\documentclass[11pt]{asaproc}

\usepackage{graphicx}
\usepackage{times}
\usepackage{eurosym}
\usepackage{caption}
\usepackage{adjustbox}
\usepackage{amsmath}
\usepackage{amsthm}
\usepackage{amsfonts}
\usepackage{amssymb}
\usepackage{calrsfs}
\usepackage{wasysym}
\usepackage{verbatim}
\usepackage{bbm}
\usepackage{color}
\usepackage[full]{harvard}
\usepackage{fancyhdr}
\usepackage{float}
\usepackage{acronym}
\usepackage{hyperref}
\usepackage{dsfont}
\usepackage{bm}
\usepackage{subcaption}

\usepackage{mathtools}
\usepackage[dvipsnames]{xcolor}

\usepackage{tikz}
\usepackage{forest}

\usetikzlibrary{arrows,shapes,positioning,shadows,trees}

\tikzset{
  basic/.style  = {draw, text width=2cm, drop shadow, font=\sffamily, rectangle},
  root/.style   = {basic, rounded corners=2pt, thin, align=center,
                   fill=green!30},
  level 2/.style = {basic, rounded corners=6pt, thin,align=center, fill=green!60,
                   text width=4em},
  level 3/.style = {basic, thin, align=left, fill=pink!60, text width=1.5em}
}
\newcommand{\relation}[3]
{
	\draw (#3.south) -- +(0,-#1) -| ($ (#2.north) $)
}

\makeatletter
\renewcommand\@biblabel[1]{}
\renewenvironment{thebibliography}[1]
      {\section*{\refname}%
       \@mkboth{\MakeUppercase\refname}{\MakeUppercase\refname}%
       \list{\@biblabel{\@arabic\c@enumiv}}%
            {\settowidth\labelwidth{\@biblabel{#1}}%
             \leftmargin\labelwidth
             \advance\leftmargin10pt
             \advance\leftmargin\labelsep
             \setlength\itemindent{-10pt}
             \@openbib@code
             \usecounter{enumiv}%
             \let\p@enumiv\@empty
             \renewcommand\theenumiv{\@arabic\c@enumiv}}%
       \sloppy
       \clubpenalty4000
       \@clubpenalty \clubpenalty
       \widowpenalty4000%
       \sfcode`\.\@m}
      {\def\@noitemerr
        {\@latex@warning{Empty `thebibliography' environment}}%
       \endlist}
\renewcommand\newblock{\hskip .11em\@plus.33em\@minus.07em}
\makeatother

\makeatletter
\def\widebreve{\mathpalette\wide@breve}
\def\wide@breve#1#2{\sbox\z@{$#1#2$}%
	\mathop{\vbox{\m@th\ialign{##\crcr
				\kern0.08em\brevefill#1{0.8\wd\z@}\crcr\noalign{\nointerlineskip}%
				$\hss#1#2\hss$\crcr}}}\limits}
\def\brevefill#1#2{$\m@th\sbox\tw@{$#1($}%
	\hss\resizebox{#2}{\wd\tw@}{\rotatebox[origin=c]{90}{\upshape(}}\hss$}
\makeatother

\makeatletter
\DeclareRobustCommand\widecheck[1]{{\mathpalette\@widecheck{#1}}}
\def\@widecheck#1#2{%
	\setbox\z@\hbox{\m@th$#1#2$}%
	\setbox\tw@\hbox{\m@th$#1%
		\widehat{%
			\vrule\@width\z@\@height\ht\z@
			\vrule\@height\z@\@width\wd\z@}$}%
	\dp\tw@-\ht\z@
	\@tempdima\ht\z@ \advance\@tempdima2\ht\tw@ \divide\@tempdima\thr@@
	\setbox\tw@\hbox{%
		\raise\@tempdima\hbox{\scalebox{1}[-1]{\lower\@tempdima\box
				\tw@}}}%
	{\ooalign{\box\tw@ \cr \box\z@}}}
\makeatother

\usepackage{array}
\usepackage{scalerel,stackengine}
\stackMath
\newcommand\reallywidehat[1]{%
\savestack{\tmpbox}{\stretchto{%
  \scaleto{%
    \scalerel*[\widthof{\ensuremath{#1}}]{\kern-.6pt\bigwedge\kern-.6pt}%
    {\rule[-\textheight/2]{1ex}{\textheight}}
  }{\textheight}%
}{0.5ex}}%
\stackon[1pt]{#1}{\tmpbox}%
}

\newcommand\norm[1]{\left\lVert#1\right\rVert}

\newcommand{\betavet}{\mathbf{\beta}}

\newcommand{\lambdavet}{\mathbf{\lambda}}
\newcommand{\avet}{\textbf{a}}
\newcommand{\bvet}{\textbf{b}}

\newcommand{\evet}{\textbf{e}}

\newcommand{\tvet}{\textbf{t}}

\newcommand{\xvet}{\textbf{x}}
\newcommand{\yvet}{\textbf{y}}
\newcommand{\vvet}{\textbf{v}}
\newcommand{\Avet}{\textbf{A}}
\newcommand{\Bvet}{\textbf{B}}
\newcommand{\Cvet}{\textbf{C}}

\newcommand{\Evet}{\textbf{E}}

\newcommand{\Gvet}{\textbf{G}}
\newcommand{\Hvet}{\textbf{H}}
\newcommand{\Ivet}{\textbf{I}}

\newcommand{\Kvet}{\textbf{K}}

\newcommand{\Mvet}{\textbf{M}}

\newcommand{\Pvet}{\textbf{P}}
\newcommand{\Qvet}{\textbf{Q}}
\newcommand{\Rvet}{\textbf{R}}
\newcommand{\Svet}{\textbf{S}}

\newcommand{\Uvet}{\textbf{U}}
\newcommand{\Wvet}{\textbf{W}}
\newcommand{\Xvet}{\textbf{X}}
\newcommand{\Yvet}{\textbf{Y}}
\newcommand{\Zvet}{\textbf{Z}}
\newcommand{\Zerovet}{\textbf{0}}
\bmdefine{\xhvet}{\mathsf{x}}
\bmdefine{\yhvet}{\mathsf{y}}
\bmdefine{\Chvet}{\mathsf{C}}
\bmdefine{\Ghvet}{\mathsf{G}}
\bmdefine{\Jhvet}{\mathsf{J}}
\bmdefine{\Shvet}{\mathsf{S}}
\bmdefine{\Uhvet}{\mathsf{U}}
\bmdefine{\Gammavet}{\mathsf{\Gamma}}
\bmdefine{\Omegavet}{\mathsf{\Omega}}
\bmdefine{\Thetavet}{\mathsf{\Theta}}
\bmdefine{\ahvet}{\mathfrak{a}}
\bmdefine{\bhvet}{\mathfrak{b}}

\DeclareMathOperator*{\argmin}{arg\,min}

\title{Cross-temporal forecast reconciliation: Optimal combination method and heuristic alternatives}

\author{Tommaso Di Fonzo\thanks{Department of Statistical Sciences, University of Padua, Italy. \textit{difonzo@stat.unipd.it}}
\and Daniele Girolimetto \thanks{Department of Statistical Sciences, University of Padua, Italy. \textit{daniele.girolimetto@studenti.unipd.it}}}

\begin{document}

\maketitle

\begin{abstract}
\hspace{-.2cm} Forecast reconciliation is a post-forecasting process aimed to improve the quality of the
base forecasts for a system of hierarchical/grouped time series (Hyndman et al., 2011).
Contemporaneous (cross-sectional) and temporal hierarchies have been
considered in the literature, but - except for Kourentzes and Athanasopoulos (2019) - generally these two features have not been fully considered together.
Adopting a notation able to simultaneously deal with both forecast reconciliation dimensions, the paper shows two new results: (i) an iterative cross-temporal forecast reconciliation procedure which extends, and overcomes some weaknesses of, the two-step procedure by Kourentzes and Athanasopoulos (2019), and (ii) the closed-form expression of the optimal (in least squares sense) point forecasts which fulfill both contemporaneous and temporal constraints.
The feasibility of the proposed procedures, along with first evaluations of their performance as compared to the most performing `single dimension' (either cross-sectional or temporal) forecast reconciliation procedures, is studied through a forecasting experiment on the 95 quarterly time series of the Australian GDP from Income and Expenditure sides considered by Athanasopoulos et al. (2019).

 \begin{keywords}
 Cross-temporal forecast reconciliation, Optimal combination, Heuristics, Hierarchical and Grouped Time Series, Quarterly Australian GDP, Income and Expenditure sides.
 \end{keywords}
\vskip1pc 
\noindent \textbf{JEL classification codes:} C22, C61, C82
\end{abstract}

\section{Introduction}
\label{sec:1}

In several operational fields, decisions to be successful require the support of accurate, detailed but also coherent forecasts. Forecasts are coherent
when the predicted values at the disaggregate and aggregate scales are equal when brought to the same level. For example, temporally coherent
monthly predictions sum up to annual ones and similarly geographically coherent regional predictions add up to country level ones.
Such a coherence is an important qualifier
for forecasts, so as to support aligned decision making across different planning units and horizons, while avoiding that different decision
making units plan on different views of the future.
For example, Kourentzes and Athanasopoulos (2019) generate forecasts for Australian domestic tourism that are
coherent across multiple geographical divisions (regions, zones, states, and whole country), but are also coherent across time
(at monthly, bi-monthly, quarterly, 4-monthly, 6-monthly, and annual levels), i.e. for different planning horizons.

As a matter of fact, in the growing literature on hierarchical forecast reconciliation the cross-sectional (contemporaneous) and temporal
coherency dimensions are mostly considered in separate ways: either `time-by-time'
cross-sectional forecast reconciliation for a $n$-dimensional time series (Hyndman et al., 2011, 2020),
or temporal coherency for the forecasts of a single variable for different time frequencies
(Athanasopoulos et al., 2017, Hyndman and Kourentzes, 2018).
The issue of getting coherent forecasts along both cross-sectional and temporal dimensions (i.e., cross-temporal coherency) has been dealt with by
Yagli et al. (2019) and Spiliotis et al. (2020). However, as far as we know, the procedure proposed by Kourentzes and Athanasopoulos (2019)
is the only one able to simultaneously fulfill both cross-sectional and temporal coherency in the final reconciled point forecasts at any considered aggregation dimension. 

Whereas the most recent forecast reconciliation procedures for each single coherence dimension are based on some optimality criterion
(van Erven and Cugliari, 2015, Wickramasuriya et al., 2019),
the cross-temporal reconciliation procedure by Kourentzes and Athanasopoulos (2019) can be considered as an heuristic with a simple and effective structure,
i.e. an approach that employs a practical method that is not guaranteed to be optimal, but which is nevertheless sufficient for reaching
an immediate goal, which in our case is the coherency along all the considered dimensions of the reconciled forecasts.
This fact is probably due to the consideration that in a cross-temporal forecast reconciliation framework the complexity and the dimensions of
the problem grow very quickly along with the requested computational time and memory
(Kourentzes and Athanasopoulos, 2019, Nystrup et al., 2020).
This is certainly true, but nevertheless an appropriate, thorough use of some well known
linear algebra tools, and the expanding computation facilities, in terms of both calculation power and memory, makes it feasible to look for the optimal solution (in least squares sense) expanding the field of application of the `forecast reconciliation methodology'
to simultaneously encompass both contemporaneous (cross-sectional) and temporal coherency dimensions.

Bottom-up and top-down approaches to forecast reconciliation are well known to both practitioners and researchers. In short, according to the bottom-up approach (Dunn et al., 1976), the forecasts at the most disaggregated level are summed up to obtain the aggregates of interest. On the contrary, in the top-down approach (Gross and Sohl, 1990) the most aggregated series is forecasted first, and then is disaggregated to provide lower level predictions (Fliedner, 2001, Athanasopoulos et al., 2009, and the references therein). A combination of these two approaches, known as middle-out (Athanasopoulos et al., 2009), selects an intermediate level of (dis)aggregation, and moves downward in a top-down fashion, and onwards according to bottom-up.

However, in the last decade there have been several contributions exploiting a regression based optimal combination approach (Hyndman et al., 2011), which has proven convincing from a mathematical-statistical point of view, flexible enough to be adapted to both cross-sectional and temporal frameworks (Wickramasuriya et al., 2019, and Athanasopoulos et al., 2017, respectively), and very effective in improving the base forecasts from many different application fields (economics, demography, energy, tourism, etc.).

In this paper we consider an optimal combination approach, which takes the base (incoherent and however obtained) forecasts of all nodes in the hierarchy, and reconcile them.
We show that all the summation constraints induced by the cross-temporal hierarchy underlying the time series, may be used to reconciliate the base forecasts through a simple projection in a well chosen linear space. At this end, we extend the optimal (in least squares sense) solutions separately proposed for each coherency dimension, developing
the optimal point forecasting procedure which - while exploiting both cross-sectional and temporal hierarchies - simultaneously fulfills
both contemporaneous and temporal constraints. We refer to this as optimal combination cross-temporal forecast reconciliation.
In addition, grounding on the existing literature on this topic (Wickramasuriya et al., 2019, Athanasopoulos et al., 2017, and Nystrup et al., 2020),
we discuss some simple approximations of the covariance matrix to be used in the statistical point forecast reconciliation, with focus on those making use of the in-sample residuals (when available) of the models used to get the base forecasts.
The strictly, and very important related issue of probabilistic forecast reconciliation (Panagiotelis et al., 2020b, Athanasopoulos et al., 2019, Hong et al., 2019, Jeon et al., 2019, Roach, 2019, Ben Taieb et al., 2020, Yang, 2020)
is not considered in this paper, and will be dealt with in the near future.

The paper is organized as follows.
We start by presenting the general framework of
point forecast reconciliation according to a projection approach (Byron, 1978, van Erven and Cugliari, 2015, Panagiotelis et al., 2020a), avoiding reference to cross-sectional or time indices (section \ref{sec:genrec}).
Hierarchical and grouped systems of multivariate time series are introduced in section \ref{sec:hts}.
The temporal hierarchies which characterize a single time series are discussed in section \ref{sec:Th}.
The cross-sectional and temporal coherency dimensions are dealt with simultaneously in section \ref{sec:CTac}, and the optimal (in least squares sense) solution to the cross-temporal forecast reconciliation problem is then developed in section \ref{sec:octforec}.
The heuristics proposed by Kourentzes and Athanasopoulos (2019) is described in section \ref{KAheu}, where some variants of that procedure are discussed. In particular, a very simple (and possibly more effective) iterative cross-temporal reconciliation procedure is proposed.
The feasibility of all the proposed procedures, along with the evaluation of their performance as compared to the most performing `single dimension' (either cross-sectional or temporal) forecast reconciliation procedures, is studied in section \ref{sec:AUSgdp} through a forecasting experiment on the 95 quarterly time series of the Australian GDP from Income and Expenditure sides considered by Athanasopoulos et al. (2019) and Bisaglia et al. (2020).
Section \ref{sec:conclusions} presents conclusions and lists some topics in this field worth to be considered for future research. Finally, the Appendix contains complementary materials, on both methodological and practical issues, not considered into the paper for length reasons. In addition, it contains supplementary tables and graphs related to the empirical application.

\clearpage
\subsection*{Symbols and notation used in the paper}
\begin{small}
\begin{adjustbox}{center}
\resizebox{1.17\linewidth}{!}{
\begin{tabular}{l|l}
	\hline
Symbols & Description \\
	\hline
$n_b$, $n_a$, $n$ & Number of bottom, upper, and total time series ($n = n_a + n_b$) \\
$m$               & Highest available sampling frequency per seasonal cycle\\
		          &(max. order of temporal aggregation) \\
$h \ge 1$         & Forecast horizon for the lowest frequency time series\\
$T$               & Number of high-frequency observations used in the\\
	              &forecasting models\\
$N$               & Number of observations at the lowest frequency: $N = \frac{T}{m}$ \\
${\cal K}$        & Set of factors of $m$ in descending order:
                    ${\cal K} = \left\{k_p, k_{p-1}, \ldots, k_2, k_1\right\}$,\\
	              & $k_p=m$, $k_1=1$\\
$k^*$             & $\displaystyle\sum_{j=1}^{p-1} k_j$ \\[.5cm]
$M_k$             & $\displaystyle\frac{m}{k}$, $k \in {\cal K}$\\       
$\bvet_t \in {\mathbb R}^{n_b}$	& vector containing the bottom time series (bts) at time $t$\\
$\avet_t \in {\mathbb R}^{n_a}$	& vector containing the upper time series (uts) at time $t$\\
$\yvet_t \in {\mathbb R}^{n}$	& vector containing the time series $\yvet_t = \left[\avet_t' \; \; \bvet_t'\right]'$ at time $t$ \\
$\Yvet, \widehat{\Yvet}, \widetilde{\Yvet} \in {\mathbb R}^{n \times h(k^* + m)}$ &
       Matrices of target, base and cross-temporally reconciled forecasts\\
$\Yvet^{[k]}, k \in {\cal K} $  & $\left(n \times hM_k\right)$
                                  matrix containing the target forecasts of the level $k$ temporally\\
                                &aggregated series. Component of matrix $\Yvet = \left[\Yvet^{[m]} \; \Yvet^{[k_{p-1}]} \; \ldots \; \Yvet^{[k_{2}]} \; \Yvet^{[1]} \right]$ \\
$\widehat{\Yvet}^{[k]}, k \in {\cal K} $ & $\left(n \times hM_k\right)$ matrix containing the base forecasts of the level $k$ temporally\\
                                         &aggregated series. Component of matrix $\widehat{\Yvet} = \left[\widehat{\Yvet}^{[m]},\widehat{\Yvet}^{[k_{p-1}]}, \ldots, \widehat{\Yvet}^{[k_{2}]}, \widehat{\Yvet}^{[1]} \right]$ \\
$\widetilde{\Yvet}^{[k]}, k \in {\cal K} $ & $\left(n \times hM_k\right)$ matrix containing the cross-temporally reconciled forecasts\\
                                           &of the level $k$ temporally aggregated series. Component of matrix\\
                                           & $\widetilde{\Yvet} = \left[\widetilde{\Yvet}^{[m]},\widetilde{\Yvet}^{[k_{p-1}]}, \ldots, \widetilde{\Yvet}^{[k_{2}]}, \widetilde{\Yvet}^{[1]} \right]$ \\
$\Avet^{[k]}$, $\Bvet^{[k]}$, $k \in {\cal K} $ & $\left(n_a \times hM_k\right)$ and $\left(n_b \times hM_k\right)$ components of matrix $\Yvet^{[k]} = \begin{bmatrix}
\Avet^{[k]} \\ \Bvet^{[k]} \end{bmatrix}$\\
$\Bvet^* \in {\mathbb R}^{n_b \times hk^*}$ & $\left[
\Bvet^{[m]} \; \Bvet^{[k_{p-1}]} \; \cdots \; \Bvet^{[k_{2}]}\right]$ \\
$\yvet, \hat{\yvet}, \tilde{\yvet} \in {\mathbb R}^{nh(k^* + m)}$ & $\yvet = \text{vec}\left(\Yvet'\right)$, $\hat{\yvet} = \text{vec}\left(\widehat{\Yvet}'\right)$, $\tilde{\yvet} = \text{vec}\left(\widetilde{\Yvet}'\right)$ \\
$\check{\yvet} \in {\mathbb R}^{nh(k^* + m)}$ & Alternative vectorization of matrix $\Yvet$, used in the cross-temporal\\
 &structural representation: $\check{\yvet} = \begin{bmatrix}
\text{vec}\left(\Avet'\right) \\ \text{vec}\left({\Bvet^*}'\right) \\ \text{vec}\left({\Bvet^{[1]}}'\right)
\end{bmatrix}$\\
$\Cvet \in {\mathbb R}^{n_a \times n_b}$ & Cross-sectional (contemporaneous) aggregation matrix \\
$\Svet \in {\mathbb R}^{n \times n_b}$ & Cross-sectional (contemporaneous) summing matrix \\
$\Uvet' \in {\mathbb R}^{n_a \times n}$ & Zero constraints cross-sectional kernel matrix:
$\Uvet'\Yvet = \Zerovet_{\left[n_a \times (k^*+m)\right]}$ \\
$\Kvet_h \in {\mathbb R}^{hk^* \times hm}$ & Temporal aggregation matrix \\
$\Rvet_h \in {\mathbb R}^{h(k^*+m) \times hm}$ & Temporal summing matrix \\
$\Zvet_h' \in {\mathbb R}^{hk^* \times h(k^*+m)}$ & Zero constraints temporal kernel matrix:
$\Zvet_h'\Yvet' = \Zerovet_{\left[hk^* \times n \right]}$ \\
$\Hvet' \in {\mathbb R}^{hn_a^* \times nh(k^*+m)}$ & Zero constraints full row-rank cross-temporal kernel matrix: $ \Hvet'\yvet = \Zerovet$\\
$\check{\Cvet} \in {\mathbb R}^{n_a^* \times n_bm}$ & Cross-temporal aggregation matrix (structural representation) for $h = 1$\\
$\check{\Svet} \in {\mathbb R}^{n(k^*+m) \times n_bm}$ & Cross-temporal summing matrix (structural representation) for $h = 1$\\
$\check{\Hvet}' \in {\mathbb R}^{hn_a^* \times nh(k^*+m)}$ & Zero constraints full row-rank cross-temporal kernel matrix\\
&valid for $\check{\yvet}$:
$ \check{\Hvet}'\check{\yvet} = \Zerovet$\\
\hline
\end{tabular}
}	
\end{adjustbox}

\end{small}


\section{Optimal point forecast reconciliation: projection approach}
\label{sec:genrec}
Forecast reconciliation is a post-forecasting process aimed to improve the quality of the {\em base} forecasts
for a system of hierarchical/grouped, and more generally linearly constrained, time series
(Hyndman et al., 2011, Panagiotelis et al., 2020a) by exploiting the constraints that the series in the system
must fulfill, whereas in general the base forecasts don't.

Let ${\yvet}$ be a ($s \times 1$) vector of target point forecasts
which are wished to satisfy the system of linearly independent constraints
\begin{equation}
\label{acconstr}
\Hvet'{\yvet} = \mathbf{0}_{(r \times 1)},
\end{equation}
where $\Hvet'$ is a ($r \times s$) matrix, with
$\text{rank}(\Hvet')=r<s$, and $\Zerovet_{(r \times 1)}$ is a ($r \times 1$)
null vector. Let $\hat{\yvet}$ be a ($s \times 1$)
vector of unbiased point forecasts, not fulfilling the linear constraints
(\ref{acconstr}) (i.e., $\Hvet'\hat{\yvet} \ne \Zerovet$).

Drawing upon Stone et al. (1942), Byron (1978), Weale (1988), Solomou and Weale (1993), and Dagum and Cholette (2006), among others, we consider a regression-based reconciliation method assuming that $\hat{\yvet}$ is
related to $\yvet$ by
\begin{equation}
\label{linmod} \hat{\yvet} = \yvet + \mathbf{\varepsilon},
\end{equation}
where $\mathbf{\varepsilon}$ is a ($s \times 1$) vector of zero mean disturbances,
with known p.d. covariance matrix  $\mathbf{W}$.
The reconciled forecasts $\tilde{\bf y}$ are found by minimizing the generalized least squares (GLS) objective function
$\left(\hat{\yvet} - \yvet\right)'\mathbf{W}^{-1}\left(\hat{\yvet} - \yvet\right)$ constrained by (\ref{acconstr}):
\[
\tilde{\bf y} = \argmin_{{\yvet}} \left({\bf y} - \hat{\bf y} \right)' {\bf W}^{-1} \left({\bf y} - \hat{\bf y} \right), \quad
\text{s.t. } {\bf H}'{\bf y} = {\bf 0}.
\]
The solution is given by (see Appendix A.1):

\begin{equation}
\label{stonest}
\tilde{\yvet}= \hat{\yvet} -
\mathbf{W}\Hvet\left(\Hvet'\mathbf{WH}\right)^{-1}\Hvet'\hat{\yvet}=
{\mathbf{M}}\hat{\yvet},
\end{equation}
where ${\mathbf{M}} = \mathbf{I}_s - \mathbf{W}\Hvet\left(\Hvet'\mathbf{WH}\right)^{-1}\Hvet'$ is a $(s \times s)$ projection matrix\footnote{A geometric interpretation of the entire hierarchical forecasting problem is given by Panagiotelis et al. (2020a).}. 
Denoting with $\mathbf{d}_{\hat{\yvet}} = \mathbf{0} - \Hvet'\hat{\yvet}$ the
$(r \times 1)$ vector containing the base forecasts' `coherency' errors, we can re-state expression (\ref{stonest}) as
\[
\tilde{\yvet}= \hat{\yvet} +
\mathbf{WH} \left(\Hvet'\mathbf{WH}\right)^{-1}\mathbf{d}_{\hat{\yvet}},
\]
which makes it clear that the reconciliation formula (\ref{stonest}) simply `adjusts' the original forecasts vector $\hat{\yvet}$ with a linear combination -- according to the smoothing matrix
$\mathbf{WH} \left(\Hvet'\mathbf{WH}\right)^{-1}$ --
of the coherency errors in the base forecasts.
In addition, if the error term of model (\ref{linmod}) is gaussian, the reconciliation error
$\tilde{\varepsilon} = \tilde{\yvet} - \yvet$ is a zero-mean gaussian vector with covariance matrix 
$$
 E \left( \tilde{\yvet} - {\yvet}\right)
\left( \tilde{\yvet} - {\yvet}\right)' =
\mathbf{W} - \mathbf{WH} \left(\Hvet'\mathbf{WH}\right)^{-1}\Hvet' = \mathbf{MW}.
$$


Hyndman et al. (2011, see also Wickramasuriya et al., 2019) propose an alternative formulation as for the reconciled estimates, equivalent to expression (\ref{stonest}) and obtained by GLS estimation of the model
\begin{equation}
\label{strucgls}
\hat{\yvet} = \Svet\betavet	+ {\bf \varepsilon},
\end{equation}
where $\Svet$ is a `structural summation matrix' describing the aggregation relationships operating on $\yvet$, and $\betavet$ is a subset of $\yvet$ containing the target forecasts of the bottom level series, such that $\yvet = \Svet\betavet$ (see section \ref{sec:hts}). Since the hypotheses on ${\bf \varepsilon}$ remain unchanged, it can be shown (see Appendix A.1) that
$$
\tilde{\betavet} = \left(\Svet'\Wvet^{-1}\Svet\right)^{-1}\Svet'\Wvet^{-1}\hat{\yvet}
$$
is the best linear unbiased estimate of $\betavet$, and that the whole reconciled forecasts vector is given by
$$
\tilde{\yvet} = \Svet\tilde{\betavet} = \Svet\Gvet\hat{\yvet},
$$
where $\Gvet = \left(\Svet'\Wvet^{-1}\Svet\right)^{-1}\Svet'\Wvet^{-1}$.

As witnessed by the huge literature on adjusting preliminary data (as the base forecasts can be considered) in order
to fulfill some externally imposed constraints, the distinctive feature
of the generalized least-squares reconciliation
approach is that it can take into account the `quality', however measured, of
the preliminary estimates, through an appropriate choice of the covariance matrix $\Wvet$.
However, for a long time these procedures have depended on the
assumption that this matrix (or any other indicators of
the estimates' accuracy) of the figures to be reconciled was
known. In many practical situation $\Wvet$ is assumed to be diagonal, and the data are adjusted in the light of their
relative variances so as to satisfy the linear restrictions. But
another - perhaps more delicate - challenge raises when either any
reliability measure is available or it can be hardly deduced by the data.
The solutions proposed in
literature for this case are basically of two types, both of which
are consistent with the least-squares approach shown so far:
\begin{enumerate}
	\item mathematical/mechanical solutions: the base forecasts are balanced
	by minimizing a penalty criterion which `induces' a covariance
	matrix (which is simply a statistical artifact);
	\item data-based solutions: the variability of the base forecasts to be reconciled is
	estimated through the models and the data used to produce the forecasts.
\end{enumerate}
As for point forecast reconciliation, in the following we will consider both approaches, with an explicit preference towards approximations of $\Wvet$ based on the in-sample residuals (when available), which appear both more convincing from a statistical point of view, and generally well-performing in practical applications. However, this topic deserves further attention (Jeon et al., 2019, p. 368, see also Kourentzes, 2017, 2018), and will be considered for future research.

\section{Hierarchical and Grouped Time Series}
\label{sec:hts}

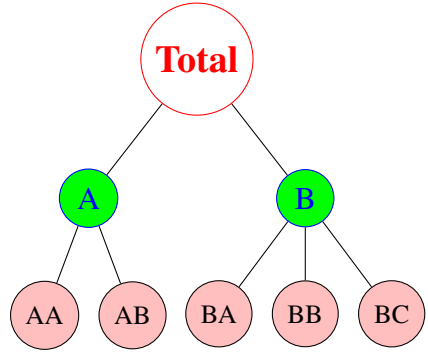
\begin{figure*}[ht]
   \centering
\begin{forest}
for tree={circle,draw, l sep=20pt}
[\textbf{\Large Total},red, 
    [\large A, blue, fill=green
      [\small AA, fill=pink]
      [\small AB, fill=pink]
    ]
    [\large B, blue, fill=green
      [\small BA, fill=pink]
      [\small BB, fill=pink]
      [\small BC, fill=pink]
  ]
]
\end{forest}
   \caption{A simple two-level hierarchical structure}
   \label{hts1}
\end{figure*}

Extending the definition of hierarchical time series given by Panagiotelis et al. (2020a), a linearly constrained time series $\yvet_t$ is a $n$-dimensional time series such that all observed values
$\yvet_1 \ldots \yvet_T$ and all future values $\yvet_{T+1}, \yvet_{T+2} \ldots$ lie in the coherent linear subspace $\cal{S}$, that is:
$\yvet_t \in \cal{S}$, $\forall t$.
In many situations, 
the time series are linked through summation constraints, which induce a hierarchy. Figure~\ref{hts1} gives an example of a hierarchical system with eight variables and three levels: the top-variable at level 1, two variables (A and B) at level 2, and five variables at level 3 (AA, AB, BA, BB, BB, BC). The temporal observations of these variables form a hierarchical time series, consisting of 5 bottom time series (bts) and 3 aggregated upper time series (uts).

Assuming that the relationship mapping the lower-level series in the hierarchy of Figure~\ref{hts1} into the higher ones always be a simple
summation\footnote{For space reasons, in this paper only simple summation for both contemporaneous and temporal aggregation
relationships is considered. Remaining in a linear framework, the extension to general linear constraints (i.e., weighted summation),
able to cover other important data features, is rather straightforward (Shang, 2017, 2019, Shang and Hyndman, 2017, Li and Hyndman, 2019, Panagiotelis et al., 2020a).},
the bottom-level series can be thought as building blocks that cannot be obtained as sum of other series in the hierarchy, while all the series at upper levels can be expressed by appropriately summing part or all of them.
For all time periods $t=1,\ldots,T$, the link between the top level series $y_t$ and the bottom level series is given by:
\begin{equation}
\label{top}
y_t = y_{AA,t} + y_{AB,t} + y_{BA,t} + y_{BB,t} + y_{BC,t}.
\end{equation}
At the same time, the nodes at the intermediate level of the hierarchy satisfy the aggregation constraints:
\begin{equation}
\label{topagg}
\begin{array}{rcl}
y_{A,t} & = & y_{AA,t} + y_{AB,t} \\
y_{B,t} & = & y_{BA,t} + y_{BB,t} + y_{BC,t} 
\end{array} .
\end{equation}
In summary, there are as many summation constraints as many nodes with leaves (3, i.e. Total, A, B).
Consider now the matrices $\Cvet$, $\Svet$, and $\Uvet'$, of dimension $(3 \times 5)$, $(8 \times 5)$, and $(3 \times 8)$, respectively:
$$
\Cvet = \left[\begin{array}{ccccc}
1 & 1 & 1 & 1 & 1 \\
1 & 1 & 0 & 0 & 0 \\
0 & 0 & 1 & 1 & 1
\end{array}\right],
\quad
\Svet = \left[\begin{array}{ccccc}
1 & 1 & 1 & 1 & 1 \\
1 & 1 & 0 & 0 & 0 \\
0 & 0 & 1 & 1 & 1 \\
1 & 0 & 0 & 0 & 0 \\
0 & 1 & 0 & 0 & 0 \\
0 & 0 & 1 & 0 & 0 \\
0 & 0 & 0 & 1 & 0 \\
0 & 0 & 0 & 0 & 1
\end{array}\right] =
\left[\begin{array}{c} \Cvet \\ \Ivet_5 \end{array}\right],
\quad
\mathbf{U}' = \left[\mathbf{I}_3 \; -\mathbf{C}  \right] ,
$$
\noindent where matrix $\Uvet'$ encodes each summation relationship in a row, with 1 at the associated node, and -1 at its leaves. 

Expressions (\ref{top}) and (\ref{topagg}) can be written in a more compact way if we define
the vectors of \textit{bottom level} ($\mathbf{b}_t$) and \textit{upper level} ($\mathbf{a}_t$) time series at time $t$ as, respectively,
$$
\mathbf{b}_t = \left[\begin{array}{c}
y_{AA,t} \\
y_{AB,t} \\
y_{BA,t} \\
y_{BA,t} \\
y_{BA,t}
\end{array}\right], \qquad
\mathbf{a}_t = \left[\begin{array}{c}
y_{t} \\
y_{A,t} \\
y_{B,t}
\end{array}\right].
$$

\noindent Denoting by $\yvet_t$ the $(8 \times 1)$ vector
$\yvet_t = \left[\mathbf{a}_t' \; \; \mathbf{b}_t'\right]'$, 
the relationships linking bottom and upper time series can be equivalently expressed as:
\begin{equation}
\label{contemp}
\mathbf{a}_t = \mathbf{Cb}_t, \quad \yvet_t = \mathbf{Sb}_t, \quad
\mathbf{U}'\yvet_t = \mathbf{0}_{(3 \times 1)}, \quad    t=1,\ldots,T.
\end{equation}
Thus, for any time index $t$, $\yvet_t$ is in the kernel of $\Uvet'$, also known as null-space of the linear transformation induced by matrix $\Uvet'$, which is given by the set of vectors $\vvet \in {\mathbb R}^7$, such that $\Uvet'\vvet = \Zerovet_{(3 \times 1)}$ (Harville, 2008, p. 591).
We call \emph{structural representation} of series $\yvet_t$ the formulation
$$
\yvet_t = \Svet\bvet_t, \quad t=1,\ldots, T ,
$$
and \emph{zero constraints kernel representation} of series $\yvet_t$ the equivalent expression
$$
\Uvet'\yvet_t = \Zerovet, \quad t=\,\ldots, T .
$$


A linearly constrained time series formed by two or more hierarchical time series sharing the same top level series, and the same bottom level series, is called grouped time series (Hyndman et al., 2011, Hyndman and Athanasopoulos, 2018). Provided matrix
$\mathbf{C}$ is appropriately designed, the definitions of matrices $\mathbf{S}$ and $\mathbf{U}'$, depending solely on matrix $\mathbf{C}$, remain unchanged.

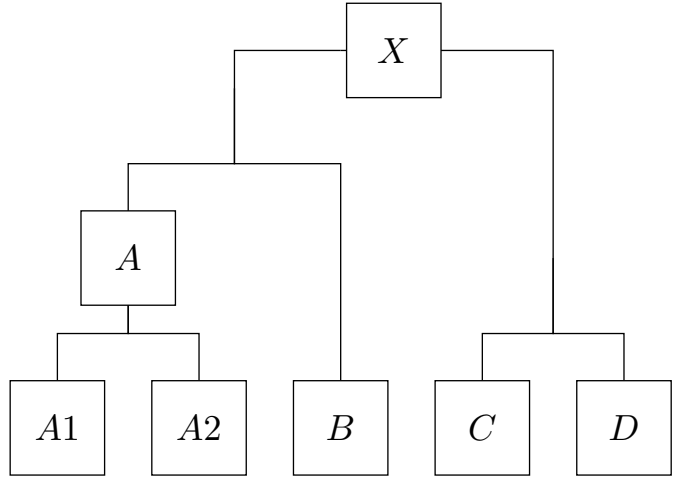
\begin{figure}[ht]
	\centering
	\resizebox{0.65\linewidth}{!}{
		\begin{tikzpicture}[baseline=(current  bounding  box.center),
	every node/.append style={shape=rectangle,
		draw=black},
	minimum width=1cm,
	minimum height=1cm]

	\node at (0, 0) (A1){$A1$};
	\node at (1.5, 0) (A2){$A2$};
	\node at (3, 0) (B){$B$};
	\node at (4.5, 0) (C){$C$};
	\node at (6, 0) (D){$D$};
	\node[draw=none, align=center,
	scale=0cm] at (0.75, 1.8) (A){};
	\node[draw=none, align=center,
	scale=0cm] at (1.875, 3.6) (invA){};
	\node[draw=none, align=center,
	scale=0cm] at (5.25, 1.8) (invC){};
	\node[draw=none, align=center,
	scale=0cm] at (3, 4) (invX){$X$};
	\relation{0.8}{A1}{A};
	\relation{0.8}{A2}{A};
	\relation{0.8}{B}{invA};
	\relation{0}{invA}{invX};
	\relation{0}{invC}{invX};
	\relation{0.8}{D}{invC};
	\relation{0.8}{C}{invC};
	\relation{0.8}{A}{invA};
	\node[fill=white] at (0.75, 1.8) (A3){$A$};
	\node[fill=white] at (3.5625, 4) (X){$X$};
	
	\end{tikzpicture}
	}
	\caption{Two hierarchies sharing the same top-level series $X$}
		\label{fig:kernelgraph}
\end{figure}

It should be noted that we can face linearly constrained time series for which the structural representaton $\yvet_t = \Svet \bvet_t$ does not give a straightforward view of the links between bottom and upper level time series. Figure \ref{fig:kernelgraph} shows two very simple hierarchies, where the variables of each hierarchy contribute (from different sides) to the same top level variable $X$, and the bottom level series of the hierarchy on the left side ($A1$,$A2$,$B$) are independent from those on the right side ($C$,$D$).

The aggregation relationships between the upper variables $X$ and $A$, and the bottom ones $A1$, $A2$, $B$, $C$, and $D$ are given by:
\begin{equation}
\label{hier2}
\begin{array}{rcl}
X & = & A1 + A2 + B \\
X & = & C + D \\
A & = & A1 + A2
\end{array} .
\end{equation}
Expression (\ref{hier2}) cannot be represented as a mapping from the bottom variables into (themselves, and) the upper variables. Nevertheless, it is possible to set up the constraints valid for all the component series in $\yvet = \left[X \; A \; A1 \; A2 \; B \; C \; D\right]'$ through the matrix 
$$
\check{\Uvet}' = \left[\begin{array}{rrrrrrr}
1 & 0 & -1 & -1 & -1 &  0 &  0 \\
1 & 0 &  0 &  0 &  0 & -1 & -1 \\
0 & 1 & -1 & -1 &  0 &  0 &  0
\end{array}\right],
$$
such that $\check{\Uvet}'\yvet = \Zerovet_{(3 \times 1)}$. After simple operations on expression (\ref{hier2}), it is found:
\begin{equation}\label{hier3}
\begin{array}{rcl}
X  & = & C + D \\
A  & = & -B + C + D \\
A1 & = & -A2 - B + C + D
\end{array} ,
\end{equation}
so we can write $\Uvet'\yvet = \Zerovet_{(3 \times 1)}$, with
$$
\Uvet' = \left[\begin{array}{rrrrrrr}
1 & 0 & 0 &  0 & 0 & -1 & -1 \\
0 & 1 & 0 &  0 & 1 & -1 & -1 \\
0 & 0 & 1 &  1 & 1 & -1 & -1
\end{array}\right]  = \left[\Ivet_{3} \; \; -\Cvet\right].
$$
While there is no practical problem in working with such constraints, it is clear that they do not conform to the visual pattern of the linearly constrained time series in Figure \ref{fig:kernelgraph}, where $A1$ appears as a `bottom variable', whereas in (\ref{hier3}) it is expressed as linear combination of series $A2$, $B$, $C$, and $D$. 

In addition, notice that the left side hierarchy of Figure \ref{fig:kernelgraph} is `unbalanced', in the sense that unlike node $A$, node $B$ has no children, and thus is located at the bottom level of the hierarchy, though it could be considered at the same level as node $A$. Situations like that, often met in practice when dealing with hierarchical/grouped time series, require an appropriate formulation of the cross-sectional aggregation matrix $\Cvet$, in order to avoid nodes' duplication  (see Appendix A.2).

\subsection{Alternative approximations of the covariance matrix for cross-sectional point forecast reconciliation}
\label{subsec:csmat}
Suppose we have the $(n \times 1)$ vector $\hat{\yvet}_h$ of unbiased base forecasts for the $n$ variables of the linearly constrained series $\yvet_t$ for the forecast horizon $h$. If the base forecasts have been independently computed, generally they do not fulfill the cross-sectional aggregation constraints, i.e. $\Uvet'\hat{\yvet}_h \ne \Zerovet_{(n \times 1)}$. By adapting the general point forecast reconciliation formula (\ref{stonest}), the vector of reconciled forecasts is given by:
\begin{equation}
\label{csstonest}
\tilde{\yvet}_h = \hat{\yvet}_h -
\Wvet_{\text{cs}}\Uvet\left(\Uvet'\Wvet_{\text{cs}}\Uvet\right)^{-1}\Uvet'\hat{\yvet}_h ,
\end{equation}
where $\Wvet_{\text{cs}}$ is a $(n \times n)$ p.d. matrix, assumed known, and suffix `cs' stands for `cross-sectional'. Alternative choices for $\Wvet_{\text{cs}}$ proposed in literature are the following:
\begin{itemize}
\item identity (cs-ols): $\Wvet_{\text{cs}} = \Ivet_{n}$ (Hyndman et al. 2011),
\item structural (cs-struc): $\Wvet_{\text{cs}} = \text{diag}\left(\Svet{\bf 1}_{n_b}\right)$ (Athanasopoulos et al., 2017),
\item series variance (cs-wls): $\Wvet_{\text{cs}} = \widehat{\Wvet}_{\text{cs-var}} = \Ivet_{n} \odot \widehat{\Wvet}_1$ (Hyndman et al., 2016),
\item MinT-shr (cs-shr):  $\Wvet_{\text{cs}} = \widehat{\Wvet}_{\text{cs-shr}} = \hat{\lambda} \widehat{\Wvet}_{\text{cs-var}} + (1 - \hat{\lambda})\widehat{\Wvet}_1$ (Wickramasuriya et al., 2019),
\item MinT-sam (cs-sam):  $\Wvet_{\text{cs}} = \widehat{\Wvet}_1$ (Wickramasuriya et al., 2019),
\end{itemize}
where the symbol $\odot$ denotes the Hadamard product, $\hat{\lambda}$ is an estimated shrinkage coefficient (Ledoit and Wolf, 2004, Sh\"{a}fer and Strimmer, 2005), $\widehat{\Wvet}_1$ is the sample covariance matrix of the one-step ahead in-sample forecast errors\footnote{Expression (\ref{Wsam1}) assumes that
$T^{-1}\displaystyle\sum_{t=1}^{T}\hat{e}_{t,i}=0$, $i=1,\ldots,n$. When this does not hold, $\widehat{\Wvet}_1$ is the sample Mean Square Error (MSE) matrix.}:
\begin{equation}
\label{Wsam1}
\widehat{\Wvet}_1 = \displaystyle\frac{1}{T}\sum_{t=1}^{T}\hat{\evet}_t\hat{\evet}_t',
\end{equation}
and $\hat{\evet}_t = \yvet_t - \hat{\yvet}_t$, $t=1,\ldots,T$, are $(n \times 1)$ vectors of in-sample forecast errors.

The first three matrices are diagonal, and in the first case the projection is orthogonal, whereas the latter two ones (cs-shr and cs-sam) have been proposed within the minimum-trace point forecast reconciliation approach by Wickramasuriya et al. (2019). It should be noted that the quality of the estimate $\widehat{\Wvet}_1$ crucially depends on the dimension of $T$. In particular, when $T < n$, matrix $\widehat{\Wvet}_1$ is singular, which prevents the matrix inversion in expression (\ref{csstonest}). The shrunk version $\widehat{\Wvet}_{\text{cs-shr}}$ is a feasible alternative, well performing in many practical situations (Wickramasuriya et al., 2019).

\subsection{Matrix representation of the cross-sectional constraints}
Let us denote with
\begin{equation}
\label{bvet}
\mathbf{b}_t = \left[b_{1t} \ldots b_{jt} \ldots b_{{n_b}t} \right]', \qquad t=1, \ldots, T,
\end{equation}
the $T$ vectors, each of dimension $\left( n_b \times 1 \right)$, containing the {\em high-frequency bottom-time series} (hf-bts),
that is the bottom series of the hierarchy/group
observed at the highest available temporal frequency.
As we shall see in section \ref{sec:CTac}, in cross-temporal hierarchies of time series the hf-bts should be considered as the `very' bottom series
of the system,
since they cannot be formed as either contemporaneous or temporal sum of other variables.
Likewise, let us denote with
\begin{equation}
\label{avet}
\avet_t = \left[ a_{1t} \ldots a_{it} \ldots a_{{n_a}t} \right]', \qquad t=1, \ldots, T,
\end{equation}
the $T$ vectors, each of dimension $\left( n_a \times 1 \right)$, containing the {\em high-frequency upper-time series} (hf-uts),
which are the cross-sectionally aggregated series of the hierarchy/group, observed at the highest temporal frequency.

At each time $t=1, \ldots, T$, the cross-sectional (contemporaneous) aggregation constraints that map the hf-bts into the hf-uts can be written as:
\begin{equation}
\label{aCb}
\mathbf{a}_t = \mathbf{Cb}_t, \quad t=1, \ldots, T,
\end{equation}
\noindent where $\mathbf{C}$ is a $\left( n_a \times n_b \right)$ {\em contemporaneous aggregation matrix}. The structural representation of the linearly constrained time series $\yvet_t$ is in turn given by (Hyndman et al., 2011):
$$
\left[\begin{array}{c} \mathbf{a}_t \\ \mathbf{b}_t \end{array} \right] = 
\left[\begin{array}{c} \mathbf{C} \\ \mathbf{I}_{n_b} \end{array} \right]\bvet_t \quad \Rightarrow \quad \yvet_t =
\mathbf{Sb}_t, \quad t=1,\ldots, T,
$$
where
$ \mathbf{S}=\left[\begin{array}{c} \mathbf{C} \\ \mathbf{I}_{n_b} \end{array} \right]$
is a $\left( n \times n_b \right)$ {\em contemporaneous summing matrix}, with $n = n_a + n_b$.
The constraints valid for $\yvet_t$ can be expressed in kernel form through the $(n_a \times n)$ \emph{zero constraints matrix}
$$
\Uvet' = \left[\Ivet_{n_a} \; -\Cvet\right],
$$
that is:
$$
\Uvet'\yvet_t = \Zerovet_{(n_a \times 1)}, \quad t=1,\ldots, T .
$$

Now, denote $\Bvet$ the $(n_b \times T)$ matrix containing the $T$ observations of the $n_b$-variate hf-bts of the system:
\[
\Bvet = \begin{bmatrix}
b_{11} & \dots & b_{1t} & \dots & b_{1T} \\
\vdots & \ddots & \vdots & \ddots & \vdots \\
b_{i1} & \dots & b_{it} & \dots & b_{iT} \\
\vdots & \ddots & \vdots & \ddots & \vdots \\
b_{n_b1} & \dots & b_{n_bt} & \dots & b_{n_bT}
\end{bmatrix}
=
\begin{bmatrix}
\bvet_1 & \dots & \bvet_t & \dots & \bvet_T
\end{bmatrix}
=
\begin{bmatrix}
\bvet_1^{\small{*\prime}} \\
\vdots \\
\bvet_i^{\small{*\prime}} \\
\vdots \\
\bvet_{n_b}^{\small{*\prime}}
\end{bmatrix}
,
\]
where $\bvet_t$ has been defined by (\ref{bvet}), and
\[
\bvet_i^{\small{*}} = \left[ b_{i1} \ldots b_{it} \ldots b_{{i}T} \right]', \quad i=1, \dots, n_b,
\]
is the $(T \times 1)$ vector containing all the observations of the $i$-th univariate hf-bts, where the asterisk
in $\bvet^{\small{*}}_i$ is used to distinguish this vector, which combines $b_{it}$ across all times for one series,
from $\bvet_t$, which combines $b_{it}$ across all series for one time.

We consider the $(n_a \times T)$ matrix $\Avet$ for the hf-uts as well:
\[
\Avet = \begin{bmatrix}
a_{11} & \dots & a_{1t} & \dots & a_{1T} \\
\vdots & \ddots & \vdots & \ddots & \vdots \\
a_{j1} & \dots & a_{jt} & \dots & a_{jT} \\
\vdots & \ddots & \vdots & \ddots & \vdots \\
a_{n_a1} & \dots & a_{n_at} & \dots & a_{n_aT}
\end{bmatrix}
=
\begin{bmatrix}
\avet_1 & \dots & \avet_t & \dots & \avet_T
\end{bmatrix}
=
\begin{bmatrix}
\avet_1^{\small{*\prime}} \\
\vdots \\
\avet_j^{\small{*\prime}} \\
\vdots \\
\avet_{n_a}^{\small{*\prime}}
\end{bmatrix}
,
\]
where $\avet_t$ was defined by (\ref{avet}), and
\[
\avet^{\small{*}}_j = \left[ a_{j1} \ldots a_{jt} \ldots a_{{j}T} \right]', \quad j=1, \dots, n_a,
\]
is the $(T \times 1)$ vector containing all the observations of the $j$-th univariate component hf-uts.

The cross-sectional (contemporaneous) aggregation relationships (\ref{aCb}) linking bottom and upper time series of $\yvet_t$
can thus be expressed in compact form, by simultaneously encompassing all $T$ time periods, for both types of data organization.
In fact, extending expression (\ref{aCb}) to the whole observation period, it is
\begin{equation}
\label{ACB}
\Avet = \Cvet \Bvet,
\end{equation}
which is equivalent to
\begin{equation}
\label{UtY}
\Uvet' \Yvet = \Zerovet_{\left(n_a \times T \right)},
\end{equation}
where
$$
\Yvet = \left[\begin{array}{c}
\Avet \\ \Bvet
\end{array}\right]
$$
is the $(n \times T)$ matrix containing the observations of all $n$ series.
It is worth noting that the cross-sectional constraints (\ref{ACB}) and (\ref{UtY}) hold at any time observation index of any temporal frequency. 
This has to be considered when dealing with cross-temporal hierarchies
(see section \ref{sec:CTac}).

Now, let us consider two vectorized forms of matrices $\Bvet$ and $\Avet$, namely:
\[
\bvet = \text{vec}\left(\Bvet\right), \quad \avet = \text{vec}\left(\Avet\right),
\]
\[
\bvet^{\small{*}} = \text{vec}\left(\Bvet'\right), \quad \avet^{\small{*}} = \text{vec}\left(\Avet'\right).
\]
Both $\bvet$ and $\bvet^{\small{*}}$ have the same dimension ($T n_b \times 1$), and this holds for $\avet$ and
$\avet^{\small{*}}$ as well, which have dimension ($T n_a \times 1$).
However, in the former case ($\bvet$ and $\avet$) the data is organized `by-time-first-and-then-by-variable':

\[
\bvet = \left[ \bvet_1' \dots \bvet_t' \dots \bvet_T' \right]', \quad
\avet = \left[ \avet_1' \dots \avet_t' \dots \avet_T' \right]',
\]
whereas in the latter ($\bvet^{\small{*}}$ and $\avet^{\small{*}}$) the data is organized `by-variable-first-and-then-by-time':
\[
\bvet^{\small{*}} = \left[ \bvet_1^{\small{*}\prime} \dots \bvet_j^{\small{*}\prime} \dots \bvet_{n_b}^{\small{*}\prime} \right]', \quad
\avet^{\small{*}} = \left[ \avet_1^{\small{*}\prime} \dots \avet_j^{\small{*}\prime} \dots \avet_{n_a}^{\small{*}\prime} \right]'.
\]

Switching between the two data representations is very simple, since vector $\bvet$ ($\avet$) can be obtained by simple transformation
of vector $\bvet^{\small{*}}$ ($\avet^{\small{*}}$) through an appropriate permutation matrix,
and {\em vice-versa} (see Appendix A.3.1).

Depending on the preferred data organization type, the cross-sectional constraints (\ref{ACB})
can be equivalently expressed in vectorized form as
(Harville, 2008, pp. 345):
\begin{equation}
\label{ContConstr}
\begin{array}{ccccccc}
\avet & = & \text{vec}\left(\Avet\right)  & = & \left(\Ivet_T \otimes \Cvet\right)\text{vec}\left(\Bvet\right) & = & \left(\Ivet_T \otimes \Cvet\right)\bvet , \\
\avet^{\small{*}} & = & \text{vec}\left(\Avet'\right) & = & \left(\Cvet \otimes \Ivet_T \right)\text{vec}\left(\Bvet'\right) & = & \left(\Cvet \otimes \Ivet_T \right)\bvet^{\small{*}} 
\end{array} ,
\end{equation}
where the symbol $\otimes$ denotes the Kronecker product. Expressions (\ref{ContConstr}) can be also formulated using matrix $\Uvet'$, as in 
(\ref{UtY}), i.e.
\begin{equation}
\label{ContConstrUt}
\left( \Ivet_T \otimes \Uvet' \right) \left[\begin{array}{c}
\avet \\ \bvet
\end{array}\right] = \Zerovet_{\left(Tn_a \times 1\right)}, \quad
\left( \Uvet' \otimes \Ivet_T \right) \left[\begin{array}{c}
\avet^* \\ \bvet^*
\end{array}\right] = \Zerovet_{\left(Tn_a \times 1\right)} .
\end{equation}
In order to avoid mistakes, one should note that, while
\begin{equation}
\label{ystar}
\yvet^* = \text{vec}\left(\Yvet'\right) = \begin{bmatrix}
\text{vec}\left(\Avet'\right) \\
\text{vec}\left(\Bvet'\right)
\end{bmatrix} = 
\begin{bmatrix}
\avet^* \\ \bvet^*
\end{bmatrix},
\end{equation}
it is in turn:
$$
\yvet = \text{vec}\left(\Yvet\right) \ne \begin{bmatrix}
\text{vec}\left(\Avet\right) \\
\text{vec}\left(\Bvet\right)
\end{bmatrix} = 
\begin{bmatrix}
\avet \\ \bvet
\end{bmatrix}.
$$
Therefore, in the following when a matrix vectorization is needed, we will generally prefer using vectorized version of matrices $\Yvet'$, $\Avet'$, and $\Bvet'$, as in (\ref{ystar}). In addition, to ease the notation, from now on the asterisk will be omitted, which means that, unlike we previously did, we denote with $\yvet$, $\avet$, and $\bvet$ the following vectors:
$$
\yvet = \text{vec}\left(\Yvet'\right), \quad \avet = \text{vec}\left(\Avet'\right), \quad
\bvet = \text{vec}\left(\Bvet'\right) .
$$

\section{Temporal hierarchies}
\label{sec:Th}

Following Athanasopoulos et al. (2017), we consider a time series $\{x_t\}_{t=1}^T$ observed at the highest available sampling frequency per seasonal cycle, say $m$ (e.g., month per year, $m=12$, quarter per year, $m=4$, hour per day, $m=24$).
Given a factor $k$ of $m$,\footnote{If $k$ is not a factor of $m$, then the seasonality of the aggregate series is non-integer, and so forecasts of the
	aggregate are more difficult to compute.} we can construct a temporally aggregated version of the time series $x_t$, through the non-overlapping sums of its $k$ successive values, which has seasonal period equal to $M_k=m/k$. To avoid ragged-edge data, we assume that the total number of observations of $x_t$ involved in the non-overlapping aggregation is a multiple of $m$, i.e. $T = N\cdot m$,
where $N$ is the length of the most temporally aggregated version of the series, i.e. the series observed at the lowest available frequency.

We denote with ${\cal{K}} = \left\{k_m, k_{p-1}, \ldots, k_{2}, k_1\right\}$ the set of the $p$ factors of $m$, in descending order, where $k_p=m$ and $k_1=1$.
The temporally aggregated series of order $k$ can be written as
\begin{equation}
\label{xk}
x^{[k]}_l = \displaystyle\sum_{t=(l-1)k+1}^{lk} x_t, \quad l=1, \ldots, \displaystyle\frac{T}{k}, \quad k \in {\cal K}.
\end{equation}
Expression (\ref{xk}) accounts also for the trivial temporal aggregation transforming $x_t$ in itself (i.e., $x_t \equiv x_l^{[1]}$, $l = t$).

Since the observation index $l$ in (\ref{xk}) varies with each aggregation level $k$, in order to express a common index for all levels, we define $\tau$ as the observation index of the most aggregated series, such that $l = \tau $ at that level, i.e.
$$
x^{[m]}_{\tau}, \quad \tau=1, \ldots, N.
$$
As for the other temporally aggregated series defined in expression (\ref{xk}), we
stack the observations for each aggregation level below $m$ in the $(M_k \times 1)$ column vectors
\begin{equation}
\label{xktau}
\xvet^{[k]}_{\tau} = \left[ x^{[k]}_{M_k(\tau-1)+1} \; x^{[k]}_{M_k(\tau-1)+2} \; \dots \; x^{[k]}_{M_k\tau} \right]', \quad
\tau = 1, \ldots, N, \quad k \in \left\{k_{p-1}, \dots , k_2, 1 \right\}.
\end{equation}
We may collect $x^{[m]}_{\tau}$ and the $p-1$ vectors defined by expression (\ref{xktau}) in a single column vector, by keeping distinct the temporally aggregated data from the high-frequency one:
$$
\xvet_{\tau} = \left[{x^{[m]}_{\tau}} \; {\xvet^{[k_{p-1}]}_{\tau}}' \; \dots \; {\xvet^{[k_2]}_{\tau}}' \; {\xvet^{[1]}_{\tau}}' \right]' =
\left[{\tvet_{x_{\tau}}}' \; \; {\xvet^{[1]}_{\tau}}'
\right]', \quad \tau = 1, \ldots, N,
$$
where
$\tvet_{x_{\tau}} = \left[{x^{[m]}_{\tau}} \; {\xvet^{[k_{p-1}]}_{\tau}}' \; \dots \; {\xvet^{[k_2]}_{\tau}}' \right]'$
is a $\left(k^* \times 1 \right)$ vector, with $k^*=\displaystyle\sum_{j=1}^{p-1}k_j$, containing all the temporally aggregated series at the observation index $\tau$, $\xvet^{[1]}_{\tau}$ is the $(m \times 1)$ vector of observations
of the time series at the highest available frequency within the complete $\tau$-th cycle, and thus each $\xvet_{\tau}$ has dimension $\left[(k^*+m) \times 1 \right]$.

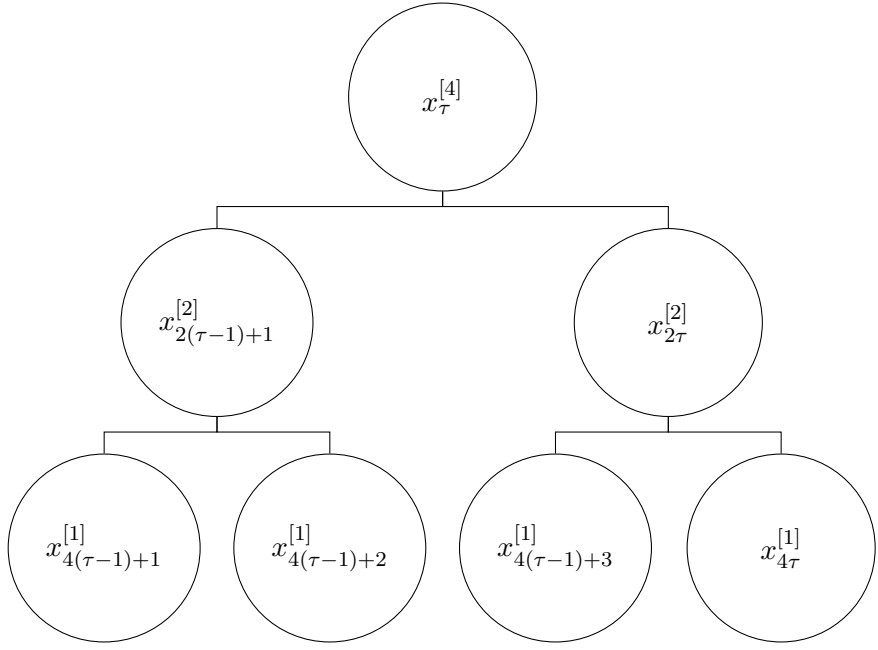
\begin{figure}[t]
	\centering
		\begin{tikzpicture}[baseline=(current  bounding  box.center),
	every node/.append style={shape=ellipse,
		draw=black},
	minimum width=2.5cm,
	minimum height=2.5cm]

	\node at (0, 0) (5Q1){$x_{4(\tau-1)+1}^{\left[ 1 \right]}$};
	\node at (3, 0) (5Q2){$x_{4(\tau-1)+2}^{\left[ 1 \right]}$};
	\node at (6, 0) (5Q3){$x_{4(\tau-1)+3}^{\left[ 1 \right]}$};
	\node at (9, 0) (5Q4){$x_{4\tau}^{\left[ 1 \right]}$};
	\node at (1.5, 3) (5SA1){$x_{2(\tau-1)+1}^{\left[ 2 \right]}$};
	\node at (7.5, 3) (5SA2){$x_{2\tau}^{\left[ 2 \right]}$};
	\node at (4.5, 6) (5A){$x_{\tau}^{\left[ 4 \right]}$};
	\relation{0.2}{5Q1}{5SA1};
	\relation{0.2}{5Q2}{5SA1};
	\relation{0.2}{5Q3}{5SA2};
	\relation{0.2}{5Q4}{5SA2};
	\relation{0.2}{5SA1}{5A};
	\relation{0.2}{5SA2}{5A};
	\end{tikzpicture}
	\caption{Temporal hierarchy for quarterly series using the common index $\tau$ for all levels of aggregation.}
	\label{temphierq-f}
\end{figure}

The relationships linking the original high-frequency series $x_t$ and its temporal aggregates can be graphically represented as a hierarchical/grouped series.
For example, for quarterly data $k \in \{4,2,1\}$,
then every four quarterly observations are aggregated up to annual and semi-annual counterparts. According to the notation so far,
for a single year the quarterly hierarchical structure can be defined as in Figure \ref{temphierq-f}, where
$x^{[4]}$, $x^{[2]}$ and $x^{[1]}$ denote annual, semi-annual, and quarterly values, respectively.

The relationships linking the nodes in the hierarchy can be expressed as we did in
(\ref{contemp}) for the cross-sectional (contemporaneous) hierarchy case:
\begin{equation}
\label{thi}
\tvet_{x_{\tau}} = \Kvet_1\xvet^{[1]}_{\tau}, \quad
\xvet_{\tau} = \Rvet_1\xvet^{[1]}_{\tau}, \quad
\Zvet_1'\xvet_{\tau} = \Zerovet_{(k^* \times 1)}, \quad \tau=1,\ldots,N,
\end{equation}
where $\Kvet_1$ is the $\left( k^* \times m\right)$ temporal aggregation matrix converting
the high-frequency observations into lower-frequency (temporally aggregated) ones,
$$
\Rvet_1 = \left[\begin{array}{c}
\Kvet_1 \\ \Ivet_m
\end{array}
\right]
$$
is the $\left[(k^*+m) \times m \right]$ \emph{temporal summing} matrix, and
$\Zvet_1' = \left[ \Ivet_{k^*} \; -\Kvet_1 \right]$ is the zero constraints kernel matrix valid for $\xvet_{\tau}$.
For example, with quarterly data it is $m=4$, $k^*=3$, and 
\[
\Kvet_1 = \begin{bmatrix}
1 & 1 & 1 & 1 \\
1 & 1 & 0 & 0 \\
0 & 0 & 1 & 1
\end{bmatrix}, \quad
\Rvet_1 = \begin{bmatrix}
1 & 1 & 1 & 1 \\
1 & 1 & 0 & 0 \\
0 & 0 & 1 & 1 \\
1 & 0 & 0 & 0 \\
0 & 1 & 0 & 0 \\
0 & 0 & 1 & 0 \\
0 & 0 & 0 & 1
\end{bmatrix}, \quad
\Zvet_1' = \left[ \Ivet_3 \; -\Kvet_1 \right].
\]

The temporal aggregation relationships can be extended to the whole time span covered by series $x_t$. Denoting by $\xvet = \left( \xvet_1' \ldots \xvet_{\tau}' \ldots \xvet_N' \right)'$
the $\left[N(k^*+m) \times 1\right]$ vector containing all the data of series $X$ at any observed temporal frequency, the complete set of temporal aggregation constraints valid for this vector is given by
\begin{equation}
\label{thivec}
\Zvet_N' \xvet = \Zerovet_{\left(Nk^* \times 1\right)},
\end{equation}
where $\Zvet_N' = \left[ \Ivet_{Nk^*} \; -\Kvet_N \right]$, and
$$
\Kvet_N = \begin{bmatrix}
\Ivet_N \otimes {\bf 1}_4' \\
\Ivet_{2N} \otimes {\bf 1}_2'
\end{bmatrix} .
$$

It is not always possible to represent the temporal aggregates of one series in a single tree\footnote{For any given positive $m >1$, there is
	a single unique temporal hierarchy only if $m=q^{\alpha}$, where $\alpha$ is a positive
	integer and $q$ is a prime number (Yang et al., 2017). A corollary is that a single unique hierarchy is only possible when there are no coprime pairs in the set
	$\left\{ k_{p-1}, \dots, k_3, k_2 \right\}$ (Athanasopoulos et al., 2017).}
such as Fig.~\ref{temphierq-f}. In Appendix A.4 the representations valid for monthly and hourly hierarchies are shown.

\subsection{Alternative approximations of the covariance matrix for point forecast reconciliation through temporal hierarchies}
\label{subsec:temprec}
Suppose we have the $\left[(k^*+m) \times 1\right]$ vector $\hat{\xvet}_h$ of unbiased base forecasts for the $p$ temporal aggregates of a single time series $X$ within a complete time cycle, i.e. at the forecast horizon $h=1$ for the lowest (most aggregated) time frequency. If the base forecasts have been independently computed, generally they do not fulfill the temporal aggregation constraints, i.e. $\Zvet_1'\hat{\xvet}_h \ne \Zerovet_{(k^* \times 1)}$. By adapting the general point forecast reconciliation formula (\ref{stonest}), and not considering suffix $h$ to simplify the notation, the vector of temporally reconciled forecasts is given by:
\begin{equation}
\label{testonest}
\tilde{\xvet} = \hat{\xvet} -
\Omegavet\Zvet_1\left(\Zvet_1'\Omegavet\Zvet_1\right)^{-1}\Zvet_1'\hat{\xvet} ,
\end{equation}
where $\Omegavet$ is a $\left[(k^*+m) \times (k^*+m)\right]$ p.d. matrix, assumed known.

In order to consider possible residual-based estimates of matrix $\Omegavet$, denote
\begin{equation}
\label{hatexk}
\hat{\evet}_{\tau}^{[k]} = \xvet^{[k]}_{\tau} - \hat{\xvet}^{[k]}_{\tau}, \quad \tau=1,\ldots,N, \quad k \in \cal{K},
\end{equation}
the $(M_k \times 1)$ vectors of the in-sample residuals at time index $\tau$ for the models used to generate the base forecasts of the temporally aggregated series of order $k$. These vectors can be organized in matrix form as
\begin{equation}
\label{hatExk}
\widehat{\Evet}^{[k]}_x = \begin{bmatrix}
\left(\hat{\evet}^{[k]}_{1}\right)'\\
\vdots \\
\left(\hat{\evet}^{[k]}_{\tau}\right)'\\
\vdots \\
\left(\hat{\evet}^{[k]}_{N}\right)'
\end{bmatrix} , \quad k \in \cal{K},
\end{equation}
where each matrix $\widehat{\Evet}^{[k]}_x$ has dimension $(N \times M_k)$, and then grouped in the $\left[N \times (k^*+m)\right]$ matrix of in-sample residuals
$$
\widehat{\Evet}_x = \begin{bmatrix}
\widehat{\Evet}^{[m]}_x \; \widehat{\Evet}^{[k_{p-1}]}_x \; \ldots \; \widehat{\Evet}^{[k_{2}]}_x \; \widehat{\Evet}^{[1]}_x
\end{bmatrix} .
$$
Each column of this matrix contains the in-sample residuals pertaining to a specific node of the temporal hierarchy, thus
the sample cross-covariance matrix of the $k^* +m$ nodes of the temporal hierarchy is given by\footnote{Expression (\ref{Wsamte}) assumes that
	$N^{-1}\displaystyle\sum_{\tau=1}^{N}\hat{e}_{\tau,l}=0$, $l=1,\ldots,k^*+m$. When this does not hold, $\reallywidehat{\Omegavet}$ is the sample Mean Square Error (MSE) matrix.}:

\begin{equation}
\label{Wsamte}
\reallywidehat{\Omegavet} = \displaystyle\frac{1}{N}\left(\widehat{\Evet}_x\right)' \widehat{\Evet}_x .
\end{equation}

\noindent This matrix is well defined if $N > (k^*+m)$, otherwise there might be singularity issues which would prevent its use in expression (\ref{testonest}) in place of matrix $\Omegavet$.

Athanasopoulos et al. (2017) and Hyndman and Kourentzes (2018) consider the following alternative choices for $\Omegavet$ (the suffix `t' stands for `temporal', to keep the `t'-procedures distinct from the `cs'-ones shown in section \ref{subsec:csmat}):
\begin{itemize}
	\item identity (t-ols): $\Omegavet = \Ivet_{k^*+m}$ ,
	\item structural (t-struc): $\Omegavet = \reallywidehat{\Omegavet}_{\text{t-struc}} = \text{diag}\left(\Rvet_1{\bf 1}_{m} \right)$ 
	\item hierarchy variance scaling (t-wlsh): $\Omegavet = \reallywidehat{\Omegavet}_{\text{t-wlsh}} = \Ivet_{k^*+m} \odot \reallywidehat{\Omegavet}$
	\item series variance scaling (t-wlsv): $\Omegavet = \reallywidehat{\Omegavet}_{\text{t-wlsv}}$
	\item MinT-shr (t-shr): $\Omegavet = \reallywidehat{\Omegavet}_{\text{t-shr}} =
	\hat{\lambda}\reallywidehat{\Omegavet}_{\text{t-wlsh}} + (1 - \hat{\lambda})\reallywidehat{\Omegavet}$
	\item MinT-sam (t-sam): $\Omegavet = \reallywidehat{\Omegavet}$
\end{itemize}
The series variance scaling matrix $\reallywidehat{\Omegavet}_{\text{t-wlsv}}$ is a diagonal matrix ``which contains estimates of the in-sample one-step-ahead error variances across each level'' (Athanasopoulos et al., 2017, p. 64), that requires a reduced number ($p$ instead of $k^*+m$) of variances to be estimated as compared to the hierarchy variance scaling matrix $ \reallywidehat{\Omegavet}_{\text{t-wlsh}}$, with increased sample size available for the estimation.

``As the purpose of temporal aggregation is to exploit important information about time series at different frequencies'', Nystrup et al. (2020) propose other formulations in order to include potential information in the autocorrelation structure. The matrices considered in this paper\footnote{For the time being, we do not consider all the newly proposed covariance matrices. The interested reader may refer to Nystrup et al. (2020).} are:  

\begin{itemize}
	\item auto-covariance scaling (t-acov): $\Omegavet = \reallywidehat{\Omegavet}_{\text{t-acov}}$
	\item structural Markov (t-strar1): $\Omegavet = \reallywidehat{\Omegavet}_{\text{t-strar1}}$
	\item series Markov (t-sar1): $\Omegavet = \reallywidehat{\Omegavet}_{\text{t-sar1}}$
	\item hierarchy Markov (t-har1): $\Omegavet = \reallywidehat{\Omegavet}_{\text{t-har1}}$
\end{itemize}

The auto-covariance scaling makes use of the estimates of the full autocovariance matrices within each aggregation level, while ignoring correlations between aggregation levels:
$$
\reallywidehat{\Omegavet}_{\text{t-acov}} = \begin{bmatrix}
\reallywidehat{\Omegavet}^{[m]} & \Zerovet  & \cdots & \Zerovet & \Zerovet \\
\Zerovet              & \reallywidehat{\Omegavet}^{[k_{p-1}]} & \cdots & \Zerovet & \Zerovet \\
\vdots & \vdots & \ddots & \vdots & \vdots \\
\Zerovet & \Zerovet & \cdots & \reallywidehat{\Omegavet}^{[k_2]} & \Zerovet \\
\Zerovet & \Zerovet & \cdots & \Zerovet & \reallywidehat{\Omegavet}^{[1]}
\end{bmatrix},
$$
where the $(M_k \times M_k)$ matrices $\reallywidehat{\Omegavet}^{[k]}$ are given by\footnote{Matrix $\reallywidehat{\Omegavet}^{[m]}$ reduces to a scalar variance.}:
\begin{equation}
\label{Wkacov}
\reallywidehat{\Omegavet}^{[k]} =
\displaystyle\frac{1}{N}\sum_{\tau=1}^{N}\hat{\evet}^{[k]}_{\tau}(\hat{\evet}^{[k]}_{\tau})'
=
\displaystyle\frac{1}{N}\left(\widehat{\Evet}^{[k]}\right)'\widehat{\Evet}^{[k]},
\quad k \in {\cal K},
\end{equation}
where vector $\hat{\evet}^{[k]}_{\tau}$ and matrix $\widehat{\Evet}_x^{[k]}$ are given by (\ref{hatexk}) and (\ref{hatExk}), respectively.
A necessary condition in order to matrix $\reallywidehat{\Omegavet}_{\text{t-acov}}$ be invertible, is $N > m$, which is less demanding than what is needed for the non-singularity of matrix $\reallywidehat{\Omegavet}$.

Because it is sometimes difficult to estimate the covariance matrix within each aggregation level without assuming that it has some special form,
Nystrup et al. (2020) propose ``an estimator that blends autocorrelation and variance information, but only requires estimation of the first order autocorrelation coefficient at each aggregation level''. They consider the Toeplitz matrix for the estimated first-order autocorrelation coefficients of the in-sample residuals for the $p-1$ levels $k = k_1, \ldots, k_{p-1}$, of the series' temporal hierarchy. Denoting these autocorrelation coefficents with $\rho_{[k]}$, it is:
$$
\Gammavet^{[m]} = 1, \qquad \Gammavet^{[k]} = \begin{bmatrix}
1 & \rho_{[k]} & \cdots & \rho_{[k]}^{M_k-1} \\[.25cm]
\rho_{[k]} & 1 & \cdots & \rho_{[k]}^{M_k-2} \\[.25cm]
\vdots & \vdots & \ddots & \vdots \\[.25cm]
\rho_{[k]}^{M_k-1} & \rho_{[k]}^{M_k-2} & \cdots & 1
\end{bmatrix}, \quad k = k_1, \ldots, k_{p-1} ,
$$
where each matrix $\Gammavet^{[k]}$, $k \in {\cal K}$, has dimension $(M_k \times M_k)$. The $p$ matrices are used to build the $\left[(k^*+m) \times (k^*+m)\right]$ matrix:
$$
\Gammavet = \begin{bmatrix}
1 & \Zerovet' & \cdots & \Zerovet' \\
\Zerovet & \Gammavet^{[k_{p-1}]} & \cdots & \Zerovet \\
\vdots & \vdots & \ddots & \vdots \\
\Zerovet & \Zerovet & \cdots & \Gammavet^{[1]}
\end{bmatrix},
$$
which can be used in three alternative estimates of matrix $\Omegavet$:
$$
\reallywidehat{\Omegavet}_{\text{t-strar1}} =
\reallywidehat{\Omegavet}^{\frac{1}{2}}_{\text{t-struc}} \Gammavet \reallywidehat{\Omegavet}^{\frac{1}{2}}_{\text{t-struc}}
$$
$$
\reallywidehat{\Omegavet}_{\text{t-sar1}} =
\reallywidehat{\Omegavet}^{\frac{1}{2}}_{\text{t-wlsv}} \Gammavet \reallywidehat{\Omegavet}^{\frac{1}{2}}_{\text{t-wlsv}}
$$
$$
\reallywidehat{\Omegavet}_{\text{t-har1}} =
\reallywidehat{\Omegavet}^{\frac{1}{2}}_{\text{t-wlsh}} \Gammavet \reallywidehat{\Omegavet}^{\frac{1}{2}}_{\text{t-wlsh}} 
$$

\section{The cross-temporal forecast reconciliation framework}
\label{sec:CTac}

\subsection{Cross-temporal aggregation constraints}
The cross sectional aggregation relationships (\ref{ContConstrUt}), linking $n$ series at a single time period, and the temporal aggregation relationships (\ref{thi}), valid for an individual variable, can be simultaneously considered, by extending (i) the cross-sectional constraints to all observation frequencies, and (ii) the temporal aggregation relationships to all variables.

Considering contemporaneous and temporal dimensions in the same framework requires to extend and adapt the notations used so far.
At this end, define
the $p$ matrices $\Yvet^{[k]}$, each of dimension $(n \times NM_k)$, as
$$
\Yvet^{[k]} = \left[\begin{array}{c}
\Avet^{[k]} \\ \Bvet^{[k]}
\end{array}\right], \quad k \in {\cal K},
$$
where
$$
{\Bvet}^{[k]} = \left[\begin{array}{c}
{{\bvet}^{[k]}_1}{'} \\
\vdots \\
{{\bvet}^{[k]}_i}{'} \\
\vdots \\
{{\bvet}^{[k]}_{n_b}}{'}
\end{array}
\right], \quad
{\Avet}^{[k]} = \left[\begin{array}{c}
{{\avet}^{[k]}_1}{'} \\
\vdots \\
{{\avet}^{[k]}_j}{'} \\
\vdots \\
{{\avet}^{[k]}_{n_a}}{'}
\end{array}
\right], \quad
k \in {\cal K} ,
$$
are the matrices containing the $k$-order temporal aggregates of the bts ($\Bvet^{[k]}$) and uts ($\Avet^{[k]}$), of dimension $(n_b \times NM_k)$ and $(n_a \times NM_k)$, respectively.

In order to be consistent with the notation so far,
$\Yvet^{[1]}$, $\Bvet^{[1]}$, and $\Avet^{[1]}$
denote the matrices containing data at the highest available sampling frequency, while
$\Yvet$, $\Bvet$, and $\Avet$ are used now to denote the
matrices containing the data at any considered temporal frequency, that is:
$$
\Yvet = \left[\begin{array}{c} \Avet \\ \Bvet \end{array}
\right] = 
\left[\begin{array}{ccccc}
\Avet^{[m]} & \Avet^{[k_{p-1}]} & \cdots & \Avet^{[k_2]} & \Avet^{[1]} \\
\Bvet^{[m]} & \Bvet^{[k_{p-1}]} & \cdots & \Bvet^{[k_2]} & \Bvet^{[1]}
\end{array}
\right] ,
$$
where $\Yvet$, $\Avet$, and $\Bvet$ have $n$, $n_a$ and $n_b$ rows, respectively, and the same number of columns, $ \left[N(k^*+m)\right]$.

\vspace{.5cm}
\noindent \textit{Cross-sectional aggregation constraints}

\noindent The cross-sectional aggregation relationships operating along all the time observation indices can be worked out by extending the latter equation in expression (\ref{ContConstrUt}):
$$
\Uvet' \Yvet^{[k]} = \Zerovet_{\left(n_a \times NM_k\right)},
\quad k \in {\cal K},
$$
which can be expressed in compact form as
\begin{equation}
\label{syscagg}
\Uvet'\Yvet =
\Zerovet_{\left[n_a \times N(k^*+m)\right]}.
\end{equation}
If we define $\yvet$ as the $(nN(k^*+m) \times 1)$ vector containing all the observations of all series at any temporal aggregation level, organized `by-variable-first-then-by-descending-aggregation-order', that is $\yvet = \text{vec}(\Yvet')$,
the cross-sectional (contemporaneous) constraints (\ref{syscagg}) can be equivalently expressed as:
\begin{equation}
\label{vecsyscagg}
\left(\Uvet' \otimes \Ivet_{N(k^*+m)}\right)\yvet =
\Zerovet_{\left[n_aN(k^* +m) \times 1\right]} .
\end{equation}

\vspace{.5cm}

\noindent \textit{Temporal aggregation constraints}

\noindent The temporal aggregation relationships (\ref{thi}), valid for a single series, can be extended to each component of the $n$-variate time series $\yvet_t$ as follows:
\begin{equation}
\label{thinseries1}
\left[\begin{array}{cc}
{\Avet^{[m]}}' & {\Bvet^{[m]}}' \\
\vdots & \vdots \\
{\Avet^{[k_2]}}' & {\Bvet^{[k_2]}}'
\end{array}
\right] =
\Kvet_N
\left[\begin{array}{cc}
{\Avet^{[1]}}' & {\Bvet^{[1]}}'
\end{array}
\right],
\end{equation}
which can be equivalently written as
$\left[\Ivet_{Nk^*} \; \; -\Kvet_N
\right]\Yvet' = \Zerovet_{\left(Nk^* \times n\right)}$, that is:
\begin{equation}
\label{thinseries2}
\Zvet_N' \Yvet' = \Zerovet_{\left(Nk^* \times n\right)}.
\end{equation}
The temporal aggregation constraints (\ref{thinseries2}) can thus be re-stated as:
\begin{equation}
\label{thinseries3}
\left( \Ivet_{n} \otimes \Zvet_N' \right)\yvet = \Zerovet_{\left(nNk^* \times 1\right)},
\end{equation}
which, for $n=1$, is equivalent to expression (\ref{thivec}).

\vspace{.5cm}
In summary, by considering expressions (\ref{vecsyscagg}) and (\ref{thinseries3}) together, the cross-temporal constraints working on the complete set of observations can be expressed as:
\begin{equation}
\label{ctvet}
\breve{\Hvet}' \yvet = \Zerovet_{\left(n^* \times 1\right)},
\end{equation}
where $n^* = n_aN(k^*+m)+nNk^*$, and
$$
\breve{\Hvet}' = \left[\begin{array}{c}
\Uvet' \otimes \Ivet_{N(k^*+m)} \\
 \Ivet_{n} \otimes \Zvet_N'
\end{array}\right]
$$
is a $\left[n^* \times nN(k^*+m)\right]$ cross-temporal zero-constraints kernel matrix.

Due to the simultaneous consideration of temporal and cross-sectional relationships linking the various time series of the system, some rows of $\breve{\Hvet}'$ are redundant, and can be eliminated if one wishes a full row-rank zero-constraints kernel matrix.
This issue is not new, since it has been encountered in the past when contemporaneous and temporal aggregation constraints are simultaneously considered for the reconciliation of a system of time series (Di Fonzo, 1990, Di Fonzo and Marini, 2011).
In detail, matrix $\breve{\Hvet}'$ consists in:
\begin{itemize}
\item $Nn_ak^*$ rows defining the cross-sectional (contemporaneous) aggregation constraints operating on the lf-uts;
\item $Nn_am$ rows defining the cross-sectional (contemporaneous) aggregation constraints operating on the hf-bts;
\item $Nn_a\left(k^* + m\right)$ rows defining the temporal aggregation constraints operating on both hf- and lf-uts;
\item $Nn_b\left(k^* + m\right)$ rows defining the temporal aggregation constraints operating on both hf- and lf-bts.
\end{itemize}
Since the first set of $Nn_ak^*$ constraints is linearly dependent from the other rows of matrix $\breve{\Hvet}'$, a full row-rank cross-temporal zero constraints kernel matrix $\Hvet'$ can be obtained by:
\begin{enumerate}
\item considering the $\left[(Nn(k^*+m) \times (Nn(k^*+m)\right]$ commutation matrix (Magnus and Neudecker, 2019, p. 54; see Appendix A.3.1) $\Pvet$ such that $\Pvet\left[\text{vec}\left(\Yvet\right)\right]=\text{vec}\left(\Yvet'\right)$;
\item defining a matrix $\Uvet^*$ as:
$$
\Uvet^* =
\begin{bmatrix}
\Zerovet_{(Nn_am \times Nnk^*)} \; \; \; \Ivet_{Nm} \otimes \Uvet'
\end{bmatrix} \Pvet';
$$
\item considering the $\left[N(n_am + nk^*) \times Nn(k^*+m)\right]$
matrix:
\begin{equation}
\label{Ht}
\Hvet' =
\begin{bmatrix}
\Uvet^* \\
\Ivet_{n} \otimes \Zvet_N'
\end{bmatrix},
\end{equation}
which has full row-rank equal to $N(n_am + nk^*) = n^* - Nn_ak^*$, and allows to re-state the complete cross-temporal constraints (\ref{ctvet}) as:
\begin{equation}
\label{ctvetfullrowrank}
\Hvet'\yvet = \Zerovet .
\end{equation}
\end{enumerate}

\vspace{.5cm}
\noindent \textit{Cross-temporal structural representation}

\noindent The cross-temporal structural representation can be seen as a generalization from a single time index $t$ to a single cycle index $\tau$ (i.e., the low-frequency time index) of the cross-sectional structural representation (see section \ref{sec:hts}), extended to cover $n(k^*+m)$ nodes instead of $n$.

Denote with $\Yvet_{\tau}$ the $\left[(n \times (k^*+m)\right]$ data matrix available at cycle $\tau$:
$$
\Yvet_{\tau} = \begin{bmatrix}
\Avet_{\tau} \\ \Bvet_{\tau}
\end{bmatrix} =
\begin{bmatrix}
\Avet^{[m]}_{\tau} \; \Avet^{[k_{p-1}]}_{\tau} \; \ldots \; \Avet^{[k_{2}]}_{\tau} \; \Avet^{[1]}_{\tau} \\[.2cm]
\Bvet^{[m]}_{\tau} \; \Bvet^{[k_{p-1}]}_{\tau} \; \ldots \; \Bvet^{[k_{2}]}_{\tau} \; \Bvet^{[1]}_{\tau} \\
\end{bmatrix}, \quad \tau = 1, \ldots N,
$$
and let $\check{\Svet}$ be the $(n^* \times n_bm)$ cross-temporal summation matrix
$$
\check{\Svet} = \begin{bmatrix}
\check{\Cvet} \\ \Ivet_{n_bm}
\end{bmatrix},
$$
where $\check{\Cvet}$ denotes a $(n_a^* \times n_bm)$ cross-temporal aggregation matrix mapping the hf-bts into the uts and lf-bts ones ($n^*_a = n_a(k^*+m)+n_bk^*$).
Denote with
$$
\avet^*_{\tau} = \begin{bmatrix}
\text{vec}\left(\Avet'_{\tau}\right) \\[0.15cm]
\text{vec}\left({\Bvet_{\tau}^*}'\right)
\end{bmatrix}, \quad \tau = 1, \ldots, N,
$$
the $\left(n^*_a \times 1\right)$ vector of `cross-temporal upper series', containing the uts and lf-bts data at the low-frequency time index $\tau$, where
$\Bvet_{\tau}^* = \left[\Bvet^{[m]}_{\tau} \; \Bvet^{[k_{p-1}]}_{\tau} \; \ldots \; \Bvet^{[k_{2}]}_{\tau}\right]$, $\tau = 1, \ldots, N$,
and with
$$
\bvet^{[1]}_{\tau} =\text{vec}\left({\Bvet_{\tau}^{[1]}}'\right), \quad \tau = 1, \ldots, N,
$$
the $(n_bm \times 1)$ vector of `cross-temporal bottom series', containing the hf-bts data. The structural representation of a cross-temporal system of $n$ time series takes the form
\begin{equation}
\label{strucrep}
\check{\yvet}_{\tau} = \check{\Svet}\bvet^{[1]}_{\tau}, \quad \tau = 1, \ldots, N,
\end{equation}
where $\check{\yvet}_{\tau}$ is a $\left[n(k^*+m) \times 1\right]$ vector where we place all the uts and the lf-bts at the top, and all the hf-bts at the bottom:
\begin{equation}
\label{yvetstar}
\check{\yvet}_{\tau} = \begin{bmatrix}
\avet_{\tau}^* \\[0.15cm]
\bvet^{[1]}_{\tau}
\end{bmatrix}, \quad \tau =1, \ldots, N .
\end{equation}

\subsection{Cross-temporal point forecast reconciliation: introduction}
\label{sec:cac}

Let us assume to have unbiased base forecasts for all the individual time series of the multivariate hierarchical/grouped time series, and for all levels of the temporal hierarchies built from the highest available sampling frequency. In addition, assume that the forecast horizon for the most
temporally aggregated time series be $h=1$,\footnote{The general case $h \ge 1$ can be dealt with in a straightforward way.}
and that the forecast horizons for the other temporally aggregated series cover the entire time cycle. This means that (i) the forecast horizon for the highest frequency time series is equal to $m$, and (ii) in general, the forecast horizon for a temporally aggregated time series of order $k$ spans from 1 to $M_k$.

The base forecasts for each bottom time series of the system form the vectors
$$
\widehat{\bvet}^{[k]}_i, \quad i=1, \ldots, n_b, \quad k \in {\cal K},
$$
where $\widehat{\bvet}^{[1]}_i = \left\{\hat{b}^{[1]}_{il}\right\}_{l=1}^{m}$ is the $(m \times 1)$ vector containing the base forecasts for the $i$-th high-frequency bottom time series (hf-bts), which are the `very' bottom time series in the cross-temporal framework, while the remaining $\widehat{\bvet}^{[k]}_i$'s (for $k \ne 1$)
contain the $M_k$ forecasts for the lower-frequency ones (lf-bts).

The base forecasts for the upper time series can be defined likeways as
$$
\widehat{\avet}^{[k]}_j, \quad j=1, \ldots, n_a, \quad k \in {\cal K},
$$
where $\widehat{\avet}^{[1]}_j = \left\{\widehat{a}^{[1]}_{jl}\right\}_{l=1}^{m}$ is the $(m \times 1)$ vector containing the base forecasts for the high-frequency $j$-th upper time series (hf-uts), and the $\widehat{\avet}^{[k]}_j$'s (for $k \ne 1$) are $(M_k \times 1)$ vectors of low-frequency upper time series (lf-uts) forecasts.

Let us collect these base forecasts in the $(n_b \times M_k)$ and $(n_a \times M_k)$, respectively, matrices
\begin{equation}
\label{basemat}
\widehat{\Bvet}^{[k]} = \left[\begin{array}{c}
{\widehat{\bvet}^{[k]}_1}{'} \\
\vdots \\
{\widehat{\bvet}^{[k]}_i}{'} \\
\vdots \\
{\widehat{\bvet}^{[k]}_{n_b}}{'}
\end{array}
\right], \quad
\widehat{\Avet}^{[k]} = \left[\begin{array}{c}
{\widehat{\avet}^{[k]}_1}{'} \\
\vdots \\
{\widehat{\avet}^{[k]}_j}{'} \\
\vdots \\
{\widehat{\avet}^{[k]}_{n_a}}{'}
\end{array}
\right], \quad
k \in {\cal K} .
\end{equation}

\noindent The matrix containing the base bts forecasts is given by:
$$
\widehat{\Bvet} = \left[
\widehat{\Bvet}^{[m]} \; \widehat{\Bvet}^{[k_{p-1}]} \; \ldots \; \widehat{\Bvet}^{[k_{2}]} \; \widehat{\Bvet}^{[1]}
\right],
$$
where $\widehat{\Bvet}$ has dimension $\left[n_b \times (k^* + m)\right]$. The base uts forecasts can be similarly arranged in the $\left[n_a \times (k^* + m)\right]$ matrix
$$
\widehat{\Avet} = \left[
\widehat{\Avet}^{[m]} \; \widehat{\Avet}^{[k_{p-1}]} \; \ldots \; \widehat{\Avet}^{[k_{2}]} \; \widehat{\Avet}^{[1]}
\right].
$$

\noindent From expression (\ref{basemat}) we can define the $p$ matrices $\widehat{\Yvet}^{[k]}$, each of dimension  $(n \times M_k)$, with $n = n_a + n_b$, containing
the base forecasts for the temporal aggregation level $k$ of both uts and bts:
$$
\widehat{\Yvet}^{[k]} = \left[\begin{array}{c}
\widehat{\Avet}^{[k]} \\
\widehat{\Bvet}^{[k]} \\
\end{array}
\right], \quad
k \in {\cal K} .
$$

\noindent Finally, denoting with $\widehat{\Yvet}$ the $\left[n \times (k^* + m)\right]$ matrix containing the base forecasts of all series and for all temporal aggregation levels, it is:
$$
\widehat{\Yvet} = \left[
\widehat{\Yvet}^{[m]} \; \widehat{\Yvet}^{[k_{p-1}]} \; \ldots \; \widehat{\Yvet}^{[k_{2}]} \; \widehat{\Yvet}^{[1]}
\right] =
\left[\begin{array}{c}
\widehat{\Avet} \\ \widehat{\Bvet}
\end{array}\right].
$$

In general, the base forecasts fulfill neither cross-sectional (contemporaneous) nor temporal aggregation constraints. That is, respectively:
$$
\Uvet' \widehat{\Yvet}  \ne \Zerovet_{\left[n_a \times (k^* +m)\right]}, \qquad
\Zvet_1' \widehat{\Yvet}' \ne \Zerovet_{\left(k^* \times n\right)} .
$$
The cross-temporal point forecast reconciliation problem can thus be stated as follows: we are looking for a reconciled point forecast matrix, say $\widetilde{\Yvet}$, which is `as-close-as-possible' (according to a pre-specified metric) to the base forecast matrix $\widehat{\Yvet}$, and simultaneously in line with the cross-sectional and temporal aggregation constraints, that is:
\begin{equation}
\label{simcon}
\Uvet' \widetilde{\Yvet}  = \Zerovet_{n_a \times (k^* +m)}, \qquad
\Zvet_1' \widetilde{\Yvet}' = \Zerovet_{(k^* \times n)} .
\end{equation}
As we have previously shown, the relationships (\ref{simcon}) can be
expressed in vectorized form as $\Hvet' \tilde{\yvet} = \Zerovet$, where $\tilde{\yvet}=\text{vec}\left(\widetilde{\Yvet}'\right)$ 
and, since $h=1$, the full row-rank matrix $\Hvet'$ in (\ref{Ht}) becomes
\begin{equation}
\label{Ht1}
\Hvet' = \begin{bmatrix}
\Uvet^* \\
\Ivet_n \otimes \Zvet_1'
\end{bmatrix} .
\end{equation}

\subsection{Bottom-up cross-temporal forecast reconciliation}
\label{subsec:buct}
Cross temporal reconciled forecasts for all series at any temporal
aggregation level can be easily computed by appropriate summation of the hf-bts base forecasts $\widehat{\bvet}^{[1]}_i$, $i=1, \ldots, n_b$, according to a bottom-up procedure like what is currently done in both the cross-sectional and temporal frameworks.

Denoting by $\ddot{\Yvet}$ the $\left[n \times (k^*+m)\right]$ matrix containing the bottom-up cross temporal reconciled forecasts, it is:
\[
\ddot{\Yvet} = \begin{bmatrix}
\ddot{\Avet} \\ \ddot{\Bvet}
\end{bmatrix} =
\begin{bmatrix}
\ddot{\Avet}^{[m]} \; \ddot{\Avet}^{[k_{p-1}]} \; \ldots \; \ddot{\Avet}^{[k_{2}]} \; \ddot{\Avet}^{[1]} \\
\ddot{\Bvet}^{[m]} \; \ddot{\Bvet}^{[k_{p-1}]} \; \ldots \; \ddot{\Bvet}^{[k_{2}]} \; \ddot{\Bvet}^{[1]}
\end{bmatrix}.
\]
Since the hf-bts reconciled forecasts are by definition equal to the hf-bts base forecasts, i.e.
$\ddot{\Bvet}^{[1]} = \widehat{\Bvet}^{[1]}$, the bottom-up forecast reconciliation procedure consists of the following steps:
\begin{enumerate}
	\item compute the hf-uts reconciled forecasts using the cross-sectional aggregation relationship (\ref{ACB}):
	\[
	\ddot{\Avet}^{[1]} = \Cvet\widehat{\Bvet}^{[1]};
	\]
	\item compute the lf-bts reconciled forecasts according to the temporal aggregation relationship (\ref{thinseries1}):
	\[
	\begin{bmatrix}
	\left(\ddot{\Bvet}^{[m]}\right)' \\  \vdots \\ \left(\ddot{\Bvet}^{[k_{2}]}\right)'
	\end{bmatrix} =  \Kvet_1 \left(\widehat{\Bvet}^{[1]}\right)' \quad
	\Rightarrow \quad
	\begin{bmatrix}
	\ddot{\Bvet}^{[m]} \; \ddot{\Bvet}^{[k_{p-1}]} \ldots \ddot{\Bvet}^{[k_2]}
	\end{bmatrix}
	= \widehat{\Bvet}^{[1]}\Kvet_1'	;
	\]
	\item compute the lf-uts reconciled forecasts by cross-sectional aggregation of the lf-bts reconciled forecasts obtained in the previous step:
	\[
	\ddot{\Avet}^{[k]} = \Cvet\ddot{\Bvet}^{[k]}, \quad k \in {\cal K} \quad
	\Rightarrow \quad 
	\begin{bmatrix}
	\ddot{\Avet}^{[m]} \; \ddot{\Avet}^{[k_{p-1}]} \ldots \ddot{\Avet}^{[k_2]}
	\end{bmatrix}
	= \Cvet\widehat{\Bvet}^{[1]}\Kvet_1' .
	\]
\end{enumerate}
In summary, the matrix containing the bottom-up reconciled forecasts, solely depending on the hf-bts base forecasts, is given by:
\begin{equation}
\label{buct}
\ddot{\Yvet} = \begin{bmatrix}
\Cvet\widehat{\Bvet}^{[1]}\Kvet_1' & \Cvet\widehat{\Bvet}^{[1]} \\
\widehat{\Bvet}^{[1]} \Kvet_1' & \widehat{\Bvet}^{[1]}
\end{bmatrix} .
\end{equation}

An equivalent, succint alternative to expression (\ref{buct}) consists in exploiting the cross-temporal structural representation (\ref{strucrep}):
\begin{equation}
\label{buctstruc}
\ddot{\check{\yvet}} = \check{\Svet} \widehat{\bvet}^{[1]},
\end{equation}
where $\widehat{\bvet}^{[1]} = \text{vec}\left(\widehat{\Bvet}^{[1]}\right)'$, keeping in mind that
the elements in $\check{\yvet}$ and in $\yvet$ are differently organized, and in general it is $\ddot{\check{\yvet}} \ne \ddot{\yvet}$, with $\ddot{\yvet} = \text{vec}\left(\ddot{\Yvet}'\right)$.
This last issue can be easily dealt with by considering the $(n^* \times n^*)$ permutation matrix such that $\yvet = \Qvet\check{\yvet}$ (see Appendix A.3.2): given the orthogonality of matrix $\Qvet$, expression (\ref{buctstruc}) can be re-stated as
$ \ddot{\yvet} = \Qvet\check{\Svet} \widehat{\bvet}^{[1]}$.
However, the formulation of matrix $\check{\Svet}$, which requires to manage linear relationships across cross-sectional and temporal dimensions, may be tedious and prone to errors, mostly for large collections of time series. In such cases, it might be preferible using formulation (\ref{buct}), where $\widehat{\Bvet}^{[1]}$, $\Cvet$ and $\Kvet_1$ are involved in simple matrix products.

Appendix A.5 describes all these features with reference to a `toy example' of a very simple two-level hierarchy with two quarterly bottom time series.


\section{Cross-temporal optimal forecast combination}
\label{sec:octforec}
Let us consider the multivariate regression model
\begin{equation}
\label{multivcs}
\widehat{\Yvet} = \Yvet + \Evet ,
\end{equation}
where the involved matrices have each dimension
$\left[n \times (k^*+m)\right]$ and contain, respectively, the base ($\widehat{\Yvet}$) and the target forecasts ($\Yvet$), and the coherency errors
($\Evet$) for the $n$ component variables of the linearly constrained time series of interest.
For each variable, $k^* + m$ base forecasts are available, pertaining to all aggregation levels of the temporal hierarchy for a complete cycle of high-frequency observation, $m$.

Consider now two vectorized versions of model (\ref{multivcs}), by transforming the matrices either in original form:
\begin{equation}
\label{vecY}
\text{vec}\left(\widehat{\Yvet}\right) = \text{vec}\left(\Yvet\right) + \text{vec}\left(\Evet\right)
\quad \Leftrightarrow \quad 
\widehat{{\bf \cal{Y}}} = {\bf \cal{Y}} + \mathbf{\varepsilon} ,
\end{equation}
\noindent or in transposed form:
\begin{equation}
\label{vecYt}
\text{vec}\left(\widehat{\Yvet}'\right) = \text{vec}\left(\Yvet'\right) + \text{vec}\left(\Evet'\right)
\quad \Leftrightarrow \quad 
\hat{\yvet} = \yvet + \mathbf{\eta} .
\end{equation}

\noindent The target forecasts must fulfill the cross-sectional (contemporaneous) constraints
$$
\Uvet'\Yvet = \Zerovet_{\left[n_a \times (k^*+m)\right]}
$$
and the temporal aggregation constraints
$$
\Zvet_1'\Yvet' = \Zerovet_{(k^* \times n)} ,
$$
that is, in vectorized form:
\begin{equation}
\label{csconstrvet}
\left(\Ivet_{k^*+m} \otimes \Uvet'\right){\bf \cal{Y}} = \Zerovet_{[n_a(k^*+m) \times 1]}
\quad \Leftrightarrow \quad
\left(\Uvet' \otimes \Ivet_{k^*+m}\right)\yvet = \Zerovet_{[n_a(k^*+m) \times 1]}
\end{equation}

\begin{equation}
\label{teconstrct}
\left( \Zvet_1' \otimes \Ivet_{n} \right){\bf \cal{Y}} = \Zerovet_{(k^*n \times 1)}
\quad \Leftrightarrow \quad
\left(\Ivet_{n} \otimes \Zvet_1'\right)\yvet = \Zerovet_{(k^*n \times 1)}
\end{equation}

Denote with $\Pvet$ the $\left[n(k^*+m) \times n(k^*+m)\right]$ commutation matrix such that
$$
\Pvet\text{vec}\left(\Yvet\right) = \text{vec}\left(\Yvet'\right) \quad
\Leftrightarrow \quad \Pvet{\bf \cal{Y}} = \yvet
$$
$$
\Pvet\text{vec}\left(\widehat{\Yvet}\right) = \text{vec}\left(\widehat{\Yvet}'\right) \quad
\Leftrightarrow \quad \Pvet\widehat{{\bf \cal{Y}}} = \hat{\yvet}
$$
$$
\Pvet\text{vec}\left(\Evet\right) = \text{vec}\left(\Evet'\right) \quad
\Leftrightarrow \quad \Pvet\mathbf{\varepsilon} = {\bf \eta}
$$
As a consequence, using the full row-rank matrix $\Hvet'$ defined by expression (\ref{Ht1}), the constraints (\ref{csconstrvet}) and (\ref{teconstrct}) can be re-stated as:
$$
\Hvet'\yvet = \Zerovet \quad \Leftrightarrow \quad \Hvet'\Pvet{\bf \cal{Y}} = \Zerovet.
$$
Let $\Wvet = E\left[\mathbf{\varepsilon\varepsilon}'\right]$ be the covariance matrix
of vector $\mathbf{\varepsilon}$,
and $\Omegavet = E\left[\mathbf{\eta\eta}'\right]$
the covariance matrix of vector $\mathbf{\eta}$.
Clearly, $\Wvet$ and $\Omegavet$ are different parameterizations of the same statistical object, i.e.
the covariance matrix of the random disturbances in the multivariate regression model
(\ref{multivcs}), for which the following relationships hold:
$$
\Omegavet = \Pvet\Wvet\Pvet', \qquad \Wvet = \Pvet' \Omegavet\Pvet .
$$

In order to apply the general point forecast reconciliation formula (\ref{stonest}) to a cross-temporal forecast reconciliation problem, we may consider either the expression
$$
\tilde{\yvet}= \hat{\yvet} -
\Omegavet\Hvet\left(\Hvet'\Omegavet\Hvet\right)^{-1}\Hvet'\hat{\yvet},
$$
where $\hat{\yvet} = \text{vec}\left(\widehat{\Yvet}'\right)$ is the row vectorization of the base forecasts matrix $\widehat{\Yvet}$, or equivalently re-state the expression above as:
$$
\label{stonevecY}
\widetilde{{\bf \cal{Y}}} = \widehat{{\bf \cal{Y}}} -
\Wvet\Pvet'\Hvet\left(\Hvet'\Pvet\Wvet\Pvet'\Hvet\right)^{-1}\Hvet'\Pvet\widehat{{\bf \cal{Y}}},
$$
by considering the column vectorization as in (\ref{vecY}).

\subsection{Simple alternative approximations of the covariance matrix for cross-temporal point forecast reconciliation}
\label{subsec:Woct}
Consider the column vectorized form of the multivariate regression (\ref{vecY}), whose random disturbances can be written as:
$$
{\bf \varepsilon} = \begin{bmatrix}
{\bf \varepsilon}^{[m]}_1\\ {\bf \varepsilon}^{[k_{p-1}]}_1\\ \vdots \\ 
{\bf \varepsilon}^{[k_{p-1}]}_{\frac{m}{k_{p-1}}}\\ \vdots \\
{\bf \varepsilon}^{[1]}_1 \\ \vdots \\ {\bf \varepsilon}^{[1]}_m 
\end{bmatrix},
$$
where each $(n \times 1)$ vector ${\bf \varepsilon}^{[k]}_l$, $k \in {\cal K}$, $l=1,\ldots,M_k$,  contains contemporaneous random disturbances, i.e. at the same observation index of any temporal aggregation order.

A simple, though rather irrealistic, generalization to the cross-temporal framework of the cross-sectional approach (see section \ref{subsec:csmat}) consists in assuming that only the disturbances at the same time index of the same temporal aggregation level are correlated, whereas no temporal dependence (either within the same series at different times, or between the $n$ series) is admitted:
$$
E\left[ \mathbf{\varepsilon}_r^{[k_i]} \left(\mbox{$\mathbf{\varepsilon}_s^{[k_j]}$}\right)'\right]
=
\left\{\begin{array}{cl}
\Wvet_{l}^{[k]} & \text{if } k_i=k_j=k, \; r=s=l \\
\Zerovet & \text{otherwise}
\end{array}
\right.
,
\begin{array}{l}
k \in {\cal K}, \\
l=1,\ldots,M_k .
\end{array}
$$

\noindent In this case, the covariance matrix $\Wvet$ has the following block-diagonal structure:
\begin{equation}
\label{Wvetcs}
\begin{array}{rcl}
\Wvet & = &\begin{bmatrix}
\Wvet^{[m]}_1 & \Zerovet          & \cdots & \Zerovet & \cdots & \Zerovet & \cdots & \Zerovet\\
\Zerovet    & \Wvet_1^{[k_{p-1}]} & \cdots & \Zerovet & \cdots & \Zerovet & \cdots & \Zerovet\\
\vdots   & \vdots               & \ddots & \vdots   & \ddots & \vdots   & \ddots & \vdots\\
\Zerovet & \Zerovet & \cdots & \Wvet_{\frac{m}{k_{p-1}}}^{[k_{p-1}]} & \cdots & \Zerovet & \cdots & \Zerovet\\
\vdots   & \vdots & \ddots & \vdots   & \ddots & \vdots & \ddots & \vdots\\
\Zerovet & \cdots & \cdots & \Zerovet & \cdots & \Wvet_{1}^{[1]} & \cdots & \Zerovet \\
\vdots   & \vdots & \ddots & \vdots   & \ddots & \vdots      & \ddots & \vdots \\
\Zerovet & \cdots & \cdots & \Zerovet & \cdots & \Zerovet    & \cdots & \Wvet_{m}^{[1]}
\end{bmatrix}
\end{array} .
\end{equation}
Furthermore, if it is assumed that within each temporal aggregation level the random disturbances follow a multivariate white noise, which means that the contemporaneous covariance matrices are constant in time (i.e., $\Wvet_{l}^{[k]} = \Wvet^{[k]}$, $k \in {\cal K}$, $l=1,\ldots,M_k$), the previous expression simplifies as follows: 
\begin{equation}
\label{Wvet}
\begin{array}{rcl}
\Wvet & = &
\begin{bmatrix}
\Wvet^{[m]} & \Zerovet          & \cdots &\Zerovet \\
\Zerovet & \left(\Ivet_{\frac{m}{k_{p-1}}} \otimes \Wvet^{[k_{p-1}]}\right) & \cdots & \Zerovet\\
\vdots & \vdots & \ddots & \vdots \\
\Zerovet & \Zerovet & \cdots & \left(\Ivet_{m} \otimes \Wvet^{[1]}\right)
\end{bmatrix}
\end{array} .
\end{equation}
From a practical point of view, each $(n \times n)$ matrix $\Wvet^{[k]}$, $k \in \cal{K}$, may be approximated like in the cross-sectional forecast reconciliation case, possibly using the in-sample residuals (see section \ref{subsec:csmat}). Expressions (\ref{Wvetcs}) and (\ref{Wvet}) can thus be seen as two simple extensions to the cross-temporal case of the approach developed in the cross-sectional framework, where no temporal dependence is accounted for both within and between the $n$ series.

We may similarly propose a simplified pattern of the disturbances covariance matrix of the multivariate regression model (\ref{multivcs}), by considering the row vectorization form (\ref{vecYt}). In this case, the random disturbances vector $\mathbf{\eta}$ can be written as
$$
\mathbf{\eta} = \begin{bmatrix}
\mathbf{\eta}_1' \; \ldots \; \mathbf{\eta}_i' \; \ldots \mathbf{\eta}_n'
\end{bmatrix}' ,
$$
where each $\left[(k^*+m) \times 1\right]$ vector $\mathbf{\eta}_i$, $i=1,\ldots,n$, contains the random disturbances at different observation indices of the various temporal aggregation levels for the same series $i$.
If we assume that the $n$ series are uncorrelated at any observation index for any temporal aggregation level (i.e. neither contemporaneous nor temporal correlation is admitted between the series, which is rather irrealistic), denoting with
$\Omegavet_{ii} = E(\mathbf{\eta}_i\mathbf{\eta}_i')$, $i=1,\ldots,n$, the $\left[(k^*+m) \times (k^* + m)\right]$ covariance matrix of the coherency errors of the temporal hierarchies of series $i$, the complete matrix
$\Omegavet$ can be written as follows:
\begin{equation}
\label{Omegavet}
\Omegavet = \begin{bmatrix}
\Omegavet_{11}  & \Zerovet & \cdots & \Zerovet \\
\Zerovet & \Omegavet_{22}  & \cdots & \Zerovet \\
\vdots   & \vdots   & \ddots & \vdots \\
\Zerovet & \Zerovet & \cdots & \Omegavet_{nn}
\end{bmatrix} ,
\end{equation}
where each matrix $\Omegavet_{ii}$, $i = 1,\ldots,n$, may be approximated as in the temporal forecast reconciliation case, possibly using the in-sample residuals (see section \ref{subsec:temprec}). Thus expression (\ref{Omegavet}) can be seen as a very simple (maybe too simple!) extension to the cross-temporal case of the approach developed in the temporal hierarchies framework, where no correlation is admitted between the random errors of the $n$ series.

Clearly, the two covariance patterns (\ref{Wvet}) and (\ref{Omegavet}) (i) are placed at opposite ends of possible ways of dealing with cross-temporal variables, and (ii) should be considered as first practical devices to make the optimal combination forecast approach feasible for the cross-temporal framework as well. This subject is undoubtedly of far greater importance between the open issues still remaining in this field, and we plan to go deep on this subject in the near future. 

Residual-based estimates of the covariance matrix $\Wvet$ (and of its re-parameterized counterpart $\Omegavet$) make use of the in-sample residuals of the models used to forecast the $n$ time series considered at any temporal aggregation level.
Denote by
$$
\widehat{\Evet}^{[k]}_{l}, \quad k \in {\cal K}, \quad l=1,\ldots,M_k,
$$
the $(n \times N)$ matrix containing the in-sample residuals for a single node of the cross-temporal hierarchy (i.e., the $i$-th row contains the residuals for the $N$ sub-periods $l$ of the model used to forecast the temporal aggregate of order $k$ of series $i$).
For each temporal aggregation level $k \in {\cal K}$, the $M_k$ matrices $\widehat{\Evet}^{[k]}_{l}$
can be grouped into the $(n \times NM_k)$ matrix
$$
\widehat{\Evet}^{[k]} = \begin{bmatrix}
\widehat{\Evet}^{[k]}_{1} \; \ldots \; \widehat{\Evet}^{[k]}_{l} \; \ldots \;
\widehat{\Evet}^{[k]}_{M_k}
\end{bmatrix} , \quad k \in {\cal K} .
$$
The $(n(k^*+m) \times N)$ matrix containing all the residuals at any time observation index and any temporal aggregation level, can in turn be written as:
$$
\widehat{\Evet} = 
\left[\begin{array}{l}
\widehat{\Evet}^{[m]}_{1}\\
\widehat{\Evet}^{[k_{p-1}]}_{1}\\ \vdots \\ \widehat{\Evet}^{[k_{p-1}]}_{\frac{m}{k_{p-1}}}\\
\vdots \\
\widehat{\Evet}^{[1]}_{1}\\ \vdots \\ \widehat{\Evet}^{[1]}_{m}
\end{array}\right] \quad = \quad
\begin{bmatrix}
\hat{\evet}_1 \; \ldots \; \hat{\evet}_{\tau} \; \ldots \; \hat{\evet}_N
\end{bmatrix},
$$
where each $\left[n(k^*+m) \times 1\right]$ vector $\hat{\evet}_{\tau}$, $\tau = 1, \ldots, N$, is given by:
$$
\hat{\evet}_{\tau} = \left[ \underbrace{\hat{e}^{[m]}_{1,\tau} \;  \ldots
	\; (\hat{\evet}^{[1]}_{1,\tau})'}_{k^*+m} \; \cdots \;
\underbrace{\hat{e}^{[m]}_{n,\tau} \;  \ldots
	\; (\hat{\evet}^{[1]}_{n,\tau})'}_{k^*+m}
\right]' .
$$
The sample residual covariance matrix\footnote{Indeed, $\reallywidehat{\Omegavet}_{\text{sam}}$ and $\reallywidehat{\Wvet}_{\text{sam}}$ are mean square error (MSE) matrices.} can be calculated according to both parameterization as:
$$
\reallywidehat{\Omegavet}_{\text{sam}} =
\displaystyle\frac{1}{N} \displaystyle\sum_{\tau=1}^{N}
\hat{\evet}_{\tau}
\left(\mbox{$\hat{\evet}_{\tau}$}\right)' =
\displaystyle\frac{1}{N} \widehat{\Evet}\widehat{\Evet}' ,
$$
$$
\reallywidehat{\Wvet}_{\text{sam}} = \Pvet'\reallywidehat{\Omegavet}_{\text{sam}}\Pvet .
$$
However, in many practical situations matrix $\widehat{\Evet}$ has a number of rows - which is equal to the number of nodes in the cross-temporal hierarchy - much larger than the number of columns, which is equal to $N=\displaystyle\frac{T}{m}$. Thus matrices $\reallywidehat{\Omegavet}_{\text{sam}}$ and $\widehat{\Wvet}_{\text{sam}}$ might not have good properties (in particular, they are not p.d. if $N \le n(k^*+m)$), and
simplified approximations must be looked for.


Two feasible alternatives are given by either the diagonalization or the shrinkage of matrix $\widehat{\Wvet}_{\text{sam}}$, that is, respectively:
$$
\widehat{\Wvet}_{\text{wlsh}} = \Ivet_{n(k^*+m)} \odot \widehat{\Wvet}_{\text{sam}} ,
$$
$$
\widehat{\Wvet}_{\text{shr}} = \hat{\lambda}\widehat{\Wvet}_{\text{wlsh}} + (1 - \hat{\lambda})\widehat{\Wvet}_{\text{sam}},
$$
where $\widehat{\Wvet}_{\text{wlsh}}$ is a diagonal matrix containing the estimates of the `hierarchy variances' for each node of the cross-temporal hierarchy, $\widehat{\Wvet}_{\text{shr}}$ is the matrix obtained by shrinkage of
$\widehat{\Wvet}_{\text{sam}}$ with target $\widehat{\Wvet}_{\text{wlsh}}$, and $\hat{\lambda}$ is an estimate of the coefficient of shrinkage intensity $\lambda$,
$0 \le \lambda \le 1$.
Both  $\widehat{\Wvet}_{\text{sam}}$ and $\widehat{\Wvet}_{\text{shr}}$ refer to all the
$n(k^*+m)$ hierarchy nodes simultaneously taken, but unlike the former matrix, the latter should not suffer for possible singularity problems.

The $(n \times n)$ matrices $\Wvet^{[k]}_l$, $k \in {\cal K}$, $l=1,\ldots,M_k$, forming the blocks on the diagonal of matrix (\ref{Wvetcs}), can be estimated both in full and shrunk version using the in-sample residuals  $\widehat{\Evet}^{[k]}_l$:
\begin{equation}
\label{Wklsam}
\widehat{\Wvet}^{[k]}_{l,\text{sam}} = \displaystyle\frac{1}{N} \widehat{\Evet}^{[k]}_l(\mbox{$\widehat{\Evet}^{[k]}_l$})' , \quad k \in {\cal K}, \quad l=1,\ldots,M_k ,
\end{equation}
$$
\widehat{\Wvet}^{[k]}_{l,\text{shr}} = \hat{\lambda}_{k,l}\left(\Ivet_{n} \odot \widehat{\Wvet}^{[k]}_{l,\text{sam}}\right) +
(1 - \hat{\lambda}_{k,l})\widehat{\Wvet}^{[k]}_{l,\text{sam}}
 , \quad k \in {\cal K}, \quad l=1,\ldots,M_k .
$$
Similarly, full and shrunk estimates of matrices $\Wvet^{[k]}$, $k \in {\cal K}$, forming the blocks on the diagonal of matrix (\ref{Wvet}), may be computed as:
\begin{equation}
\label{Wksam}
\widehat{\Wvet}^{[k]}_{\text{sam}} = \displaystyle\frac{1}{NM_k} \widehat{\Evet}^{[k]}(\mbox{$\widehat{\Evet}^{[k]}$})' , \quad k \in {\cal K} ,
\end{equation}
\begin{equation}
\label{Wkshr}
\widehat{\Wvet}^{[k]}_{\text{shr}} = \hat{\lambda}_{k}\left(\Ivet_{n} \odot \widehat{\Wvet}^{[k]}_{\text{sam}}\right) +
(1 - \hat{\lambda}_{k})\widehat{\Wvet}^{[k]}_{\text{sam}} , \quad k \in {\cal K} .
\end{equation}
While expression (\ref{Wklsam}) always requires $N > n$ in order to have good properties, formula (\ref{Wksam}) makes it clear that - except $\widehat{\Wvet}^{[m]}_{\text{sam}}$, which is calculated with the same $N$ residuals for each series as $\widehat{\Wvet}^{[k]}_{1,\text{sam}}$ - the estimates are based on more data, and the necessary condition to have a p.d. matrix is $NM_k > n$. However, in order to have all the $p$ matrices $\widehat{\Wvet}^{[k]}_{\text{sam}}$ well defined, the more restrictive condition $N > n$ should be met.
Matrices (\ref{Wklsam}) - (\ref{Wkshr}) can be used to approximate $\Wvet$ 
as follows:
$$
\begin{array}{rcl}
\widehat{\Wvet}^{BD}_{\text{hsam}} & = &\begin{bmatrix}
\widehat{\Wvet}^{[m]}_{1,\text{sam}} & \Zerovet          & \cdots & \Zerovet & \cdots & \Zerovet & \cdots & \Zerovet\\
\Zerovet    & \widehat{\Wvet}_{1,\text{sam}}^{[k_{p-1}]} & \cdots & \Zerovet & \cdots & \Zerovet & \cdots & \Zerovet\\
\vdots   & \vdots               & \ddots & \vdots   & \ddots & \vdots   & \ddots & \vdots\\
\Zerovet & \Zerovet & \cdots & \widehat{\Wvet}_{\frac{m}{k_{p-1}},\text{sam}}^{[k_{p-1}]} & \cdots & \Zerovet & \cdots & \Zerovet\\
\vdots   & \vdots & \ddots & \vdots   & \ddots & \vdots & \ddots & \vdots\\
\Zerovet & \cdots & \cdots & \Zerovet & \cdots & \widehat{\Wvet}_{1,\text{sam}}^{[1]} & \cdots & \Zerovet \\
\vdots   & \vdots & \ddots & \vdots   & \ddots & \vdots      & \ddots & \vdots \\
\Zerovet & \cdots & \cdots & \Zerovet & \cdots & \Zerovet    & \cdots & \widehat{\Wvet}_{m,\text{sam}}^{[1]}
\end{bmatrix}
\end{array} .
$$
$$
\begin{array}{rcl}
\widehat{\Wvet}^{BD}_{\text{hshr}} & = &\begin{bmatrix}
\widehat{\Wvet}^{[m]}_{1,\text{shr}} & \Zerovet          & \cdots & \Zerovet & \cdots & \Zerovet & \cdots & \Zerovet\\
\Zerovet    & \widehat{\Wvet}_{1,\text{shr}}^{[k_{p-1}]} & \cdots & \Zerovet & \cdots & \Zerovet & \cdots & \Zerovet\\
\vdots   & \vdots               & \ddots & \vdots   & \ddots & \vdots   & \ddots & \vdots\\
\Zerovet & \Zerovet & \cdots & \widehat{\Wvet}_{\frac{m}{k_{p-1}},\text{shr}}^{[k_{p-1}]} & \cdots & \Zerovet & \cdots & \Zerovet\\
\vdots   & \vdots & \ddots & \vdots   & \ddots & \vdots & \ddots & \vdots\\
\Zerovet & \cdots & \cdots & \Zerovet & \cdots & \widehat{\Wvet}_{1,\text{shr}}^{[1]} & \cdots & \Zerovet \\
\vdots   & \vdots & \ddots & \vdots   & \ddots & \vdots      & \ddots & \vdots \\
\Zerovet & \cdots & \cdots & \Zerovet & \cdots & \Zerovet    & \cdots & \widehat{\Wvet}_{m,\text{shr}}^{[1]}
\end{bmatrix}
\end{array} .
$$

$$
\reallywidehat{\Wvet}^{BD}_{\text{sam}} = \begin{bmatrix}
\reallywidehat{\Wvet}^{[m]}_{\text{sam}} & \Zerovet & \ldots & \Zerovet\\
\Zerovet & \Ivet_{\frac{m}{k_{p-1}}} \otimes \reallywidehat{\Wvet}^{[k_{p-1}]}_{\text{sam}} & \ldots & \Zerovet\\
\vdots & \vdots & \ddots & \vdots \\
\Zerovet & \Zerovet & \ldots & \Ivet_{m} \otimes \reallywidehat{\Wvet}^{[1]}_{\text{sam}}	
\end{bmatrix}
$$

$$
\reallywidehat{\Wvet}^{BD}_{\text{shr}} = \begin{bmatrix}
\reallywidehat{\Wvet}^{[m]}_{\text{shr}} & \Zerovet & \ldots & \Zerovet\\
\Zerovet & \Ivet_{\frac{m}{k_{p-1}}} \otimes \reallywidehat{\Wvet}^{[k_{p-1}]}_{\text{shr}} & \ldots & \Zerovet\\
\vdots & \vdots & \ddots & \vdots \\
\Zerovet & \Zerovet & \ldots & \Ivet_{m} \otimes \reallywidehat{\Wvet}^{[1]}_{\text{shr}}	
\end{bmatrix}
$$

Most of the alternative choices for $\Wvet$ (or $\Omegavet$) shown so far are simple extensions to the cross-temporal framework of the approximations for $\Wvet$ (or $\Omegavet$) considered either in cross-sectional or in temporal forecast reconciliation. For the time being, we are considering the following approximations (`oct' stands for `optimal cross-temporal'):
\begin{itemize}
\item identity (oct-ols): $\Wvet = \Omegavet = \Ivet_{n(k^*+m)}$
\item structural (oct-struc):
$\Wvet = \widehat{\Wvet}_{\text{struc}} = \Pvet'\left[\text{diag}\left(\Qvet\check{\Svet}{\bf 1}_{n_bm}\right)\Pvet\right] =
\text{diag}\left(\Pvet'\Qvet\check{\Svet}{\bf 1}_{n_bm}\right)$
(see section \ref{subsec:buct}, and appendix A.3.2)
\item hierarchy variance scaling (oct-wlsh): $\Wvet = \widehat{\Wvet}_{\text{wlsh}}$
\item series variance scaling (oct-wlsv): $\Wvet = \reallywidehat{\Wvet}_{\text{wlsv}} = \Pvet'\reallywidehat{\Omegavet}_{\text{wlsv}}\Pvet$, where $\reallywidehat{\Omegavet}_{\text{wlsv}}$ is a straightforward extension of $\reallywidehat{\Omegavet}_{\text{t-wlsv}}$ (see section \ref{subsec:temprec})
\item block-diagonal shrunk cross-covariance scaling (oct-bdshr): $\Wvet = \reallywidehat{\Wvet}^{BD}_{\text{shr}}$
\item block-diagonal cross-covariance scaling (oct-bdsam): $\Wvet = \reallywidehat{\Wvet}^{BD}_{\text{sam}}$
\item auto-covariance scaling (acov): $\Wvet = \reallywidehat{\Wvet}_{\text{acov}} = \Pvet'\reallywidehat{\Omegavet}_{\text{acov}}\Pvet$, where
$\reallywidehat{\Omegavet}_{\text{acov}}$ is a straightforward extension of $\reallywidehat{\Omegavet}_{\text{t-acov}}$ (see section \ref{subsec:temprec})
\item MinT-shr (oct-shr):  $\Wvet = \widehat{\Wvet}_{\text{shr}}$
\item MinT-sam (oct-sam):  $\Wvet = \widehat{\Wvet}_{\text{sam}}$
\end{itemize}

\section{An heuristic cross-temporal reconciliation procedure}
\label{KAheu}
Kourentzes and Athanasopoulos (2019), henceforth KA, have proposed a cross-temporal reconciliation procedure that can be viewed as an ensemble forecasting procedure which exploits the simple averaging of different forecasts.
The procedure consists in the following steps (it is assumed $h=1$):


\noindent{\bf Step 1}

\noindent For each individual variable, compute the temporally reconciled forecasts
and collect them in the $\left[n \times (k^* + m)\right]$ matrix $\widecheck{\Yvet}$:
$$
\widehat{\Yvet} \quad \rightarrow \quad \widecheck{\Yvet} .
$$
This result can be obtained by applying the point forecast reconciliation formula (\ref{testonest}) to each column of matrix $\widehat{\Yvet}'$, which can be written as:
$$
\widehat{\Yvet}' = \left[\begin{array}{cccccc}
\hat{\tvet}_{a_1} & \cdots & \hat{\tvet}_{a_{n_a}} &
\hat{\tvet}_{b_1} & \cdots & \hat{\tvet}_{b_{n_b}} \\[.25cm]
\hat{\avet}^{[1]}_1 & \cdots & \hat{\avet}^{[1]}_{n_a} &
\hat{\bvet}^{[1]}_1 & \cdots & \hat{\bvet}^{[1]}_{n_b}
\end{array}
\right].
$$

\noindent The $n_a$ vectors of temporally reconciled forecasts of the uts can be obtained as:
$$
\left[\begin{array}{c}
\check{\tvet}_{a_j} \\[0.25cm] \check{\avet}^{[1]}_{j}
\end{array}
\right] = \Mvet_{a_j}
\left[\begin{array}{c}
\hat{\tvet}_{a_j} \\[0.25cm] \hat{\avet}^{[1]}_{j}
\end{array}
\right], \quad
\Mvet_{a_j} = \Ivet_{k^*+m} - \Omegavet_{a_j}\Zvet_1\left(\Zvet_1'\Omegavet_{a_j}\Zvet_1\right)^{-1}\Zvet_1',
\quad j=1,\ldots,n_a.
$$
Likeways, the $n_b$ vectors of temporally reconciled forecasts of the bts are given by:
$$
\left[\begin{array}{c}
\check{\tvet}_{b_i} \\[0.25cm] \check{\bvet}^{[1]}_{i}
\end{array}
\right] = \Mvet_{b_i}
\left[\begin{array}{c}
\hat{\tvet}_{b_i} \\[0.25cm] \hat{\bvet}^{[1]}_{i}
\end{array}
\right], \quad
\Mvet_{b_i} = \Ivet_{k^*+m} - \Omegavet_{b_i}\Zvet_1\left(\Zvet_1'\Omegavet_{b_i}\Zvet_1\right)^{-1}\Zvet_1',
\quad i=1,\ldots,n_b,
$$
where the $n_a + n_b$ matrices $\Mvet_{a_j}$ and $\Mvet_{b_i}$ have dimension
$\left[(k^*+m) \times (k^*+m)\right]$, and $\Omegavet_{a_j}$, $j=1,\ldots, n_a$, and $\Omegavet_{b_i}$, $i=1,\ldots,n_b$, are known p.d. $\left[(k^*+m) \times (k^*+m) \right]$ matrices.

The mapping performing the transformation of the base forecasts into the temporally reconciled ones can be expressed in compact form as:
$$
\text{vec}\left(\widecheck{\Yvet}'\right) =
\left[\begin{array}{cccccc}
\Mvet_{a_1} & \cdots & \Zerovet & \Zerovet & \cdots & \Zerovet \\
\vdots    & \ddots & \vdots   & \vdots   & \ddots & \vdots \\
\Zerovet  & \cdots & \Mvet_{a_{n_a}} & \Zerovet & \cdots & \Zerovet \\
\Zerovet  & \cdots & \Zerovet & \Mvet_{b_{1}} & \cdots & \Zerovet \\
\vdots    & \ddots & \vdots   & \vdots   & \ddots & \vdots \\
\Zerovet  & \cdots & \Zerovet & \Zerovet & \cdots & \Mvet_{b_{n_b}}
\end{array}
\right] \text{vec}\left(\widehat{\Yvet}'\right),
$$
and then matrix $\widecheck{\Yvet}'$ can be re-stated as:
$$
\widecheck{\Yvet}' = \left[\begin{array}{cccccc}
\check{\tvet}_{a_1} & \cdots & \check{\tvet}_{a_{n_a}} &
\check{\tvet}_{b_1} & \cdots & \check{\tvet}_{b_{n_b}} \\[.25cm]
\check{\avet}^{[1]}_1 & \cdots & \check{\avet}^{[1]}_{n_a} &
\check{\bvet}^{[1]}_1 & \cdots & \check{\bvet}^{[1]}_{n_b}
\end{array}
\right] =
\begin{bmatrix}
(\widecheck{\Avet}^{[m]})' & (\widecheck{\Bvet}^{[m]})' \\
\vdots & \vdots \\
(\widecheck{\Avet}^{[k_2]})' & (\widecheck{\Bvet}^{[k_2]})' \\
(\widecheck{\Avet}^{[1]})' & (\widecheck{\Bvet}^{[1]})'
\end{bmatrix} .
$$
These reconciled forecasts are in line with the temporal aggregation constraints, i.e.
$\Zvet_1' \widecheck{\Yvet}' = \Zerovet_{(k^* \times n)}$, but in general they are not in line
with the cross-sectional (contemporaneous) constraints, that is:
$\Uvet' \widecheck{\Yvet}  \ne \Zerovet_{\left[n_a \times (k^* +m)\right]}$.

\vspace{.5cm}

\noindent{\bf Step 2}

\noindent Transform $\widecheck{\Yvet}$ by computing time-by-time cross-sectional reconciled forecasts for all the temporal aggregation levels, and collect them in the $\left[n \times (k^* + m)\right]$ matrix $\widebreve{\Yvet}$:
$$
\widecheck{\Yvet} \quad \rightarrow \quad \widebreve{\Yvet} .
$$
Matrix $\widecheck{\Yvet}$ can be written as
$$
\widecheck{\Yvet} = \begin{bmatrix}
\widecheck{\Yvet}^{[m]} \; \widecheck{\Yvet}^{[k_{p-1}]} \ldots
\widecheck{\Yvet}^{[k_2]} \; \widecheck{\Yvet}^{[1]}
\end{bmatrix},
$$
where $\widecheck{\Yvet}^{[k]}$, $k \in {\cal K}$, has dimension $\left(n \times M_k\right)$.
Thus, the cross-sectionally reconciled forecasts can be computed by transforming each
$\widecheck{\Yvet}^{[k]}$ as: 
$$
\widebreve{\Yvet}^{[k]} = \Mvet^{[k]} \widecheck{\Yvet}^{[k]}, \quad
k \in {\cal K},
$$
where $\Mvet^{[k]}$ denotes the $(n \times n)$ projection matrix used to reconcile forecasts of $k$-level temporally aggregated time series:
$$
\Mvet^{[k]} = \Ivet_{n} - \Wvet^{[k]}\Uvet\left(\Uvet'\Wvet^{[k]}\Uvet\right)^{-1}\Uvet', \quad
k \in {\cal K},
$$
and $\Wvet^{[k]}$ is a $(n \times n)$ known p.d. matrix. Since it is
$\Uvet'\Mvet^{[k]} = \Zerovet_{(n_a \times n)}$, $k \in {\cal K}$,
the reconciled forecasts are cross-sectionally coherent, i.e.
$\Uvet' \widebreve{\Yvet}  = \Zerovet_{\left[n_a \times (k^* +m)\right]}$, but not
temporally:
$\Zvet_1' \widebreve{\Yvet}' \ne \Zerovet_{(k^* \times n)}$.

\clearpage

\noindent{\bf Step 3}

\noindent Transform again the step 1 forecasts $\widecheck{\Yvet}$, by computing time-by-time cross-sectional reconciled forecasts for all the temporal aggregation levels using the $(n \times n)$ matrix $\overline{\Mvet}$, given by the average of the matrices 
$\Mvet^{[k]}$
obtained at step 2:
$$
\widecheck{\Yvet} \quad \rightarrow \quad \widetilde{\Yvet}^{\small KA} .
$$
Matrix $\overline{\Mvet}$ can be expressed as:
$$
\overline{\Mvet}=\displaystyle\frac{1}{p}\sum_{k \in {\cal K}} \Mvet^{[k]},
$$
and the final cross-temporal reconciled forecasts are given by:
\begin{equation}
\label{KArec}
\widetilde{\Yvet}^{\small KA} =  \overline{\Mvet} \widecheck{\Yvet}.
\end{equation}
Since
	$\Uvet'\overline{\Mvet} = \displaystyle\frac{1}{p}\displaystyle\sum_{k \in {\cal K}} \Uvet'\Mvet^{[k]} =\Zerovet_{\left(n_a \times n\right)}$,
	and
	$\Zvet_1'\widecheck{\Yvet}' = \Zerovet_{(k^* \times n)}$,
the reconciled forecasts (\ref{KArec}) fulfill both cross-sectional and temporal aggregation constraints:
$$
\Uvet' \widetilde{\Yvet}^{\small KA}  = \Uvet'\overline{\Mvet}\widebreve{\Yvet} = \Zerovet_{\left[n_a \times (k^* +m)\right]},
$$
$$
\Zvet_1'\left(\widetilde{\Yvet}^{\small KA}\right)' = \Zvet_1'\widebreve{\Yvet}'\overline{\Mvet}'=
\Zerovet_{(k^* \times n)}.
$$

\subsection{Some remarks}
To perform step 1, 
KA consider two alternatives as for the
$\left[(k^*+m) \times (k^*+m)\right]$ matrices $\Omegavet_{a_j}$ and $\Omegavet_{b_i}$ needed for computing the transformation matrices $\Mvet_{a_j}$ and $\Mvet_{b_i}$, respectively.
The former is t-struc, while
the latter is t-wlsv 
(see section \ref{subsec:temprec}).
As for step 2, KA use either cs-wls or cs-shr (see section \ref{subsec:csmat}).

\vspace{.25cm}
\noindent\emph{Remark 1}

\noindent These two steps 
can be seen as the successive applications of two distinct multivariate reconciliation procedures, each characterized by different covariance matrix and constraints.
For, in the first step it is solved a quadratic linear problem, where only temporal aggregation constraints are considered:

\[
\check{\bf y} = \argmin_{{\yvet}} \left({\bf y} - \hat{\bf y} \right)' \Omegavet^{-1} \left({\bf y} - \hat{\bf y} \right), \quad
\text{s.t. } \left(\Ivet_n \otimes \Zvet_1'\right)\yvet = \Zerovet,
\]
where $\Omegavet$ is the block-diagonal matrix in (\ref{Omegavet}).
The solution is given by:

$$
\check{\yvet}= \hat{\yvet} -
\Omegavet\left(\Ivet_n \otimes \Zvet_1\right)
\left[\left(\Ivet_n \otimes \Zvet_1'\right)\Omegavet\left(\Ivet_n \otimes \Zvet_1\right)\right]^{-1}
\left(\Ivet_n \otimes \Zvet_1'\right)\hat{\yvet}=
\Mvet\hat{\yvet},
$$
where
$$
\Mvet = \Ivet -
\Omegavet\left(\Ivet_n \otimes \Zvet_1\right)
\left[\left(\Ivet_n \otimes \Zvet_1'\right)\Omegavet\left(\Ivet_n \otimes \Zvet_1\right)\right]^{-1}
\left(\Ivet_n \otimes \Zvet_1'\right)
$$
is the
$\left[n(k^*+m) \times n(k^*+m)\right]$
projection matrix
$$
\Mvet = \begin{bmatrix}
\Mvet_1 & \Zerovet & \cdots & \Zerovet \\
\Zerovet & \Mvet_2 & \cdots & \Zerovet \\
\vdots & \vdots & \ddots & \vdots\\
\Zerovet & \Zerovet & \cdots & \Mvet_n
\end{bmatrix},
$$
with
$$
\Mvet_i = \Ivet_{k^*+m} - \Omegavet_{ii}\Zvet_1\left(\Zvet_1'\Omegavet_{ii}\Zvet_1\right)^{-1}\Zvet_1',
\quad i=1, \ldots, n .
$$

The second step consists in another quadratic minimization problem, where only cross-sectional (contemporaneous) constraints are considered:
\[
\widebreve{{\bf \cal{Y}}} = \argmin_{{\bf \cal{Y}}} \left({\bf \cal{Y}} - \widecheck{\bf \cal{Y}} \right)' \Wvet^{-1} \left({\bf \cal{Y}} - \widecheck{\bf \cal{Y}} \right), \quad
\text{s.t. } \left(\Ivet_{k^*+m} \otimes \Uvet' \right){\bf \cal{Y}} = \Zerovet ,
\]
where $\Wvet$ is the block-diagonal matrix in (\ref{Wvet}), and whose solution is given by:
\begin{equation}
\label{stonecs}
\widebreve{{\bf \cal{Y}}} = 
\widecheck{{\bf \cal{Y}}} -
\Wvet \left(\Ivet_{k^*+m} \otimes \Uvet \right)
\left[ \left(\Ivet_{k^*+m} \otimes \Uvet'\right) \Wvet
\left(\Ivet_{k^*+m} \otimes \Uvet \right)\right]^{-1}
\left(\Ivet_{k^*+m} \otimes \Uvet' \right) \widecheck{{\bf \cal{Y}}} =
{\bf \cal{M}}\widecheck{{\bf \cal{Y}}}
\end{equation}
where
$$
{\bf \cal{M}} = \Ivet -
\Wvet\left(\Ivet_{k^*+m} \otimes \Uvet \right)
\left[ \left(\Ivet_{k^*+m} \otimes \Uvet' \right) \Wvet
\left(\Ivet_{k^*+m} \otimes \Uvet \right)\right]^{-1}
\left(\Ivet_{k^*+m} \otimes \Uvet' \right)
$$
is the $\left[n(k^*+m) \times n(k^*+m)\right]$ projection matrix
$$
\begin{array}{rcl}
{\bf \cal{M}} & = &\begin{bmatrix}
\Mvet^{[m]} & \Zerovet          & \cdots & \Zerovet & \cdots & \Zerovet & \cdots & \Zerovet\\
\Zerovet    & \Mvet^{[k_{p-1}]} & \cdots & \Zerovet & \cdots & \Zerovet & \cdots & \Zerovet\\
\vdots   & \vdots               & \ddots & \vdots   & \ddots & \vdots   & \ddots & \vdots\\
\Zerovet & \Zerovet & \cdots & \Mvet^{[k_{p-1}]} & \cdots & \Zerovet & \cdots & \Zerovet\\
\vdots   & \vdots & \ddots & \vdots   & \ddots & \vdots & \ddots & \vdots\\
\Zerovet & \cdots & \cdots & \Zerovet & \cdots & \Mvet^{[1]} & \cdots & \Zerovet \\
\vdots   & \vdots & \ddots & \vdots   & \ddots & \vdots      & \ddots & \vdots \\
\Zerovet & \cdots & \cdots & \Zerovet & \cdots & \Zerovet    & \cdots & \Mvet^{[1]}
\end{bmatrix} \\
\\
& = &
\begin{bmatrix}
\Mvet^{[m]} & \Zerovet          & \cdots &\Zerovet \\
\Zerovet & \left(\Ivet_{\frac{m}{k_{p-1}}} \otimes \Mvet^{[k_{p-1}]}\right) & \cdots & \Zerovet\\
\vdots & \vdots & \ddots & \vdots \\
\Zerovet & \Zerovet & \cdots & \left(\Ivet_{m} \otimes \Mvet^{[1]}\right)
\end{bmatrix}
\end{array}
$$
with
$$
\Mvet^{[k]} = \Ivet_n - \Wvet^{[k]}\Uvet\left(\Uvet'\Wvet^{[k]}\Uvet\right)^{-1}\Uvet',
\quad k \in  \cal{K} .
$$
It is worth noting that the reconciled forecasts (\ref{stonecs}) can be expressed according to the alternative vectorization, as:
$$
\breve{\yvet} = \Pvet'\breve{{\bf \cal{Y}}} .
$$

\vspace{0.25cm}
\noindent\emph{Remark 2}

\noindent The cross-sectional reconciliation performed at step 2 of the KA procedure involves the transformations of $k^*+m$
vector of forecasts. More precisely, each transformation matrix $\Mvet^{[k]}$, $k \in {\cal K}$,
is applied to $M_k$ different $(n \times 1)$ vectors. Thus, a sensible alternative to the KA proposal might be considering the weighted average of the transformation matrices:
$$
\overline{\Mvet}^*=\displaystyle\frac{1}{k^*+m}\sum_{k \in {\cal K}} M_k\Mvet^{[k]}.
$$

\vspace{0.25cm}
\noindent\emph{Remark 3}

\noindent In general the final result of the reconciliation procedure would change if the user invert the order of application of the two reconciliation steps. In Appendix A.6 the  `cross-sectional-first-then-temporal' reconciliation procedure is shown, along with the relevant $\overline{\Mvet}$ matrix, which in this case is obtained through an average of the projection matrices used for the reconciliation of the $n$ series according to temporal hierarchies. Since the differences between the reconciled point forecasts according to these two approaches could be not negligible (see section \ref{subsec:itekah}), in our view this is a weakness of the procedure, and calls for a decision rule about the final reconciled forecasts to retain. A practical way of doing could be choosing the reconciled forecasts which are the `closest' (according to a given metric) to the base forecasts between the two alternatives.

\vspace{0.25cm}
\noindent\emph{Remark 4}

\noindent The calculation of the average matrix $\overline{\Mvet}$ in the final step of the procedure, needed to recover the cross-temporal coherency across the point forecasts, requires the availability of the projection matrices used in the second step. This poses no problem when closed form reconciliation formulae can be used. Unfortunately, this is not the case when the user is interested in considering more general linear constraints (e.g., non-negativity of the final reconciled estimates), that should be treated with appropriate numerical procedures (Kourentzes and Athanasopoulos, 2020a, Wickramasuriya et al., 2020)\footnote{This issue is currently under study, in order to develop a procedure which, by exploiting some distinctive features of a hierarchical/grouped time series, be able to produce non-negative reconciled forecasts with a reduced computational effort.}.

\vspace{0.2cm}

In the next subsection we extend the heuristic KA procedure in such a way that these issues can be overcome in a simple and effective manner.

\subsection{An iterative heuristic cross-temporal reconciliation procedure}
\label{subsec:itekah}
Taking inspiration from the heuristic KA reconciliation procedure, we consider an iterative procedure which produces cross-temporally reconciled forecasts by alternating forecast reconciliation along one single dimension (either cross-sectional or temporal) at each iteration step.

Each iteration consists in the first two steps of the heuristic KA procedure, so the forecasts are reconciled by alternating cross-sectional (contemporaneous) reconciliation, and reconciliation through temporal hierarchies in a cyclic fashion.

Starting from the base forecasts $\widehat{\Yvet}$, denote with $d_{cs}$ and $d_{te}$, respectively, the cross-sectional and temporal gross discrepancies, given by:
$$
d_{cs} = \norm{\Uvet'\widehat{\Yvet}}_1 \qquad d_{te} = \norm{\Zvet_1'\widehat{\Yvet}'}_1
$$
where $\norm{\Xvet}_1 = \sum_{i,j}|x_{i,j}|$. Since the base forecasts are not in line with either type of constraints, in general both $d_{cs}$ and $d_{te}$ are greater than zero.

The iterative procedure can be described as follows:
\begin{enumerate}
	\item Start the iterations by calculating the temporally reconciled forecasts
$\widetilde{\Yvet}^{(1)}$, such that $\Zvet_1'\left(\widetilde{\Yvet}^{(1)}\right)'=\Zerovet$, and
$d_{cs}^{(1)} = \norm{\Uvet'\widetilde{\Yvet}^{(1)}}_1 \ge 0$.
\item The point forecasts in matrix $\widetilde{\Yvet}^{(1)}$ are then cross-sectionally reconciled,  obtaining $\widetilde{\Yvet}^{(2)}$, which is such that $\Uvet'\widetilde{\Yvet}^{(2)} = \Zerovet$,
and $d_{te}^{(1)} = \norm{\Zvet_1'\left(\widetilde{\Yvet}^{(2)}\right)'}_1 \ge 0$.
\item The updates in steps 1. and 2. are performed at each iteration $j$, $j=1,2,\ldots$, until a convergence criterion is met, that is
$d_{te}^{(j)} < \delta$, where $\delta$ is a positive tolerance value (e.g., $\delta = 10^{-6})$,
and matrix $\widetilde{\Yvet}^{(2j)}$ contains the final cross-temporal reconciled forecasts.
\end{enumerate}
The above procedure can be seen as an extension of the well known iterative proportional fitting procedure (Deming and Stephan, 1940, Johnston and Pattie, 1993), also known as RAS method (Miller and Blair, 2009), to adjust the internal cell values of a two-dimensional matrix iteratively until they sum to some predetermined row and column totals. In that case the adjustment follows a proportional adjustment scheme, whereas in the cross-temporal reconciliation framework each adjustment step is made according to the penalty function associated to the single-dimension reconciliation procedure adopted.

\begin{figure}[htb]
	\begin{center}
		\includegraphics[scale=0.78]{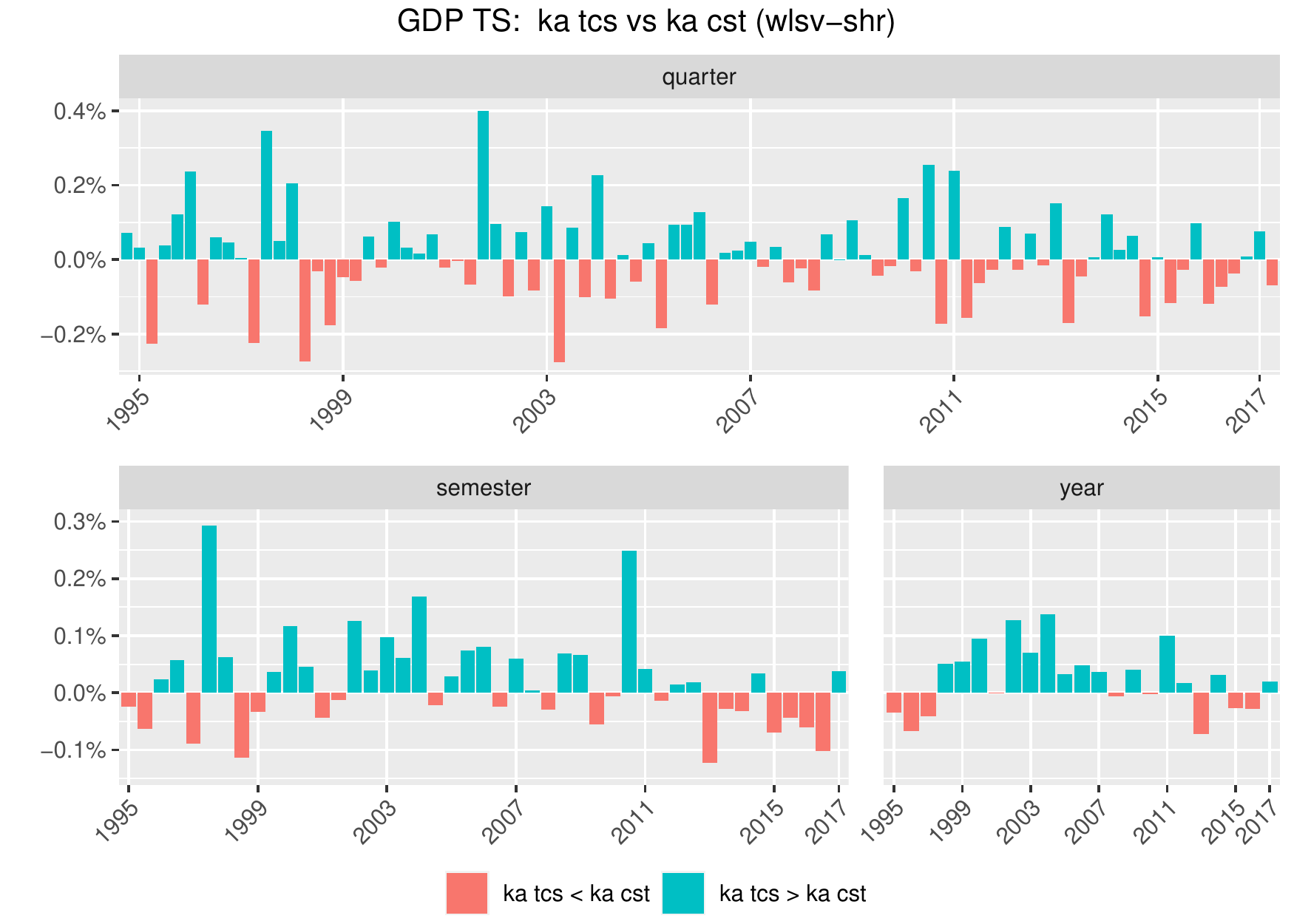}
		\vspace{-0.5cm}
		\caption{\scriptsize Quarterly, semi-annual and annual Australian GDP one-step-ahead reconciled forecasts according to the Kourentzes and Athanasopoulos (2019)
			cross-temporal reconciliation approach (t-wlsv for the temporal step, cs-shr for the cross-sectional step) by alternating the constraint dimensions to be fulfilled:
			percentage differences between the reconciled forecasts obtained through
			(i) temporal-then-cross-sectional reconciliation, and (ii) cross-sectional-then-temporal
			reconciliation. The differences between the two reconciled forecasts are divided by their arithmetic mean.}
		\label{fig:GDPdiff}
	\end{center}
\end{figure}

Indeed, the choice of the dimension along with the first reconciliation step in each iteration is performed is up to the user, and there is no particular reason why one should perform the temporal reconciliation first, and the cross-sectional reconciliation then.
Figure \ref{fig:GDPdiff} shows the percentage discrepancies in the Australian GDP at current prices one-step-ahead forecasts for any temporal aggregation level (quarterly, semi-annual, annual, see section \ref{sec:AUSgdp}), when the cross-temporal reconciliation is performed according to either the KA approach, or to the analogous procedure where the cross-sectional constraints are considered first, and then the temporal dimension is accounted for. Percentage differences in the reconciled forecasts for this single, very important variable, are visually evident, though bounded within (-0.3\% -- +0.4\%).

\begin{figure}[ht]
	\begin{center}
		\includegraphics[scale=0.85]{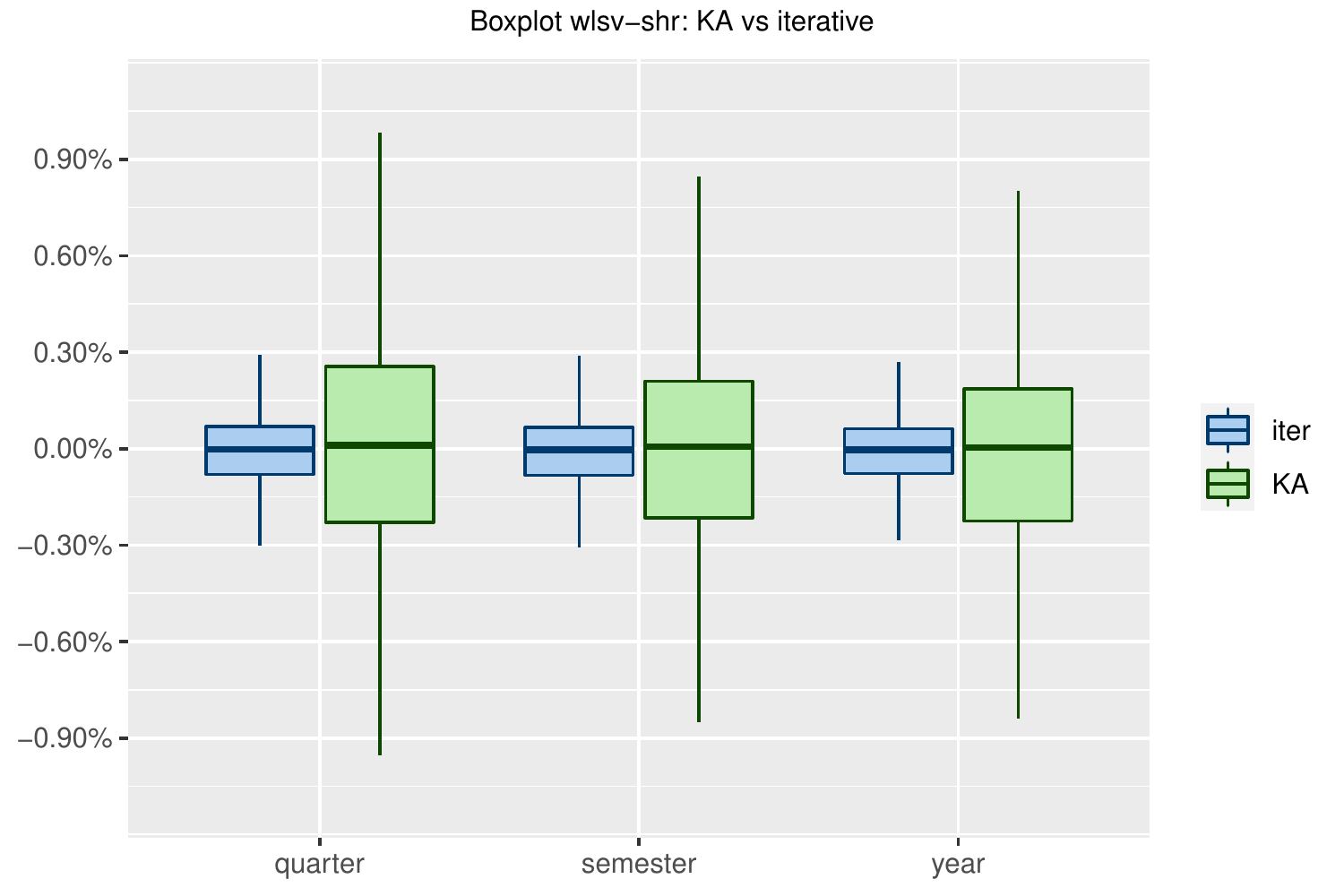}
		\caption{\scriptsize Quarterly, semi-annual and annual one-step-ahead reconciled forecasts of 79 out of 95 times series of the Australian GDP from Income and Expenditure sides using both the original KA
			cross-temporal reconciliation procedure (t-wlsv for the temporal step, and cs-shr for the cross-sectional one), and its iterative variant:
			boxplots of the percentage differences between the reconciled forecasts obtained through
			(i) temporal-then-cross-sectional reconciliation, and (ii) cross-sectional-then-temporal
			reconciliation. The differences between each pair of reconciled forecasts are divided by their arithmetic mean.}
		\label{fig:tcs-cst1}
	\end{center}
\end{figure}

\vspace{-0.1cm}
Figure \ref{fig:tcs-cst1} completes the results shown so far, by considering the forecasts of the strictly positive 79 (out of 95) variables from both Income and Expenditure sides, cross-temporally reconciled  according to the KA procedure and its iterative variant. The boxplots show the distributions of the percentage discrepancies between the reconciled forecasts obtained using temporal reconciliation first, and cross-sectional reconciliation then, \emph{vis-\`{a}-vis} the results obtained by inverting the order of application of the two reconciliation procedures. It clearly appears that the iterative variant of the original KA proposal
produces less pronounced discrepancies\footnote{Temporal reconciliation has been done using t-wlsv, while cross-sectional reconciliation was performed using cs-shr. However, this result does not seem to depend on the reconciliation procedures considered: t-struc+cs-wls, t-struc+cs-shr, and t-wlsv+cs-wls give very similar results, here not presented for space reasons, but available on request from the authors.}.

\begin{figure}[ht]
\begin{center}
\includegraphics[width=1\linewidth]{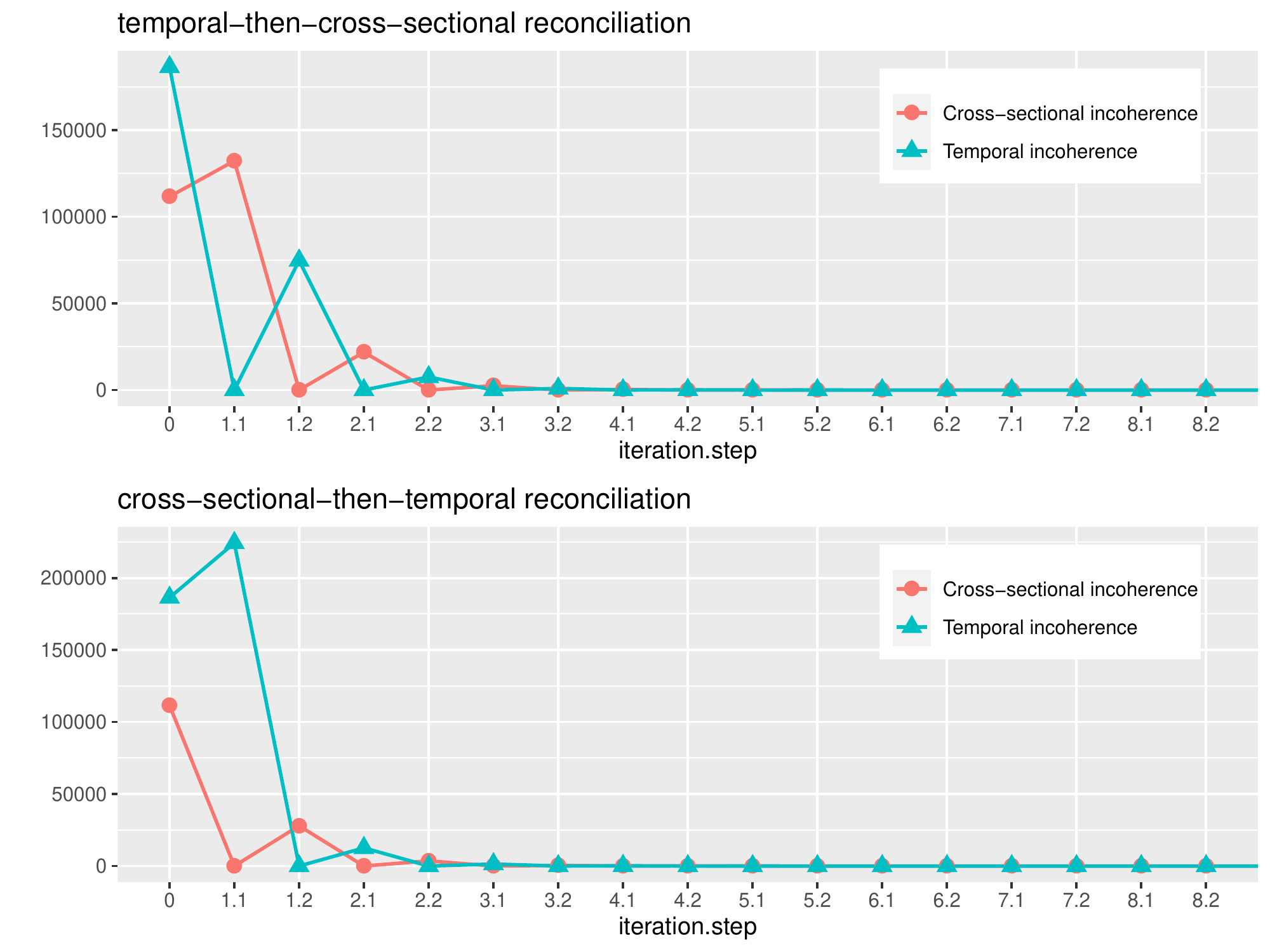}
\caption{\small Cross-sectional and temporal gross incoherence at each iteration step of the iterative cross-temporal forecast reconciliation procedure (t-wlsv + cs-shr) for the Australian GDP time series, at the first forecast origin 1994:Q3.}
\label{fig:KAiter}
\end{center}
\vspace{-.75cm}
\end{figure}

It must also be said that the convergence speed of the iterative procedure does not seem to be affected by the choice of the first dimension to be fulfilled when the iteration starts.
Figure \ref{fig:KAiter} shows an example of the convergence speed of the iterative procedure either starting with the cross-sectional 
(bottom panel) or temporal (top panel) reconciliation procedure for the Australian GDP forecasts. In both cases, the convergence is achieved very quickly: fixing  $\delta = 10^{-6}$, 15 (14) iterates are needed when starting from the temporal (cross-sectional) dimension. Furthermore, from the fourth iteration onwards the constraints are practically fulfilled in both cases.
Nevertheless, since the final reconciled values depend on this choice,
it would be useful having an ex-ante `choice rule' between the two alternatives. We are currently working on this issue, however in the rest of the paper, when considering heuristic cross-temporal forecast reconciliation procedures, for ease of presentation we maintain the original choice made by KA, performing temporal forecast reconciliation first, and  cross-sectional reconciliation then.

\section[Cross-temporal reconciliation of the Australian GDP forecasts]{Cross-temporal reconciliation of the Australian GDP forecasts from Income and Expenditure sides}
\label{sec:AUSgdp}
\begin{figure}[ht]
\begin{center}
\includegraphics[scale=0.325]{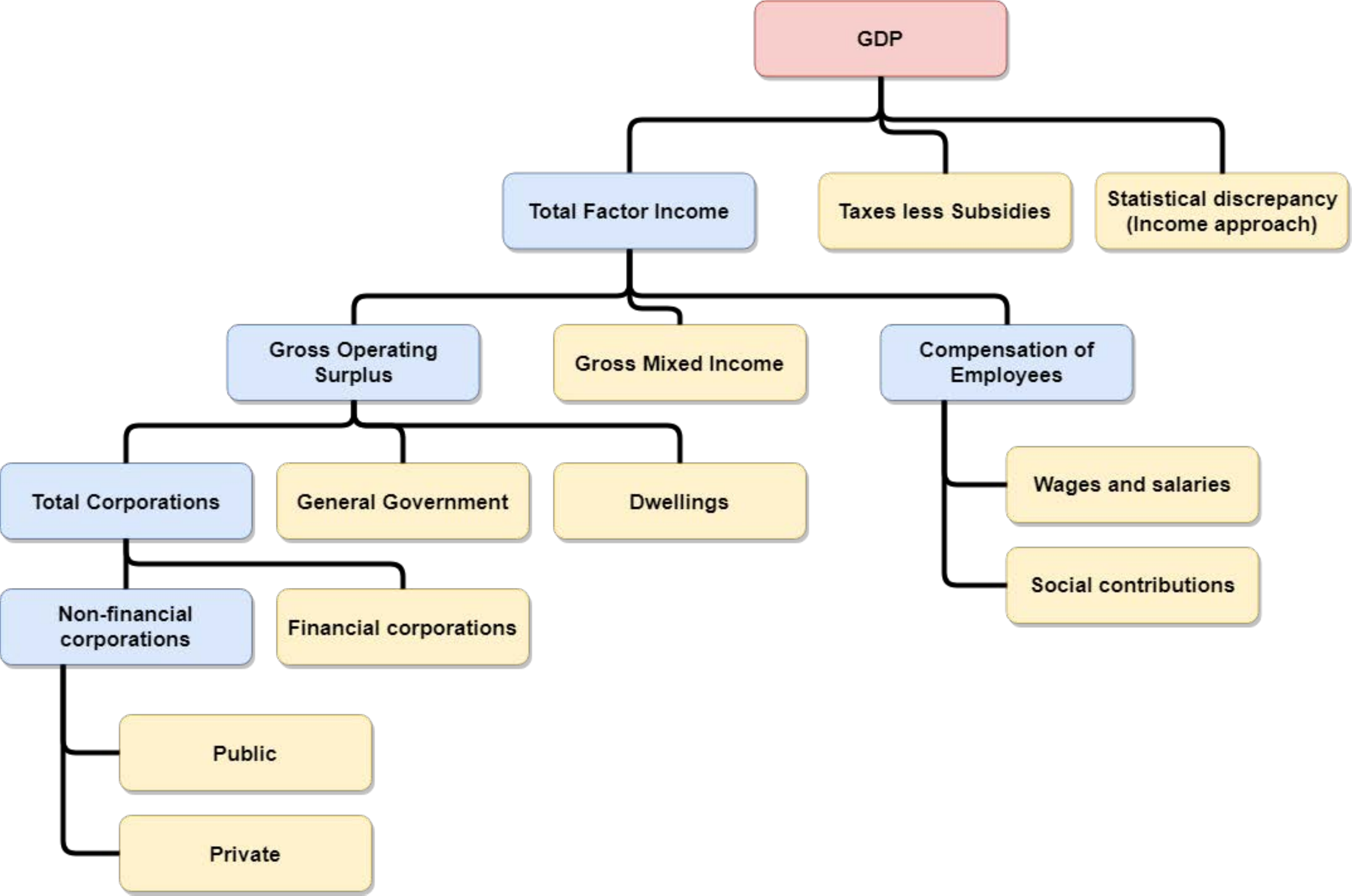}
\caption{\footnotesize Hierarchical structure of the income approach for Australian GDP.
The pink cell contains the most aggregate series. The blue cell contain intermediate-level series and
the yellow cells correspond to the most disaggregate bottom-level series.
Source: Athanasopoulos et al., 2019, p. 702.}
\label{fig:AUSINC}
\end{center}
\vspace{-.75cm}
\end{figure}

In a recent paper, Athanasopoulos et al. (2019, p. 690) propose ``the application of state-of-the-art forecast reconciliation methods
to macroeconomic forecasting'' in order to perform aligned decision making and to improve forecast accuracy.
In their empirical study they consider the cross-sectional forecast reconciliation for 95 Australian Quarterly National Accounts time series, describing the
Gross Domestic Product ($GDP$) at current prices from
Income and Expenditure sides, interpreted as two 
distinct
hierarchical structures.
In the former case (Income), $GDP$ is on the top of 15 lower level aggregates (figure \ref{fig:AUSINC}),
while in the latter (Expenditure), $GDP$ is the top level aggregate of a hierarchy of 79 time series (see figures 21.5-21.7 in
Athanasopoulos et al., 2019, pp. 703-705).

By managing the complete set of 95 time series following the approach described in section \ref{sec:hts}, Bisaglia et al. (2020) have extended the results of Athanasopoulos et al. (2019),
showing that fully reconciled forecasts of $GDP$, coherent with all the
reconciled forecasts from both Expenditure and Income sides, can be obtained through the projection approach described in section \ref{sec:genrec}.
According to the notation adopted so far, the  $(33 \times 95)$ kernel matrix accounting for the cross-sectional zero constraints is given by (Bisaglia et al., 2020):
$$
{\bf U}' = \left[\begin{array}{ccccc}
1 & {\bf 0}_{(5 \times 1)}'             & -{\bf 1}_{(10 \times 1)}'         & {\bf 0}_{(26 \times 1)}'         &  {\bf 0}_{(53 \times 1)}'         \\
1 & {\bf 0}_{(5 \times 1)}'             &  {\bf 0}_{(10 \times 1)}'         & {\bf 0}_{(26 \times 1)}'         & -{\bf 1}_{(53 \times 1)}'         \\
{\bf 0}_{(5 \times 1)} & {\bf I}_5              & -{\bf C}^I             & {\bf 0}_{(5 \times 26)} &  {\bf 0}_{(5 \times 53)} \\
{\bf 0}_{(26 \times 1)} & {\bf 0}_{(26 \times 5)}  & {\bf 0}_{(26 \times 10)} & {\bf I}_{26}          &  -{\bf C}^E
\end{array}\right] .
$$
In what follows, cross-temporal forecast reconciliation is applied within the same forecasting experiment designed by Athanasopoulos et al. (2019), extended in order to consider semi-annual and annual forecasts as well: for the available time series span (1984:Q4 - 2018:Q1),
quarterly base forecasts from 1 up to 4 quarters ahead have been obtained for the $n=95$ separate time series through simple univariate
ARIMA models
selected using the {\tt auto.arima} function of the \texttt{R}-package {\tt forecast} (Hyndman et al., 2020). The forecasting experiment uses a recursive training sample with expanding window length, where the first 
training sample is set from 1984:Q4 to 1994:Q3  and the last ends on 2017:Q1, for a total of 91 forecast origins\footnote{The \texttt{R} scripts, the data and the results of the paper by Athanasopoulos et al. (2019)
	are available in the github repository located at \url{https://github.com/PuwasalaG/Hierarchical-Book-Chapter}. We did not change this first, crucial step in the forecast reconciliation workflow, since 
	the focus is on the potential of cross-temporal forecast reconciliation. However, Athanasopoulos et al. (2019) point out that this fast and flexible approach performs
well in forecasting Australian GDP aggregates, even compared to other more complex methods.}.
Likeways, in the same automatic fashion we have computed (i) one and two-step ahead forecasts for the time series obtained by temporal aggregation of two successive quarters, and (ii) one-step-ahead forecasts for the time series obtained by temporal aggregation of four successive quarters.

\subsection{Performance measures for multiple comparisons}

We evaluate the performance of multiple (say, $J>1$) forecast reconciliation procedures through forecast accuracy indices
calculated on the forecast error\footnote{Sagaert et al. (2019) warn practitioners that this could be `a myopic choice as (the accuracy metrics) consider solely the first moment of the error distribution and ignore higher moments, which can have significant implications for decision making'. This important issue will be dealt with in the near future.}
$$
\hat{e}_{i,j,t}^{[k],h} = y_{i,t+h}^{[k]} - \hat{y}_{i,j,t}^{[k],h}, \quad
\begin{array}{l}
i=1,\ldots,95, \\
j=0,\ldots,J,
\end{array} \;  t=1,\ldots, 91 , \; 
\begin{array}{l}
k \in {\cal K}, \\
h=1,\ldots,h_k ,
\end{array}
$$
where $y$ and $\hat{y}$ are the observed and forecasted values, respectively, $i$ denotes the series ($i=1,\ldots,32$, for the uts, $i=33,\ldots,95$, for the bts), $j=0$ denotes the base forecasts, $t$ is the forecast origin ($t=1$ corresponds to 1994:Q3), ${\cal K} = \{4,2,1\}$, and $h_4=1$, $h_2=2$, $h_1=4$, are the forecast horizons for annual, semi-annual, and quarterly time series, respectively.

The accuracy is evaluated using 
the Average Relative Mean Square Error (AvgRelMSE, Davydenko and Fildes, 2013; Kourentzes and Athanasopoulos, 2019, 2020b), obtained by transforming the MSE index, given by the average across all 91 forecasts origins of the squared forecast errors:
\begin{equation}
\label{MSEijkh}
\text{MSE}_{i,j}^{[k],h} = \frac{1}{91} \displaystyle\sum_{t=1}^{91}\left(\hat{e}_{i,j,t}^{[k],h}\right)^2 , \quad
\begin{array}{l}
i=1,\ldots,95, \\
j=0,\ldots,J,
\end{array} \; 
\begin{array}{l}
k \in {\cal K}, \\
h=1,\ldots,h_k .
\end{array}
\end{equation}


\noindent The AvgRelMSE is the geometric mean across all 95 series of the MSE ratio\footnote{Davydenko and Fildes (2013) develop the Average Relative MAE (AvgRelMAE), based on the Mean Absolute Error of the forecasts, but suggest that this formulation `If required (...) can also be extended to other measures of dispersion or loss functions', as the AvgRelMSE in (\ref{AvgRelMSEjkh}) and
	the AvgRelRMSE, based on the Root Mean Square Error (Sagaert et al., 2019, Kourentzes and Athanasopoulos, 2020a).}  
of a forecast over a benchmark given by the base, incoherent ARIMA forecasts, across all evaluation samples, for a given horizon $h$:
\begin{equation}
\label{AvgRelMSEjkh}
\text{AvgRelMSE}_{j}^{[k],h} =
\left(\displaystyle\prod_{i=1}^{95} \text{rMSE}_{i,j}^{[k],h}\right)^{\frac{1}{95}} , \; j=0,\ldots,J, \;
\begin{array}{l}
k \in {\cal K}, \\
h=1,\ldots,h_k ,
\end{array}
\end{equation}
where $\text{rMSE}_{i,j}^{[k],h}$ is the relative MSE:
$$
\text{rMSE}_{i,j}^{[k],h} = \displaystyle\frac{\text{MSE}_{i,j}^{[k],h}}{\text{MSE}_{i,0}^{[k],h}}  , \quad
\begin{array}{l}
i=1,\ldots,95, \\
j=0,\ldots,J,
\end{array} \; 
\begin{array}{l}
k \in {\cal K}, \\
h=1,\ldots,h_k .
\end{array}
$$
If a forecast outperforms the base forecasts, then the AvgRelMSE becomes smaller than one and vice-versa, and the percentage improvement in accuracy over the benchmark can be calculated as $\left(1-\text{AvgRelMSE}_{j}^{[k],h}\right) \times 100$.

Expression (\ref{AvgRelMSEjkh}), which refers to all 95 time series, can be re-stated for (i) groups of variables (e.g., bts and uts), (ii) multiple forecast horizons (e.g., $h=1-4$ for quarterly forecasts, $k=1$; $h=1-2$ for semi-annual forecasts, $k=2$), (iii) different temporal aggregation levels over the whole forecast horizon (e.g., accuracy indices for the whole temporal hierarchy of each series)\footnote{On this last point, Kourentzes and Athanasopoulos (2020b, pp. 17-18) raise an important issue by claiming that `in contrast to common practice, we believe that there is limited benefit in an empirical evaluation setting, to report average accuracy measures across all levels of the hierarchy (...). It is very improbable that this reflects a realistic situation. 
Hence, it is paramount that the modeller attempts to establish a strong connection between the objectives of the forecasts and the evaluation'.
In the forecasting experiment of this paper, where only three temporal aggregation levels are in order, it could be sensible not to consider as strategic the semi-annual time frequency, which is instrumentally used to improve the accuracy of quarterly and annual forecasts, but is probably of reduced practical usefulness to analysts and decision makers.}. In Appendix A.7 we show the expressions used to compute forecast accuracy indices in a rolling forecast experiment, like the one we are dealing with, for selected combinations of variables/time frequencies/forecast horizons.

In order to give a complete picture of the evaluation results, in the next subsection we show and discuss the MSE-based accuracy indices, at multiple timescales and forecast horizons, for a set of selected forecast reconciliation procedures. Appendix A.8 reports the indices based on MAE as well, and several tables and graphs of the accuracy indices (for both MSE and MAE) for all the forecast reconciliation procedures described in the previous sections, by keeping distinct one-dimension (either cross-sectional or temporal) forecast reconciliation procedures from cross-temporal heuristic and optimal combination procedures.

Furthermore, we use the non-parametric Friedman and post-hoc Nemenyi tests (see also Koning et al., 2005, and Hibon et al., 2012), as implemented in the \texttt{R}-package \texttt{tsutils} (Kourentzes, 2019), to establish if the differences in the forecasts produced by the considered procedures are significant. According to Kourentzes and Athanasopoulos (2019, p. 402)
``the Friedman test first establishes whether at least one of the forecasts is
significantly different from the rest. If this is the case, we use the Nemenyi test to identify groups of forecasts for which there is no
evidence of statistically significant differences. The advantage of this testing approach is that it does not impose any distributional
assumptions and does not require multiple pairwise testing between forecasts, which would distort the outcome of the tests''.


\subsection{The considered forecast reconciliation procedures}

The empirical application mainly aims to evaluate the performance of the most convincing new cross-temporal reconciliation procedures, which basically are those using residual-based approximations of the covariance matrix, as compared to the state-of-the-art point forecast reconciliation procedures. More precisely, we consider five selected procedures recently proposed in the hierarchical forecasting literature:
\begin{itemize}
\item cs-shr (Wickramasuriya, et al. 2019),
\item t-wlsv (Kourentzes et al., 2017),
\item t-acov (Nystrup et al., 2020),
\item t-sar1 (Nystrup et al., 2020),
\item kah-wlsv-shr (Kourentzes and Athanasopoulos, 2019),
\end{itemize}
five (two-step and iterative) variants of the KA approach:
\begin{itemize}
	\item tcs-acov-shr, i.e. two-step t-acov + cs-shr,
	\item tcs-sar1-shr, i.e. two-step t-sar1 + cs-shr,
	\item ite-wlsv-shr, i.e. iterative t-wlsv + cs-shr (see section \ref{subsec:itekah}),
	\item ite-acov-shr, i.e. iterative t-acov + cs-shr (see section \ref{subsec:itekah}),
	\item ite-sar1-shr, i.e. iterative t-sar1 + cs-shr (see section \ref{subsec:itekah}),
\end{itemize}
and finally, three optimal combination forecast procedures:
\begin{itemize}
	\item oct-wlsv, i.e. $\Wvet = \widehat{\Wvet}_{\text{wlsv}}$ (see section \ref{subsec:Woct}),
	\item oct-bdshr,  i.e. $\Wvet = \widehat{\Wvet}^{BD}_{\text{shr}}$ (see section \ref{subsec:Woct}),
	\item oct-acov,  i.e. $\Wvet = \widehat{\Wvet}_{\text{acov}}$ (see section \ref{subsec:Woct}).
\end{itemize}

The first five procedures have proven well performing in the empirical applications where they have been used (Athanasopoulos et al., 2017, 2019, Wickramasuriya et al., 2019, Bisaglia et al., 2020, Nystrup et al., 2020, among others). Clearly, the one-dimension reconciliation procedures (cs-shr, t-wlsv, t-acov, and t-sar1) do not give fully coherent forecasts. Rather, as far as it is expected that they improve on the base forecasts, the best-practice one-dimension procedures should be viewed as stricter benchmarks for the cross-temporal forecast reconciliation procedures, which are requested to give accurate one-number-forecasts as well.

In summary, the forecasting experiment was designed to evaluate the capability of the cross-temporal forecast reconciliation procedures to improve the forecast accuracy as compared (i) to the base forecasts, and (ii) to the most performing one-dimension forecast reconciliation procedures. In addition, the experiment should help in assessing (iii) the performance of both KA-variants (two-step and iterative procedures) and optimal combination forecasts as compared to the original proposal by KA, and (iv) the feasibility and the accuracy of the optimal combination cross-temporal reconciliation procedures, which for the time being - even when they are computed using the in-sample residuals - are based on rather simple/unrealistic approximations of the covariance matrix (see section \ref{subsec:Woct}).
As for this last point, we are interested in understading if there is any significant difference between the reconciled forecasts produced by the most performing heuristic and optimal combination forecast procedures.

\subsection{Main results}
Table \ref{Table1} presents the AvgRelMSE's obtained for the forecasting techniques (base + 13 reconciliation procedures) listed in the previuos sub-section. We provide results for all 95 component time series, and for the 32 upper-level and the 63 bottom-level time series separately. The results are shown by level of temporal aggregation and forecast horizon. At each column, the lowest error is highlighted in red boldface, while values greater than one, which mean that the reconciled forecasts are worse than the base ones, are highlighted in black boldface.


\begin{table}[!ht]
	\centering
	\caption{AvgRelMSE at any temporal aggregation level and any forecast horizon.}
	\resizebox{0.95\linewidth}{!}{
		\begin{tabular}{c|c|c|c|c|c|c|c|c|c|c}
			\hline
			& \multicolumn{5}{c|}{\textbf{Quarterly}} & \multicolumn{3}{c|}{\textbf{Semi-annual}} & \multicolumn{1}{c|}{\textbf{Annual}} & \textbf{All}\\
			\textbf{Procedure} & 1 & 2 & 3 & 4 & 1-4 & 
			1 & 2 & 1-2 & 1 & \\
   			\hline
   			\multicolumn{11}{c}{\addstackgap[10pt]{\emph{all 95 series}}} \\
base & 1 & 1 & 1 & 1 & 1 & 1 & 1 & 1 & 1 & 1 \\ 
  cs-shr & 0.9583 & 0.9701 & 0.9757 & 0.9824 & 0.9716 & 0.9526 & 0.9781 & 0.9652 & 0.9657 & 0.9689 \\ 
  t-wlsv & \textbf{1.0017} & 0.9994 & 0.9875 & 0.9853 & 0.9934 & 0.8444 & 0.9316 & 0.8869 & 0.7729 & 0.9279 \\ 
  t-acov & 0.9780 & 0.9912 & 0.9986 & 0.9888 & 0.9891 & 0.8253 & 0.9353 & 0.8786 & 0.7694 & 0.9225 \\ 
  t-sar1 & \textbf{1.0018} & 0.9994 & 0.9875 & 0.9854 & 0.9935 & 0.8445 & 0.9317 & 0.8870 & 0.7729 & 0.9279 \\ 
  kah-wlsv-shr & 0.9684 & 0.9697 & 0.9596 & \textbf{\textcolor{red}{0.9603}} & 0.9645 & 0.8175 & \textbf{\textcolor{red}{0.9085}} & 0.8618 & 0.7518 & 0.9013 \\ 
  tcs-acov-shr & 0.9453 & \textbf{\textcolor{red}{0.9583}} & 0.9710 & 0.9626 & 0.9592 & 0.7977 & 0.9117 & 0.8528 & 0.7481 & 0.8952 \\ 
  tcs-sar1-shr & 0.9684 & 0.9697 & 0.9597 & 0.9603 & 0.9645 & 0.8175 & 0.9086 & 0.8619 & 0.7518 & 0.9013 \\ 
  ite-wlsv-shr & 0.9611 & 0.9680 & \textbf{\textcolor{red}{0.9587}} & 0.9604 & 0.9620 & 0.8148 & 0.9091 & 0.8606 & 0.7512 & 0.8995 \\ 
  ite-acov-shr & \textbf{\textcolor{red}{0.9398}} & 0.9583 & 0.9709 & 0.9653 & \textbf{\textcolor{red}{0.9585}} & \textbf{\textcolor{red}{0.7957}} & 0.9127 & \textbf{\textcolor{red}{0.8522}} & \textbf{\textcolor{red}{0.7476}} & \textbf{\textcolor{red}{0.8945}} \\ 
  ite-sar1-shr & 0.9613 & 0.9683 & 0.9588 & 0.9605 & 0.9622 & 0.8151 & 0.9092 & 0.8609 & 0.7514 & 0.8997 \\ 
  oct-wlsv & 0.9692 & 0.9719 & 0.9622 & 0.9631 & 0.9666 & 0.8203 & 0.9125 & 0.8652 & 0.7562 & 0.9042 \\ 
  oct-bdshr & 0.9838 & 0.9798 & 0.9618 & 0.9665 & 0.9730 & 0.8297 & 0.9144 & 0.8710 & 0.7573 & 0.9095 \\ 
  oct-acov & 0.9553 & 0.9648 & 0.9767 & 0.9707 & 0.9668 & 0.8013 & 0.9185 & 0.8579 & 0.7531 & 0.9016 \\ 
			\multicolumn{11}{c}{\addstackgap[10pt]{\emph{32 upper series}}} \\
base & 1 & 1 & 1 & 1 & 1 & 1 & 1 & 1 & 1 & 1 \\ 
  cs-shr & \textbf{\textcolor{red}{0.9157}} & \textbf{\textcolor{red}{0.927}} & 0.9300 & 0.9315 & \textbf{\textcolor{red}{0.926}} & 0.9174 & 0.9387 & 0.928 & 0.9232 & 0.9262 \\ 
  t-wlsv & \textbf{1.0064} & \textbf{1.0091} & 0.9909 & 0.9920 & 0.9996 & 0.8556 & 0.9386 & 0.8961 & 0.7684 & 0.9331 \\ 
  t-acov & \textbf{1.0018} & \textbf{1.0146} & 0.9922 & 0.9934 & \textbf{1.0004} & 0.8537 & 0.9382 & 0.8950 & 0.7683 & 0.9332 \\ 
  t-sar1 & \textbf{1.0066} & \textbf{1.0093} & 0.9908 & 0.9921 & 0.9997 & 0.8560 & 0.9386 & 0.8963 & 0.7684 & 0.9333 \\ 
  kah-wlsv-shr & 0.9398 & 0.9467 & 0.9281 & 0.9302 & 0.9362 & 0.7996 & 0.8769 & 0.8373 & 0.7151 & 0.8726 \\ 
  tcs-acov-shr & 0.9411 & 0.9435 & 0.9307 & 0.9331 & 0.9371 & 0.7956 & 0.8779 & 0.8357 & 0.7146 & 0.8725 \\ 
  tcs-sar1-shr & 0.9399 & 0.9464 & 0.9280 & 0.9301 & 0.9361 & 0.7995 & 0.8767 & 0.8372 & 0.7149 & 0.8725 \\ 
  ite-wlsv-shr & 0.9253 & 0.9420 & 0.9224 & 0.9274 & 0.9292 & 0.7932 & 0.8739 & 0.8326 & 0.7114 & \textbf{\textcolor{red}{0.8668}} \\ 
  ite-acov-shr & 0.9283 & 0.9398 & 0.9259 & 0.9314 & 0.9313 & \textbf{\textcolor{red}{0.7893}} & 0.8754 & \textbf{\textcolor{red}{0.8313}} & \textbf{\textcolor{red}{0.7111}} & 0.8675 \\ 
  ite-sar1-shr & 0.9256 & 0.9424 & \textbf{\textcolor{red}{0.9223}} & \textbf{\textcolor{red}{0.9274}} & 0.9294 & 0.7936 & \textbf{\textcolor{red}{0.8738}} & 0.8327 & 0.7114 & 0.8669 \\ 
  oct-wlsv & 0.9411 & 0.9506 & 0.9316 & 0.9326 & 0.939 & 0.8032 & 0.8811 & 0.8412 & 0.7198 & 0.8760 \\ 
  oct-bdshr & 0.9453 & 0.9559 & 0.9246 & 0.9340 & 0.9399 & 0.8091 & 0.8791 & 0.8433 & 0.7174 & 0.8767 \\ 
  oct-acov & 0.9388 & 0.9498 & 0.9353 & 0.9371 & 0.9402 & 0.7984 & 0.8844 & 0.8403 & 0.7193 & 0.8763 \\ 
  \multicolumn{11}{c}{\addstackgap[10pt]{\emph{63 bottom series}}} \\
			base & 1 & 1 & 1 & 1 & 1 & 1 & 1 & 1 & 1 & 1 \\ 
  cs-shr & 0.9806 & 0.9928 & 0.9998 & \textbf{1.0094} & 0.9956 & 0.9709 & 0.9987 & 0.9847 & 0.9880 & 0.9914 \\ 
  t-wlsv & 0.9992 & 0.9945 & 0.9858 & 0.9819 & 0.9903 & 0.8387 & 0.9281 & 0.8823 & 0.7752 & 0.9252 \\ 
  t-acov & 0.9661 & 0.9796 & \textbf{1.0019} & 0.9864 & 0.9834 & 0.8112 & 0.9338 & 0.8704 & 0.7699 & 0.9171 \\ 
  t-sar1 & 0.9994 & 0.9944 & 0.9858 & 0.9820 & 0.9904 & 0.8388 & 0.9282 & 0.8824 & 0.7752 & 0.9253 \\ 
  kah-wlsv-shr & 0.9832 & 0.9817 & \textbf{\textcolor{red}{0.976}} & \textbf{\textcolor{red}{0.9759}} & 0.9792 & 0.8267 & \textbf{\textcolor{red}{0.9250}} & 0.8745 & 0.7712 & 0.9163 \\ 
  tcs-acov-shr & 0.9474 & \textbf{\textcolor{red}{0.9659}} & 0.9921 & 0.9780 & \textbf{\textcolor{red}{0.9707}} & \textbf{\textcolor{red}{0.7988}} & 0.9294 & \textbf{\textcolor{red}{0.8616}} & \textbf{\textcolor{red}{0.7658}} & \textbf{\textcolor{red}{0.9069}} \\ 
  tcs-sar1-shr & 0.9832 & 0.9818 & 0.9762 & 0.9761 & 0.9793 & 0.8268 & 0.9253 & 0.8746 & 0.7713 & 0.9164 \\ 
  ite-wlsv-shr & 0.9798 & 0.9814 & 0.9776 & 0.9776 & 0.9791 & 0.8259 & 0.9275 & 0.8753 & 0.7723 & 0.9166 \\ 
  ite-acov-shr & \textbf{\textcolor{red}{0.9457}} & 0.9679 & 0.9945 & 0.9830 & 0.9726 & 0.7989 & 0.9323 & 0.8631 & 0.7669 & 0.9086 \\ 
  ite-sar1-shr & 0.9800 & 0.9817 & 0.9779 & 0.9778 & 0.9793 & 0.8262 & 0.9278 & 0.8755 & 0.7725 & 0.9169 \\ 
  oct-wlsv & 0.9837 & 0.9828 & 0.9782 & 0.9789 & 0.9809 & 0.8292 & 0.9288 & 0.8776 & 0.7754 & 0.9188 \\ 
  oct-bdshr & \textbf{1.0040} & 0.9922 & 0.9813 & 0.9835 & 0.9902 & 0.8404 & 0.9329 & 0.8854 & 0.7784 & 0.9267 \\ 
  oct-acov & 0.9639 & 0.9725 & 0.9984 & 0.9881 & 0.9806 & 0.8028 & 0.9363 & 0.8670 & 0.7709 & 0.9147 \\ 
			\hline
	\end{tabular}}
	\label{Table1}
\end{table}

\begin{figure}[!ht]
	\centering
	\includegraphics[width=\linewidth]{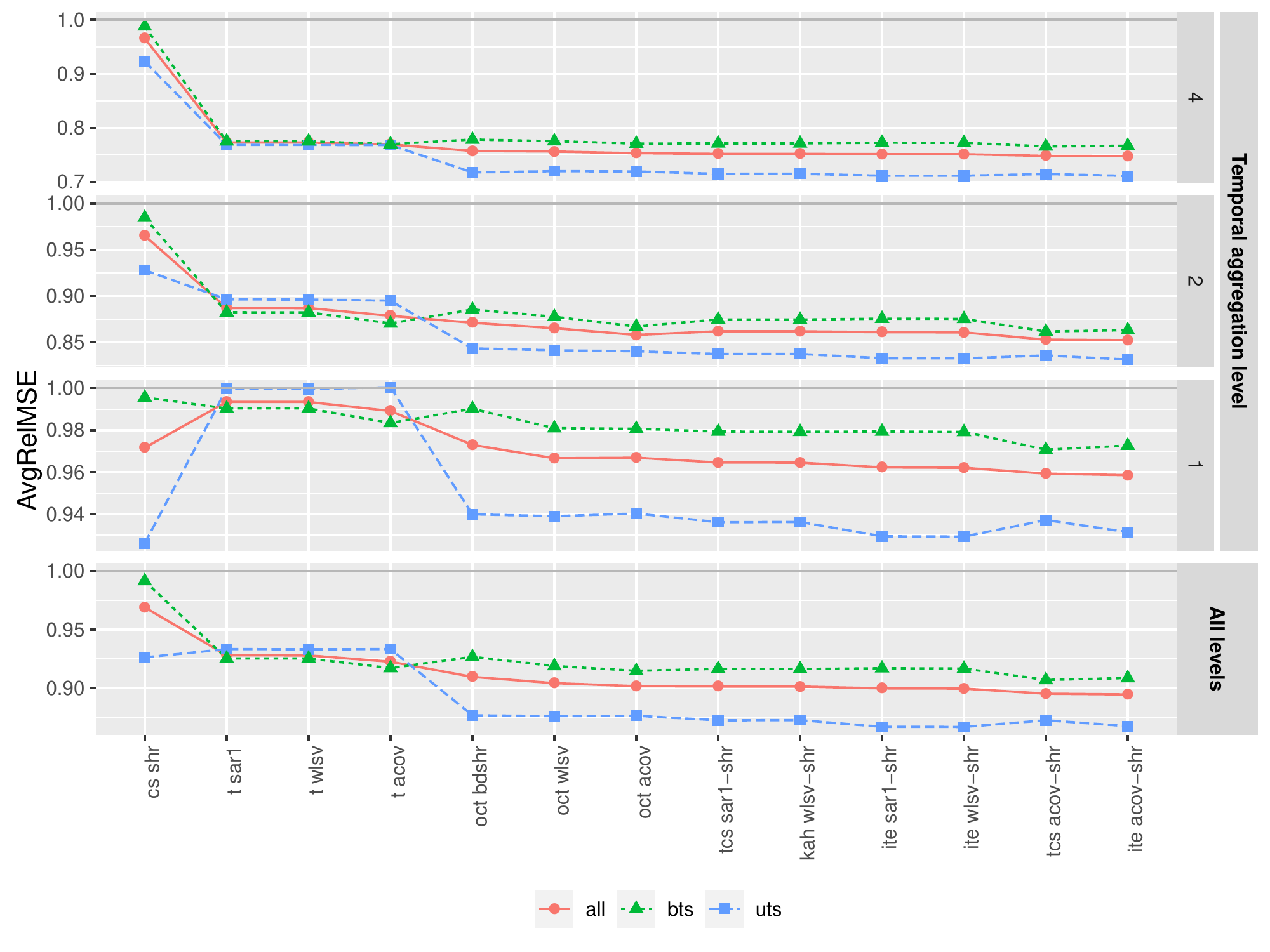}
	\includegraphics{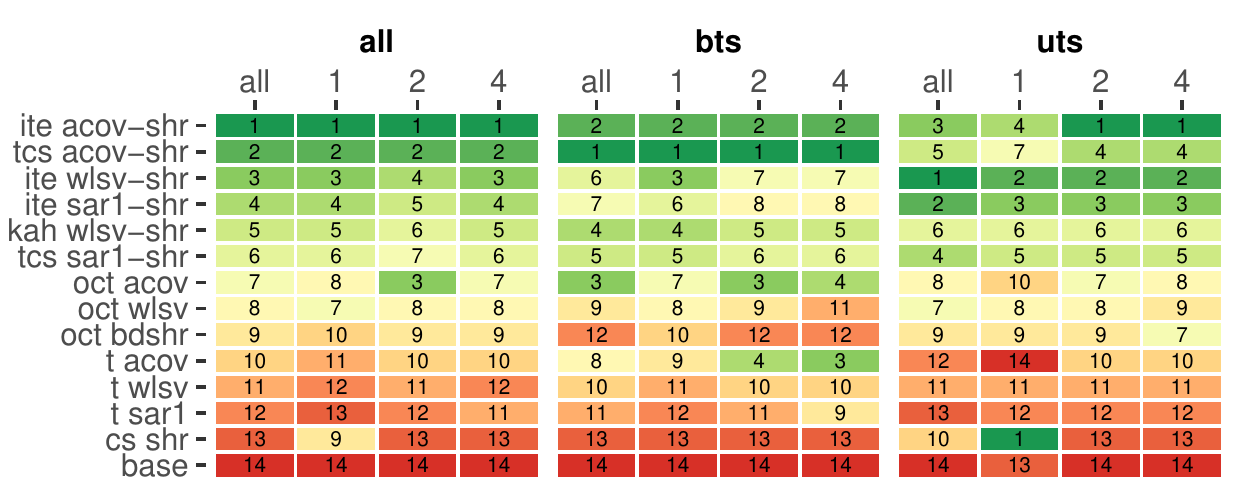}
	\caption{Top panel: Average Relative MSE across all series and forecast horizons, by frequency of observation. Bottom panel: Rankings by frequency of observation and forecast horizon.}
	\label{fig:best_mse}
\end{figure}

\begin{figure}[!ht]
	\centering
	\includegraphics[width=\linewidth]{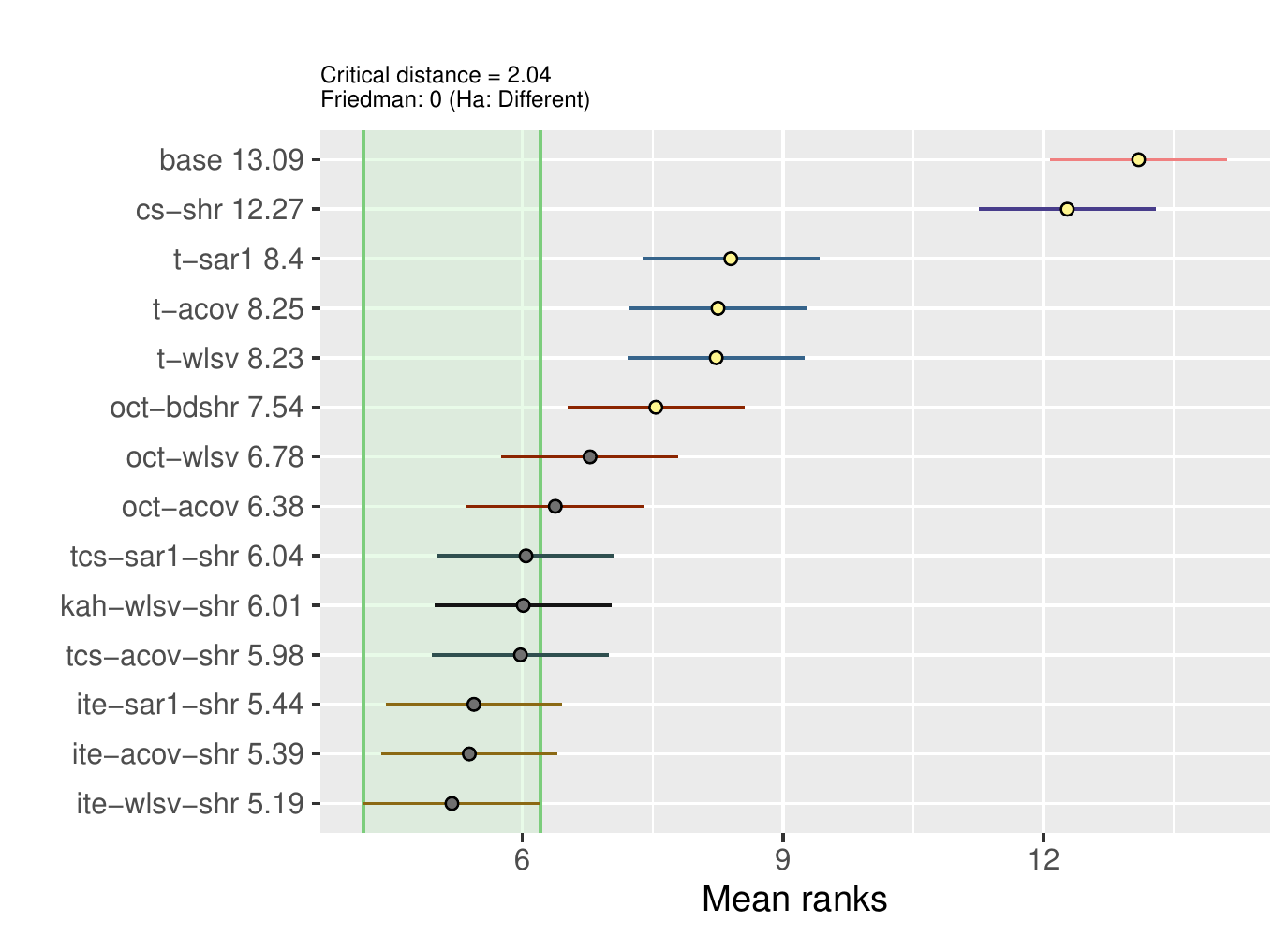}
	\includegraphics[width=\linewidth]{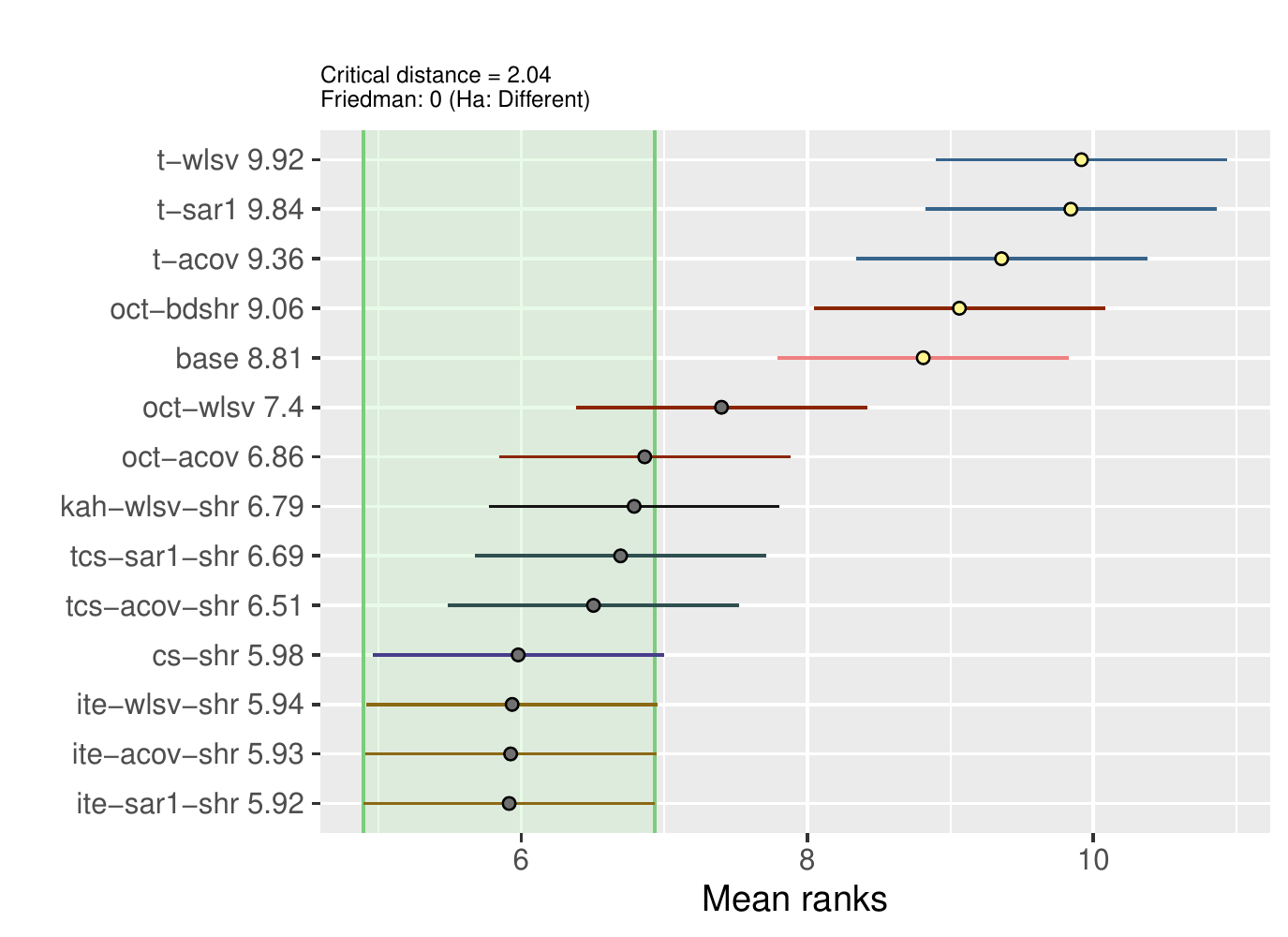}
	\caption{Nemenyi test results at 5\% significance level for all 95 series. The reconciliation procedures are sorted vertically according to the MSE mean rank (i) across all time frequencies and forecast horizons (top), and (ii) for one-step-ahead quarterly forecasts (bottom).}
	\label{fig:nem_best_mse}
\end{figure}

Most of the data in the table are represented in the top panel of Figure \ref{fig:best_mse}, containing the graphs of the AvgRelMSE's for the considered procedures, across all forecast horizons, by temporal aggregation level of the forecasted series. The ranks of these indices are reported in the bottom panel of the same figure, with colours in background chosen to highlight the procedures' performance, from best (green) to worst (red). In this figure the procedures have been put in the order given by the overall AvgRelMSE, which seems a good compromise to represent such a multiple comparison. Figure \ref{fig:nem_best_mse} shows the Multiple Comparison with the Best Nemenyi test, after that the Friedman test has shown that the forecasts given by the considered procedures are different both when all temporal aggregation levels and forecast horizons (top panel), and when only one-step-ahead quarterly forecasts (bottom panel), are considered.

\clearpage

The main results found on this dataset can be summarized as follows:
\begin{itemize}
\item as compared to both base forecasts and one-dimension reconciliation procedures, using cross-temporal hierarchies provides a clear decrease in the AvgRelMSE for the uts (likely the most important variables for the decision maker, e.g. GDP) at any temporal aggregation level and any forecast horizon;
\item this accuracy improvement is less marked, though yet visually evident, for the bottom level series, as compared to the reconciled forecasts through temporal hierarchies alone, which however are cross-sectionally incoherent;
\item each iterative procedure performs better than its two-step counterpart;
\item within the cross-temporal procedures, the heuristic procedures provide better results than the optimal combination ones.
\end{itemize}

Looking at the performances of each procedure, it's worth noting that cs-shr scores first as for the quarterly forecasts of the uts, and almost always improves on the base forecasts' accuracy, regardless of series' group, temporal aggregation level and forecast horizon\footnote{The only exception is an AvgRelMSE greater than 1 (1.0094) for the bts quarterly forecasts at horizon 4.}. In addition, from the bottom panel of Figure \ref{fig:nem_best_mse} we observe that, when considered only on quarterly basis, the one-step-ahead forecasts for all series provided by cs-shr are (temporally incoherent and) not significantly different from those provided by the best procedure (which in this case is ite-acov-shr). However, since the temporal dimension is not accounted for by this reconciliation procedure, the relative performance worsens (i.e., the cross-temporal procedures improve on the base forecasts more than cs-shr) as the temporal aggregation level increases.

Overall, ite-acov-shr always scores best for all series and all forecast horizons, and second-best for the bts series and all forecast horizons, while tcs-acov-shr scores second and first, in turn. However, ite-acov-shr shows good results for the uts forecasts as well. In this case, the best performances are given by ite-sar1-shr and ite-wlsv-shr. Figure \ref{fig:nem_best_mse} shows that the differences in the forecasts produced by all the considered heuristic procedures are not statistically significant at any temporal aggregation level and forecast horizon\footnote{Figure \ref{fig:nem_best_mse} reports only the test results across all temporal aggregation levels and forecast horizons (top), and for $k=1$ and $h=1$ (bottom). The graphs of the Nemenyi test for each temporal aggregation level and each forecast horizon are provided in Appendix A.8.}. 
Furthermore, two optimal combination procedures (oct-acov and oct-wlsv) produce reconciled forecasts not significantly different from the best procedure according to the Nemenyi test (see Figure \ref{fig:nem_best_mse}), while oct-bdshr is significantly (worse and) different from the best forecast reconciliation procedure.

Finally, in Table \ref{Table2} the AvgRelMSE's for selected upper time series and reconciliation procedures are shown\footnote{The results for all 95 series, and AvgRelMAE as well, are available in Appendix A.8.}. The series we analyze (see Figure \ref{fig:seriesB}) come from the first three levels of both the Income and Expenditure sides hierarchies:
\begin{itemize}
\item Gross Domestic Product
\item Total Factor Income (Income side)
\item Gross Operating Surplus (Income side)
\item Compensation of Employees (Income side)
\item Gross National Expenditure (Expenditure side)
\item Domestic Final Demand (Expenditure side)
\item Changes in Inventories (Expenditure sides)
\item Final Consumption Expenditures (Expenditure side)
\item Gross Fixed Capital Formation (Expenditure side)
\end{itemize}

\begin{table}[!ht]
	\centering
	\caption{AvgRelMSE at any temporal aggregation level and any forecast horizon \\ for selected upper time series and reconciliation procedures.}
	\resizebox{0.92\linewidth}{!}{
		\begin{tabular}{c|c|c|c|c|c|c|c|c|c|c}
			\hline
			& \multicolumn{5}{c|}{\textbf{Quarterly}} & \multicolumn{3}{c|}{\textbf{Semi-annual}} & \multicolumn{1}{c|}{\textbf{Annual}} & \textbf{All}\\
			\textbf{Procedure} & 1 & 2 & 3 & 4 & 1-4 & 
			1 & 2 & 1-2 & 1 & \\
			\hline
			\multicolumn{11}{c}{\rule{0pt}{4ex}\emph{Gross Domestic Product}} \\ 
cs-shr & \textbf{\textcolor{red}{0.9740}} & \textbf{\textcolor{red}{0.9397}} & 0.9028 & \textbf{\textcolor{red}{0.8924}} & \textbf{\textcolor{red}{0.9267}} & 0.8382 & 0.8713 & 0.8546 & 0.7116 & 0.8719 \\
t-acov & \textbf{1.0883} & \textbf{1.0356} &\textbf{1.0108} & 1.0000 & \textbf{1.0331} & 0.6539 & 0.8717 & 0.7549 & 0.6047 & 0.8750 \\
kah-wlsv-shr & \textbf{1.1249} & 0.9876 & 0.9068 & 0.8719 & 0.9681 & 0.6510 & \textbf{\textcolor{red}{0.7698}} & 0.7079 & 0.5485 & 0.8163 \\
ite-acov-shr & \textbf{1.0503} & 0.9808 & 0.9027 & 0.8853 & 0.9526 & 0.6281                          & 0.7730 & 0.6968 & 0.5427 & 0.8039 \\
oct-acov & \textbf{1.0696} & 0.9689 & \textbf{\textcolor{red}{0.8975}} & 0.8926 & 0.9545 & \textbf{\textcolor{red}{0.6245}} & 0.7745 & \textbf{\textcolor{red}{0.6954}} & \textbf{\textcolor{red}{0.5402}} & \textbf{\textcolor{red}{0.8039}} \\

\multicolumn{11}{c}{\rule{0pt}{4ex}\emph{Total Factor Income}} \\ 

cs-shr & \textbf{\textcolor{red}{0.8316}}	& \textbf{\textcolor{red}{0.9002}} & 0.8769 & 0.8335 & \textbf{\textcolor{red}{0.8600}} & 0.8232 & 0.8760 & 0.8492 & 0.7162 & 0.8348 \\
t-acov & \textbf{1.0434} & \textbf{1.0927} & 0.9971 & 0.9818 & \textbf{1.0279} & 0.7174 & 0.9408 & 0.8215 & 0.6636 & 0.9057 \\
kah-wlsv-shr & 0.9598 & 0.9523 & 0.8696 & 0.7984 & 0.8925 & 0.6353 & \textbf{\textcolor{red}{0.7870}} & 0.7071 & 0.5680 & 0.7829 \\
ite-acov-shr & 0.8995 & 0.9428 & \textbf{\textcolor{red}{0.8663}} & \textbf{\textcolor{red}{0.8134}} & 0.8792 & 0.6141 & 0.7909 & 0.6969 & 0.5642 & 0.7722 \\
oct-acov & 0.8819 & 0.9335 & 0.8635 & 0.8131 & 0.8719 & \textbf{\textcolor{red}{0.6078}} & 0.7907 & \textbf{\textcolor{red}{0.6932}} & \textbf{\textcolor{red}{0.5603}} & \textbf{\textcolor{red}{0.7666}} \\

\multicolumn{11}{c}{\rule{0pt}{4ex}\emph{Gross Operating Surplus}} \\ 

cs-shr & \textbf{\textcolor{red}{0.9170}} & \textbf{\textcolor{red}{0.8834}} & 0.9140 & 0.9008 & \textbf{\textcolor{red}{0.9037}} & \textbf{1.0425} & \textbf{1.0489} & \textbf{1.0457} & 0.9354 & 0.9468 \\
t-acov & \textbf{1.0180} & 0.9768 & 0.9760 & 0.9459 & 0.9789 & 0.8958 & \textbf{1.1015} & 0.9933 & 0.8807 & 0.9682 \\
kah-wlsv-shr & 0.9867 & 0.9139 & 0.8988 & \textbf{\textcolor{red}{0.8717}} & 0.9168 & 0.8572 & \textbf{1.0133} & 0.9320 & 0.8134 & 0.9055 \\
ite-acov-shr & 0.9673 & 0.8943 & \textbf{\textcolor{red}{0.8985}} & 0.8810 & 0.9097 & \textbf{\textcolor{red}{0.8338}} & \textbf{1.0147} & \textbf{\textcolor{red}{0.9199}} & \textbf{\textcolor{red}{0.8083}} & \textbf{\textcolor{red}{0.8973}} \\
oct-acov & 0.9524 & 0.9233 & 0.9181 & 0.8826 & 0.9187 & 0.8534 & \textbf{1.0301} & 0.9376 & 0.8262 & 0.9102 \\

\multicolumn{11}{c}{\rule{0pt}{4ex}\emph{Compensation of Employees}} \\ 

cs-shr & \textbf{\textcolor{red}{0.9416}} & \textbf{\textcolor{red}{0.9880}} & \textbf{1.0172} & \textbf{1.0112} & \textbf{\textcolor{red}{0.9891}} & \textbf{1.0519} & \textbf{1.0820} & \textbf{1.0669} & \textbf{1.0488} & \textbf{1.0192} \\
t-acov & \textbf{1.0635} & \textbf{1.0506} & \textbf{1.0593} & \textbf{1.0365} & \textbf{1.0524} & 0.7474 & 0.8618 & \textbf{\textcolor{red}{0.8026}} & 0.5876 & 0.8962 \\
kah-wlsv-shr & \textbf{1.0893} & \textbf{1.0739} & \textbf{1.0886} & \textbf{1.0330} & \textbf{1.0709} & 0.7663 & 0.8726 & 0.8177 & 0.5932 & 0.9112 \\
ite-acov-shr & \textbf{1.0060} & \textbf{1.0417} & \textbf{1.0778} & \textbf{1.0668} & \textbf{1.0477} & \textbf{\textcolor{red}{0.7326}} & 0.8859 & 0.8056 & 0.5931 & 0.8961 \\
oct-acov & \textbf{1.0585} & \textbf{1.0662} & \textbf{1.0576} & \textbf{1.0251} & \textbf{1.0517} & 0.7560 & \textbf{\textcolor{red}{0.8567}} & 0.8048 & \textbf{\textcolor{red}{0.5853}} & \textbf{\textcolor{red}{0.8960}} \\

\multicolumn{11}{c}{\rule{0pt}{4ex}\emph{Gross National Expenditure}} \\ 

cs-shr & \textbf{\textcolor{red}{0.9243}} & \textbf{\textcolor{red}{0.9407}} & 0.9212 & \textbf{\textcolor{red}{0.8897}} & \textbf{\textcolor{red}{0.9188}} & 0.9865 & 0.8728 & 0.9280 & 0.9302 & 0.9230 \\
t-acov & \textbf{1.0197} & \textbf{1.0367} & \textbf{1.0113} & \textbf{1.0060} & \textbf{1.0184} & 0.8447 & 0.9008 & 0.8723 & 0.6630 & 0.9164 \\
kah-wlsv-shr & 0.9966 & 0.9959 & 0.9265 & 0.9017 & 0.9542 & 0.8284 & 0.8113 & 0.8198 & 0.6156 & 0.8583 \\
ite-acov-shr & 0.9723 & 0.9925 & \textbf{\textcolor{red}{0.9155}} & 0.9094 & 0.9467 & 0.8244 & 0.8158 & 0.8201 & 0.6156 & \textbf{\textcolor{red}{0.8545}} \\
oct-acov & \textbf{1.0071} & \textbf{1.0002} & 0.9278 & 0.9031 & 0.9585 & \textbf{\textcolor{red}{0.8193}} & \textbf{\textcolor{red}{0.8075}} & \textbf{\textcolor{red}{0.8133}} & \textbf{\textcolor{red}{0.6064}} & 0.8567 \\

\multicolumn{11}{c}{\rule{0pt}{4ex}\emph{Domestic Final Demand}} \\ 

cs-shr & \textbf{\textcolor{red}{0.8713}} & \textbf{\textcolor{red}{0.9737}} & \textbf{1.0182} & 0.9958 & \textbf{\textcolor{red}{0.9631}} & 0.9787 & \textbf{1.0192} & 0.9988 & \textbf{1.0038} & 0.9789 \\
t-acov & 0.9844 & \textbf{1.0002} & \textbf{1.0031} & \textbf{\textcolor{red}{0.9851}} & 0.9932 & 0.8656 & \textbf{\textcolor{red}{0.9421}} & 0.9030 & 0.6745 & 0.9146 \\
kah-wlsv-shr & 0.9112 & \textbf{1.0152} & \textbf{1.0136} & \textbf{1.0039} & 0.9850 & 0.8184 & 0.9562 & 0.8846 & 0.6747 & 0.9049 \\
ite-acov-shr & 0.8843 & \textbf{1.0014} & \textbf{1.0049} & \textbf{1.0161} & 0.9751 & \textbf{\textcolor{red}{0.8119}} & 0.9662 & 0.8857 & 0.6758 & 0.9003 \\
oct-acov & 0.9274 & \textbf{1.0114} & \textbf{1.0088} & 0.9956 & 0.9852 & 0.8142 & 0.9428 & \textbf{\textcolor{red}{0.8761}} & \textbf{\textcolor{red}{0.6608}} & \textbf{\textcolor{red}{0.8999}} \\

\multicolumn{11}{c}{\rule{0pt}{4ex}\emph{Changes in Inventories}} \\ 

cs-shr & \textbf{1.0791} & \textbf{1.0228} & \textbf{1.0412} & \textbf{\textcolor{red}{0.9250}} & \textbf{1.0154} & 0.7215 & 0.8134 & 0.7661 & 0.8811 & 0.9181 \\
t-acov & \textbf{1.0382} & \textbf{1.0609} & \textbf{1.0032} & 0.9999 & \textbf{1.0253} & 0.6886 & 0.7098 & 0.6991 & 0.8996 & 0.9020 \\
kah-wlsv-shr & \textbf{1.0204} & \textbf{1.0339} & \textbf{1.0163} & 0.9467 & \textbf{1.0037} & \textbf{\textcolor{red}{0.6644}} & 0.6795 & \textbf{\textcolor{red}{0.6719}} & \textbf{\textcolor{red}{0.8369}} & 0.8720 \\
ite-acov-shr & \textbf{1.0317} & \textbf{1.0239} & \textbf{\textcolor{red}{0.9908}} & 0.9285 & \textbf{\textcolor{red}{0.9929}} & 0.6776 & \textbf{\textcolor{red}{0.6676}} & 0.6726 & 0.8407 & \textbf{\textcolor{red}{0.8674}} \\
oct-acov & \textbf{1.0074} & \textbf{1.0401} & \textbf{1.0087} & 0.9540 & \textbf{1.0021} & 0.6813 & 0.7051 & 0.6931 & 0.9084 & 0.8894 \\

\multicolumn{11}{c}{\rule{0pt}{4ex}\emph{Final Consumption Expenditures}} \\ 

cs-shr & \textbf{\textcolor{red}{0.8826}} & \textbf{\textcolor{red}{0.8184}} & \textbf{\textcolor{red}{0.8216}} & \textbf{\textcolor{red}{0.8223}} & \textbf{\textcolor{red}{0.8358}} & 0.9482 & 0.9741 & 0.9611 & 0.9988 & 0.8923 \\
t-acov & 0.9956 & \textbf{1.0268} & 0.9982 & \textbf{1.0199} & \textbf{1.0100} & 0.8978 & 0.9456 & 0.9214 & 0.7540 & 0.9436 \\
kah-wlsv-shr & 0.9370 & 0.9094 & 0.9000 & 0.8956 & 0.9104 & 0.8077 & 0.8395 & 0.8234 & 0.6708 & 0.8469 \\
ite-acov-shr & 0.9263 & 0.8804 & 0.8963 & 0.8913 & 0.8984 & \textbf{\textcolor{red}{0.7861}} & \textbf{\textcolor{red}{0.8331}} & \textbf{\textcolor{red}{0.8093}} & \textbf{\textcolor{red}{0.6593}} & \textbf{\textcolor{red}{0.8343}} \\
oct-acov & 0.9691 & 0.9489 & 0.9310 & 0.9373 & 0.9464 & 0.8277 & 0.8673 & 0.8473 & 0.6888 & 0.8763 \\

\multicolumn{11}{c}{\rule{0pt}{4ex}\emph{Gross Fixed Capital Formation}} \\ 

cs-shr & \textbf{\textcolor{red}{0.9442}} & \textbf{\textcolor{red}{0.9828}} & \textbf{1.0156} & \textbf{1.0096} & 0.9876 & \textbf{1.0225} & \textbf{1.0185} & \textbf{1.0205} & 0.9719 & 0.9946 \\
t-acov & 0.9875 & \textbf{1.0066} & 0.9967 & 0.9653 & 0.9889 & 0.8881 & \textbf{1.0002} & 0.9425 & 0.7258 & 0.9333 \\
kah-wlsv-shr & 0.9875 & 0.9790 & 0.9768 & 0.9663 & 0.9774 & 0.8480 & 0.9829 & 0.9130 & 0.7052 & 0.9149 \\
ite-acov-shr & 0.9524 & 0.9827 & 0.9651 & 0.9859 & 0.9714 & 0.8453 & 0.9973 & 0.9182 & 0.7097 & 0.9140 \\
oct-acov & 0.9498 & 0.9511 & \textbf{\textcolor{red}{0.9539}} & \textbf{\textcolor{red}{0.9426}} & \textbf{\textcolor{red}{0.9448}} & \textbf{\textcolor{red}{0.8149}} & \textbf{\textcolor{red}{0.9476}} & \textbf{\textcolor{red}{0.8787}} & \textbf{\textcolor{red}{0.6726}} & \textbf{\textcolor{red}{0.8816}} \\

\hline
\end{tabular}}
\label{Table2}
\end{table}

The ability of cs-shr to improve on short-term (1 or 2-quarter ahead) base forecasts clearly emerges, with the only exception of the forecasts of the Change in Inventories series, where most indices at quarterly level are greater than 1. However, this bad performance is shared by the other reconciliation procedures as well, and is likely due to the low quality of the base forecasts as compared to the other considered series (see Figure \ref{fig:seriesB}).

To conclude, the general improvement registered on average (last column of Table \ref{Table2}) by the cross-temporal reconciliation procedures may be considered a positive outcome, which combines an acceptable forecasting performance at quarterly level with a good performance at semi-annual and annual-levels, with the additional feature that the complete system of forecasts is internally and temporally coherent.



\begin{figure}[!t]
	\centering
	\includegraphics[width=\linewidth]{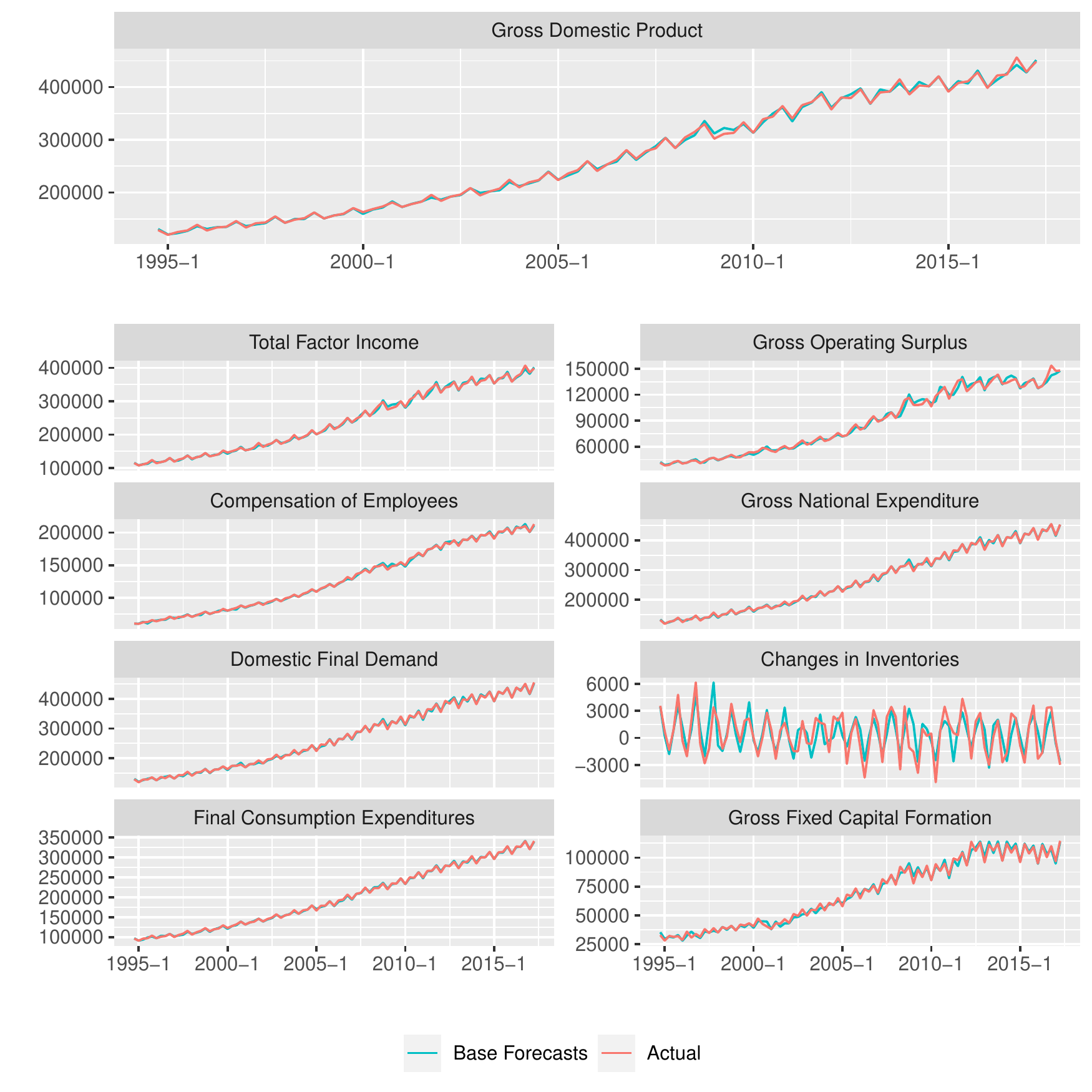}
	\caption{Quarterly GDP and selected time series from both Income and Expenditure sides: actual values and one-step-ahead base forecasts during the testing period (1994:Q4 - 2018:Q1)}
	\label{fig:seriesB}
\end{figure}
\clearpage

\section{Conclusions}
\label{sec:conclusions}
The hierarchical framework is currently considered as an effective way to improve the accuracy of forecasts in many different fields of application.
In this paper we give some contributions and extensions to a topic which has been widely studied in the last decade, by connecting it to the widespread
literature on least-squares adjustment of preliminary data (Stone et al., 1942, Byron, 1978), with focus on a projection approach which
\emph{de facto} encompasses and extends the modelling framework by Hyndman et al. (2011) (see Wickramasuriya et al., 2019, and Panagiotelis et al., 2020a).
However, we do agree with Jeon et al. (2019, p. 368) that a ``shortcoming of many of the approaches above, including WLS with structural scaling, is that
the weights (...) are a function of in-sample errors and are not directly determined with reference to an objective function ultimately used to asses
forecast quality''.
This problem, yet present for cross-temporal hierarchies, is added to the dimensionality issues which generally characterize these structures,
whose number of nodes is considerably larger than the relevant single-dimension hierarchies, and calls for alternative estimation strategies,
based for example on cross validation, as proposed by Jeon et al. (2019), or - when enough data is available - on Machine Learning techniques
(Mancuso et al., 2020, Spiliotis et al., 2020).

Nevertheless, cross-temporal point forecast reconciliation seems to be a promising theme, which is worth considering for future research. In particular, we
developed an \texttt{R} package offering classical and new optimal and heuristic combination forecast reconciliation procedures (\texttt{FoReco} - \texttt{Fo}recast \texttt{Rec}onciliation, Di Fonzo and Girolimetto, 2020). In addition, we plan to perform simulation experiments to better understand behaviour, potentiality, and possible shortcomings of the proposed procedures. Other topics in our research agenda are:
\begin{itemize}
\item looking for more realistic (and hopefully effective) approximations of the covariance matrices for cross-temporal reconciliation, (i) by building on Jeon et al. (2019), (ii) by deepening some ideas by Kourentzes (2017, 2018), and (iii) by extending/adapting some proposals by Nystrup et al. (2020) to the cross-temporal framework;
\item extending the cross-temporal framework to the reconciliation of probabilistic forecasts (Panagiotelis et al., 2020b, Jeon et al., 2019, Ben Taieb et al., 2020), and for bayesian (Eckert et al., 2020) and fast (Ashouri et al., 2019) forecast reconciliation procedures;
\item extending the cross-temporal optimal combination approach to the case of intermittent demand forecasts (Petropoulos and Kourentzes, 2015), with the related non-negativity issues
(Kourentzes and Athanasopoulos, 2020a, Wickramasuriya et al., 2020), and possible consideration of `soft' constraints (Danilov and Magnus, 2008).
\end{itemize}


\clearpage
\setcounter{figure}{0}
\setcounter{table}{0}
\setcounter{equation}{0}
\renewcommand{\thepage}{\arabic{page}}
\renewcommand\theequation{A.\arabic{equation}}  
\renewcommand\thefigure{A.\arabic{figure}}  
\renewcommand\thetable{A.\arabic{table}} 
\renewcommand*{\thefootnote}{\fnsymbol{footnote}}
\setcounter{footnote}{0}
\section*{\large Appendix\\ Cross-temporal forecast reconciliation: Optimal combination method and heuristic alternatives}
\addcontentsline{toc}{section}{Appendix}
\markright{Appendix}
\begin{center}
Tommaso Di Fonzo\footnote{Department of Statistical Sciences, University of Padua, Italy. \textit{difonzo@stat.unipd.it}} \hskip0.5cm 
Daniele Girolimetto \footnote{Department of Statistical Sciences, University of Padua, Italy. \textit{daniele.girolimetto@studenti.unipd.it}}
\end{center}
\renewcommand*{\thefootnote}{\arabic{footnote}}
\setcounter{footnote}{0}
\vskip1.5cm
\begin{table*}[h]
	\centering
\begin{tabular}%
  {>{\raggedright\arraybackslash}p{0.03\textwidth}
   >{\raggedright\arraybackslash}p{0.8\textwidth}%
   >{\raggedleft\arraybackslash}p{0.07\textwidth}%
  }
  \textbf{A.1} & \textbf{Alternative, equivalent formulations of the solution to the optimal point forecast reconciliation problem} & \textbf{\pageref{A1}}\\
&&\\
  \textbf{A.2} & \textbf{Balanced and unbalanced hierarchies} & \textbf{\pageref{A2}}\\
  &&\\
  \textbf{A.3} & \textbf{Commutation matrix and the relationships linking vectors and matrices of bottom and upper time series} & \textbf{\pageref{A3}}\\
  &&\\
  \textbf{A.4} & \textbf{Monthly and hourly temporal hierarchies} & \textbf{\pageref{A4}}\\
  &&\\
  \textbf{A.5} & \textbf{Cross-temporal hierarchy: a toy example} & \textbf{\pageref{A5}}\\
  &&\\
\textbf{A.6} & \textbf{An alternative heuristic cross-temporal reconciliation procedure} & \textbf{\pageref{A6}}\\
&&\\
\textbf{A.7} & \textbf{Average relative accuracy indices for selected groups of variables/time frequencies/forecast horizons, in a rolling forecast experiment} & \textbf{\pageref{A7}}\\
&&\\
\textbf{A.8} & \textbf{Forecast reconciliation experiment: supplementary tables and graphs} & \textbf{\pageref{A8}}
\end{tabular}
\end{table*}
\vfill
\clearpage
\section*{A.1 Alternative, equivalent formulations of the solution to the optimal point forecast reconciliation problem}
\label{A1}
Given the model
$$
\hat{\yhvet} = \Svet\betavet + \mathbf{\varepsilon}, \quad E\left(\mathbf{\varepsilon}\right)=\Zerovet, \quad
 E\left(\mathbf{\varepsilon}\mathbf{\varepsilon}'\right)=\Wvet,
$$
the $GLS$ estimator of vector $\betavet$ is given by
$$
\tilde{\betavet}=\left(\Svet'\Wvet^{-1}\Svet\right)^{-1}\Svet'\Wvet^{-1}\hat{\yhvet}
$$
and then the vector containing all reconciled forecasts is given by
\begin{equation}
\label{A1}
\tilde{\yhvet}=\Svet\tilde{\betavet}=\Svet\left(\Svet'\Wvet^{-1}\Svet\right)^{-1}\Svet'\Wvet^{-1}\hat{\yhvet} = \Svet\Gvet\hat{\yhvet},
\end{equation}
where $\Gvet = \left(\Svet'\Wvet^{-1}\Svet\right)^{-1}\Svet'\Wvet^{-1}$.

Now we show that solution (\ref{A1}) is equivalent to the one we obtain considering the following model
and its subsequent formulation in terms of constrained quadratic minimization:
$$
\hat{\yhvet} = \yhvet + \mathbf{\varepsilon}, \quad E\left(\mathbf{\varepsilon}\right)=\Zerovet, \quad
E\left(\mathbf{\varepsilon}\mathbf{\varepsilon}'\right)=\Wvet, \text{ s.t. }
\Hvet'\yhvet = \Zerovet,
$$
where $\Hvet = \left[ \begin{array}{cc}
\Ivet_{n_a} & -\Chvet
\end{array}\right]$
is a matrix of dimension $\left[n_a \times \left(n_a + n_b\right)\right]$.

In this case the following constrained minimization problem must be solved:
$$
\min_{\yhvet} \left(\yhvet - \hat{\yhvet}\right)'\Wvet^{-1}\left(\yhvet - \hat{\yhvet}\right), \text{ s.t. }
\Hvet'\yhvet = \Zerovet
$$
Let's consider the lagrangean function
$$
{\cal{L}} = \left(\yhvet - \hat{\yhvet}\right)'\Wvet^{-1}\left(\yhvet - \hat{\yhvet}\right) +
2 \lambdavet'\Hvet'\yhvet=
\yhvet'\Wvet^{-1}\yhvet - 2\hat{\yhvet}'\Wvet^{-1}\yhvet + 2\lambdavet'\Hvet'\yhvet,
$$
where $\lambdavet$ is a $\left(n_a \times 1\right)$ vector of Lagrange multipliers.

Differentiating ${\cal{L}}$ wrt $\yhvet$ and $\lambdavet$ and then equating to zero (first order conditions), we get the linear system
$$
\begin{array}{rcl}2\Wvet^{-1}\yhvet + 2\Hvet\lambdavet & = & 2\Wvet^{-1}\hat{\yhvet} \\
\Hvet'\yhvet & = & \Zerovet
\end{array}
$$
that is:
$$
\left[
\begin{array}{cc}
\Wvet^{-1} & \Hvet \\
\Hvet' & \Zerovet
\end{array}
\right]
\left[
\begin{array}{c}
\yhvet \\ \lambdavet
\end{array}
\right] =
\left[
\begin{array}{c}
\Wvet^{-1}\hat{\yhvet} \\ \Zerovet
\end{array}
\right].
$$
According to the lemma of inversion of a block-partitioned matrix (Lou and Shiou, 2002), it is:
$$
\left[
\begin{array}{cc}
\Wvet^{-1} & \Hvet \\
\Hvet' & \Zerovet
\end{array}
\right]^{-1} =
\left[
\begin{array}{cc}
\Wvet - \Wvet\Hvet\left(\Hvet'\Wvet\Hvet\right)^{-1}\Hvet'\Wvet & \Wvet\Hvet\left(\Hvet'\Wvet\Hvet\right)^{-1} \\
\left(\Hvet'\Wvet\Hvet\right)^{-1}\Hvet'\Wvet & -\left(\Hvet'\Wvet\Hvet\right)^{-1}
\end{array}
\right],
$$
and thus
$$
\tilde{\yhvet} = \left[
\Ivet_n - \Wvet\Hvet\left(\Hvet'\Wvet\Hvet\right)^{-1}\Hvet'
\right] \hat{\yhvet}.
$$
Now, let us consider matrix $\Jhvet$, defined as
$$
\Jhvet = \left[\begin{array}{cc}
\Zerovet_{n_b\times n_a} & \Ivet_{n_b}
\end{array}\right],
$$
which has dimension $\left[n_b \times n\right]$, and is such that when applied to a
$(n \times 1)$ vector, `extracts' its last $n_b$ elements.
In other words, by denoting
$\tilde{\betavet} = \Jhvet\tilde{\yhvet}$, it is:
$$
\tilde{\yhvet} = \Svet\tilde{\betavet}=
\left[\Jhvet - \Jhvet\Wvet\Hvet\left(\Hvet'\Wvet\Hvet\right)^{-1}\Hvet'
\right]\hat{\yhvet}=\Gvet\hat{\yhvet},
$$
from which we can conclude that
$$
\Gvet = \left(\Svet'\Wvet^{-1}\Svet\right)^{-1}\Svet'\Wvet^{-1} =
        \left[ \Jhvet - \Jhvet\Wvet\Hvet\left(\Hvet'\Wvet\Hvet\right)^{-1}\Hvet'\right].
$$
In the former case, the expression requires the inversion of a $(n \times n)$ matrix, $\Wvet$,
and of a $\left(n_b \times n_b \right)$ matrix, $\left(\Svet'\Wvet^{-1}\Svet\right)$.
In the latter case the matrix to be inverted, $\left(\Hvet'\Wvet\Hvet\right)$ has dimension $\left(n_a \times n_a \right)$.

\section*{A.2 Balanced and unbalanced hierarchies}
\label{A2}
\begin{figure}[ht]
	\centering
	\begin{tabular}{ccc}
		\begin{tikzpicture} [baseline=(current  bounding  box.center),
		every node/.append style={shape=rectangle,
			draw=black,
			minimum size=1cm}
		]
		\node at (-5, 0) (AA){$AA$};
		\node at (-3.75, 0) (AB){$AB$};
		\node at (-2.5, 0) (BA){$BA$};
		\node at (-1.25, 0) (BB){$BB$};
		\node at (0, 0) (CA){$C$};
		\node at (-4.375, 2) (A){$A$};
		\node at (-1.875, 2) (B){$B$};
		\node at (0, 2) (C){$C$};
		\node at (-1.875, 4) (T){$Tot.$};
		\relation{0.2}{C}{T};
		\relation{0.2}{B}{T};
		\relation{0.2}{A}{T};
		\relation{0.2}{CA}{C};
		\relation{0.2}{BA}{B};
		\relation{0.2}{BB}{B};
		\relation{0.2}{AA}{A};
		\relation{0.2}{AB}{A};
		\end{tikzpicture} & ~ &
		\begin{tikzpicture} [baseline=(current  bounding  box.center),
		every node/.append style={shape=rectangle,
			draw=black,
			minimum size=1cm}
		]
		\node at (-5, 0) (AA){$AA$};
		\node at (-3.75, 0) (AB){$AB$};
		\node at (-2.5, 0) (BA){$BA$};
		\node at (-1.25, 0) (BB){$BB$};
		\node at (0, 0) (CA){$C$};
		\node at (-4.375, 2) (A){$A$};
		\node at (-1.875, 2) (B){$B$};
		\node at (-1.875, 4) (T){$Tot.$};
		\relation{0.2}{B}{T};
		\relation{0.2}{A}{T};
		\relation{0.2}{CA}{T};
		\relation{0.2}{BA}{B};
		\relation{0.2}{BB}{B};
		\relation{0.2}{AA}{A};
		\relation{0.2}{AB}{A};
		\end{tikzpicture}
	\end{tabular}
	\caption{A simple unbalanced hierarchy (right) and its balanced version (left)}
	\label{unbal}
\end{figure}
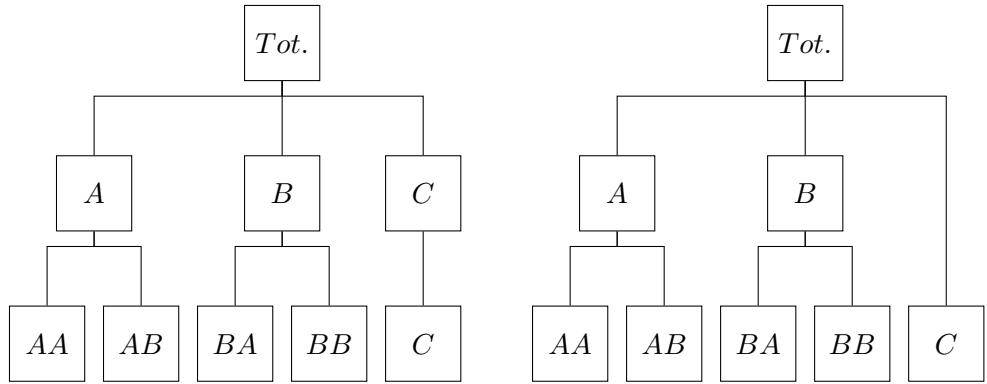

A simple three-level hierarchy is shown in the right panel of figure \ref{unbal}, where variable $C$ at the second
level of the hierarchy
has no `children', and thus is considered as a bottom variable too, at level three of the hierarchy.

The left panel shows the `balanced version' of the same hierarchy, where variable $C$ is (duplicated and)
present at both levels two and three.

The aggregation relationships linking the component series can be expressed as follows:
$$
\begin{array}{rcl}
y_{Tot} & = & y_{AA} + y_{AB} + y_{BA} + y_{BB} + y_{C} \\
y_{A}   & = & y_{AA} + y_{AB} \\
y_{B}   & = & y_{BA} + y_{BB} \\
y_{C}   & = & y_{C}
\end{array},
$$
where the last equality has merely the function of making the hierarchy balanced.
The corresponding contemporaneous aggregation matrix $\Cvet$ is given by:
$$
\Cvet = \left[\begin{array}{ccccc}
1 & 1 & 1 & 1 & 1 \\
1 & 1 & 0 & 0 & 0 \\
0 & 0 & 1 & 1 & 0 \\
0 & 0 & 0 & 0 & 1
\end{array}\right].
$$
The redundant relationship ($y_C = y_C$) makes the last row of matrix $\Cvet$
equal to the last row of the contemporaneous summing matrix
$\Svet = \left[\begin{array}{c} \Cvet \\ \Ivet_5 \end{array}\right]$.
This redundancy can be easily eliminated by considering
the new contemporaneous aggregation matrix $\tilde{\Cvet}$, which in the case of an unbalanced hierarchy
has clearly one row less than in the balanced version:
$$
\tilde{\Cvet} = \left[\begin{array}{ccccc}
1 & 1 & 1 & 1 & 1 \\
1 & 1 & 0 & 0 & 0 \\
0 & 0 & 1 & 1 & 0
\end{array}\right].
$$
The new contemporaneous summing matrix is thus given by
$\tilde{\Svet} = \left[\begin{array}{c} \tilde{\Cvet} \\ \Ivet_5 \end{array}\right]$,
which has dimension $(8 \times 5)$ instead of $(9 \times 5)$ as for matrix $\Svet$.
In a complex hierarchy, mostly when contemporaneous and temporal hierarchies are
simultaneously considered,
this fact should be carefully considered in order to save memory
space and computing time.

The R package \texttt{hts} (Hyndman et al., 2020) manages only balanced hierarchy, and
thus builds matrix $\Svet$ instead of $\tilde{\Svet}$. Due to this fact, large cross-sectional hierarchies might require
computational efforts larger than necessary, and could face numerical problems when more sofisticated reconciliation strategies are applied.
For example, the grouped time series of the Australian Tourism Demand (Visitor Nights) analyzed by
Wickramasuriya et al. (2019) (see also Ashouri et al., 2019; Bertani et al., 2020; Wickramasuriya et al., 2020),
contains 30 duplicated time series, since it comes from two unbalanced hierarchies with only 525 `unique'
time series (304 bts and 221 uts),
as compared to the 555 time series of the balanced version.
Similar, though less pronounced cases found in literature are (i) the reduced version (an unbalanced geographical hierarchy of 105 `unique' time series out of 111) of the Australian Visitor Nights dataset analyzed	by Kourentzes and Athanasopoulos (2019), and (ii) the Australian Tourism Demand (Overnight Trips) dataset considered by Panagiotelis et al. (2020a), which consists of 104 `unique' time series out of 110 for the balanced hierarchy.

\section*{A.3 Commutation matrix and the relationships linking vectors and matrices of bottom and upper time series}
\label{A3}
Given an $(r \times c)$ matrix $\Xvet$, denote with $\Cvet_{r,c}$ the $(rc \times rc)$ commutation matrix (Magnus and Neudecker, 2019) which maps $\text{vec}\left(\Xvet\right)$ into $\text{vec}\left(\Xvet'\right)$:
$$
\Cvet_{r,c}\text{vec}\left(\Xvet\right) = \text{vec}\left(\Xvet'\right).
$$
This matrix is a special type of permutation matrix, obtained by simple exchanges of rows of the
identity matrix, and is therefore orthogonal, that is:
$$
\Cvet_{r,c}^{-1} = \Cvet_{r,c}' =  \Cvet_{c,r}.
$$

\subsection*{A.3.1 Cross-sectional case: the permutation matrices linking vectors $\bvet^*$ to $\bvet$ and $\avet^*$ to $\avet$}

Denoting
$\bvet = \text{vec}\left(\Bvet\right)$,
$\bvet^* = \text{vec}\left(\Bvet'\right)$,
$\avet = \text{vec}\left(\Avet\right)$,
$\avet^* = \text{vec}\left(\Avet'\right)$,
the mappings of $\bvet$ into $\bvet^*$ and $\avet$ into $\avet^*$, respectively, can be expressed as
\[
\Pvet_b \bvet = \bvet^*, \quad  \Pvet_a \avet = \avet^*,
\]
where $\Pvet_b = \Cvet_{n_bT,n_bT}$ and $\Pvet_a = \Cvet_{n_aT,n_aT}$
are $(n_bT \times n_bT)$ and $(n_aT \times n_aT)$, respectively, commutation matrices.
Since both $\Pvet_b$ and $\Pvet_a$ are orthogonal, 
it is:
\[\bvet = \Pvet_b' \bvet^{\small{*}}, \quad \avet = \Pvet_a'\avet^{\small{*}} .
\]
The index $k$, $k=1, \dots, n_bT$, of the generic element of vector $\bvet^*$ can be expressed
in terms of the row and column indices of the corresponding element of matrix $\Bvet'$:
\[
\text{vec}(\Bvet') = \bvet^* = \left\{ b^*_k \right\}, \quad b^*_k = b_{ti}, \mbox{ with } k=t+(i-1)T .
\]
As for the index $l$, $l=1, \dots, n_aT$, of the generic element of vector $\avet^*$, we have:
\[
\text{vec}(\Avet') = \avet^* = \left\{ a^*_l \right\}, \quad a^*_l = a_{tj}, \mbox{ with } l=t+(j-1)T .
\]

\clearpage
\noindent{\em A numerical example}

\noindent Assuming that $n=2$ variables and $T=3$ time periods are considered,
matrix $\Xvet = \begin{bmatrix} 11 & 12 & 13 \\ 21 & 22 & 23 \end{bmatrix}$ can be vectorized either as
\[
\text{vec}(\Xvet) = \xvet = \begin{bmatrix} 11 & 21 & 12 & 22 & 13 & 23 \end{bmatrix}'
\]
or
\[
\text{vec}(\Xvet') = \xvet^* = \begin{bmatrix} 11 & 12 & 13 & 21 & 22 & 23 \end{bmatrix}' .
\]

\noindent In this case, the permutation matrix $\Pvet$ mapping $\xvet^*$ into $\xvet$, such that $\xvet = \Pvet \xvet^*$
(and $\xvet^* = \Pvet' \xvet$),
is given by:
\[
\Pvet = \begin{bmatrix}
1 & 0 & 0 & 0 & 0 & 0\\
0 & 0 & 0 & 1 & 0 & 0\\
0 & 1 & 0 & 0 & 0 & 0\\
0 & 0 & 0 & 0 & 1 & 0\\
0 & 0 & 1 & 0 & 0 & 0\\
0 & 0 & 0 & 0 & 0 & 1
\end{bmatrix} .
\]

\noindent The following \textbf{\textsf{R}} script performs the calculation of matrix $\Pvet$:
\footnotesize
\begin{verbatim}
n <- 2;
t <- 3;
I <- matrix(1:(n*t), n, t, byrow = T) 
I <- as.vector(I) # vectorize the required indices
P <- diag(n*t);   # Initialize an identity matrix
P <- P[I,]        # Re-arrange the rows of the identity matrix

# A numerical example
X <- matrix(c(11,12,13,21,22,23), byrow=T, nrow=2) # (2 x 3) matrix 
Xt <- t(X) 
x <- as.vector(X)      # x = vec(X) 
xstar <- as.vector(Xt) # xstar = vec(X')
xnew <- P%*%xstar      # vector xstar is mapped into vector xnew
norm(x - xnew)         # check: the norm of the difference should be zero
xstarnew <- t(P)*x;    # vector x is mapped into vector xstarnew
norm(xstar - xstarnew) # check: the norm of the difference should be zero
\end{verbatim}

\normalsize
%

\subsection*{A.3.2 Cross-temporal case: the relationship between $\check{\yvet}$ and $\text{vec}\left(\Yvet'\right)$}

Assuming $h=1$, denote with 
$\Yvet = \begin{bmatrix}
\Avet \\ \Bvet
\end{bmatrix}$
the $\left[n \times (k^*+m)\right]$ matrix of the target forecasts at any temporal frequency.
The $\left[n_b \times (k^*+m)\right]$ submatrix $\Bvet$, which contains the target forecasts of the bottom time series, can be written as:
$$
\Bvet = \left[\Bvet^{[m]} \; \Bvet^{[k_{p-1}]} \;  \ldots \; \Bvet^{[k_2]} \; \Bvet^{[1]}\right]
=
 \left[\Bvet^* \; \Bvet^{[1]}\right],
$$
where the $(n_b \times k^*)$ matrix
$\Bvet^* = \left[\Bvet^{[m]} \; \Bvet^{[k_{p-1}]} \;  \ldots \; \Bvet^{[k_2]}\right]$,
and matrix $\Bvet^{[1]}$ contain the target forecasts for, respectively, the temporally aggregated time series (lf-bts) and the high-frequency ones (hf-bts).
The following relationships hold:
\[
\Cvet_{n_b,(k^*+m)}\left[\text{vec}\left(\Bvet\right)\right] = \text{vec}\left(\Bvet'\right),
\]

\[
\Cvet_{n_b,k^*}\left[\text{vec}\left(\Bvet^*\right)\right] = \text{vec}\left[(\Bvet^*)'\right],
\]

\[
\Cvet_{n_b,m}\left[\text{vec}\left(\Bvet^{[1]}\right)\right] = \text{vec}\left[(\Bvet^{[1]})'\right].
\]
Since
$\text{vec}\left(\Bvet\right) = \begin{bmatrix}
\text{vec}\left({\Bvet^*}\right) \\[.25cm]
\text{vec}\left({\Bvet^{[1]}}\right)
\end{bmatrix}$, we can write:
\[
\Cvet_{n_b,(k^*+m)}\text{vec}\left(\Bvet\right) =
\Cvet_{n_b,(k^*+m)}
\begin{bmatrix}\Cvet_{k^*,n_b} & \Zerovet_{(n_bk^* \times n_bm)} \\[.2cm]
\Zerovet_{(n_bm \times n_bk^*)} & \Cvet_{m,n_b}
\end{bmatrix}
\begin{bmatrix}
\text{vec}\left[(\Bvet^*)'\right] \\[.25cm]
\text{vec}\left[(\Bvet^{[1]})'\right]
\end{bmatrix},
\]
that is:
$$
\text{vec}\left(\Bvet'\right) = \widetilde{\Qvet}\begin{bmatrix}
\text{vec}\left[(\Bvet^*)'\right] \\[.25cm]
\text{vec}\left[(\Bvet^{[1]})'\right]
\end{bmatrix},
$$
where
\begin{equation}
\widetilde{\Qvet}=
\Cvet_{n_b,(k^*+m)}
\begin{bmatrix}\Cvet_{k^*,n_b} & \Zerovet_{(n_bk^* \times n_bm)} \\[.2cm]
\Zerovet_{(n_bm \times n_bk^*)} & \Cvet_{m,n_b}
\end{bmatrix}.
\end{equation}
According to expression (\ref{yvetstar}), the $\left[n(k^*+m) \times 1\right]$ vector $\check{\yvet}$
can be written as:
$$
\check{\yvet} = \begin{bmatrix}
\text{vec}(\Avet') \\[.2cm] \text{vec}\left[(\Bvet^*)'\right] \\[.2cm] \text{vec}\left[(\Bvet^{[1]})'\right]
\end{bmatrix} .
$$
Then, $\text{vec}\left(\Yvet'\right)$ can be expressed in terms of $\check{\yvet}$ as:
$$
\label{vecYty}
\text{vec}\left(\Yvet'\right) = \Qvet\check{\yvet},
$$
where
$$
\Qvet=
\begin{bmatrix}\Ivet_{n_a(k^*+m)} & \Zerovet_{\left[n_a(k^*+m) \times n_bm\right]} \\[.2cm]
\Zerovet_{\left[n_bm \times n_a(k^*+m)\right]} & \widetilde{\Qvet}
\end{bmatrix}.
$$
\section*{A.4 Monthly and hourly temporal hierarchies}
\label{A4}

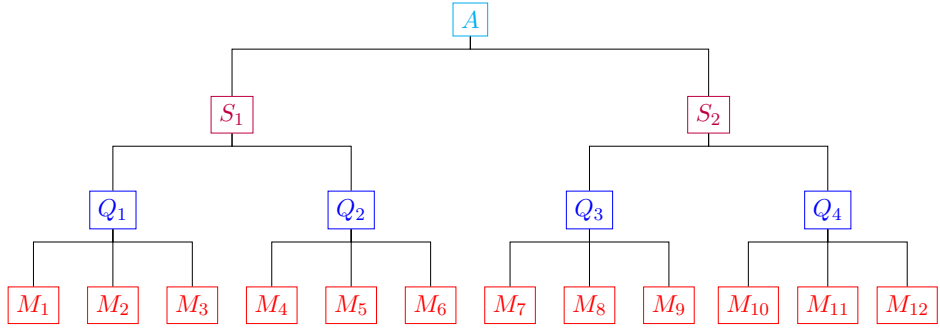
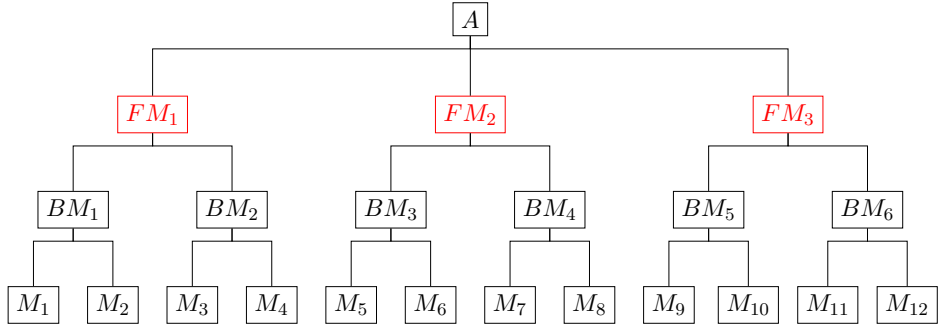
\begin{figure}[ht]
	\centering
	\begin{subfigure}{0.9\textwidth}
			\resizebox{\linewidth}{!}{
			\begin{tikzpicture}[baseline=(current  bounding  box.center),
			every node/.append style={shape=rectangle,
				draw=black},
			minimum size=0.01cm]
			\node[color=red] at (0, 0)     (M01){$M_1$};
			\node[color=red] at (1.25, 0)  (M02){$M_2$};
			\node[color=red] at (2.5, 0)   (M03){$M_3$};
			\node[color=red] at (3.75, 0)  (M04){$M_4$};
			\node[color=red] at (5, 0)     (M05){$M_5$};
			\node[color=red] at (6.25, 0)  (M06){$M_6$};
			\node[color=red] at (7.5, 0)   (M07){$M_7$};
			\node[color=red] at (8.75, 0)  (M08){$M_8$};
			\node[color=red] at (10, 0)    (M09){$M_9$};
			\node[color=red] at (11.25, 0) (M10){$M_{10}$};
			\node[color=red] at (12.5, 0)  (M11){$M_{11}$};
			\node[color=red] at (13.75, 0) (M12){$M_{12}$};
			
			\node[color=blue] at (1.25, 1.5) (Q1){$Q_1$};
			\node[color=blue] at (5.00, 1.5) (Q2){$Q_2$};
			\node[color=blue] at (8.75, 1.5) (Q3){$Q_3$};
			\node[color=blue] at (12.5, 1.5) (Q4){$Q_4$};
			
			\node[color=purple] at (3.125, 3)  (S1){$S_1$};
			\node[color=purple] at (10.625, 3) (S2){$S_2$};
			
			\node[color=cyan] at (6.875, 4.5) (A){$A$};
			\relation{0.2}{M01}{Q1};
			\relation{0.2}{M02}{Q1};
			\relation{0.2}{M03}{Q1};
			\relation{0.2}{M04}{Q2};
			\relation{0.2}{M05}{Q2};
			\relation{0.2}{M06}{Q2};
			\relation{0.2}{M07}{Q3};
			\relation{0.2}{M08}{Q3};
			\relation{0.2}{M09}{Q3};
			\relation{0.2}{M10}{Q4};
			\relation{0.2}{M11}{Q4};
			\relation{0.2}{M12}{Q4};
			\relation{0.2}{Q1}{S1};
			\relation{0.2}{Q2}{S1};
			\relation{0.2}{Q3}{S2};
			\relation{0.2}{Q4}{S2};
			\relation{0.2}{S1}{A};
			\relation{0.2}{S2}{A};
			\end{tikzpicture}
		}
		\caption{Monthly - Quarterly - Semi-Annual - Annual frequencies}
		\label{fig:demo1}
		\vspace*{6mm}
	\end{subfigure}
	\begin{subfigure}{0.9\textwidth}
					\resizebox{1\linewidth}{!}{
				\begin{tikzpicture}[baseline=(current  bounding  box.center),
				every node/.append style={shape=rectangle,
					draw=black},
				minimum size=0.01cm]
				\node at (0, 0)     (M01){$M_1$};
				\node at (1.25, 0)  (M02){$M_2$};
				\node at (2.5, 0)   (M03){$M_3$};
				\node at (3.75, 0)  (M04){$M_4$};
				\node at (5, 0)     (M05){$M_5$};
				\node at (6.25, 0)  (M06){$M_6$};
				\node at (7.5, 0)   (M07){$M_7$};
				\node at (8.75, 0)  (M08){$M_8$};
				\node at (10, 0)    (M09){$M_9$};
				\node at (11.25, 0) (M10){$M_{10}$};
				\node at (12.5, 0)  (M11){$M_{11}$};
				\node at (13.75, 0) (M12){$M_{12}$};
				
				\node at (0.625, 1.5)  (BM1){$BM_1$};
				\node at (3.125, 1.5)  (BM2){$BM_2$};
				\node at (5.625, 1.5)  (BM3){$BM_3$};
				\node at (8.125, 1.5)  (BM4){$BM_4$};
				\node at (10.625, 1.5) (BM5){$BM_5$};
				\node at (13.125, 1.5) (BM6){$BM_6$};
				
				\node[color=red] at (1.875, 3)  (FM1){$FM_1$};
				\node[color=red] at (6.875, 3)  (FM2){$FM_2$};
				\node[color=red] at (11.875, 3) (FM3){$FM_3$};
				
				\node at (6.875, 4.5) (A){$A$};
				\relation{0.2}{M01}{BM1};
				\relation{0.2}{M02}{BM1};
				\relation{0.2}{M03}{BM2};
				\relation{0.2}{M04}{BM2};
				\relation{0.2}{M05}{BM3};
				\relation{0.2}{M06}{BM3};
				\relation{0.2}{M07}{BM4};
				\relation{0.2}{M08}{BM4};
				\relation{0.2}{M09}{BM5};
				\relation{0.2}{M10}{BM5};
				\relation{0.2}{M11}{BM6};
				\relation{0.2}{M12}{BM6};
				\relation{0.2}{BM1}{FM1};
				\relation{0.2}{BM2}{FM1};
				\relation{0.2}{BM3}{FM2};
				\relation{0.2}{BM4}{FM2};
				\relation{0.2}{BM5}{FM3};
				\relation{0.2}{BM6}{FM3};
				\relation{0.2}{FM1}{A};
				\relation{0.2}{FM2}{A};
				\relation{0.2}{FM3}{A};
				\end{tikzpicture}
			}
		\caption{Monthly - Bi-Monthly - Four-Monthly - Annual frequencies}
		\label{fig:demo3}
		\vspace*{3mm}
	\end{subfigure}
	\caption{The two temporal hierarchies induced by a monthly time series.}
	\label{fig:4}
\end{figure}

For monthly data, the aggregates of interest are for $k \in \left\{12, 6, 4, 3, 2, 1 \right\}$.
Hence the monthly observations are aggregated to annual, semi-annual, four-monthly, quarterly and bi-monthly observations.
These can be represented in two separate hierarchies, as shown in Fig. \ref{fig:4},
which means that the temporal hierarchies form a grouped series, sharing the `top level' (annual) aggregate, and the same twelve
`bottom' nodes, one for each month of the original temporally disaggregated time series.


However, the $(16 \times 12)$ temporal aggregation matrix $\Kvet_1$ for this case is easily obtained:
\[
\Kvet_1 = \left[\begin{array}{cccccccccccc}
1 & 1 & 1 & 1 & 1 & 1 & 1 & 1 & 1 & 1 & 1 & 1 \\
1 & 1 & 1 & 1 & 1 & 1 & 0 & 0 & 0 & 0 & 0 & 0 \\
0 & 0 & 0 & 0 & 0 & 0 & 1 & 1 & 1 & 1 & 1 & 1 \\
1 & 1 & 1 & 1 & 0 & 0 & 0 & 0 & 0 & 0 & 0 & 0 \\
0 & 0 & 0 & 0 & 1 & 1 & 1 & 1 & 0 & 0 & 0 & 0 \\
0 & 0 & 0 & 0 & 0 & 0 & 0 & 0 & 1 & 1 & 1 & 1 \\
1 & 1 & 1 & 0 & 0 & 0 & 0 & 0 & 0 & 0 & 0 & 0 \\
0 & 0 & 0 & 1 & 1 & 1 & 0 & 0 & 0 & 0 & 0 & 0 \\
0 & 0 & 0 & 0 & 0 & 0 & 1 & 1 & 1 & 0 & 0 & 0 \\
0 & 0 & 0 & 0 & 0 & 0 & 0 & 0 & 0 & 1 & 1 & 1 \\
1 & 1 & 0 & 0 & 0 & 0 & 0 & 0 & 0 & 0 & 0 & 0 \\
0 & 0 & 1 & 1 & 0 & 0 & 0 & 0 & 0 & 0 & 0 & 0 \\
0 & 0 & 0 & 0 & 1 & 1 & 0 & 0 & 0 & 0 & 0 & 0 \\
0 & 0 & 0 & 0 & 0 & 0 & 1 & 1 & 0 & 0 & 0 & 0 \\
0 & 0 & 0 & 0 & 0 & 0 & 0 & 0 & 1 & 1 & 0 & 0 \\
0 & 0 & 0 & 0 & 0 & 0 & 0 & 0 & 0 & 0 & 1 & 1
\end{array}\right],
\]
\noindent and thus $\Rvet_1 = \left[\begin{array}{c}
\Kvet_1 \\ \Ivet_{12} \end{array}\right]$,
$ \Zvet_1' = \left[ \Ivet_{16} \; -\Kvet_1 \right]$, 
$\xvet_{\tau} = \Rvet_1\xvet^{[1]}_{\tau}$, 
and $\Zvet_1'\xvet_{\tau}= \Zerovet$, $\tau = 1, \ldots, N$,
where $\xvet_{\tau} = \left[{x^{[12]}_{\tau}}, {\xvet^{[6]}_{\tau}}', {\xvet^{[4]}_{\tau}}',
       {\xvet^{[3]}_{\tau}}', {\xvet^{[2]}_{\tau}}', {\xvet^{[1]}_{\tau}}' \right]'$
       is the $(28 \times 1)$ vector containing all temporal aggregates of variable $X$ at the observation index $\tau$ (i.e., within the complete $\tau$-th cycle).

Let's conclude with considering the case of an hourly time series with diurnal periodicity. In this case it is $m=24$, $k^*=36$,
and $\Kvet_N$ is the $(36N \times 24N)$ matrix
$$
\Kvet_N = \left[\begin{array}{c}
\Ivet_{N} \otimes \mathbf{1}'_{24} \\
\Ivet_{2N} \otimes \mathbf{1}'_{12} \\
\Ivet_{3N} \otimes \mathbf{1}'_{8} \\
\Ivet_{4N} \otimes \mathbf{1}'_{6} \\
\Ivet_{6N} \otimes \mathbf{1}'_{4} \\
\Ivet_{8N} \otimes \mathbf{1}'_{3} \\
\Ivet_{12N} \otimes \mathbf{1}'_{2}
\end{array}\right],
$$
which converts single hour values into the sum of 2, 3, 4, 6, 8, 12, and 24 hours data,
respectively, and $ \Zvet_N' = \left[ \Ivet_{36N} \; -\Kvet_N \right]$ is a full row-rank $(36N \times 60N)$ matrix.
\section*{A.5 Cross-temporal hierarchy: a toy example}
\label{A5}
\begin{figure}[t]
	\centering
	\resizebox{0.8\linewidth}{!}{
		\begin{tikzpicture}[baseline=(current  bounding  box.center),
	every node/.append style={shape=ellipse,
		draw=black},
	minimum width=1.2cm,
	minimum height=1.2cm]

	\node at (0, 0) (5Q1){$w^{[1]}_1$};
	\node at (1.5, 0) (5Q2){$w^{[1]}_2$};
	\node at (3, 0) (5Q3){$w^{[1]}_3$};
	\node at (4.5, 0) (5Q4){$w^{[1]}_4$};
	\node at (0.75, 1.8) (5SA1){$w^{[2]}_1$};
	\node at (3.75, 1.8) (5SA2){$w^{[2]}_2$};
	\node at (2.25, 3.6) (5A){$w^{[4]}_1$};
	\relation{0.2}{5Q1}{5SA1};
	\relation{0.2}{5Q2}{5SA1};
	\relation{0.2}{5Q3}{5SA2};
	\relation{0.2}{5Q4}{5SA2};
	\relation{0.2}{5SA1}{5A};
	\relation{0.2}{5SA2}{5A};
	\node[draw=none, align=center] at (0.25,3.6) {\Large $W$};
	
	\node at (11.5, 0) (6Q1){$z^{[1]}_1$};
	\node at (13, 0) (6Q2){$z^{[1]}_2$};
	\node at (14.5, 0) (6Q3){$z^{[1]}_3$};
	\node at (16, 0) (6Q4){$z^{[1]}_4$};
	\node at (12.25, 1.8) (6SA1){$z^{[2]}_1$};
	\node at (15.25, 1.8) (6SA2){$z^{[2]}_2$};
	\node at (13.75, 3.6) (6A){$z^{[4]}_1$};
	\relation{0.2}{6Q1}{6SA1};
	\relation{0.2}{6Q2}{6SA1};
	\relation{0.2}{6Q3}{6SA2};
	\relation{0.2}{6Q4}{6SA2};
	\relation{0.2}{6SA1}{6A};
	\relation{0.2}{6SA2}{6A};
	\node[draw=none, align=center] at (11.75,3.6) {\Large $Z$};
	
	\node at (5.75, 10) (7Q1){$x^{[1]}_1$};
	\node at (7.25, 10) (7Q2){$x^{[1]}_2$};
	\node at (8.75, 10) (7Q3){$x^{[1]}_3$};
	\node at (10.25, 10) (7Q4){$x^{[1]}_4$};
	\node at (6.5, 11.8) (7SA1){$x^{[2]}_1$};
	\node at (9.5, 11.8) (7SA2){$x^{[2]}_2$};
	\node at (8, 13.6) (7A){$x^{[4]}_1$};
	\relation{0.2}{7Q1}{7SA1};
	\relation{0.2}{7Q2}{7SA1};
	\relation{0.2}{7Q3}{7SA2};
	\relation{0.2}{7Q4}{7SA2};
	\relation{0.2}{7SA1}{7A};
	\relation{0.2}{7SA2}{7A};
	\node[draw=none, align=center] at (6,13.6) {\Large $X$};
	
	\node[draw=black, fit={(5Q1) (5Q2) (5Q3) (5Q4) (5SA1) (5SA2) (5A)},
	minimum width=8cm,
	minimum height=8cm](U1){};
	\node[draw=black, fit={(6Q1) (6Q2) (6Q3) (6Q4) (6SA1) (6SA2) (6A)},
	minimum width=8cm,
	minimum height=8cm](U2){};
	\node[draw=black, fit={(7Q1) (7Q2) (7Q3) (7Q4) (7SA1) (7SA2) (7A)},
	minimum width=8cm,
	minimum height=8cm](T){};
	\relation{0.5}{U1}{T};
	\relation{0.5}{U2}{T};
	\end{tikzpicture}
	}
	\caption{A two level cross-temporal hierarchy with quarterly data}
        \label{toyquarterlyhierarchy}
\end{figure}
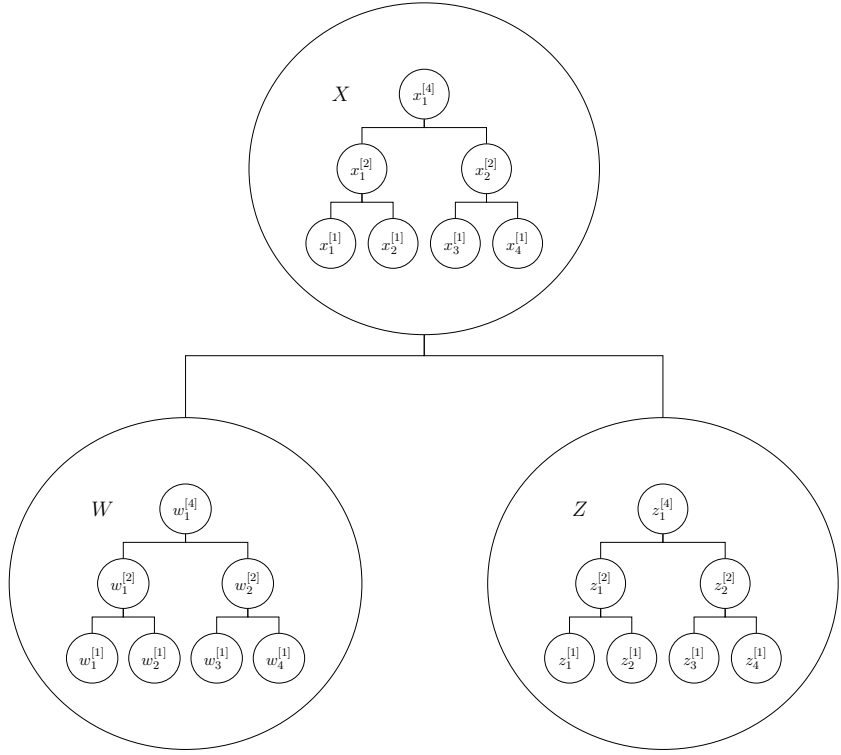

Let us consider the relationships linking all the variables implied by a cross-temporal hierarchy for the very simple case
of a total quarterly series observed for one year, $X$, obtained as the sum of two component variables, $W$ and $Z$, respectively.
The contemporaneous (cross-sectional) constraint, $X = W + Z$,
must hold at any observation index of all temporal frequencies considered in the temporal hierarchy of Figure \ref{temphierq-f} (annual, semi-annual and quarterly), as shown in Figure \ref{toyquarterlyhierarchy}, which gives a graphical view of the the way in which the two dimensions (cross-sectional and temporal) are combined within a complete time cycle (one year).

All the nodes in the cross-temporal hierarchy can be expressed in terms of the quarterly bottom time series $w^{[1]}_t$ and $z^{[1]}_t$, $t=1,\ldots,4$, according to the structural representation:

\begin{footnotesize}
$$
\underbrace{
\begin{bmatrix}
x^{[4]}_{1}\\[.1cm]x^{[2]}_{1}\\[.1cm]x^{[2]}_{2}\\[.1cm]x^{[1]}_{1}\\[.1cm]x^{[1]}_{2}\\[.1cm]x^{[1]}_{3}\\[.1cm]x^{[1]}_{4}\\[.1cm]
w^{[4]}_{1}\\[.1cm]w^{[2]}_{1}\\[.1cm]w^{[2]}_{2}\\[.1cm]
z^{[4]}_{1}\\[.1cm]z^{[2]}_{1}\\[.1cm]z^{[2]}_{2}\\[.1cm]
w^{[1]}_{1}\\[.1cm]w^{[1]}_{2}\\[.1cm]w^{[1]}_{3}\\[.1cm]w^{[1]}_{4}\\[.1cm]
z^{[1]}_{1}\\[.1cm]z^{[1]}_{2}\\[.1cm]z^{[1]}_{3}\\[.1cm]z^{[1]}_{4}
\end{bmatrix}}_{{\Large{\check{\yvet}}}}
 = \underbrace{\left[\begin{array}{ccccccccccccccccccccc}
1 & 1 & 1 & 1 & 1 & 1 & 1 & 1 \\
1 & 1 & 0 & 0 & 1 & 1 & 0 & 0 \\
0 & 0 & 1 & 1 & 0 & 0 & 1 & 1 \\
1 & 0 & 0 & 0 & 1 & 0 & 0 & 0 \\
0 & 1 & 0 & 0 & 0 & 1 & 0 & 0 \\
0 & 0 & 1 & 0 & 0 & 0 & 1 & 0 \\
0 & 0 & 0 & 1 & 0 & 0 & 0 & 1 \\
1 & 1 & 1 & 1 & 0 & 0 & 0 & 0 \\
1 & 1 & 0 & 0 & 0 & 0 & 0 & 0 \\
0 & 0 & 1 & 1 & 0 & 0 & 0 & 0 \\
0 & 0 & 0 & 0 & 1 & 1 & 1 & 1 \\
0 & 0 & 0 & 0 & 1 & 1 & 0 & 0 \\
0 & 0 & 0 & 0 & 0 & 0 & 1 & 1 \\
\multicolumn{8}{c}{\Ivet_8}
\end{array}\right]}_{\check{\Svet}}
\underbrace{\begin{bmatrix}
w^{[1]}_{1}\\[.1cm]w^{[1]}_{2}\\[.1cm]w^{[1]}_{3}\\[.1cm]w^{[1]}_{4}\\[.1cm]
z^{[1]}_{1}\\[.1cm]z^{[1]}_{2}\\[.1cm]z^{[1]}_{3}\\[.1cm]z^{[1]}_{4}
\end{bmatrix}}_{\bvet},
$$
\end{footnotesize}

\noindent where 
$\avet = \left[x^{[4]}_{1} \; x^{[2]}_{1} \; x^{[2]}_{2} \; x^{[1]}_{1} \; x^{[1]}_{2} \; x^{[1]}_{3} \; x^{[1]}_{4} \right]'$,
$\bvet = \left[\
w^{[1]}_{1} \; w^{[1]}_{2} \; w^{[1]}_{3} \; w^{[1]}_{4} \; 
z^{[1]}_{1} \; z^{[1]}_{2} \; z^{[1]}_{3} \; z^{[1]}_{4}
\right]'$,
$\check{\yvet} = \begin{bmatrix}
\avet \\ \bvet
\end{bmatrix}$,
$\check{\Svet} =\begin{bmatrix} \check{\Cvet} \\ \Ivet_8 \end{bmatrix}$, and $\check{\Cvet}$ is the $(13 \times 8)$ matrix:

$$
\check{\Cvet} =
\begin{bmatrix}
1 & 1 & 1 & 1 & 1 & 1 & 1 & 1 \\
1 & 1 & 0 & 0 & 1 & 1 & 0 & 0 \\
0 & 0 & 1 & 1 & 0 & 0 & 1 & 1 \\
1 & 0 & 0 & 0 & 1 & 0 & 0 & 0 \\
0 & 1 & 0 & 0 & 0 & 1 & 0 & 0 \\
0 & 0 & 1 & 0 & 0 & 0 & 1 & 0 \\
0 & 0 & 0 & 1 & 0 & 0 & 0 & 1 \\
1 & 1 & 1 & 1 & 0 & 0 & 0 & 0 \\
1 & 1 & 0 & 0 & 0 & 0 & 0 & 0 \\
0 & 0 & 1 & 1 & 0 & 0 & 0 & 0 \\
0 & 0 & 0 & 0 & 1 & 1 & 1 & 1 \\
0 & 0 & 0 & 0 & 1 & 1 & 0 & 0 \\
0 & 0 & 0 & 0 & 0 & 0 & 1 & 1 
\end{bmatrix}.
$$

\noindent The zero constraints valid for the nodes of the cross-temporal hierarchy can be represented through the $(13 \times 21)$ matrix $\check{\Hvet}' = \left[\Ivet_{13} \; \; -\check{\Cvet}\right]$, which has full row-rank, and is such that:
\begin{equation}
\label{strucconstr}
\check{\Hvet}'\check{\yvet} = \Zerovet_{(13\times1)}.
\end{equation}


\noindent According to the notation used so far, it is
$n_a=1$, $n_b=2$, $T=m=4$, $N=1$, $p=3$, and ${\cal K} = \left\{ 4,2,1 \right\}$.
The contemporaneous aggregation matrix $\Cvet$, mapping bts into uts, is simply a $(1 \times 2)$ row vector of ones: $\Cvet = [1 \; 1]$, and thus $\Uvet'$ is the $(1 \times 3)$ row vector $\Uvet' = [1 \; -1 \; -1]$. Furthermore, the $(3 \times 4)$ temporal aggregation matrix $\Kvet_1$ mapping a quarterly series into its semi-annual and annual counterparts, and the related $(3 \times 7)$ matrix $\Zvet_1' = \left[\Ivet_3 \; -\Kvet_1 \right]$, are given by:
$$
\Kvet_1 = \begin{bmatrix} 1 & 1 & 1 & 1 \\ 1 & 1 & 0 & 0 \\ 0 & 0 & 1 & 1 \end{bmatrix},
\qquad
\Zvet_1' = \begin{bmatrix}
1 & 0 & 0 & -1 & -1 & -1 & -1 \\
0 & 1 & 0 & -1 & -1 & 0 & 0 \\
0 & 0 & 1 & 0 & 0 & -1 & -1 
\end{bmatrix} .
$$
The $(3 \times 7)$ matrix $\Yvet$, collecting all the time series at any observation frequency, is given by:
$$
\Yvet = \begin{bmatrix}
x^{[4]}_1 & x^{[2]}_{1} & x^{[2]}_{2} & x^{[1]}_{1} & x^{[1]}_{2} & x^{[1]}_{3} & x^{[1]}_{4} \\[.2cm]
w^{[4]}_1 & w^{[2]}_{1} & w^{[2]}_{2} & w^{[1]}_{1} & w^{[1]}_{2} & w^{[1]}_{3} & w^{[1]}_{4} \\[.2cm]
z^{[4]}_1 & z^{[2]}_{1} & z^{[2]}_{2} & z^{[1]}_{1} & z^{[1]}_{2} & z^{[1]}_{3} & z^{[1]}_{4}
\end{bmatrix},
$$
and then $\yvet = \text{vec}\left(\Yvet'\right)$ is the $(21 \times 1)$ vector
\begin{small}
$$
\yvet = \left[
x^{[4]}_{1} \; x^{[2]}_{1} \; x^{[2]}_{2} \;
x^{[1]}_{1} \; x^{[1]}_{2} \; x^{[1]}_{3} \; x^{[1]}_{4} \;
w^{[4]}_{1} \; w^{[2]}_{1} \; w^{[2]}_{2} \;
w^{[1]}_{1} \; w^{[1]}_{2} \; w^{[1]}_{3} \; w^{[1]}_{4} \; 
z^{[4]}_{1} \; z^{[2]}_{1} \; z^{[2]}_{2} \;
z^{[1]}_{1} \; z^{[1]}_{2} \; z^{[1]}_{3} \; z^{[1]}_{4}
\right]' ,
$$
\end{small}
\noindent which is differently organized as compared to $\check{\yvet}$. However, it is easy to show that $\yvet = \Qvet \check{\yvet}$, where $\Qvet$ is the $(21 \times 21)$ permutation matrix
$$
\Qvet = \left[\begin{array}{ccccccccccccccccccccc}
\multicolumn{10}{c}{\Ivet_{10}} & \multicolumn{11}{c}{\Zerovet_{(10 \times 11)}} \\
0 & 0 & 0 & 0 & 0 & 0 & 0 & 0 & 0 & 0 & 0 & 0 & 0 & 1 & 0 & 0 & 0 & 0 & 0 & 0 & 0 \\
0 & 0 & 0 & 0 & 0 & 0 & 0 & 0 & 0 & 0 & 0 & 0 & 0 & 0 & 1 & 0 & 0 & 0 & 0 & 0 & 0 \\
0 & 0 & 0 & 0 & 0 & 0 & 0 & 0 & 0 & 0 & 0 & 0 & 0 & 0 & 0 & 1 & 0 & 0 & 0 & 0 & 0 \\
0 & 0 & 0 & 0 & 0 & 0 & 0 & 0 & 0 & 0 & 0 & 0 & 0 & 0 & 0 & 0 & 1 & 0 & 0 & 0 & 0 \\
0 & 0 & 0 & 0 & 0 & 0 & 0 & 0 & 0 & 0 & 1 & 0 & 0 & 0 & 0 & 0 & 0 & 0 & 0 & 0 & 0 \\
0 & 0 & 0 & 0 & 0 & 0 & 0 & 0 & 0 & 0 & 0 & 1 & 0 & 0 & 0 & 0 & 0 & 0 & 0 & 0 & 0 \\
0 & 0 & 0 & 0 & 0 & 0 & 0 & 0 & 0 & 0 & 0 & 0 & 1 & 0 & 0 & 0 & 0 & 0 & 0 & 0 & 0 \\
\multicolumn{17}{c}{\Zerovet_{(4 \times 17)}} & \multicolumn{4}{c}{\Ivet_4} 
\end{array}\right] .
$$
Given the orthogonality of matrix $\Qvet$, it is $ \check{\yvet} = \Qvet'\yvet$, and then the constraints (\ref{strucconstr}) can be re-stated as
$\check{\Hvet}'\Qvet'\yvet = \Zerovet_{(13 \times 1)}$, that is
$$
\Hvet'\yvet = \Zerovet_{(13 \times 1)},
$$
where $\Hvet' = \left(\Qvet\check{\Hvet}\right)'$ is a $(13 \times 21)$ full row-rank matrix.

The cross-temporal constraints can be formulated according to expression (\ref{ctvet}) as well, where $\breve{\Hvet}'$ is the $(16 \times 21)$ matrix
$$
\breve{\Hvet}' = \begin{bmatrix}
\Uvet' \otimes \Ivet_7 \\
\Ivet_3 \otimes \Zvet_1'
\end{bmatrix} =
\begin{bmatrix}
\Ivet_7 & -\Ivet_7 & -\Ivet_7 \\
\Zvet_1'  & \Zerovet & \Zerovet_1 \\
\Zerovet & \Zvet_1' & \Zerovet_1 \\
\Zerovet & \Zerovet & \Zvet_1'
\end{bmatrix}.
$$
The rank of $\breve{\Hvet}'$ is 13, which means that the matrix is not full row-rank.
The choice of the rows to remove is not unique\footnote{For example, in this simple example a different full row-rank $\Hvet'$ can be obtained either by removing the last three rows, or by eliminating rows 5, 10 and 15 from matrix $\breve{\Hvet}'$.}, and in real life applications the elimination of linear dependent relationships from the cross-temporal constraint set might be not as simple as in this toy example. In general, the computation of $\Hvet'$ as in (\ref{Ht}) seems rather quick and effective. In this toy example, the resulting $\Hvet'$ matrix according to that procedure is simply matrix $\breve{\Hvet}'$ without the first three rows, that is:
$$
\Hvet' = 
\begin{bmatrix}
\Ivet^* & -\Ivet^* & -\Ivet^* \\
\Zvet_1'  & \Zerovet & \Zerovet \\
\Zerovet & \Zvet_1' & \Zerovet \\
\Zerovet & \Zerovet & \Zvet_1'
\end{bmatrix},
$$
where $\Ivet^*$ is the $(4 \times 7)$ matrix
$$
\Ivet^* =
\begin{bmatrix}
0 & 0 & 0 & 1 & 0 & 0 & 0 \\
0 & 0 & 0 & 0 & 1 & 0 & 0 \\
0 & 0 & 0 & 0 & 0 & 1 & 0 \\
0 & 0 & 0 & 0 & 0 & 0 & 1
\end{bmatrix} .
$$
\section*{A.6 An alternative heuristic cross-temporal reconciliation procedure}
\label{A6}
Let us consider a cross-temporal reconciliation procedure based on the reversal of the order in which the one-dimension forecast reconciliation procedures are applied by KA.
The procedure consists in the following steps (it is assumed $h=1$):

\vspace{.5cm}

\noindent{\bf Step 1}

\noindent Transform $\widehat{\Yvet}$ by computing time-by-time cross-sectional reconciled forecasts $\widebreve{\Yvet}$  for all the temporal aggregation levels:
$$
\widehat{\Yvet} \quad \rightarrow \quad \widebreve{\Yvet} .
$$
The $\left[n \times (k^*+m)\right]$ matrix $\widehat{\Yvet}$ can be re-written also as:
$$
\widehat{\Yvet} = \begin{bmatrix}
\widehat{\Yvet}^{[m]} \; \widehat{\Yvet}^{[k_{p-1}]} \ldots
\widehat{\Yvet}^{[k_2]} \; \widehat{\Yvet}^{[1]}
\end{bmatrix},
$$
where $\widehat{\Yvet}^{[k]}$, $k \in {\cal K}$, has dimension $\left(n \times M_k\right)$.
Cross-sectionally reconciled forecasts can be computed by transforming each
$\widehat{\Yvet}^{[k]}$ as: 
$$
\widebreve{\Yvet}^{[k]} = \Mvet^{[k]} \widehat{\Yvet}^{[k]}, \quad
k \in {\cal K},
$$
where $\Mvet^{[k]}$ are $p$ transformation matrices, each of dimension $(n \times n)$, given by:
$$
\Mvet^{[k]} = \Ivet_{n} - \Wvet^{[k]}\Uvet\left(\Uvet'\Wvet^{[k]}\Uvet\right)^{-1}\Uvet', \quad
k \in {\cal K},
$$
and $\Wvet^{[k]}$ is a $(n \times n)$ p.d. known matrix. Since it is
$\Uvet'\Mvet^{[k]} = \Zerovet_{(n_a \times n)}$, $k \in {\cal K}$,
the reconciled forecasts are cross-sectionally coherent, i.e.
$\Uvet' \widebreve{\Yvet}  = \Zerovet_{\left[n_a \times (k^* +m)\right]}$, but not
temporally:
$\Zvet_1' \widebreve{\Yvet}' \ne \Zerovet_{(k^* \times n)}$.

\vspace{.5cm}

\noindent{\bf Step 2}

\noindent For each individual variable, compute the temporally reconciled forecasts $\widecheck{\Yvet}$:
$$
\widebreve{\Yvet} \quad \rightarrow \quad \widecheck{\Yvet} .
$$
This result can be obtained by applying the point forecast reconciliation formula according to temporally hierarchies (\ref{testonest}) to each column of matrix $\widebreve{\Yvet}'$. In fact, using the notation of section 4, it is
$$
\widebreve{\Yvet}' = \left[\begin{array}{cccccc}
\breve{\tvet}_{a_1} & \cdots & \breve{\tvet}_{a_{n_a}} &
\breve{\tvet}_{b_1} & \cdots & \breve{\tvet}_{b_{n_b}} \\[.25cm]
\breve{\avet}^{[1]}_1 & \cdots & \breve{\avet}^{[1]}_{n_a} &
\breve{\bvet}^{[1]}_1 & \cdots & \breve{\bvet}^{[1]}_{n_b}
\end{array}
\right].
$$

\noindent The $n_a$ vectors of temporally reconciled forecasts of the uts can be obtained as:
$$
\left[\begin{array}{c}
\check{\tvet}_{a_j} \\[0.25cm] \check{\avet}^{[1]}_{j}
\end{array}
\right] = \Mvet_{a_j}
\left[\begin{array}{c}
\breve{\tvet}_{a_j} \\[0.25cm] \breve{\avet}^{[1]}_{j}
\end{array}
\right], \quad
\Mvet_{a_j} = \Ivet_{k^*+m} - \Omegavet_{a_j}\Zvet_1\left(\Zvet_1'\Omegavet_{a_j}\Zvet_1\right)^{-1}\Zvet_1',
\quad j=1,\ldots,n_a.
$$
Likeways, the $n_b$ vectors of temporally reconciled forecasts of the bts are given by:
$$
\left[\begin{array}{c}
\check{\tvet}_{b_i} \\[0.25cm] \check{\bvet}^{[1]}_{i}
\end{array}
\right] = \Mvet_{b_i}
\left[\begin{array}{c}
\breve{\tvet}_{b_i} \\[0.25cm] \breve{\bvet}^{[1]}_{i}
\end{array}
\right], \quad
\Mvet_{b_i} = \Ivet_{k^*+m} - \Omegavet_{b_i}\Zvet_1\left(\Zvet_1'\Omegavet_{b_i}\Zvet_1\right)^{-1}\Zvet_1',
\quad i=1,\ldots,n_b,
$$
where the $n_a + n_b$ matrices $\Mvet_{a_j}$ and $\Mvet_{b_i}$ have dimension
$\left[(k^*+m) \times (k^*+m)\right]$, and each $\Omegavet_{a_j}$, $j=1,\ldots, n_a$, and $\Omegavet_{b_i}$, $i=1,\ldots,n_b$, respectively, is a known p.d. $\left[(k^*+m) \times (k^*+m) \right]$ matrix.

The mapping performing the transformation of the base forecasts into the temporally reconciled ones can be expressed in compact form as:
$$
\text{vec}\left(\widecheck{\Yvet}'\right) =
\left[\begin{array}{cccccc}
\Mvet_{a_1} & \cdots & \Zerovet & \Zerovet & \cdots & \Zerovet \\
\vdots    & \ddots & \vdots   & \vdots   & \ddots & \vdots \\
\Zerovet  & \cdots & \Mvet_{a_{n_a}} & \Zerovet & \cdots & \Zerovet \\
\Zerovet  & \cdots & \Zerovet & \Mvet_{b_{1}} & \cdots & \Zerovet \\
\vdots    & \ddots & \vdots   & \vdots   & \ddots & \vdots \\
\Zerovet  & \cdots & \Zerovet & \Zerovet & \cdots & \Mvet_{b_{n_b}}
\end{array}
\right] \text{vec}\left(\widebreve{\Yvet}'\right).
$$
The temporally reconciled forecasts can be then collected in the matrix $\widecheck{\Yvet}'$:
$$
\widecheck{\Yvet}' = \left[\begin{array}{cccccc}
\check{\tvet}_{a_1} & \cdots & \check{\tvet}_{a_{n_a}} &
\check{\tvet}_{b_1} & \cdots & \check{\tvet}_{b_{n_b}} \\[.25cm]
\check{\avet}^{[1]}_1 & \cdots & \check{\avet}^{[1]}_{n_a} &
\check{\bvet}^{[1]}_1 & \cdots & \check{\bvet}^{[1]}_{n_b}
\end{array}
\right] =
\begin{bmatrix}
(\widecheck{\Avet}^{[m]})' & (\widecheck{\Bvet}^{[m]})' \\
\vdots & \vdots \\
(\widecheck{\Avet}^{[k_2]})' & (\widecheck{\Bvet}^{[k_2]})' \\
(\widecheck{\Avet}^{[1]})' & (\widecheck{\Bvet}^{[1]})'
\end{bmatrix} ,
$$
which is in line with the temporal aggregation constraints, i.e.
$\Zvet_1' \widecheck{\Yvet}' = \Zerovet_{(k^* \times n)}$, but in general it is not in line
with the cross-sectional (contemporaneous) constraints:
$\Uvet' \widecheck{\Yvet}  \ne \Zerovet_{n_a \times (k^* +m)}$.

\vspace{.5cm}

\noindent{\bf Step 3}

\noindent Transform again the step 1 forecasts $\widebreve{\Yvet}$, by computing temporally reconciled forecasts for all $n$ variables using the $\left[(k^*+m) \times (k^*+m)\right]$ matrix $\overline{\Mvet}^{\text{cst}}$, where `cst' stands for `cross-sectional-then-temporal', given by the average of the matrices 
$\Mvet_i$
obtained at step 2:
$$
\widebreve{\Yvet} \quad \Rightarrow \widetilde{\Yvet}^{\text{cst}} .
$$
Matrix $\overline{\Mvet}^{\text{cst}}$ can be expressed as:
$$
\overline{\Mvet}^{\text{cst}}=\displaystyle\frac{1}{n}\sum_{i=1}^{n} \Mvet_i.
$$
The final cross-temporal reconciled forecasts are given by:
\begin{equation}
\label{NEWrec}
\widetilde{\Yvet}^{\text{cst}} = \left(\overline{\Mvet}^{\text{cst}} \widebreve{\Yvet}'\right)' =
\widebreve{\Yvet}(\overline{\Mvet}^{\text{cst}})'.
\end{equation}
Since
$\Uvet'\widebreve{\Yvet} =  \Zerovet_{\left[ n_a \times (k^*+m)\right]}$, and
$\Zvet_1'\overline{\Mvet}^{\text{cst}} = n^{-1}\sum_{i=1}^{n}\Zvet_1'\Mvet_i = \Zerovet_{\left[k^* \times (k^*+m)\right]}$,
the reconciled forecasts (\ref{NEWrec}) fulfill both cross-sectional and temporal aggregation constraints:
$$
\Uvet' \widetilde{\Yvet}^{\text{cst}}  = \Uvet'\widebreve{\Yvet}(\overline{\Mvet}^{\text{cst}})' = \Zerovet_{\left[n_a \times (k^* +m)\right]},
$$
$$
\Zvet_1'\left(\widetilde{\Yvet}^{\text{cst}}\right)' = \Zvet_1'\overline{\Mvet}^{\text{cst}}\widebreve{\Yvet}'=
\Zerovet_{(k^* \times n)}.
$$

\section*{A.7 Average relative accuracy indices for selected groups of variables/time frequencies/forecast horizons, in a rolling forecast experiment}
\label{A7}
Let
$$
\hat{e}_{i,j,t}^{[k],h} = y_{i,t+h}^{[k]} - \hat{y}_{i,j,t}^{[k],h}, \quad
\begin{array}{l}
i=1,\ldots,n, \\
j=0,\ldots,J,
\end{array} \;  t=1,\ldots, q , \;
\begin{array}{l}
k \in {\cal K}, \\
h=1,\ldots,h_k ,
\end{array}
$$
be the forecast error, where $y$ and $\hat{y}$ are the actual and the forecasted values, respectively, suffix $i$ denotes the variable of interest, $j$ is the forecasting technique, where $j=0$ is the benchmark forecasting procedure, $t$ is the forecast origin, ${\cal K}$ is the set of the time frequencies at which the series is observed, and $h$ is the forecast horizon, whose lead time depends on the time frequency $k$.

Denote by A$_{i,j}^{[k],h}$ the forecasting accuracy of the technique $j$, computed across $q$ forecast origins, for the $h$-step-ahead forecasts of the variable $i$ at the temporal aggregation level $k$. For example, A$_{i,j}^{[k],h} = MSE_{i,j}^{[k],h}$, as defined in expression (\ref{MSEijkh}), otherwise we might have
A$_{i,j}^{[k],h} = MAE_{i,j}^{[k],h}$ or A$_{i,j}^{[k],h} = RMSE_{i,j}^{[k],h}$,
where
$$
\begin{array}{rcl}
MAE_{i,j}^{[k],h} & = & \displaystyle\frac{1}{q}\displaystyle\sum_{t=1}^{q} \left|\hat{e}_{i,j}^{[k],h}\right| \\[.75cm]
RMSE_{i,j}^{[k],h} & = & \sqrt{\displaystyle\frac{1}{q}\displaystyle\sum_{t=1}^{q} \left(\hat{e}_{i,j}^{[k],h}\right)^2}
\end{array}
$$

\noindent In any case, we consider the relative version of the accuracy index $A_{i,j}^{[k],h}$, given by:
$$
r_{i,j}^{[k],h} = \displaystyle\frac{A_{i,j}^{[k],h}}{A_{i,0}^{[k],h}} ,
\quad i=1,\ldots,n, \quad  j=0,\ldots, J, \quad k \in {\cal K}, \quad h=1, \ldots, h_k,
$$
and use it to compute the Average relative accuracy index of the forecasting procedure $j$, for given $k$ and $h$, through the geometric mean:
$$
\text{AvgRelA}_{j}^{[k],h} = \left(\displaystyle\prod_{i=1}^{n} r_{i,j}^{[k],h} \right)^{\frac{1}{n}} , \quad j=0,\ldots,J .
$$

\noindent We may consider the following average relative accuracy indices for selected groups of variables/time frequencies and forecast horizons:

\vspace{.5cm}
\noindent \textbf{Average relative accuracy indices for a single variable at a given time frequency, for multiple forecast horizons}

$$
\label{AvgRelAikh1h2}
\text{AvgRelA}_{i,j}^{[k],q_1:q_2} = \left(\prod_{h=q_1}^{q_2} r_{i,j}^{[k],h}\right)^{\frac{1}{q_2 - q_1 + 1}}, \;
\begin{array}{l}
	i=1,\ldots,n, \\
	j=0,\ldots,J,
\end{array} \;
\begin{array}{l}
	k \in {\cal K}, \\
	1 \le q_1 \le q_2 \le h_k
\end{array} .
$$

\vspace{.5cm}
\noindent \textbf{Average relative accuracy indices for a group of variables (either all, or selected groups, e.g. a: uts, b: bts) at a given time frequency, either for a single forecast horizon or across them}
$$
\begin{array}{rcll}
\text{AvgRelA}^{[k],h}_j & = &
\left(\displaystyle\prod_{i=1}^{n} r_{i,j}^{[k],h}\right)^{\frac{1}{n}}, &
 j=0,\ldots,J , \; k \in {\cal K}, \; h=1,\ldots,h_k \\[.5cm]
\text{AvgRelA}^{[k],h}_{a,j} & = & \left(\displaystyle\prod_{i=1}^{n_a} r_{i,j}^{[k],h}\right)^{\frac{1}{n_a}}, &
 j=0,\ldots, J, \; k \in {\cal K} \\[.5cm]
\text{AvgRelA}^{[k],h}_{b,j} & = & \left(\displaystyle\prod_{i=n_a+1}^{n} r_{i,j}^{[k],h}\right)^{\frac{1}{n_b}}, &
j=0,\ldots, J, \; k \in {\cal K} \\[.5cm]
\text{AvgRelA}^{[k]}_j & = & \left(\displaystyle\prod_{i=1}^{n} \prod_{h=1}^{h_k} r_{i,j}^{[k],h}\right)^{\frac{1}{n h_k}}, &
\; j=0,\ldots,J, \; k \in {\cal K} \\[.5cm]
\text{AvgRelA}^{[k]}_{a,j} & = &\left(\displaystyle\prod_{i=1}^{n_a} \prod_{h=1}^{h_k} r_{i,j}^{[k],h}\right)^{\frac{1}{n_a h_k}}, &
j=0,\ldots, J, \; k \in {\cal K} \\[.5cm]
\text{AvgRelA}^{[k]}_{b,j} & = &\left(\displaystyle\prod_{i=n_a+1}^{n} \prod_{h=1}^{h_k} r_{i,j}^{[k],h}\right)^{\frac{1}{n_b h_k}}, &
 j=0,\ldots, J, \; k \in {\cal K}
\end{array}
$$

\vspace{.5cm}
\noindent \textbf{Average relative accuracy indices for a single variable or for a group of variables (all, a: uts, b: bts), across all time frequencies and forecast horizons}
$$
\begin{array}{rcll}
\text{AvgRelA}_{i,j} & = & \left(\displaystyle\prod_{k \in {\cal K}} \prod_{h=1}^{h_k} \text{r}_{i,j}^{[k],h}\right)^{\frac{1}{k^*+m}}, &
\begin{array}{l}
i=1,\ldots,n  \\ j=0,\ldots,J
\end{array} \\[.5cm]
\text{AvgRelA}_j & = & \left(\displaystyle\prod_{i=1}^{n} \prod_{k \in {\cal K}} \prod_{h=1}^{h_k} \text{r}_{i,j}^{[k],h}\right)^{\frac{1}{n(k^*+m)}}  , & j=0,\ldots,J \\[.5cm]
\text{AvgRelA}_{a,j} & = & \left(\displaystyle\prod_{i=1}^{n_a} \prod_{k \in {\cal K}} \prod_{h=1}^{h_k} \text{r}_{i,j}^{[k],h}\right)^{\frac{1}{n_a(k^*+m)}}  , & j=0,\ldots,J \\[.5cm]
\text{AvgRelA}_{b,j} & = & \left(\displaystyle\prod_{i=n_a+1}^{n} \prod_{k \in {\cal K}} \prod_{h=1}^{h_k} \text{r}_{i,j}^{[k],h}\right)^{\frac{1}{n_b(k^*+m)}} , & j=0,\ldots,J
\end{array}
$$


\clearpage
\section*{A.8 Forecast reconciliation experiment: supplementary tables and graphs}
\label{A8}
\subsection*{A.8.1 Selected forecast reconciliation procedures: performance results using AvgRelMAE}
\begin{table}[ht]
	\centering
	\caption{AvgRelMAE at any temporal aggregation level and any forecast horizon.}
	\resizebox{0.92\linewidth}{!}{
		\begin{tabular}{c|c|c|c|c|c|c|c|c|c|c}
			\hline
			& \multicolumn{5}{c|}{\textbf{Quarterly}} & \multicolumn{3}{c|}{\textbf{Semi-annual}} & \multicolumn{1}{c|}{\textbf{Annual}} & \textbf{All}\\
			\hline
			\textbf{Procedure} & 1 & 2 & 3 & 4 & 1-4 & 
			1 & 2 & 1-2 & 1 & \\
			\hline
			\multicolumn{11}{c}{\rule{0pt}{4ex}\emph{all 95 series}} \\ [2ex]
			base & 1 & 1 & 1 & 1 & 1 & 1 & 1 & 1 & 1 & 1 \\ 
  cs-shr & 0.9769 & 0.9842 & 0.9863 & 0.9893 & 0.9842 & 0.9733 & 0.9871 & 0.9802 & 0.9840 & 0.9830 \\ 
  t-wlsv & 0.9997 & 0.9988 & 0.9967 & 0.9953 & 0.9976 & 0.9212 & 0.9617 & 0.9412 & 0.8714 & 0.9624 \\ 
  t-acov & 0.9893 & 0.9951 & \textbf{1.002} & 0.9971 & 0.9959 & 0.9106 & 0.9632 & 0.9365 & 0.8697 & 0.9598 \\ 
  t-sar1 & 0.9997 & 0.999 & 0.9964 & 0.9953 & 0.9976 & 0.9213 & 0.9615 & 0.9412 & 0.8712 & 0.9624 \\ 
  kah-wlsv-shr & 0.9796 & 0.9829 & 0.9797 & 0.9790 & 0.9803 & 0.9046 & 0.9459 & 0.9250 & 0.8572 & 0.9459 \\ 
  tcs-acov-shr & 0.9698 & \textbf{\textcolor{red}{0.9764}} & 0.9852 & 0.9805 & 0.9780 & 0.8936 & 0.9476 & \textbf{\textcolor{red}{0.9202}} & 0.8555 & 0.9429 \\ 
  tcs-sar1-shr & 0.9797 & 0.9830 & 0.9796 & 0.9790 & 0.9803 & 0.9048 & \textbf{\textcolor{red}{0.9459}} & 0.9251 & 0.8572 & 0.9459 \\ 
  ite-wlsv-shr & 0.9750 & 0.9815 & 0.9783 & \textbf{\textcolor{red}{0.9784}} & 0.9783 & 0.9035 & 0.9459 & 0.9245 & 0.8562 & 0.9444 \\ 
  ite-acov-shr & \textbf{\textcolor{red}{0.9672}} & 0.9770 & 0.9849 & 0.9812 & \textbf{\textcolor{red}{0.9776}} & \textbf{\textcolor{red}{0.8936}} & 0.9481 & 0.9205 & \textbf{\textcolor{red}{0.8547}} & \textbf{\textcolor{red}{0.9426}} \\ 
  ite-sar1-shr & 0.9751 & 0.9819 & \textbf{\textcolor{red}{0.9781}} & 0.9784 & 0.9784 & 0.9038 & 0.9459 & 0.9246 & 0.8563 & 0.9445 \\ 
  oct-wlsv & 0.9813 & 0.9858 & 0.9829 & 0.9830 & 0.9832 & 0.9078 & 0.9506 & 0.9289 & 0.8620 & 0.9494 \\ 
  oct-bdshr & 0.9858 & 0.9880 & 0.9809 & 0.9833 & 0.9845 & 0.9112 & 0.9499 & 0.9304 & 0.8620 & 0.9505 \\ 
  oct-acov & 0.9762 & 0.9831 & 0.9904 & 0.9879 & 0.9844 & 0.8965 & 0.9541 & 0.9248 & 0.8600 & 0.9485 \\ 
			\multicolumn{11}{c}{\rule{0pt}{4ex}\emph{32 upper series}} \\ [2ex]
base & 1 & 1 & 1 & 1 & 1 & 1 & 1 & 1 & 1 & 1 \\ 
  cs-shr & 0.9484 & \textbf{\textcolor{red}{0.9628}} & \textbf{\textcolor{red}{0.9595}} & \textbf{\textcolor{red}{0.9652}} & \textbf{\textcolor{red}{0.959}} & 0.9521 & 0.9679 & 0.9600 & 0.9691 & 0.9607 \\ 
  t-wlsv & 0.9947 & \textbf{1.0034} & 0.9994 & \textbf{1.0006} & 0.9995 & 0.9273 & 0.9628 & 0.9448 & 0.8689 & 0.9641 \\ 
  t-acov & 0.9965 & \textbf{1.0061} & \textbf{1.0019} & \textbf{1.0011} & \textbf{1.0014} & 0.9270 & 0.9627 & 0.9447 & 0.8688 & 0.9651 \\ 
  t-sar1 & 0.9947 & \textbf{1.0034} & 0.9993 & \textbf{1.0005} & 0.9995 & 0.9275 & 0.9626 & 0.9449 & 0.8686 & 0.9640 \\ 
  kah-wlsv-shr & 0.9538 & 0.9713 & 0.9646 & 0.9685 & 0.9645 & 0.8912 & 0.9265 & 0.9087 & 0.8319 & 0.9284 \\ 
  tcs-acov-shr & 0.9595 & 0.9679 & 0.9655 & 0.9700 & 0.9657 & 0.8895 & 0.9269 & 0.9080 & 0.8314 & 0.9288 \\ 
  tcs-sar1-shr & 0.9539 & 0.9711 & 0.9643 & 0.9684 & 0.9644 & 0.8913 & 0.9263 & 0.9086 & 0.8318 & 0.9283 \\ 
  ite-wlsv-shr & \textbf{\textcolor{red}{0.9466}} & 0.9700 & 0.9614 & 0.9665 & 0.9611 & 0.8897 & 0.9247 & 0.9071 & 0.8299 & \textbf{\textcolor{red}{0.9257}} \\ 
  ite-acov-shr & 0.9528 & 0.9674 & 0.9627 & 0.9684 & 0.9628 & \textbf{\textcolor{red}{0.8879}} & 0.9257 & \textbf{\textcolor{red}{0.9066}} & \textbf{\textcolor{red}{0.8297}} & 0.9265 \\ 
  ite-sar1-shr & 0.9469 & 0.9704 & 0.9611 & 0.9665 & 0.9612 & 0.8901 & \textbf{\textcolor{red}{0.9246}} & 0.9072 & 0.8300 & 0.9258 \\ 
  oct-wlsv & 0.9589 & 0.9773 & 0.9712 & 0.9752 & 0.9706 & 0.8969 & 0.9339 & 0.9152 & 0.8404 & 0.9350 \\ 
  oct-bdshr & 0.9552 & 0.9790 & 0.9632 & 0.9719 & 0.9673 & 0.8983 & 0.9288 & 0.9134 & 0.8364 & 0.9320 \\ 
  oct-acov & 0.9631 & 0.9756 & 0.9729 & 0.9764 & 0.9720 & 0.8933 & 0.9356 & 0.9142 & 0.8383 & 0.9351 \\ 
			\multicolumn{11}{c}{\rule{0pt}{4ex}\emph{63 bottom series}} \\ [2ex]
base & 1 & 1 & 1 & 1 & 1 & 1 & 1 & 1 & 1 & 1 \\ 
  cs-shr & 0.9917 & 0.9953 & \textbf{1.0002} & \textbf{1.0018} & 0.9972 & 0.9842 & 0.997 & 0.9906 & 0.9917 & 0.9945 \\ 
  t-wlsv & \textbf{1.0022} & 0.9965 & 0.9953 & 0.9926 & 0.9967 & 0.9181 & 0.9611 & 0.9393 & 0.8727 & 0.9615 \\ 
  t-acov & 0.9856 & 0.9896 & \textbf{1.0021} & 0.9951 & 0.9931 & 0.9023 & 0.9635 & 0.9324 & 0.8702 & 0.9571 \\ 
  t-sar1 & \textbf{1.0023} & 0.9968 & 0.9950 & 0.9926 & 0.9967 & 0.9183 & 0.9609 & 0.9393 & 0.8726 & 0.9615 \\ 
  kah-wlsv-shr & 0.9930 & 0.9888 & 0.9875 & 0.9844 & 0.9884 & 0.9115 & \textbf{\textcolor{red}{0.9559}} & 0.9334 & 0.8703 & 0.9549 \\ 
  tcs-acov-shr & 0.9751 & \textbf{\textcolor{red}{0.9807}} & 0.9953 & 0.9859 & \textbf{\textcolor{red}{0.9842}} & \textbf{\textcolor{red}{0.8957}} & 0.9582 & \textbf{\textcolor{red}{0.9265}} & 0.8680 & \textbf{\textcolor{red}{0.9502}} \\ 
  tcs-sar1-shr & 0.9930 & 0.9891 & 0.9875 & 0.9844 & 0.9885 & 0.9117 & 0.9559 & 0.9335 & 0.8704 & 0.9550 \\ 
  ite-wlsv-shr & 0.9898 & 0.9874 & 0.9869 & \textbf{\textcolor{red}{0.9844}} & 0.9871 & 0.9106 & 0.9568 & 0.9334 & 0.8699 & 0.9541 \\ 
  ite-acov-shr & \textbf{\textcolor{red}{0.9746}} & 0.9819 & 0.9963 & 0.9878 & 0.9851 & 0.8965 & 0.9597 & 0.9276 & \textbf{\textcolor{red}{0.8677}} & 0.9509 \\ 
  ite-sar1-shr & 0.9897 & 0.9879 & \textbf{\textcolor{red}{0.9869}} & 0.9844 & 0.9872 & 0.9109 & 0.9569 & 0.9336 & 0.8700 & 0.9542 \\ 
  oct-wlsv & 0.9929 & 0.9901 & 0.9888 & 0.9870 & 0.9897 & 0.9133 & 0.9591 & 0.9360 & 0.8732 & 0.9568 \\ 
  oct-bdshr & \textbf{1.0018} & 0.9925 & 0.9900 & 0.9892 & 0.9933 & 0.9178 & 0.9609 & 0.9391 & 0.8753 & 0.9600 \\ 
  oct-acov & 0.9830 & 0.9869 & 0.9994 & 0.9937 & 0.9907 & 0.8981 & 0.9636 & 0.9303 & 0.8712 & 0.9554 \\ 
			\hline
	\end{tabular}}
\end{table}

\begin{figure}[ht]
	\centering
	\includegraphics[width=\linewidth]{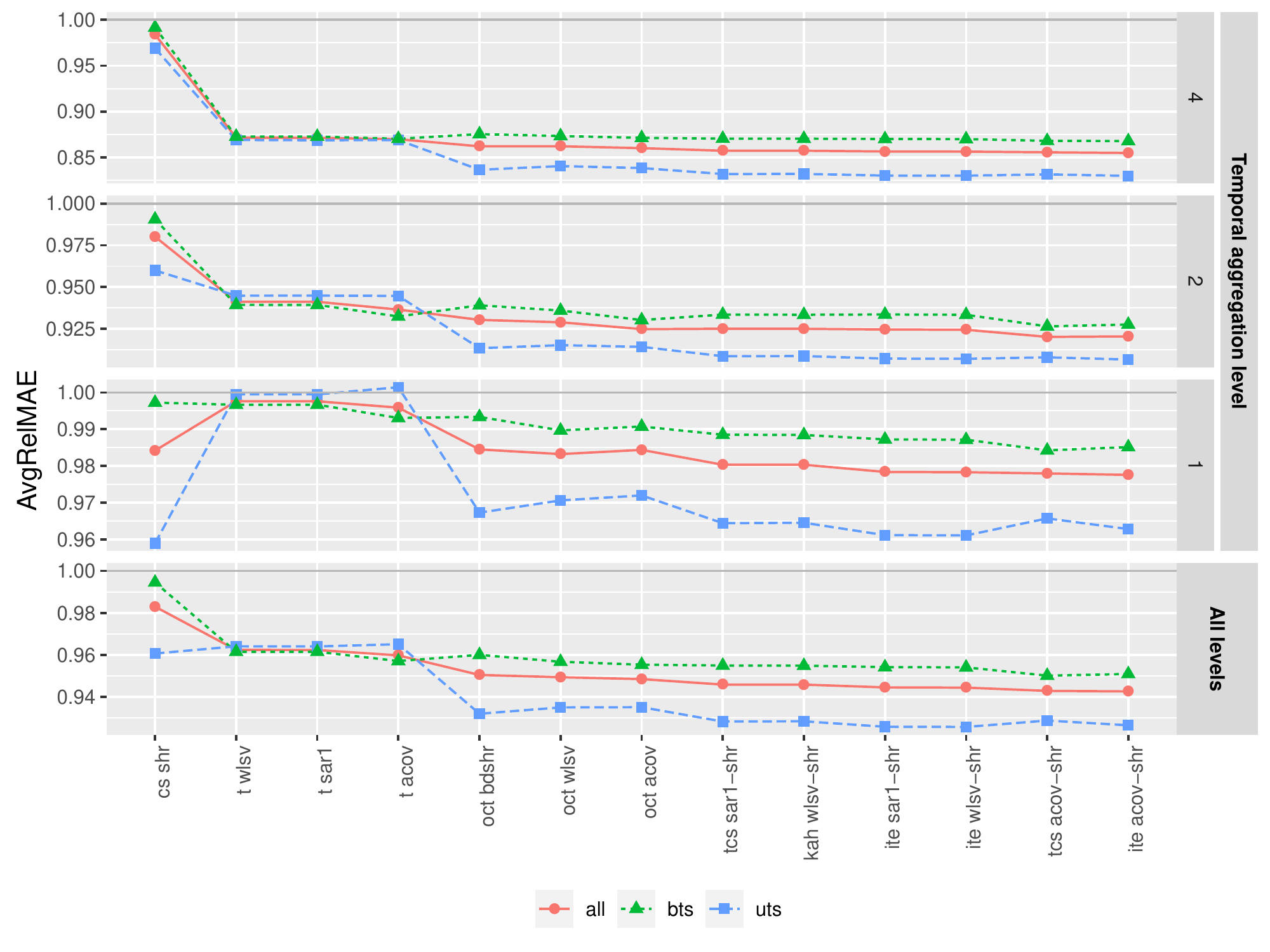}
	\includegraphics{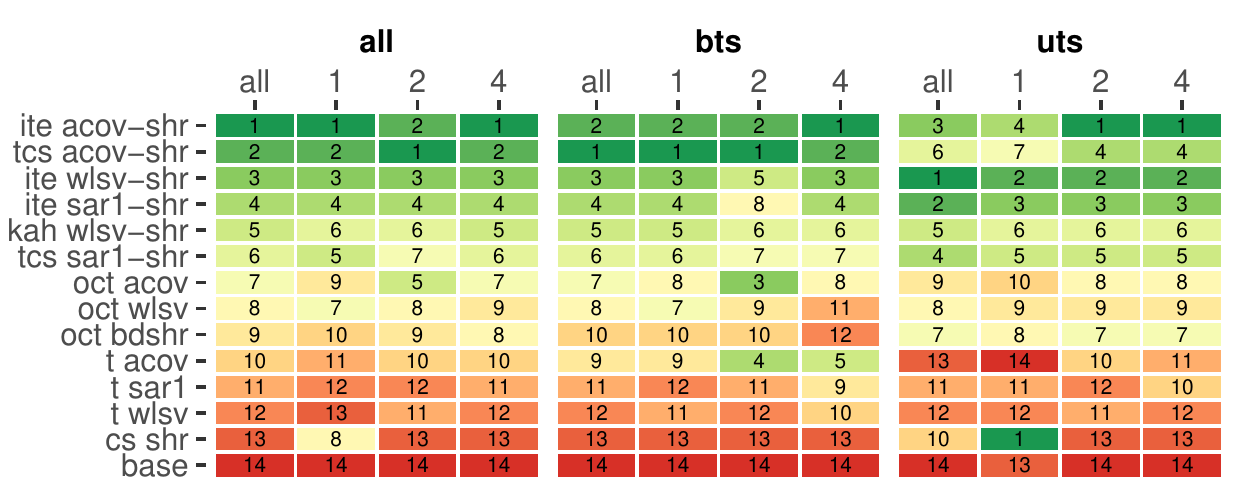}
	\caption{Top panel: Average Relative MAE across all series and forecast horizons, by frequency of observation. Bottom panel: Rankings by frequency of observation and forecast horizon.}
	\label{fig:best_mae}
\end{figure}

\begin{figure}[ht]
	\centering
	\includegraphics[width=\linewidth]{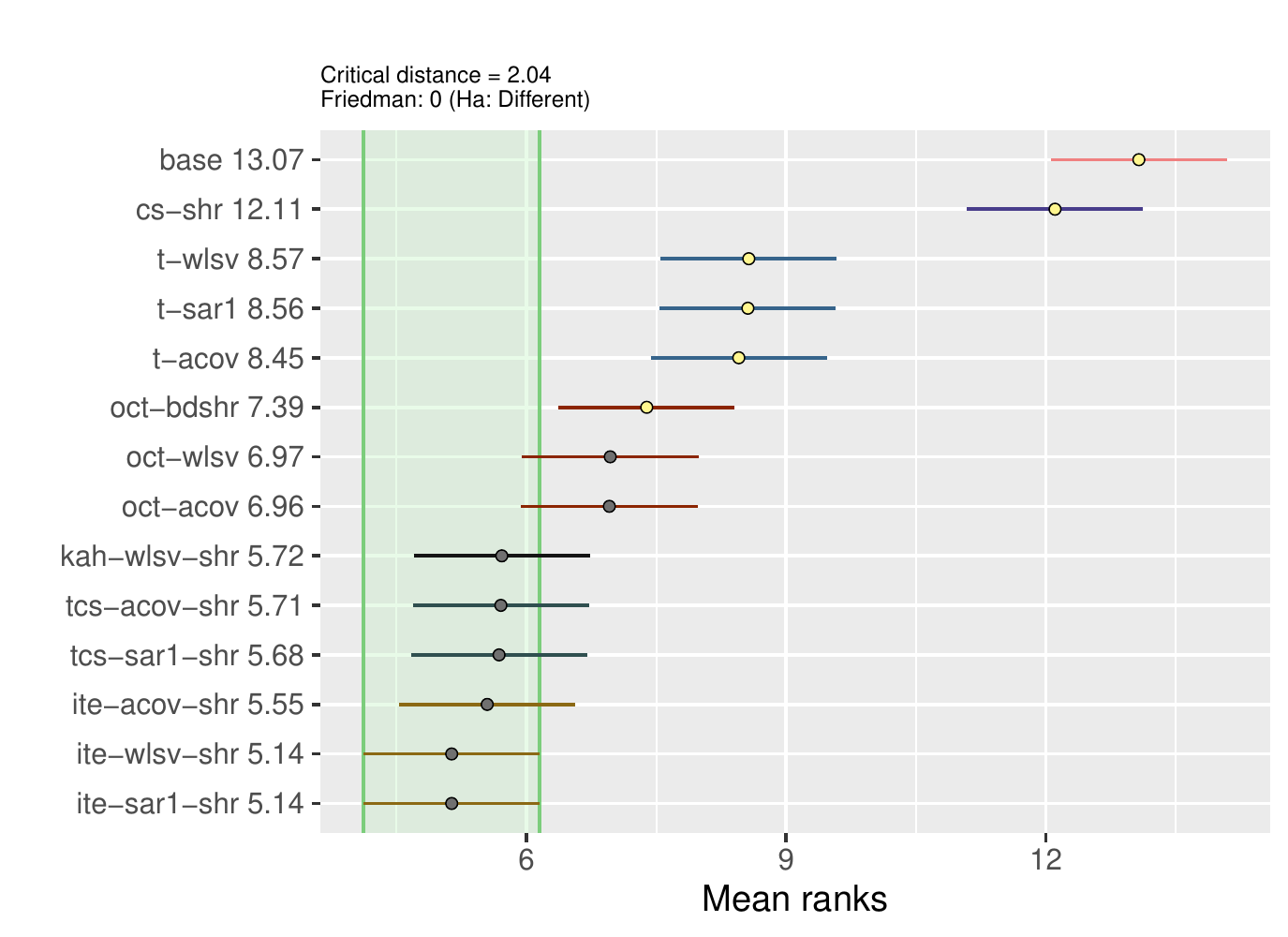}
	\includegraphics[width=\linewidth]{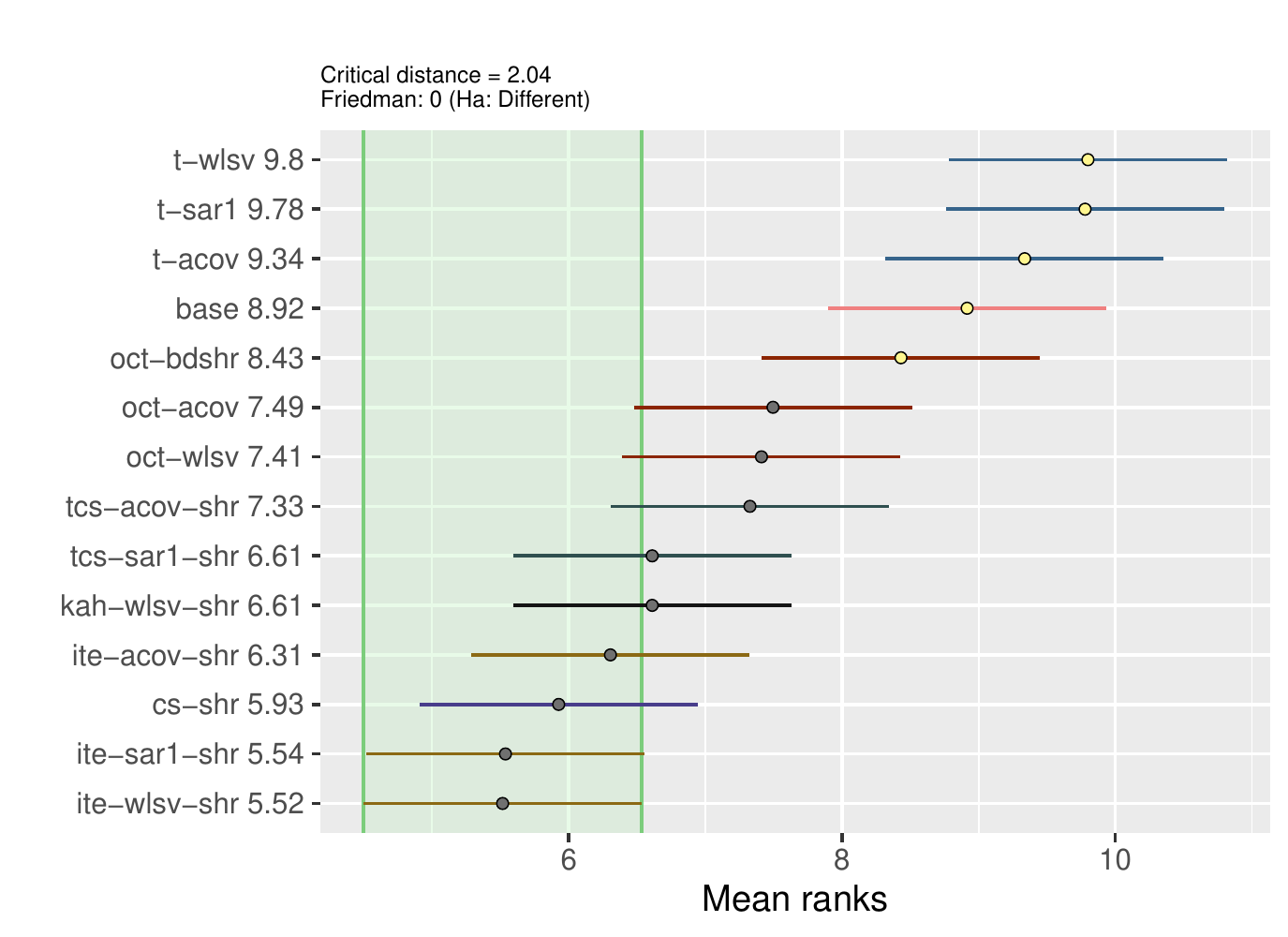}
	\caption{Nemenyi test results at 5\% significance level for all 95 series. The reconciliation procedures are sorted vertically according to the MAE mean rank (ii) across all time frequencies and forecast horizons (top), and (ii) for one-step-ahead quarterly forecasts (bottom).}
	\label{fig:nem_best_mae}
\end{figure}

\begin{figure}[ht]
	\centering
	\includegraphics[width=\linewidth]{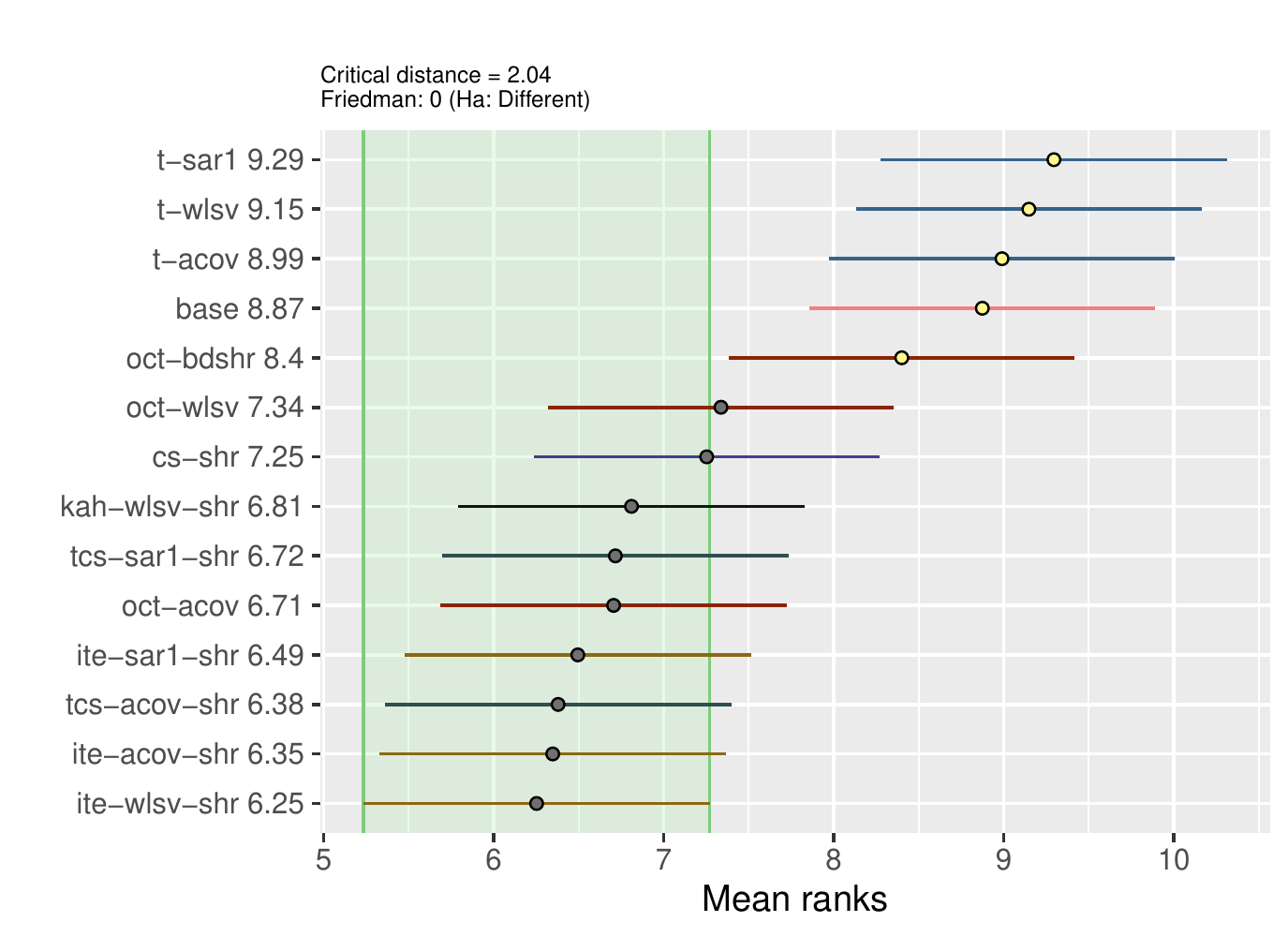}
	\includegraphics[width=\linewidth]{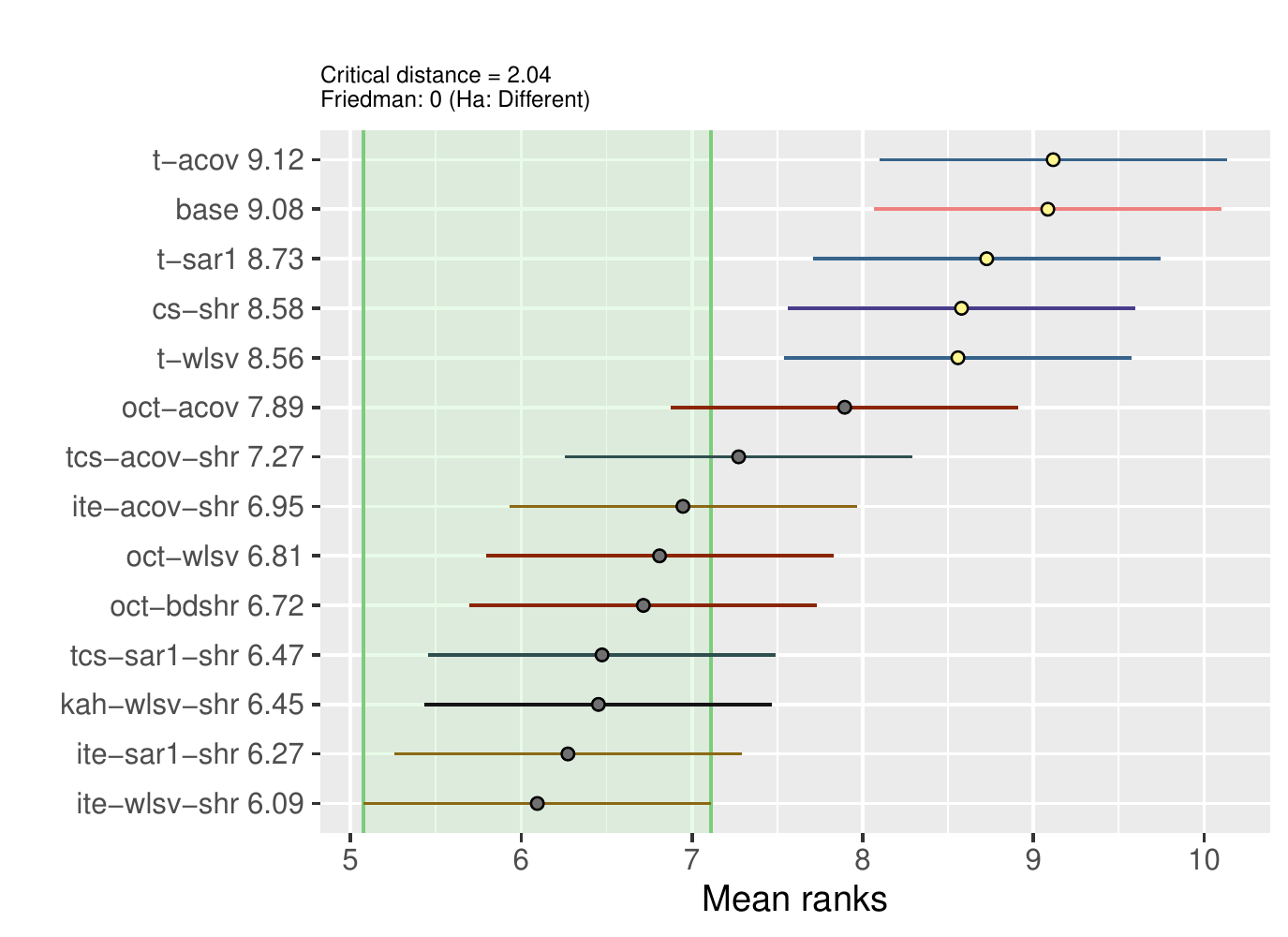}
	\caption{Nemenyi test results at 5\% significance level for all 95 series. The reconciliation procedures are sorted vertically according to the MSE mean rank for two-step-ahead (top) and three-step-ahead (bottom) quarterly forecasts.}
	\label{fig:nem_best_mae}
\end{figure}

\begin{figure}[ht]
	\centering
	\includegraphics[width=\linewidth]{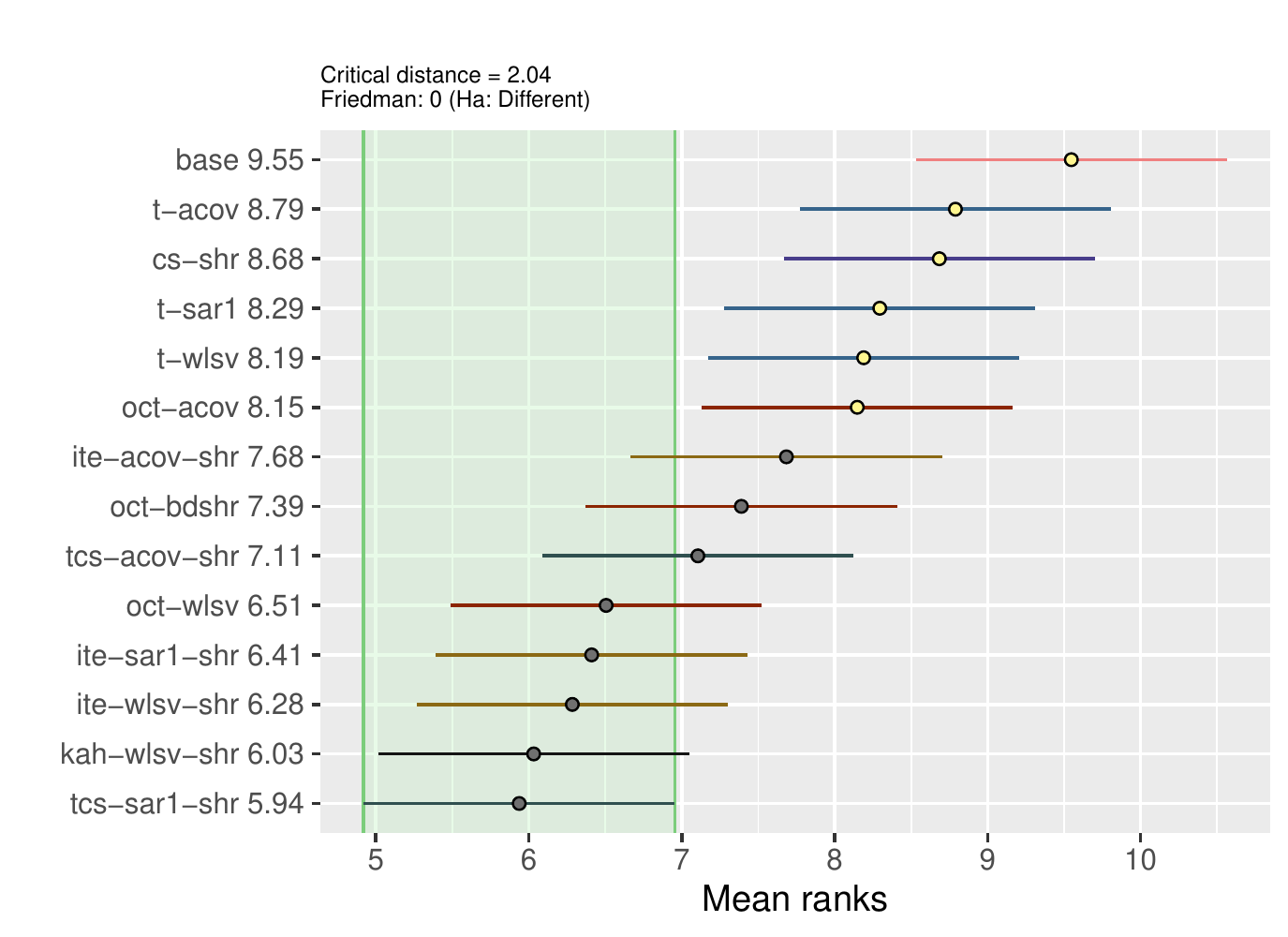}
	\includegraphics[width=\linewidth]{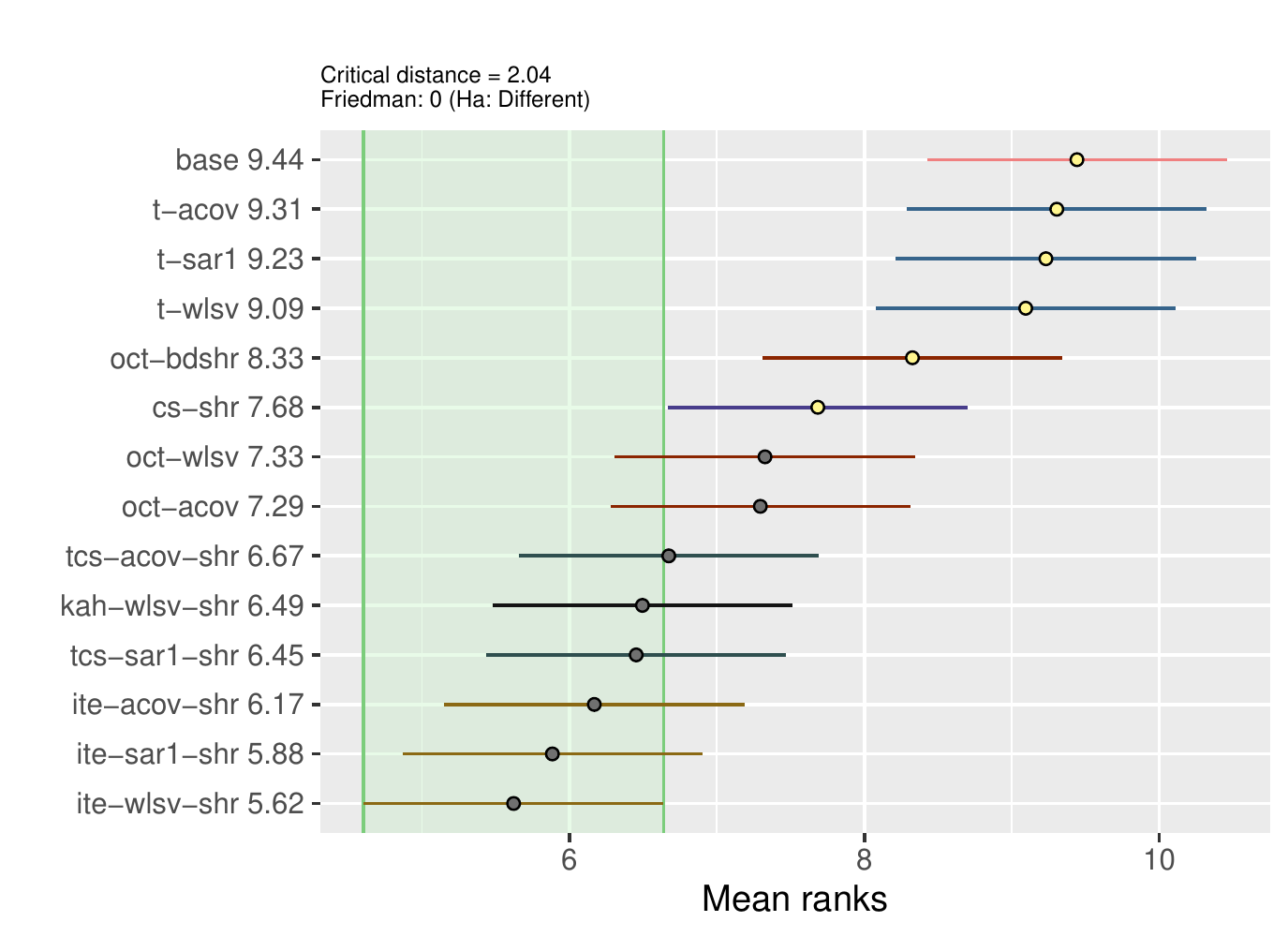}
	\caption{Nemenyi test results at 5\% significance level for all 95 series. The reconciliation procedures are sorted vertically according to the MSE mean rank for four-step-ahead (top) and one-to-four-step-ahead (bottom) quarterly forecasts.}
	\label{fig:nem_best_mae}
\end{figure}

\begin{figure}[ht]
	\centering
	\includegraphics[width=\linewidth]{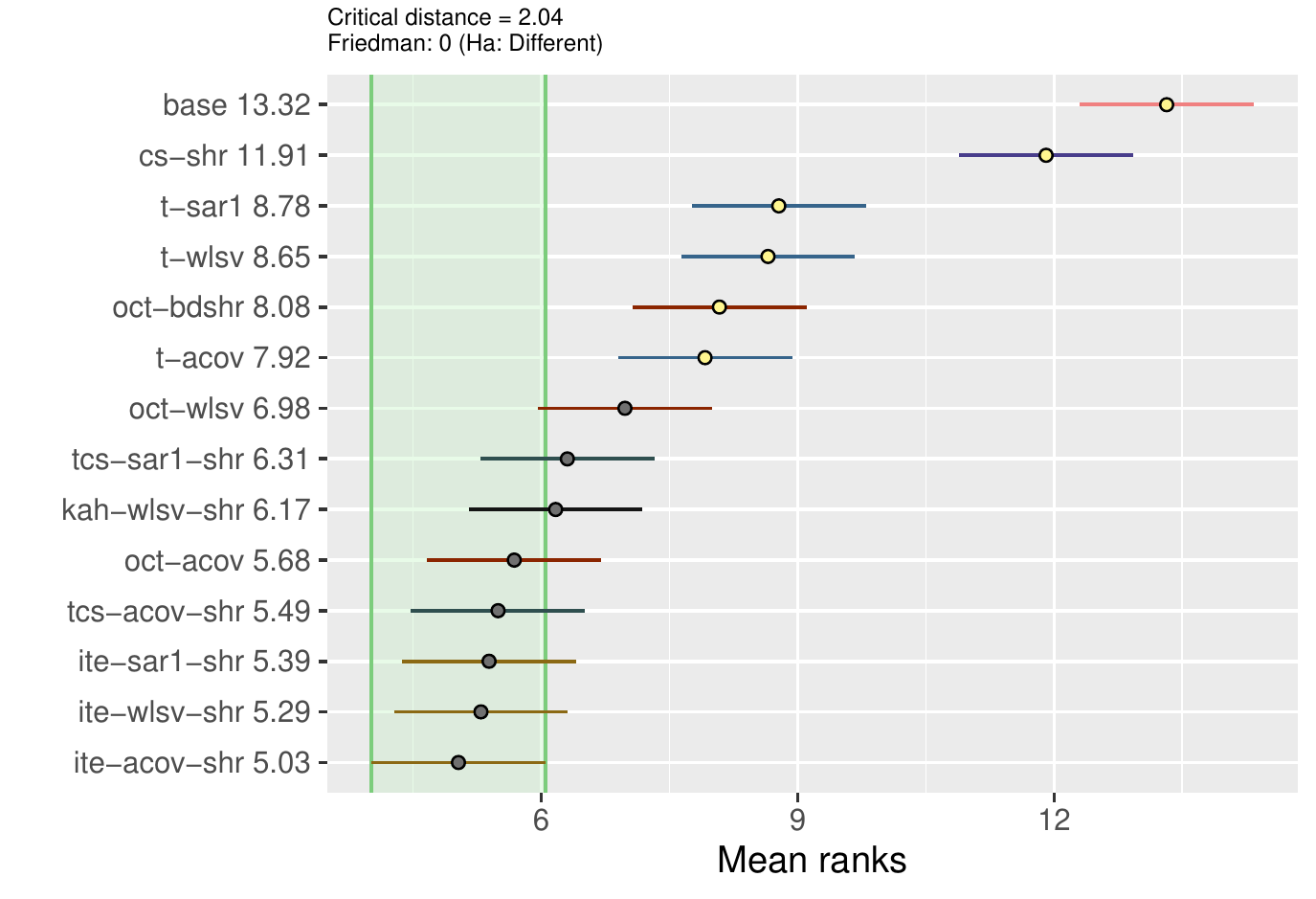}
	\includegraphics[width=\linewidth]{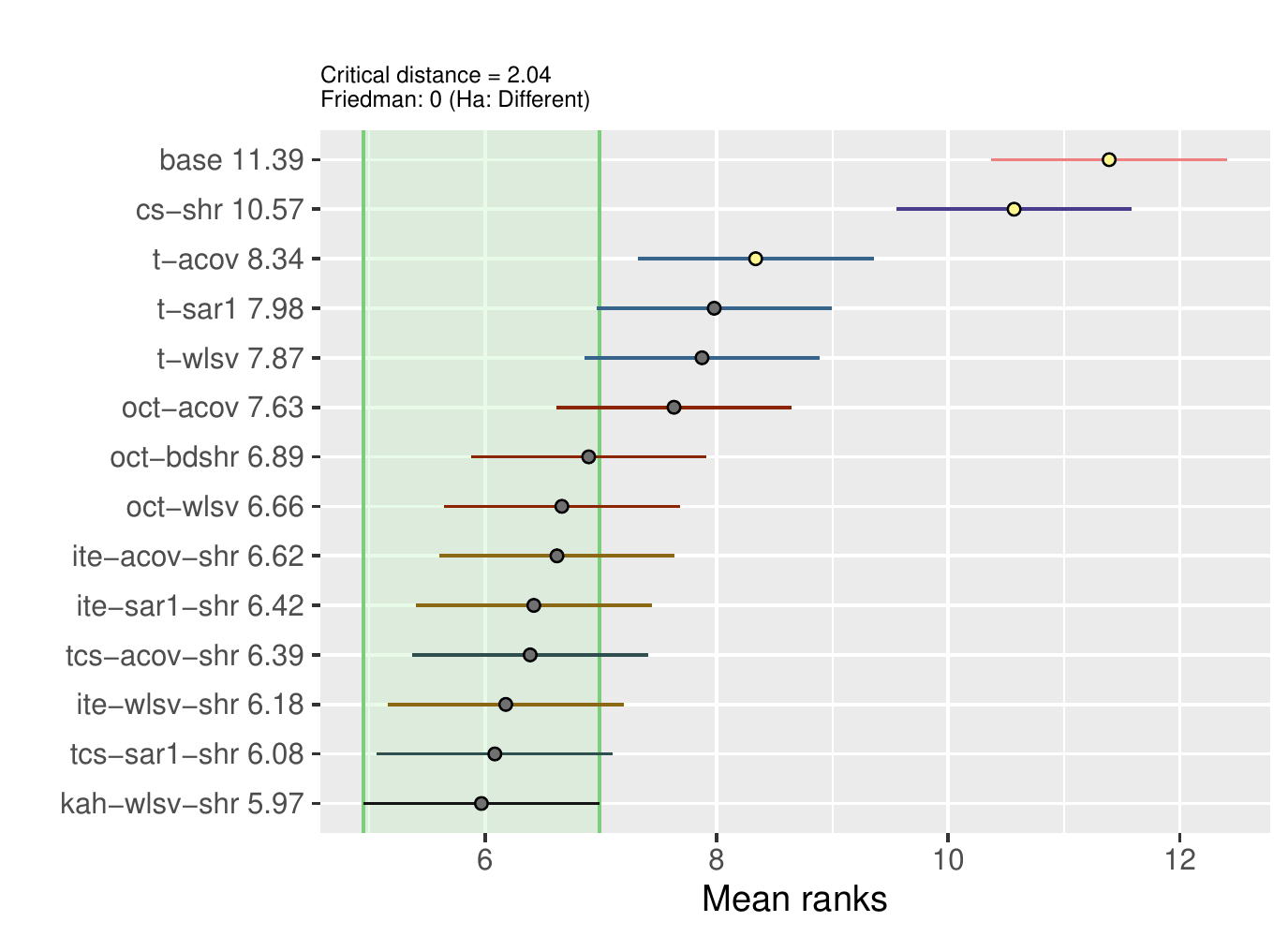}
	\caption{Nemenyi test results at 5\% significance level for all 95 series. The reconciliation procedures are sorted vertically according to the MSE mean rank for one-step-ahead (top) and two-step-ahead (bottom) six-months forecasts.}
	\label{fig:nem_best_mae}
\end{figure}

\begin{figure}[ht]
	\centering
	\includegraphics[width=\linewidth]{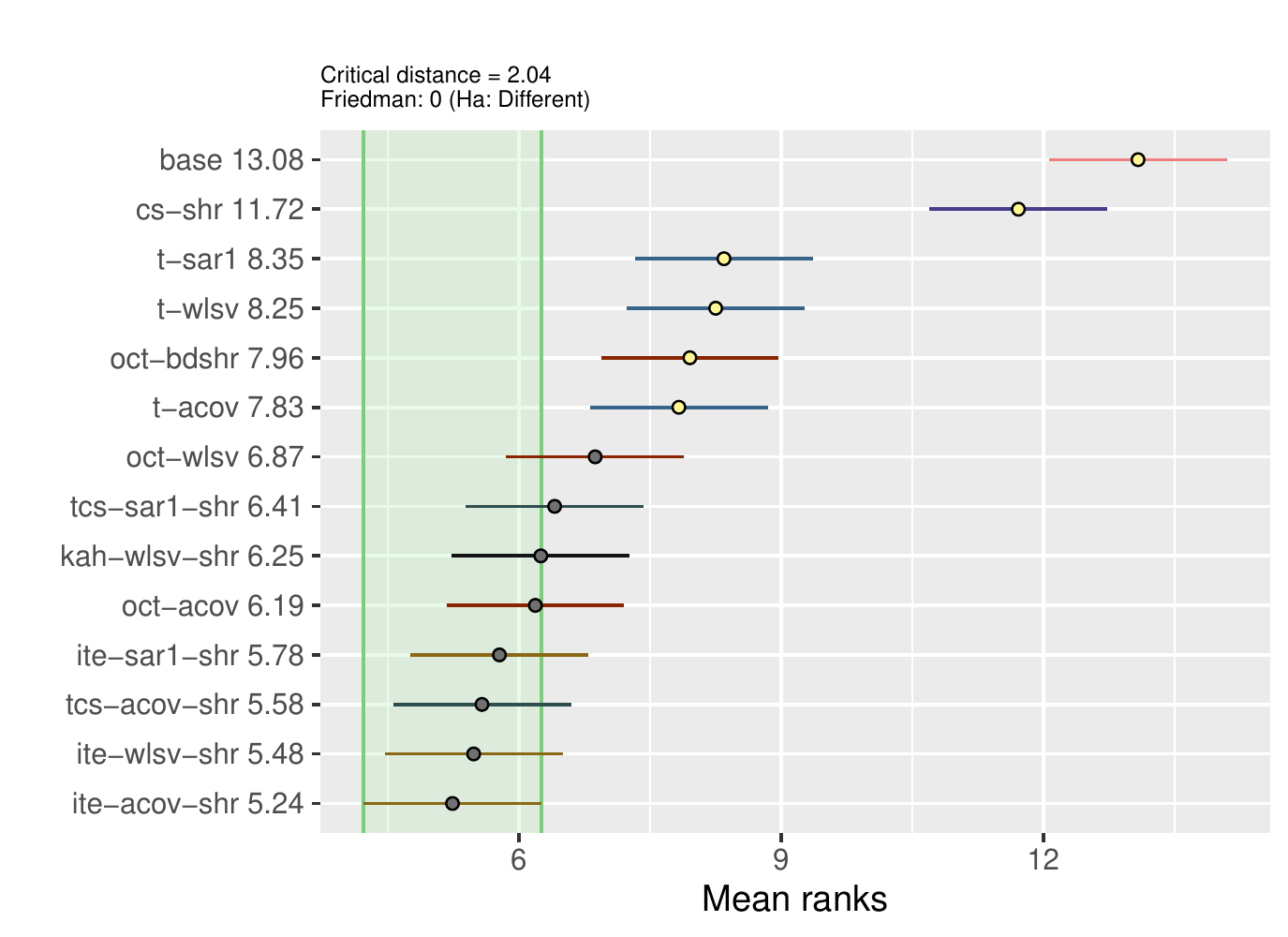}
	\includegraphics[width=\linewidth]{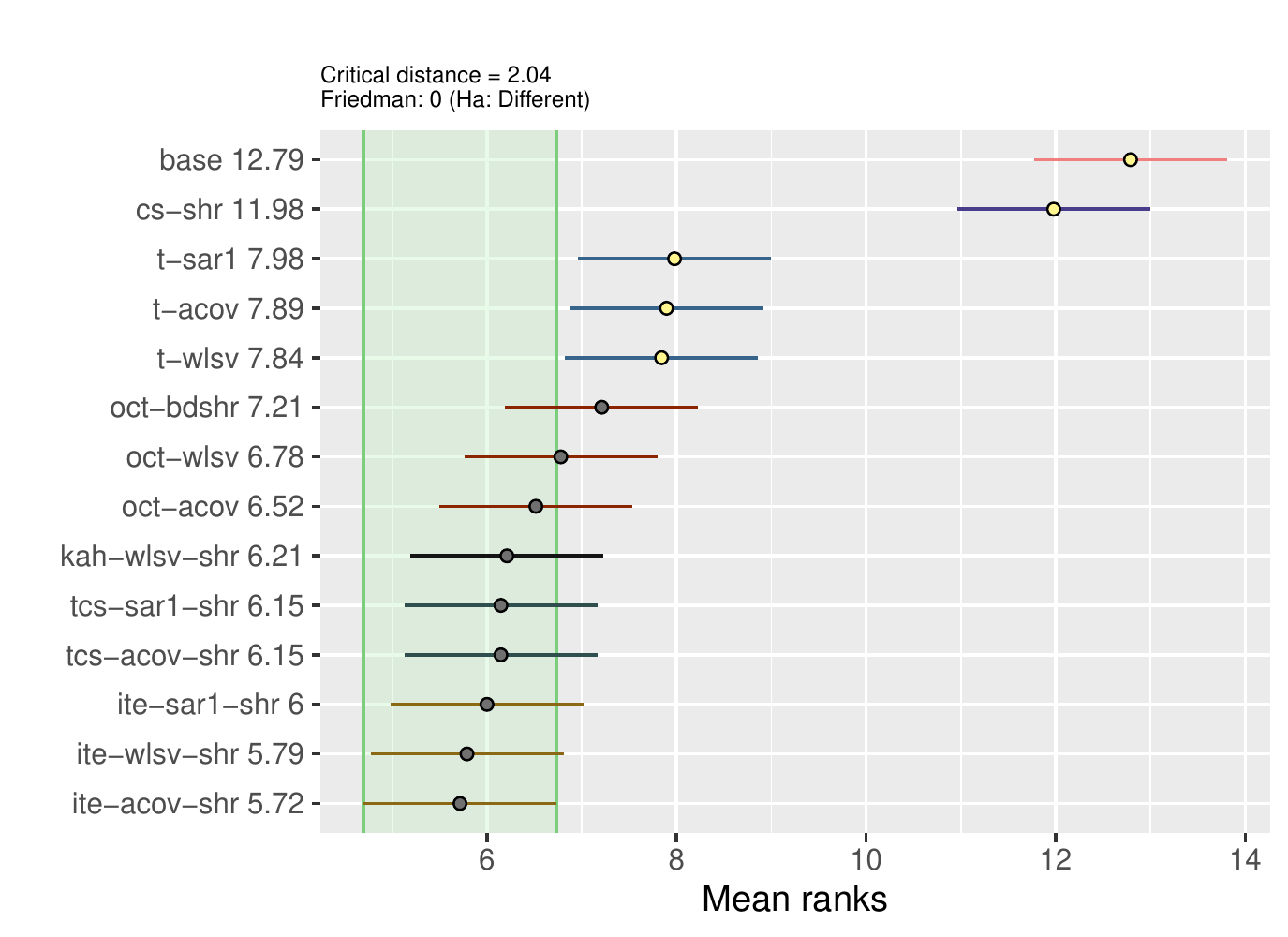}
	\caption{Nemenyi test results at 5\% significance level for all 95 series. The reconciliation procedures are sorted vertically according to the MSE mean rank for one-to-two-step-ahead (top) six-months forecasts and one-step-ahead twelve-months forecasts (bottom).}
	\label{fig:nem_best_mae}
\end{figure}

\begin{figure}[ht]
	\centering
	\includegraphics[width=\linewidth]{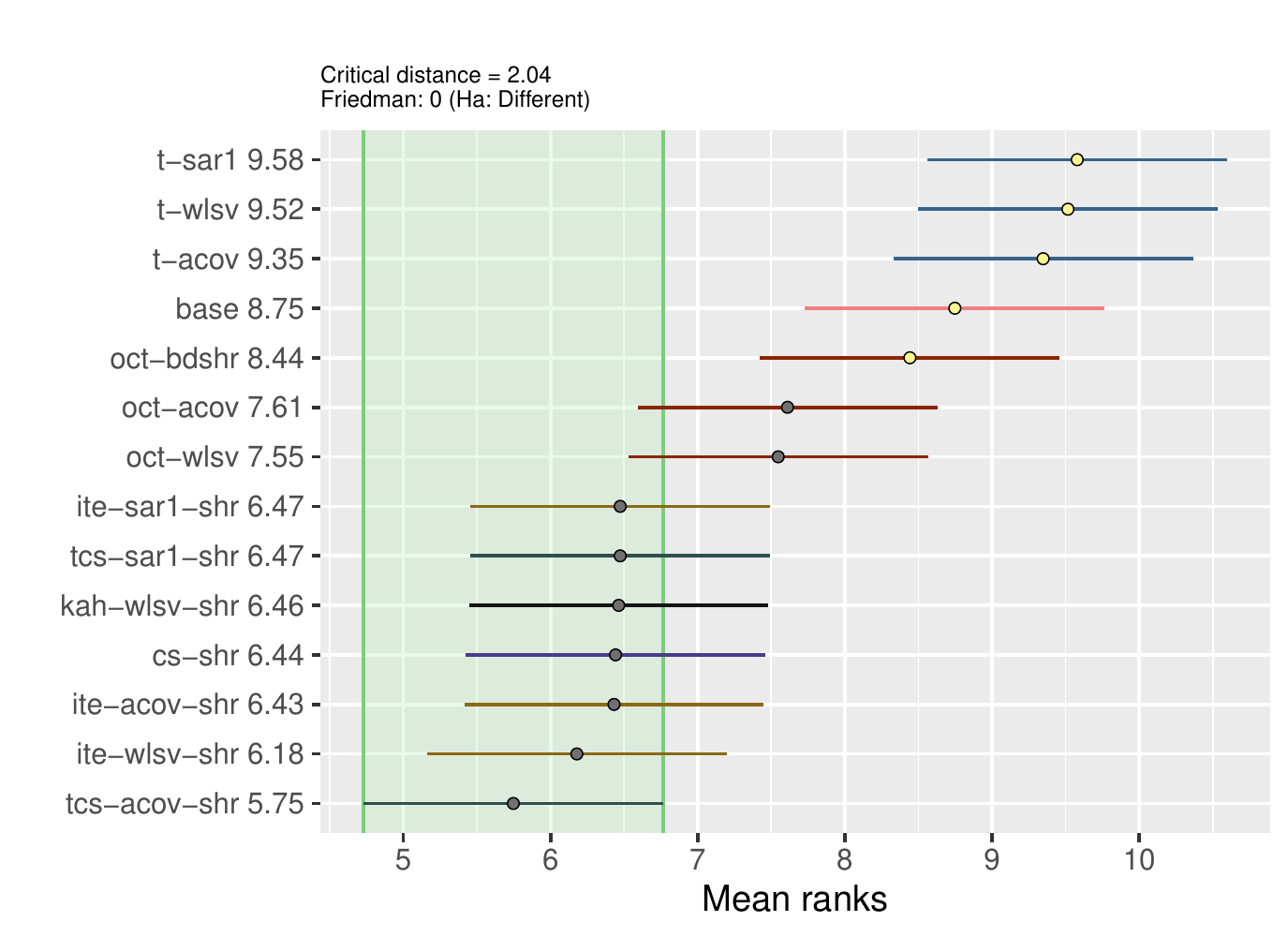}
	\includegraphics[width=\linewidth]{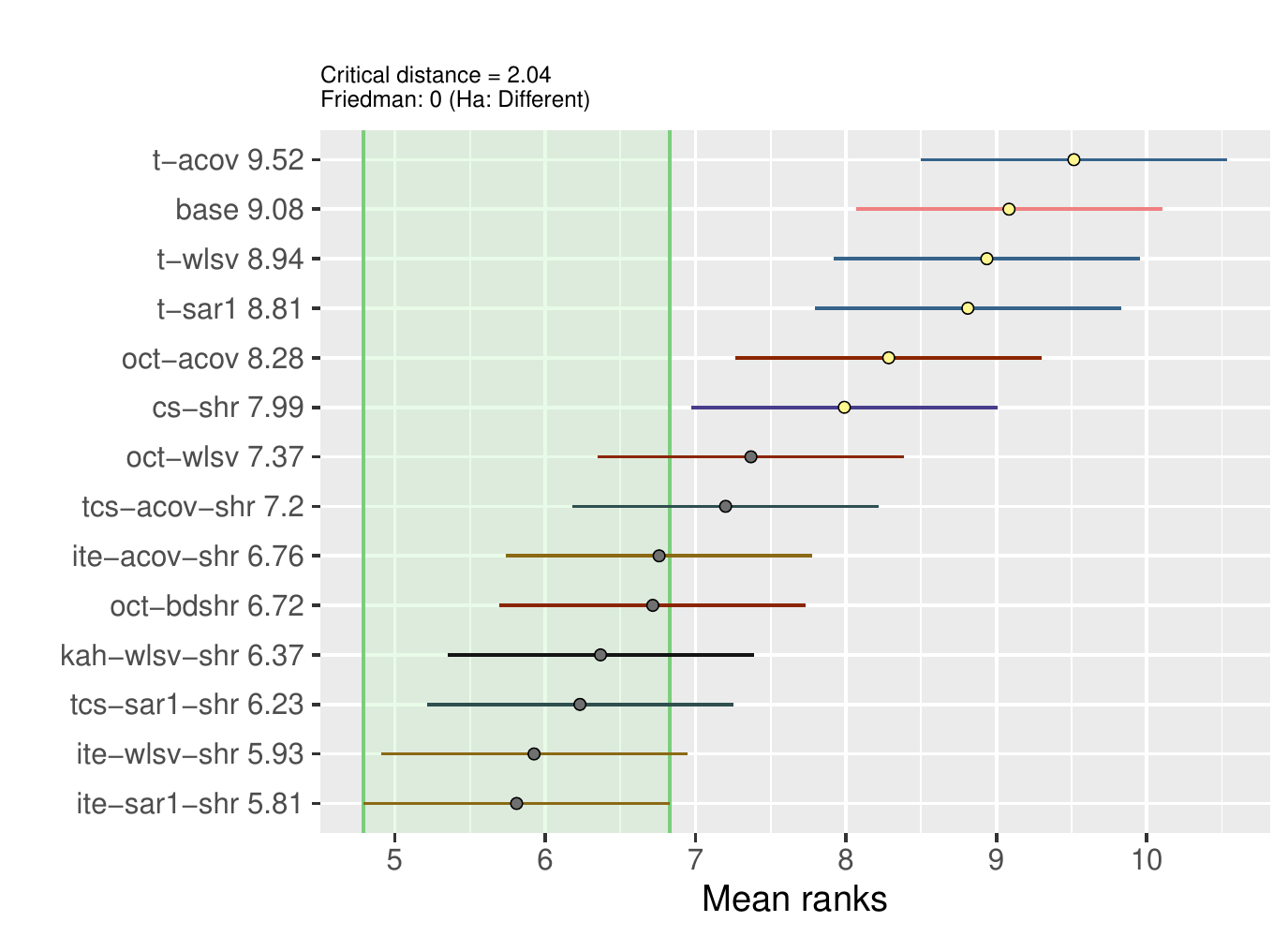}
	\caption{Nemenyi test results at 5\% significance level for all 95 series. The reconciliation procedures are sorted vertically according to the MAE mean rank for two-step-ahead (top) and three-step-ahead (bottom) quarterly forecasts.}
	\label{fig:nem_best_mae}
\end{figure}

\begin{figure}[ht]
	\centering
	\includegraphics[width=\linewidth]{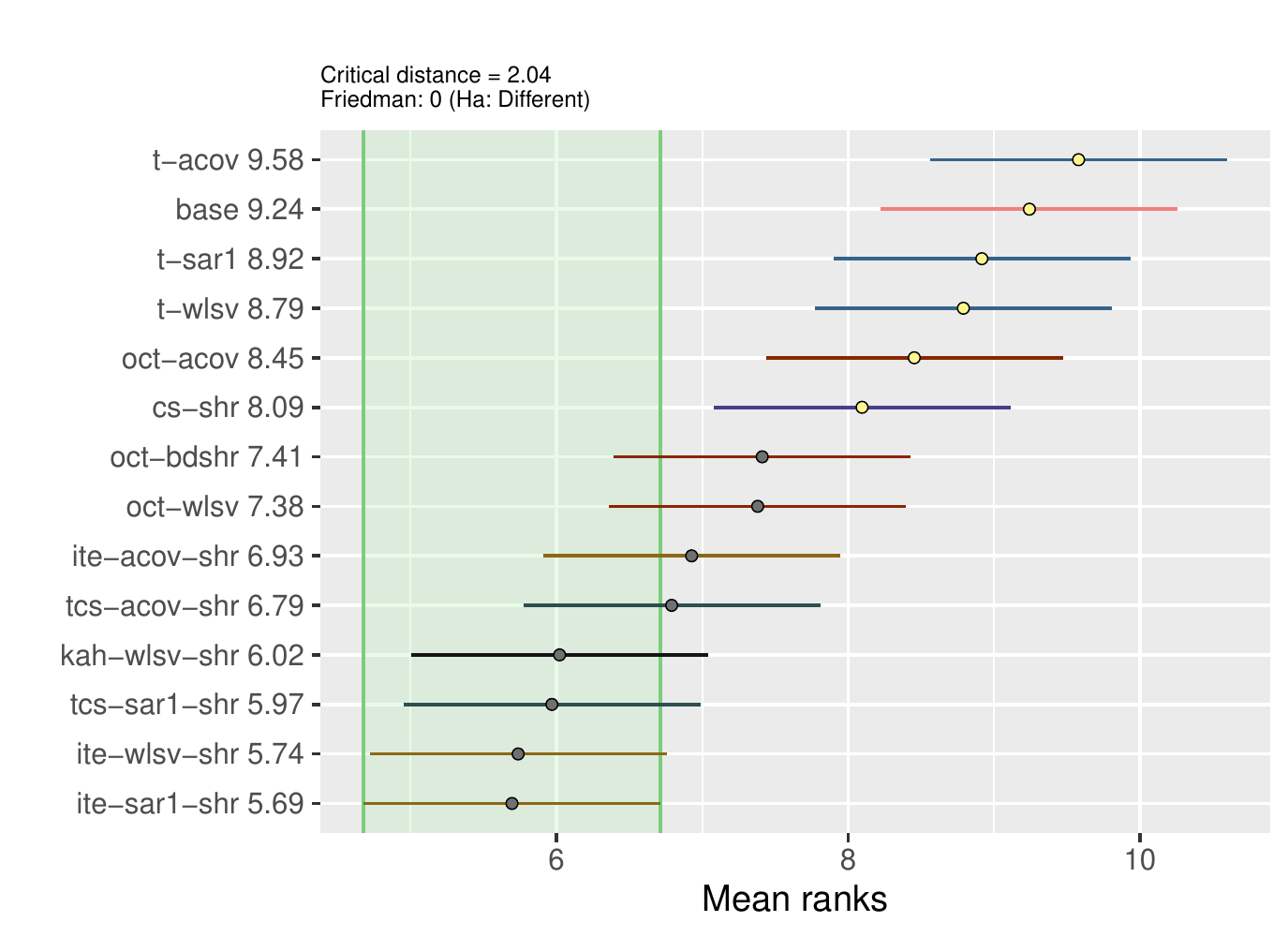}
	\includegraphics[width=\linewidth]{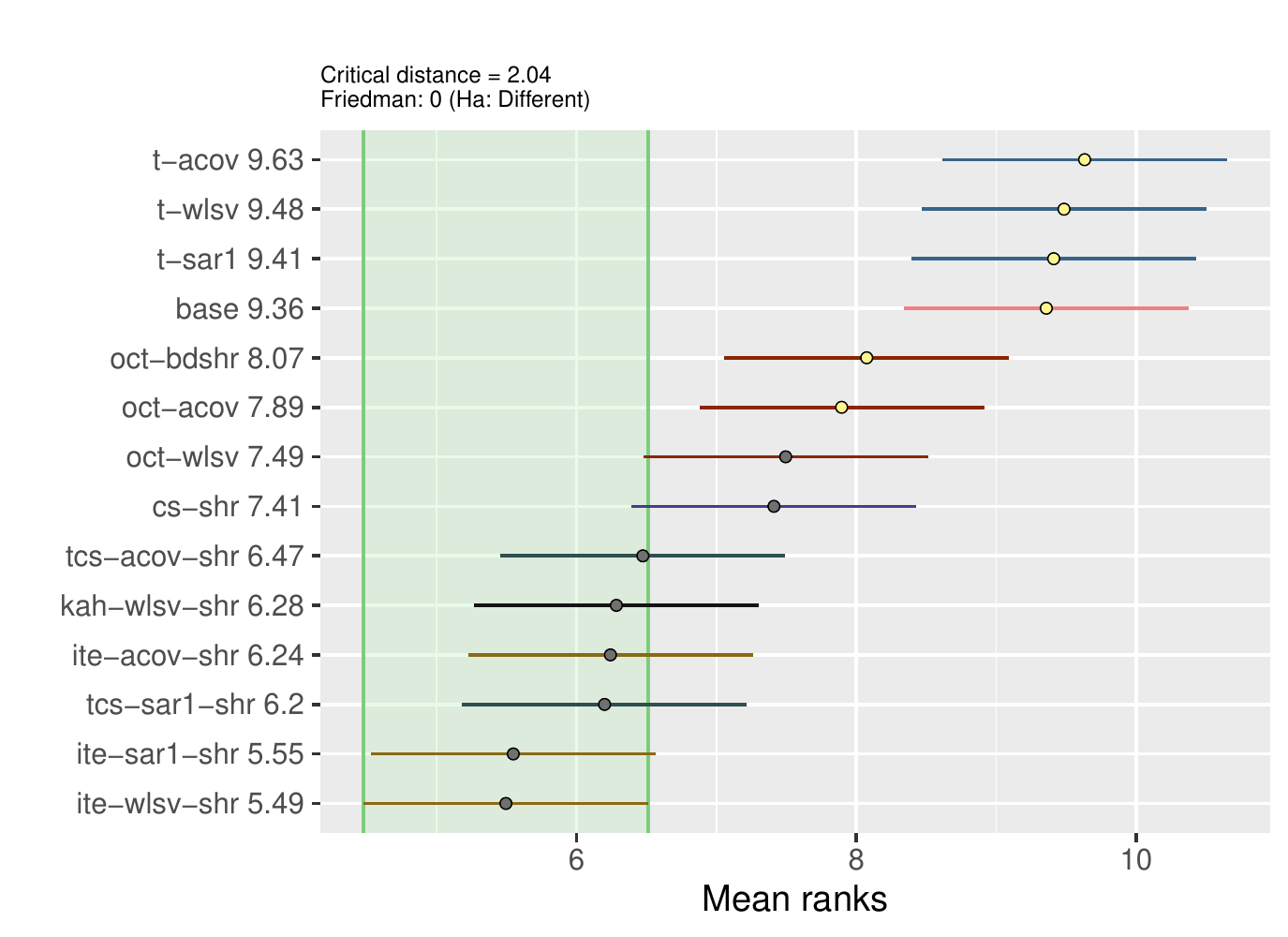}
	\caption{Nemenyi test results at 5\% significance level for all 95 series. The reconciliation procedures are sorted vertically according to the MAE mean rank for four-step-ahead (top) and one-to-four-step-ahead (bottom) quarterly forecasts.}
	\label{fig:nem_best_mae}
\end{figure}

\begin{figure}[ht]
	\centering
	\includegraphics[width=\linewidth]{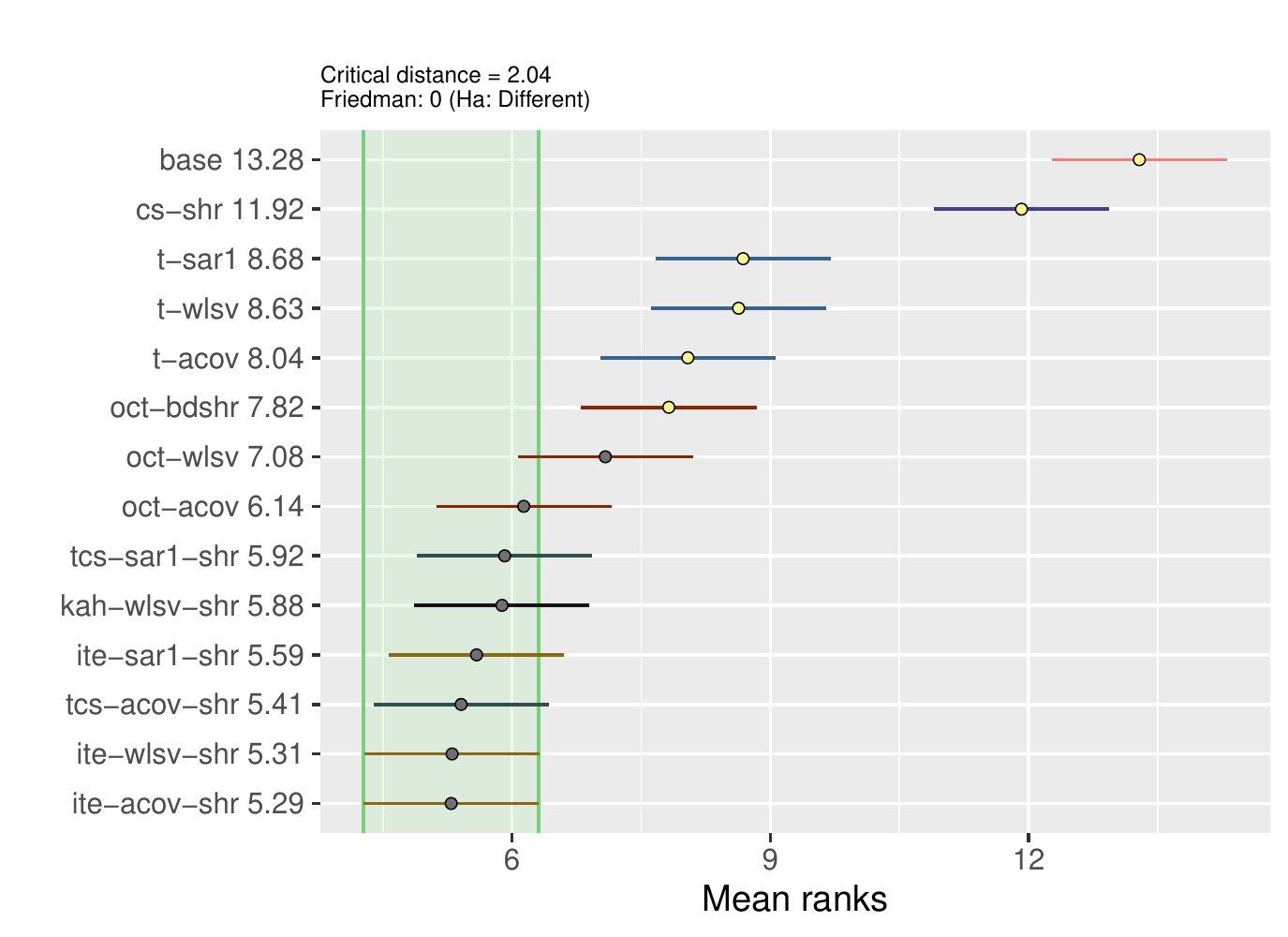}
	\includegraphics[width=\linewidth]{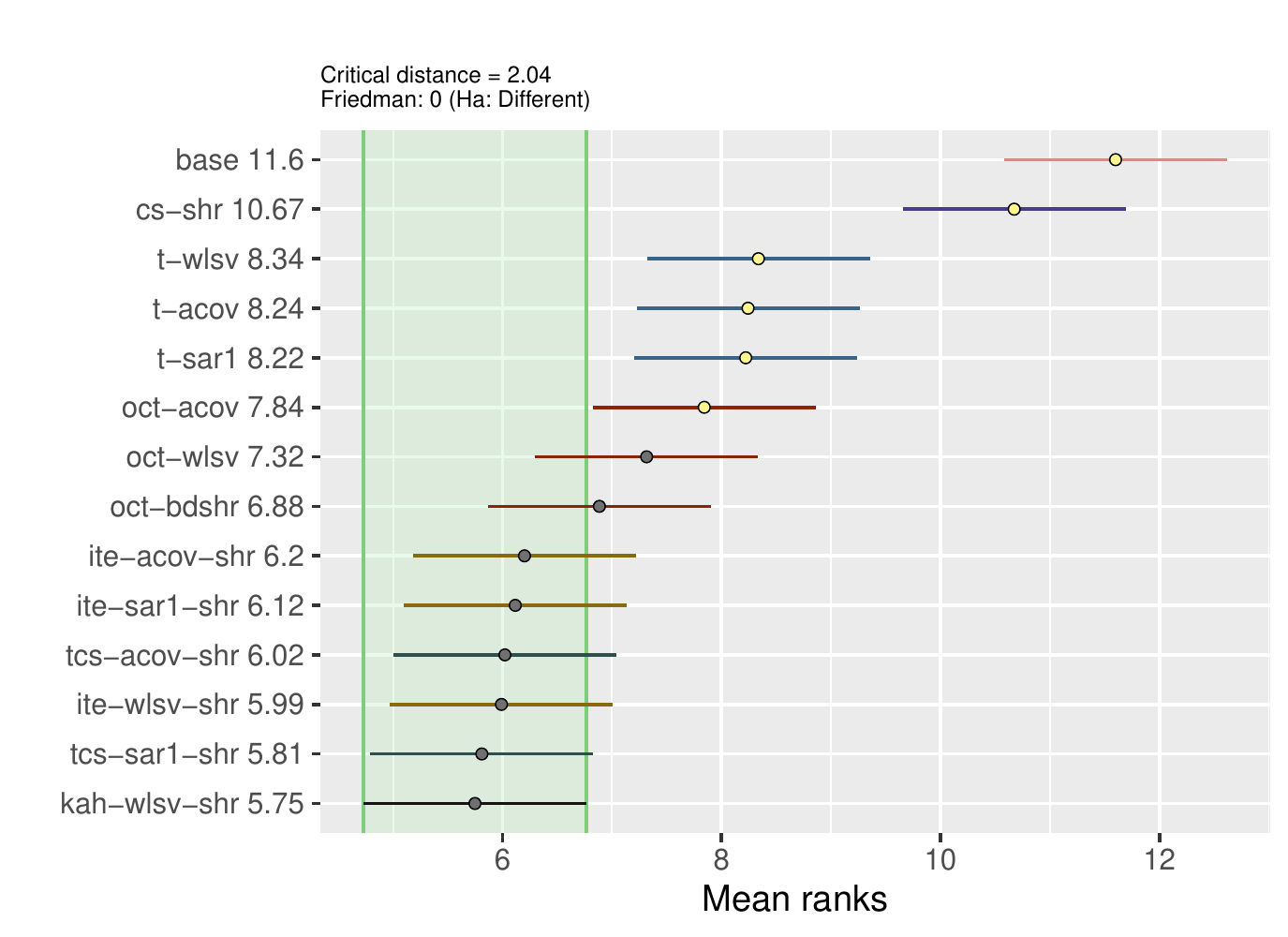}
	\caption{Nemenyi test results at 5\% significance level for all 95 series. The reconciliation procedures are sorted vertically according to the MAE mean rank for one-step-ahead (top) and two-step-ahead (bottom) six-months forecasts.}
	\label{fig:nem_best_mae}
\end{figure}

\begin{figure}[ht]
	\centering
	\includegraphics[width=\linewidth]{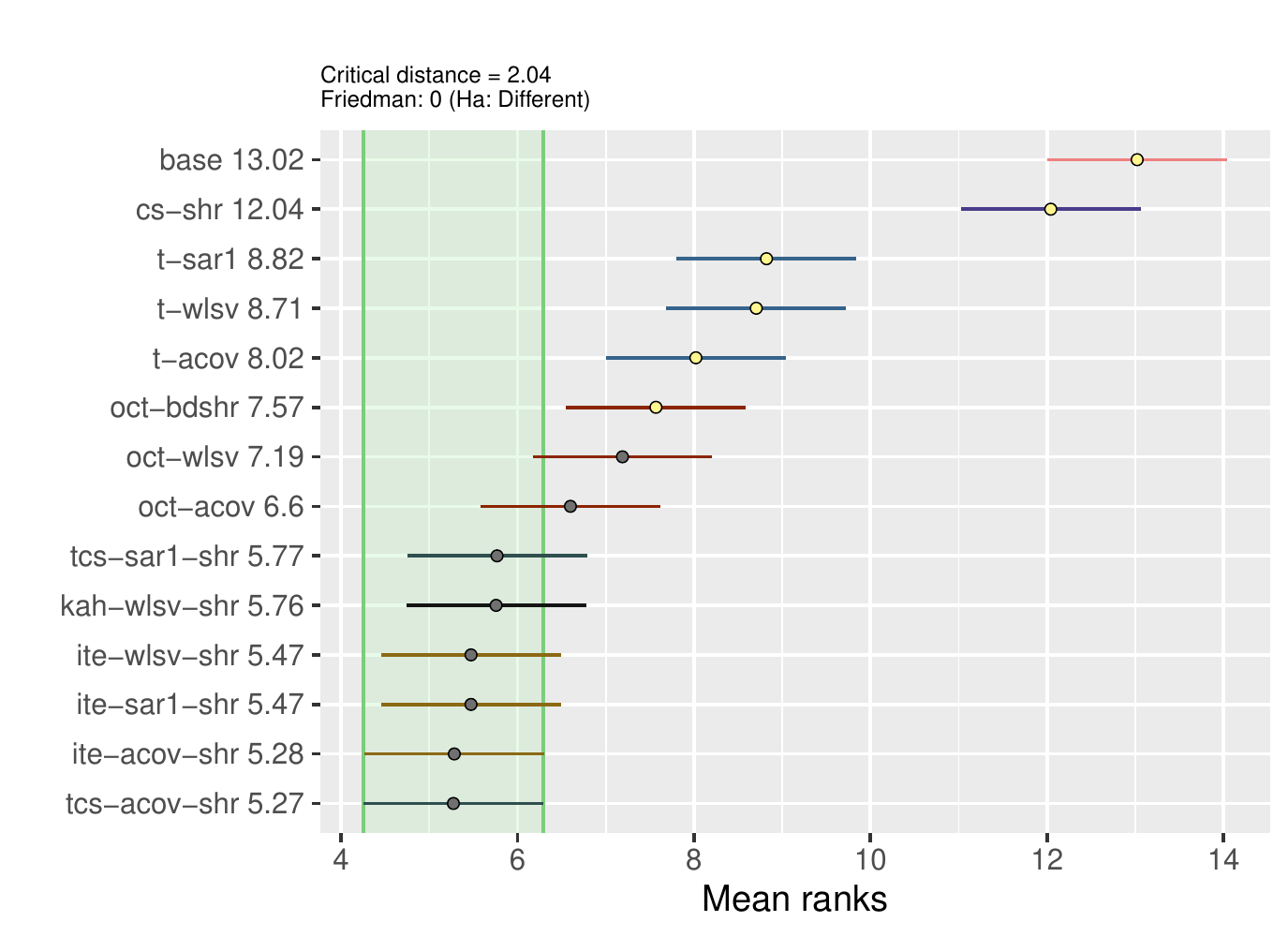}
	\includegraphics[width=\linewidth]{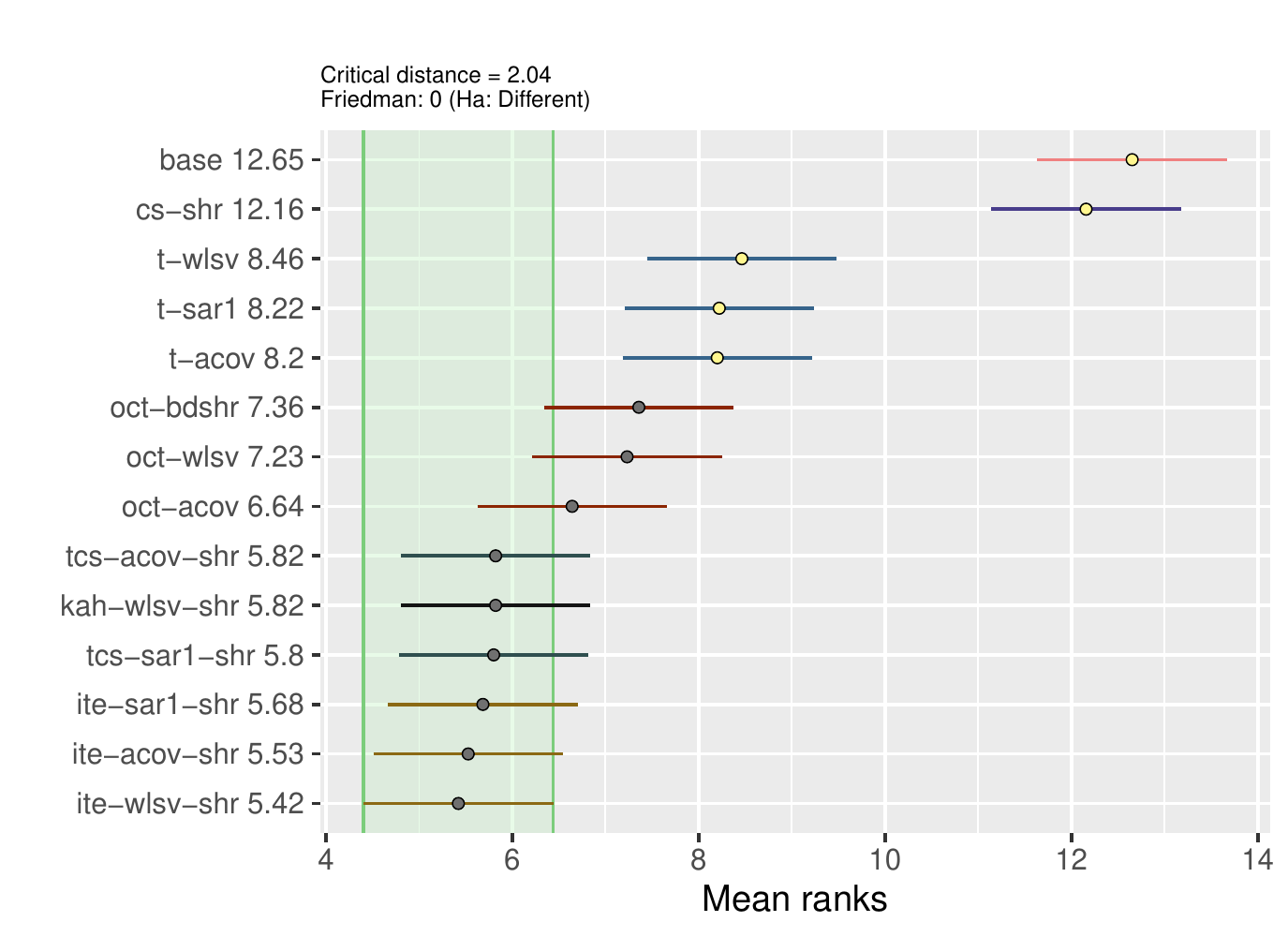}
	\caption{Nemenyi test results at 5\% significance level for all 95 series. The reconciliation procedures are sorted vertically according to the MAE mean rank for one-to-two-step-ahead (top) six-months forecasts and one-step-ahead twelve-months forecasts (bottom).}
	\label{fig:nem_best_mae}
\end{figure}

\clearpage
\begin{table}[ht]
\centering
\caption{AvgRelMSE at any temporal aggregation level and any forecast horizon for all 95 time series and selected reconciliation procedures.} 
\label{hseriesMSE}
\resizebox{\linewidth}{!}{
\begin{tabular}{r|cccccccccc}
\hline
			& \multicolumn{5}{c|}{\textbf{Quarterly}} & \multicolumn{3}{c|}{\textbf{Semi-annual}} & \multicolumn{1}{c|}{\textbf{Annual}} & \textbf{All}\\
			\hline
			\textbf{Series} & 1 & 2 & 3 & 4 & 1-4 & 
			1 & 2 & 1-2 & 1 & \\
  \hline
  \multicolumn{11}{c}{\addstackgap[10pt]{\emph{cs-shr}}} \\
Gdp & 0.9740 & 0.9397 & 0.9028 & 0.8924 & 0.9267 & 0.8382 & 0.8713 & 0.8546 & 0.7116 & 0.8719 \\ 
  Tfi & 0.8316 & 0.9002 & 0.8769 & 0.8335 & 0.8600 & 0.8232 & 0,8760 & 0.8492 & 0.7162 & 0.8348 \\ 
  TfiGos & 0.9170 & 0.8834 & 0.9140 & 0.9008 & 0.9037 & \textbf{1.0425} & \textbf{1.0489} & \textbf{1.0457} & 0.9354 & 0.9468 \\ 
  TfiCoe & 0.9416 & 0.9880 & \textbf{1.0172} & \textbf{1.0112} & 0.9891 & \textbf{1.0519} & \textbf{1.0820} & \textbf{1.0669} & \textbf{1.0488} & \textbf{1.0192} \\ 
  TfiGosCop & 0.8994 & 0.9722 & \textbf{1.0203} & \textbf{1.0302} & 0.9791 & 0.9818 & \textbf{1.0060} & 0.9938 & \textbf{1.0113} & 0.9879 \\
\end{tabular}}
\end{table}

\begin{table}[ht]
\centering
\resizebox{\linewidth}{!}{
\begin{tabular}{r|cccccccccc}
  GneDfdFceHfcAbtCig & 0.8864 & 0.9089 & 0.9201 & 0.9864 & 0.9247 & 0.9083 & 0.9672 & 0.9373 & 0.9889 & 0.9372 \\ 
  GneDfdFceHfcMisOgd & 0.9932 & 0.9826 & 0.9538 & 0.9366 & 0.9663 & \textbf{1.0094} & 0.9896 & 0.9994 & \textbf{1.0034} & 0.9809 \\ 
  GneDfdFceHfcMisOsv & \textbf{1.0009} & 0.9946 & 0.9920 & 0.9839 & 0.9929 & 0.9380 & \textbf{1.0020} & 0.9695 & \textbf{1.0061} & 0.9880 \\ 
  GneDfdFceHfcMisIfs & 0.9113 & 0.9297 & 0.8367 & 0.8769 & 0.8879 & 0.8905 & 0.9017 & 0.8961 & 0.8831 & 0.8895 \\ 
  GneDfdFceHfcTptTsv & 0.9585 & 0.9832 & 0.9686 & 0.9521 & 0.9655 & \textbf{1.0157} & \textbf{1.0761} & \textbf{1.0455} & \textbf{1.0196} & 0.9954 \\ 
  GneDfdFceHfcTptPvh & 0.9766 & 0.9906 & 0.9845 & \textbf{1.0074} & 0.9897 & 0.9297 & \textbf{1.1746} & \textbf{1.0450} & \textbf{1.0152} & \textbf{1.0089} \\ 
  GneDfdFceHfcTptOvh & 0.9588 & 0.9439 & 0.9461 & 0.9333 & 0.9455 & \textbf{1.0326} & \textbf{1.1102} & \textbf{1.0707} & \textbf{1.0268} & 0.9913 \\ 
  GneDfdFceHfcHcrAsv & \textbf{1.0063} & 0.9681 & 0.9700 & 0.9930 & 0.9842 & 0.9942 & 0.9783 & 0.9862 & 0.9517 & 0.9801 \\
\end{tabular}}
\end{table}
\begin{table}[ht]
\centering
\resizebox{\linewidth}{!}{
\begin{tabular}{r|cccccccccc}
  TfiGosGvt & 0.4874 & 0.7062 & \textbf{1.3241} & \textbf{1.0692} & 0.8355 & 0.244 & 0.7633 & 0.4316 & 0.2685 & 0.5883 \\ 
  TfiGosDwl & \textbf{1.0092} & 0.9711 & 0.9639 & \textbf{1.0151} & 0.9896 & 0.8787 & 0.9649 & 0.9208 & 0.5876 & 0.8998 \\ 
  TfiGmi & \textbf{1.0412} & 0.9778 & 0.9906 & 0.9929 & \textbf{1.0004} & 0.8936 & 0.9638 & 0.9280 & 0.8520 & 0.9569 \\ 
  TfiCoeWns & \textbf{1.0779} & \textbf{1.0581} & \textbf{1.0743} & \textbf{1.0515} & \textbf{1.0654} & 0.7286 & 0.8438 & 0.7841 & 0.5622 & 0.8908 \\ 
  TfiCoeEsc & \textbf{1.0415} & \textbf{1.0569} & \textbf{1.0257} & \textbf{1.0039} & \textbf{1.0318} & 0.7958 & 0.9304 & 0.8604 & 0.6295 & 0.9129 \\ 
  Tsi & 0.9932 & \textbf{1.0096} & \textbf{1.0240} & \textbf{1.0014} & \textbf{1.0070} & 0.8515 & 0.8890 & 0.8700 & 0.7989 & 0.9344 \\ 
  Sdi & \textbf{1.0270} & \textbf{1.0139} & \textbf{1.0706} & \textbf{1.0304} & \textbf{1.0353} & 0.9412 & 0.9247 & 0.9329 & 0.9245 & 0.9888 \\ 
  GneDfdFceGvtNatNdf & 0.9154 & 0.9423 & 0.9137 & 0.9653 & 0.9339 & 0.9567 & 0.9529 & 0.9548 & 0.8692 & 0.9303 \\
\end{tabular}}
\end{table}
\begin{table}[ht]
\centering
\resizebox{\linewidth}{!}{
\begin{tabular}{r|cccccccccc}
  GneDfd & 0.9112 & \textbf{1.0152} & \textbf{1.0136} & \textbf{1.0039} & 0.9850 & 0.8184 & 0.9562 & 0.8846 & 0.6747 & 0.9049 \\ 
  GneCii & \textbf{1.0204} & \textbf{1.0339} & \textbf{1.0163} & 0.9467 & \textbf{1.0037} & 0.6644 & 0.6795 & 0.6719 & 0.8369 & 0,8720 \\ 
  GneDfdFce & 0.9370 & 0.9094 & 0.9000 & 0.8956 & 0.9104 & 0.8077 & 0.8395 & 0.8234 & 0.6708 & 0.8469 \\ 
  GneDfdGfc & 0.9875 & 0.9790 & 0.9768 & 0.9663 & 0.9774 & 0.8480 & 0.9829 & 0.9130 & 0.7052 & 0.9149 \\ 
  GneCiiPnf & 0.8501 & 0.8479 & 0.8625 & 0.9379 & 0.8738 & 0.7933 & 0.7834 & 0.7883 & 0.7799 & 0.8348 \\ 
  GneDfdFceGvt & 0.8485 & 0.8657 & 0.8652 & 0.9246 & 0.8755 & 0.8663 & 0.8580 & 0.8621 & 0.7005 & 0.8443 \\ 
  GneDfdFceHfc & \textbf{1.0250} & \textbf{1.0672} & \textbf{1.0255} & 0.9787 & \textbf{1.0236} & 0.7785 & 0.7895 & 0.7840 & 0.5944 & 0.8776 \\ 
  GneDfdGfcPub & 0.8643 & 0.9285 & 0.9176 & 0.9409 & 0.9124 & 0.7842 & 0.8682 & 0.8251 & 0.8702 & 0.8806 \\
\end{tabular}}
\end{table}
\begin{table}[ht]
\centering
\resizebox{\linewidth}{!}{
\begin{tabular}{r|cccccccccc}
  GneDfdFceHfcHweRnt & \textbf{1.1249} & \textbf{1.0070} & 0.9527 & 0.9535 & \textbf{1.0072} & 0.7317 & 0.7403 & 0.7360 & 0.3396 & 0.7884 \\ 
  GneDfdFceHfcHweWsc & 0.9278 & 0.8946 & 0.8863 & 0.9039 & 0,9030 & 0.9905 & \textbf{1.0043} & 0.9973 & 0.7529 & 0.9052 \\ 
  GneDfdFceHfcHweEgf & 0.9421 & 0.9792 & 0.9894 & 0.9650 & 0.9687 & 0.8533 & 0.9184 & 0.8853 & 0.8144 & 0.9210 \\ 
  GneDfdFceHfcFud & \textbf{1.0435} & \textbf{1.0365} & 0.9907 & 0.9714 & \textbf{1.0101} & 0.7503 & \textbf{1.0102} & 0.8706 & 0.8632 & 0.9466 \\ 
  GneDfdFceHfcCnf & 0.9746 & \textbf{1.0060} & 0.9876 & 0.9926 & 0.9901 & 0.9570 & 0.9684 & 0.9627 & 0.8526 & 0.9614 \\ 
  GneDfdFceHfcRnc & 0.9749 & 0.9896 & 0.9583 & 0.9592 & 0.9704 & 0.8803 & 0.9830 & 0.9302 & 0.8437 & 0.9398 \\ 
  GneDfdFceHfcEdc & \textbf{1.0082} & 0.9805 & 0.9827 & \textbf{1.0081} & 0.9948 & 0.9367 & \textbf{1.0012} & 0.9684 & 0.8615 & 0.9671 \\ 
  GneDfdFceHfcCom & \textbf{1.0925} & \textbf{1.0582} & \textbf{1.0683} & \textbf{1.0551} & \textbf{1.0684} & 0.8563 & 0.8699 & 0.8631 & 0.7969 & 0.9640 \\
\end{tabular}}
\end{table}
\begin{table}[ht]
\centering
\resizebox{\linewidth}{!}{
\begin{tabular}{r|cccccccccc}
  GneDfdGfcPvtPbiIprRnd & 0.7186 & 0.8211 & \textbf{1.3404} & \textbf{1.0933} & 0.9643 & 0.3203 & 0.8051 & 0.5078 & 0.2956 & 0.6781 \\ 
  GneDfdGfcPvtPbiIprMnp & 0.9702 & \textbf{1.0647} & \textbf{1.0146} & \textbf{1.0348} & \textbf{1.0205} & 0.9229 & 0.9941 & 0.9579 & 0.7514 & 0.9593 \\ 
  GneDfdGfcPvtPbiIprCom & 0.6086 & 0.8125 & \textbf{1.2119} & \textbf{1.0191} & 0.8840 & 0.2733 & 0.7282 & 0.4461 & 0.3397 & 0.6343 \\ 
  GneDfdGfcPvtPbiIprArt & 0.7032 & 0.7705 & \textbf{1.1453} & 0.9671 & 0.8802 & 0.3217 & 0.6735 & 0.4655 & 0.2882 & 0.6255 \\ 
  GneDfdGfcPvtPbiNdcNbd & \textbf{1.0485} & \textbf{1.1286} & \textbf{1.1367} & \textbf{1.0777} & \textbf{1.0973} & 0.8485 & \textbf{1.0544} & 0.9459 & 0.7934 & \textbf{1.0041} \\ 
  GneDfdGfcPvtPbiNdcNec & 0.8956 & 0.8537 & 0.8668 & 0.8709 & 0.8716 & 0.8109 & 0.9229 & 0.8651 & 0.5003 & 0.8034 \\ 
  GneDfdGfcPvtPbiNdcSha & 0.9375 & \textbf{1.0458} & \textbf{1.0399} & \textbf{1.1169} & \textbf{1.0330} & 0.8758 & 0.9709 & 0.9221 & 0.9583 & 0.9894 \\ 
  GneDfdGfcPvtPbiNdmNew & 0.9321 & 0.9600 & 0.9147 & 0.9322 & 0.9346 & 0.8370 & 0.9358 & 0.8850 & 0.7396 & 0.8899 \\
\end{tabular}}
\end{table}
\begin{table}[th]
\centering
\resizebox{\linewidth}{!}{
\begin{tabular}{r|cccccccccc}
  GneDfdFceHfcFhe & 0.9782 & 0.9749 & 0.9617 & 0.9142 & 0.9569 & 0.8415 & 0.7836 & 0.8121 & 0.7530 & 0.8823 \\ 
  GneDfdFceHfcHwe & 0.7829 & 0.8061 & 0.8020 & 0.8759 & 0.8160 & 0.5803 & 0.7330 & 0.6522 & 0.5549 & 0.7244 \\ 
  GneDfdGfcPubGvtNat & 0.8693 & 0.8991 & 0.8749 & 0.8987 & 0.8854 & 0.9940 & 0.9754 & 0.9846 & 0.8661 & 0.9098 \\ 
  GneDfdGfcPvtPbiIpr & 0.7368 & 0.7365 & 0.8195 & 0.8122 & 0.7752 & 0.629 & 0.7903 & 0.7051 & 0.4687 & 0.7022 \\ 
  GneDfdGfcPvtPbiNdc & 0.9218 & 0.8003 & 0.8633 & 0.8812 & 0.8655 & 0.7415 & 0.8047 & 0.7725 & 0.5741 & 0.7901 \\ 
  GneDfdGfcPvtPbiNdm & 0.9173 & 0.9770 & 0.9587 & 0.9495 & 0.9504 & 0.9245 & 0.9733 & 0.9486 & 0.7686 & 0.9215 \\ 
  TfiGosCopNfnPub & \textbf{1.0016} & \textbf{1.0412} & \textbf{1.0607} & \textbf{1.0340} & \textbf{1.0341} & 0.9171 & 0.9816 & 0.9488 & 0.7800 & 0.9692 \\ 
  TfiGosCopNfnPvt & 0.9434 & \textbf{1.0018} & 0.9856 & \textbf{1.0337} & 0.9906 & 0.8472 & 0.9728 & 0.9078 & 0.8974 & 0.9527 \\
\hline
\end{tabular}}
\end{table}

\begin{table}[ht]
\centering
\caption{AvgRelMAE at any temporal aggregation level and any forecast horizon for all 95 time series and selected reconciliation procedures.} 
\label{hseriesMAE}
\resizebox{\linewidth}{!}{
\begin{tabular}{r|cccccccccc}
\hline
			& \multicolumn{5}{c|}{\textbf{Quarterly}} & \multicolumn{3}{c|}{\textbf{Semi-annual}} & \multicolumn{1}{c|}{\textbf{Annual}} & \textbf{All}\\
			\hline
			\textbf{Series} & 1 & 2 & 3 & 4 & 1-4 & 
			1 & 2 & 1-2 & 1 & \\
  \hline
  \multicolumn{11}{c}{\addstackgap[10pt]{\emph{cs-shr}}} \\
Gdp & \textbf{1.0083} & 0.9940 & \textbf{1.0029} & 0.9939 & 0.9997 & 0.9670 & 0.9911 & 0.9790 & 0.9444 & 0.9857 \\ 
  Tfi & 0.9229 & \textbf{1.0063} & 0.9720 & 0.9519 & 0.9628 & 0.9513 & 0.9818 & 0.9664 & 0.9044 & 0.9553 \\ 
  TfiGos & 0.9642 & 0.9437 & 0.9695 & 0.9398 & 0.9542 & \textbf{1.0514} & \textbf{1.0135} & \textbf{1.0323} & 0.9691 & 0.9781 \\ 
  TfiCoe & 0.9369 & 0.9572 & 0.9447 & 0.9439 & 0.9456 & \textbf{1.0014} & 0.9931 & 0.9972 & 0.9514 & 0.9609 \\ 
  TfiGosCop & 0.9392 & 0.9828 & \textbf{1.0053} & 0.9777 & 0.9760 & \textbf{1.0058} & 0.9988 & \textbf{1.0023} & 0.9806 & 0.9841 \\
\end{tabular}}
\end{table}

\begin{table}[ht]
\centering
\resizebox{\linewidth}{!}{
\begin{tabular}{r|cccccccccc}
GneDfdFceHfcAbtCig & 0.9480 & 0.9524 & 0.9609 & 0.9875 & 0.9621 & 0.9481 & 0.9702 & 0.9591 & \textbf{1.0085} & 0.9677 \\ 
  GneDfdFceHfcMisOgd & 0.9874 & 0.9974 & 0.9657 & 0.9562 & 0.9765 & 0.9873 & 0.9655 & 0.9763 & 0.9889 & 0.9782 \\ 
  GneDfdFceHfcMisOsv & \textbf{1.0072} & 0.9851 & 0.9911 & 0.9944 & 0.9944 & 0.9661 & \textbf{1.0123} & 0.9889 & \textbf{1.0146} & 0.9957 \\ 
  GneDfdFceHfcMisIfs & 0.9320 & 0.9580 & 0.9120 & 0.9525 & 0.9384 & 0.9246 & 0.9243 & 0.9245 & 0.9291 & 0.9331 \\ 
  GneDfdFceHfcTptTsv & 0.9825 & 0.9934 & 0.9894 & 0.9798 & 0.9862 & \textbf{1.0017} & \textbf{1.0187} & \textbf{1.0102} & 0.9813 & 0.9923 \\ 
  GneDfdFceHfcTptPvh & 0.9944 & 0.9979 & 0.9811 & \textbf{1.0111} & 0.9961 & 0.9879 & \textbf{1.0586} & \textbf{1.0226} & \textbf{1.0050} & \textbf{1.0049} \\ 
  GneDfdFceHfcTptOvh & 0.9909 & 0.9799 & 0.9655 & 0.9624 & 0.9746 & \textbf{1.0117} & \textbf{1.0374} & \textbf{1.0245} & 0.9976 & 0.9919 \\ 
  GneDfdFceHfcHcrAsv & \textbf{1.0096} & 0.9904 & 0.9763 & 0.9856 & 0.9904 & 0.9845 & 0.9686 & 0.9765 & 0.9620 & 0.9823 \\
\end{tabular}}
\end{table}
\begin{table}[ht]
\centering
\resizebox{\linewidth}{!}{
\begin{tabular}{r|cccccccccc}
  TfiGosGvt & 0.7047 & 0.8383 & \textbf{1.1628} & \textbf{1.0574} & 0.9232 & 0.5274 & 0,8760 & 0.6797 & 0.4881 & 0.7723 \\ 
  TfiGosDwl & \textbf{1.0242} & \textbf{1.0012} & 0.9841 & \textbf{1.0124} & \textbf{1.0053} & 0.9317 & 0.9834 & 0.9572 & 0.7493 & 0.9506 \\ 
  TfiGmi & \textbf{1.0362} & 0.9838 & \textbf{1.0045} & 0.9933 & \textbf{1.0043} & 0.9156 & 0.9609 & 0.9380 & 0.9220 & 0.9729 \\ 
  TfiCoeWns & \textbf{1.0187} & \textbf{1.0129} & \textbf{1.0455} & \textbf{1.0207} & \textbf{1.0244} & 0.8766 & 0.9120 & 0.8941 & 0.7372 & 0.9401 \\ 
  TfiCoeEsc & \textbf{1.0035} & \textbf{1.0112} & \textbf{1.0170} & 0.9927 & \textbf{1.0061} & 0.9310 & 0.9693 & 0.9499 & 0.8353 & 0.9637 \\ 
  Tsi & \textbf{1.0031} & \textbf{1.0070} & \textbf{1.0135} & \textbf{1.0114} & \textbf{1.0088} & 0.8993 & 0.9551 & 0.9268 & 0.9153 & 0.9710 \\ 
  Sdi & \textbf{1.0271} & \textbf{1.0203} & \textbf{1.0434} & \textbf{1.0213} & \textbf{1.0280} & 0.9388 & 0.9533 & 0.9460 & 0.9908 & 0.9986 \\ 
  GneDfdFceGvtNatNdf & 0.9647 & 0.9748 & 0.9561 & 0.9899 & 0.9713 & 0.9388 & 0.9748 & 0.9566 & 0.9018 & 0.9569 \\
\end{tabular}}
\end{table}
\begin{table}[ht]
\centering
\resizebox{\linewidth}{!}{
\begin{tabular}{r|cccccccccc}
  GneDfd & \textbf{1.0008} & \textbf{1.0197} & \textbf{1.0155} & \textbf{1.0708} & \textbf{1.0264} & 0.8925 & 0.9881 & 0.9391 & 0.8350 & 0.9716 \\ 
  GneCii & 0.9889 & \textbf{1.0053} & \textbf{1.0034} & 0.9549 & 0.9879 & 0.8202 & 0.8146 & 0.8174 & 0.8531 & 0.9164 \\ 
  GneDfdFce & 0.9782 & 0.9560 & 0.9549 & 0.9311 & 0.9549 & 0.9102 & 0.8950 & 0.9025 & 0.8345 & 0.9217 \\ 
  GneDfdGfc & 0.9800 & 0.9599 & 0.9481 & 0.9660 & 0.9634 & 0.9344 & 0.9394 & 0.9369 & 0.7774 & 0.9269 \\ 
  GneCiiPnf & 0.9085 & 0.9122 & 0.9439 & 0.9983 & 0.9400 & 0.9023 & 0.8496 & 0.8755 & 0.8323 & 0.9053 \\ 
  GneDfdFceGvt & 0.9257 & 0.9368 & 0.9518 & 0.9602 & 0.9435 & 0.9134 & 0.9095 & 0.9114 & 0.8531 & 0.9209 \\ 
  GneDfdFceHfc & 0.9753 & \textbf{1.0719} & \textbf{1.0115} & 0.9981 & \textbf{1.0135} & 0.8714 & 0.8591 & 0.8652 & 0.7629 & 0.9302 \\ 
  GneDfdGfcPub & 0.9402 & 0.9617 & 0.9587 & \textbf{1.0199} & 0.9697 & 0.8874 & 0.9642 & 0.9250 & 0.9476 & 0.9536 \\
\end{tabular}}
\end{table}
\begin{table}[ht]
\centering
\resizebox{\linewidth}{!}{
\begin{tabular}{r|cccccccccc}
    GneDfdFceHfcHweRnt & \textbf{1.0676} & 0.9723 & 0.9575 & 0.9531 & 0.9865 & 0.8284 & 0.8146 & 0.8215 & 0.5911 & 0.8702 \\ 
  GneDfdFceHfcHweWsc & 0.9515 & 0.9515 & 0.9426 & 0.9567 & 0.9506 & 0.9474 & \textbf{1.0013} & 0.9740 & 0.7869 & 0.9317 \\ 
  GneDfdFceHfcHweEgf & \textbf{1.0002} & 0.9899 & \textbf{1.0007} & 0.9671 & 0.9894 & 0.9237 & 0.9227 & 0.9232 & 0.8407 & 0.9477 \\ 
  GneDfdFceHfcFud & \textbf{1.0401} & \textbf{1.0267} & \textbf{1.0087} & 0.9959 & \textbf{1.0177} & 0.8425 & \textbf{1.0180} & 0.9261 & 0.9670 & 0.9834 \\ 
  GneDfdFceHfcCnf & 0.9997 & \textbf{1.0190} & \textbf{1.0145} & \textbf{1.0019} & \textbf{1.0087} & 0.9688 & 0.9817 & 0.9752 & 0.9043 & 0.9836 \\ 
  GneDfdFceHfcRnc & 0.9581 & \textbf{1.0061} & 0.9939 & 0.9984 & 0.9889 & 0.9204 & \textbf{1.0134} & 0.9658 & 0.9291 & 0.9735 \\ 
  GneDfdFceHfcEdc & \textbf{1.0022} & 0.9763 & 0.9895 & 0.9859 & 0.9884 & 0.9590 & 0.9871 & 0.9730 & 0.9174 & 0.9736 \\ 
  GneDfdFceHfcCom & \textbf{1.0619} & \textbf{1.0446} & \textbf{1.0476} & \textbf{1.0363} & \textbf{1.0476} & 0.9469 & 0.9583 & 0.9526 & 0.8570 & 0.9907 \\
\end{tabular}}
\end{table}
\begin{table}[ht]
\centering
\resizebox{\linewidth}{!}{
\begin{tabular}{r|cccccccccc}
  GneDfdGfcPvtPbiIprRnd & 0.8281 & 0.9210 & \textbf{1.1872} & \textbf{1.0664} & 0.9913 & 0.5936 & 0.9026 & 0.7320 & 0.5028 & 0.8250 \\ 
  GneDfdGfcPvtPbiIprMnp & \textbf{1.0104} & \textbf{1.0405} & \textbf{1.0238} & \textbf{1.0392} & \textbf{1.0284} & 0.8877 & 0.9769 & 0.9312 & 0.8274 & 0.9691 \\ 
  GneDfdGfcPvtPbiIprCom & 0.7822 & 0.9018 & \textbf{1.1189} & \textbf{1.0363} & 0.9510 & 0.5528 & 0.8376 & 0.6804 & 0.5956 & 0.8084 \\ 
  GneDfdGfcPvtPbiIprArt & 0.8207 & 0.9123 & \textbf{1.0975} & \textbf{1.0104} & 0.9546 & 0.6157 & 0.8012 & 0.7024 & 0.5123 & 0.8001 \\ 
  GneDfdGfcPvtPbiNdcNbd & \textbf{1.0234} & \textbf{1.0313} & \textbf{1.0481} & \textbf{1.0254} & \textbf{1.0320} & 0.9596 & \textbf{1.0079} & 0.9835 & 0.8134 & 0.9839 \\ 
  GneDfdGfcPvtPbiNdcNec & 0.9861 & 0.9836 & 0.9868 & 0.9549 & 0.9778 & \textbf{1.0131} & 0.9744 & 0.9936 & 0.7389 & 0.9437 \\ 
  GneDfdGfcPvtPbiNdcSha & 0.9848 & \textbf{1.0751} & \textbf{1.0850} & \textbf{1.0697} & \textbf{1.0529} & 0.9576 & \textbf{1.0008} & 0.9789 & 0.9723 & \textbf{1.0195} \\ 
  GneDfdGfcPvtPbiNdmNew & 0.9953 & 0.9995 & 0.9837 & 0.9783 & 0.9892 & 0.9125 & 0.9716 & 0.9416 & 0.8667 & 0.9571 \\
\end{tabular}}
\end{table}
\begin{table}[th]
\centering
\resizebox{\linewidth}{!}{
\begin{tabular}{r|cccccccccc}
  GneDfdFceHfcFhe & 0.9850 & \textbf{1.0022} & 0.9836 & 0.9649 & 0.9838 & 0.9251 & 0.8991 & 0.9120 & 0.8471 & 0.9424 \\ 
  GneDfdFceHfcHwe & 0.8912 & 0.9129 & 0.9134 & 0.9780 & 0.9233 & 0.7517 & 0,8710 & 0.8092 & 0.7237 & 0.8587 \\ 
  GneDfdGfcPubGvtNat & 0.9581 & 0.9644 & 0.9755 & 0.9467 & 0.9611 & 0.9693 & 0.9390 & 0.9541 & 0.9266 & 0.9541 \\ 
  GneDfdGfcPvtPbiIpr & 0.8702 & 0,8720 & 0.9239 & 0.9372 & 0.9003 & 0.7509 & 0.8775 & 0.8118 & 0.6742 & 0.8387 \\ 
  GneDfdGfcPvtPbiNdc & 0.9266 & 0.8953 & 0.9375 & 0.9078 & 0.9167 & 0.8570 & 0.8921 & 0.8744 & 0.7253 & 0.8746 \\ 
  GneDfdGfcPvtPbiNdm & 0.9614 & \textbf{1.0097} & 0.9957 & 0.9993 & 0.9913 & 0.9775 & \textbf{1.0001} & 0.9887 & 0.8832 & 0.9744 \\ 
  TfiGosCopNfnPub & 0.9640 & \textbf{1.0273} & \textbf{1.0036} & 0.9987 & 0.9982 & 0.9426 & 0.9759 & 0.9591 & 0.8974 & 0.9720 \\ 
  TfiGosCopNfnPvt & \textbf{1.0012} & 0.9884 & 0.9941 & \textbf{1.0062} & 0.9975 & 0.9475 & 0.9736 & 0.9605 & 0.9127 & 0.9743 \\
\hline
\end{tabular}}
\end{table}

\begin{table}
	\caption{Variables, series IDs and their descriptions for the Expenditure Approach}
	\centering
	\footnotesize
\begin{tabular*}{\columnwidth}[width = \textwidth]{llp{7cm}}
		\hline
		\textbf{Variable} & \textbf{Series ID} & \textbf{Description}\\
		\hline
	Gdp            & A2302467A & GDP - Gross Domestic Product \\
    Sde                   & A2302566J & Statistical Discrepancy(E)\\
    Exp                   & A2302564C & Exports of goods and services\\
    Imp                   & A2302565F & Imports of goods and services\\
    Gne                   & A2302563A & Gross national exp.\\[2ex]
    GneDfdFceGvtNatDef    & A2302523J & Gen. gov. - National; Final consumption exp. - Defence\\
    GneDfdFceGvtNatNdf    & A2302524K & Gen. gov. - National; Final consumption exp. - Non-defence\\
\hline
	\end{tabular*}
	\label{Tab:Expenditure-hierarchy-1}
\begin{flushleft}{Source: Athanasopoulos et al. (2019).}\end{flushleft}
\end{table}

\begin{table}
	\caption{Variables, series IDs and their descriptions for Household Final Consumption - Expenditure Approach}
	\centering
	\footnotesize
\begin{tabular*}{\columnwidth}[width = \textwidth]{llp{8cm}}
		\hline
		\textbf{Variable} & \textbf{Series ID} & \textbf{Description}\\
		\hline
    GneDfdHfc          & A2302254W & Household Final Consumption Expenditure\\
    GneDfdFceHfcFud    & A2302237V & Food\\
    GneDfdFceHfcAbt    & A3605816F & Alcoholic beverages and tobacco\\
    GneDfdFceHfcAbtCig & A2302238W & Cigarettes and tobacco\\
    GneDfdFceHfcAbtAlc & A2302239X & Alcoholic beverages\\[1ex]
    GneDfdFceHfcCnf    & A2302240J & Clothing and footwear\\
    GneDfdFceHfcHwe    & A3605680F & Housing, water, electricity, gas and other fuels\\
\hline
	\end{tabular*}
	\label{Tab:Expenditure-hierarchy-3}
\begin{flushleft}{Source: Athanasopoulos et al. (2019).}\end{flushleft}
\end{table}

\clearpage
\null
\vfill
\begin{table}[h]
	\caption{Variables, series IDs and their descriptions for Changes in Inventories - Expenditure Approach}
	\centering
	\footnotesize
\begin{tabular*}{\columnwidth}[width = \textwidth]{llp{8cm}}
		\hline
		\textbf{Variable} & \textbf{Series ID} & \textbf{Description}\\
		\hline
    GneCii       & A2302562X  & Changes in Inventories\\
    GneCiiPfm    & A2302560V  & Farm\\
    GneCiiPba    & A2302561W  & Public authorities\\
    GneCiiPnf    & A2302559K  & Private; Non-farm Total\\
    GneCiiPnfMin & A83722619L & Private; Mining (B)\\[1ex]
    GneCiiPnfMan & A3348511X  & Private; Manufacturing (C)\\
    GneCiiPnfWht & A3348512A  & Private; Wholesale trade (F)\\
    GneCiiPnfRet & A3348513C  & Private; Retail trade (G)\\
    GneCiiPnfOnf & A2302273C  & Private; Non-farm; Other non-farm industries\\
		\hline
	\end{tabular*}
	\label{Tab:Expenditure-hierarchy-2}
\begin{flushleft}{Source: Athanasopoulos et al. (2019).}\end{flushleft}
\end{table}

\begin{table}[!h]
	\caption{Variables, series IDs and their descriptions for the Income approach}
	\centering
	\footnotesize
\begin{tabular*}{\columnwidth}[width = \textwidth]{llp{8cm}}
		\hline
		\textbf{Variable} & \textbf{Series ID} & \textbf{Description}\\
		\hline
    Sdi             & A2302413V & Statistical discrepancy (I)\\
    Tsi             & A2302412T & Taxes less subsidies (I)\\
    TfiCoeWns       & A2302399K & Compensation of employees; Wages and salaries\\
    TfiCoeEsc       & A2302400J & Compensation of employees; Employers' social contributions\\[2ex]
    TfiCoe          & A2302401K & Compensation of employees\\
    TfiGosCopNfnPvt & A2323369L & Private non-financial corporations; Gross operating surplus\\
    TfiGosCopNfnPub & A2302403R & Public non-financial corporations; Gross operating surplus\\
    TfiGosCopNfn    & A2302404T & Non-financial corporations; Gross operating surplus\\
    TfiGosCopFin    & A2302405V & Financial corporations; Gross operating surplus\\[1ex]
    TfiGosCop       & A2302406W & Total corporations; Gross operating surplus\\
    TfiGosGvt       & A2298711F & General government; Gross operating surplus\\
    TfiGosDwl       & A2302408A & Dwellings owned by persons; Gross operating surplus\\
    TfiGos          & A2302409C & All sectors; Gross operating surplus\\
    TfiGmi          & A2302410L & Gross mixed income\\
    Tfi             & A2302411R & Total factor income\\
		\hline
	\end{tabular*}
	\label{Tab: Income-hierarchy}
\begin{flushleft}{Source: Athanasopoulos et al. (2019).}\end{flushleft}
\end{table}
\vfill

\clearpage
\subsection*{A.8.2 Cross-sectional \& temporal reconciliation procedures}
\begin{table}[ht]
\centering
\caption{AvgRelMSE at any temporal aggregation level and any forecast horizon.}
\resizebox{0.86\linewidth}{!}{
\begin{tabular}{c|c|c|c|c|c|c|c|c|c|c}
  \hline
 & \multicolumn{5}{c|}{\textbf{Quarterly}} & \multicolumn{3}{c|}{\textbf{Semi-annual}} & \multicolumn{1}{c|}{\textbf{Annual}} & \textbf{All}\\
	 \hline
\textbf{Procedure} & 1 & 2 & 3 & 4 & 1-4 & 
            1 & 2 & 1-2 & 1 & \\
     \hline
 \multicolumn{11}{c}{\rule{0pt}{4ex}\emph{all 95 series}} \\[2ex]
base & 1 & 1 & 1 & 1 & 1 & 1 & 1 & 1 & 1 & 1 \\ 
  cs-ols & \textbf{1.2189} & \textbf{1.1663} & \textbf{1.1631} & \textbf{1.1510} & \textbf{1.1745} & \textbf{1.1427} & \textbf{1.1428} & \textbf{1.1427} & \textbf{1.2238} & \textbf{1.1722} \\ 
  cs-struc & \textbf{1.1911} & \textbf{1.1363} & \textbf{1.1248} & \textbf{1.1049} & \textbf{1.1388} & \textbf{1.1058} & \textbf{1.0733} & \textbf{1.0894} & \textbf{1.1046} & \textbf{1.1196} \\ 
  cs-wls & 0.9619 & \textbf{\textcolor{red}{0.9698}} & \textbf{\textcolor{red}{0.9730}} & \textbf{\textcolor{red}{0.9778}} & \textbf{\textcolor{red}{0.9706}} & 0.9613 & 0.9741 & 0.9677 & 0.9724 & 0.9700 \\ 
  cs-shr & \textbf{\textcolor{red}{0.9583}} & 0.9701 & 0.9757 & 0.9824 & 0.9716 & 0.9526 & 0.9781 & 0.9652 & 0.9657 & 0.9689 \\
\hline
\end{tabular}}
\end{table}

\begin{table}[ht]
\centering
\caption{AvgRelMAE at any temporal aggregation level and any forecast horizon.}
\resizebox{0.86\linewidth}{!}{
\begin{tabular}{c|c|c|c|c|c|c|c|c|c|c}
  \hline
 & \multicolumn{5}{c|}{\textbf{Quarterly}} & \multicolumn{3}{c|}{\textbf{Semi-annual}} & \multicolumn{1}{c|}{\textbf{Annual}} & \textbf{All}\\
	 \hline
\textbf{Procedure} & 1 & 2 & 3 & 4 & 1-4 & 
            1 & 2 & 1-2 & 1 & \\
     \hline
 \multicolumn{11}{c}{\rule{0pt}{4ex}\emph{all 95 series}} \\[2ex]
base & 1 & 1 & 1 & 1 & 1 & 1 & 1 & 1 & 1 & 1 \\ 
  cs-ols & \textbf{1.1168} & \textbf{1.0846} & \textbf{1.0806} & \textbf{1.0716} & \textbf{1.0883} & \textbf{1.0722} & \textbf{1.0621} & \textbf{1.0672} & \textbf{1.1032} & \textbf{1.0843} \\ 
  cs-struc & \textbf{1.0997} & \textbf{1.0679} & \textbf{1.0600} & \textbf{1.0497} & \textbf{1.0692} & \textbf{1.0473} & \textbf{1.0306} & \textbf{1.0389} & \textbf{1.0536} & \textbf{1.0582} \\ 
  cs-wls & 0.9793 & 0.9846 & 0.9866 & \textbf{\textcolor{red}{0.9893}} & 0.9850 & 0.9788 & 0.9857 & 0.9823 & 0.9907 & 0.9850 \\ 
  cs-shr & \textbf{\textcolor{red}{0.9769}} & \textbf{\textcolor{red}{0.9842}} & \textbf{\textcolor{red}{0.9863}} & 0.9893 & \textbf{\textcolor{red}{0.9842}} & 0.9733 & 0.9871 & 0.9802 & 0.9840 & 0.9830 \\
\hline
\end{tabular}}
\end{table}

\begin{figure}[ht]
	\centering
	\includegraphics[width=\linewidth]{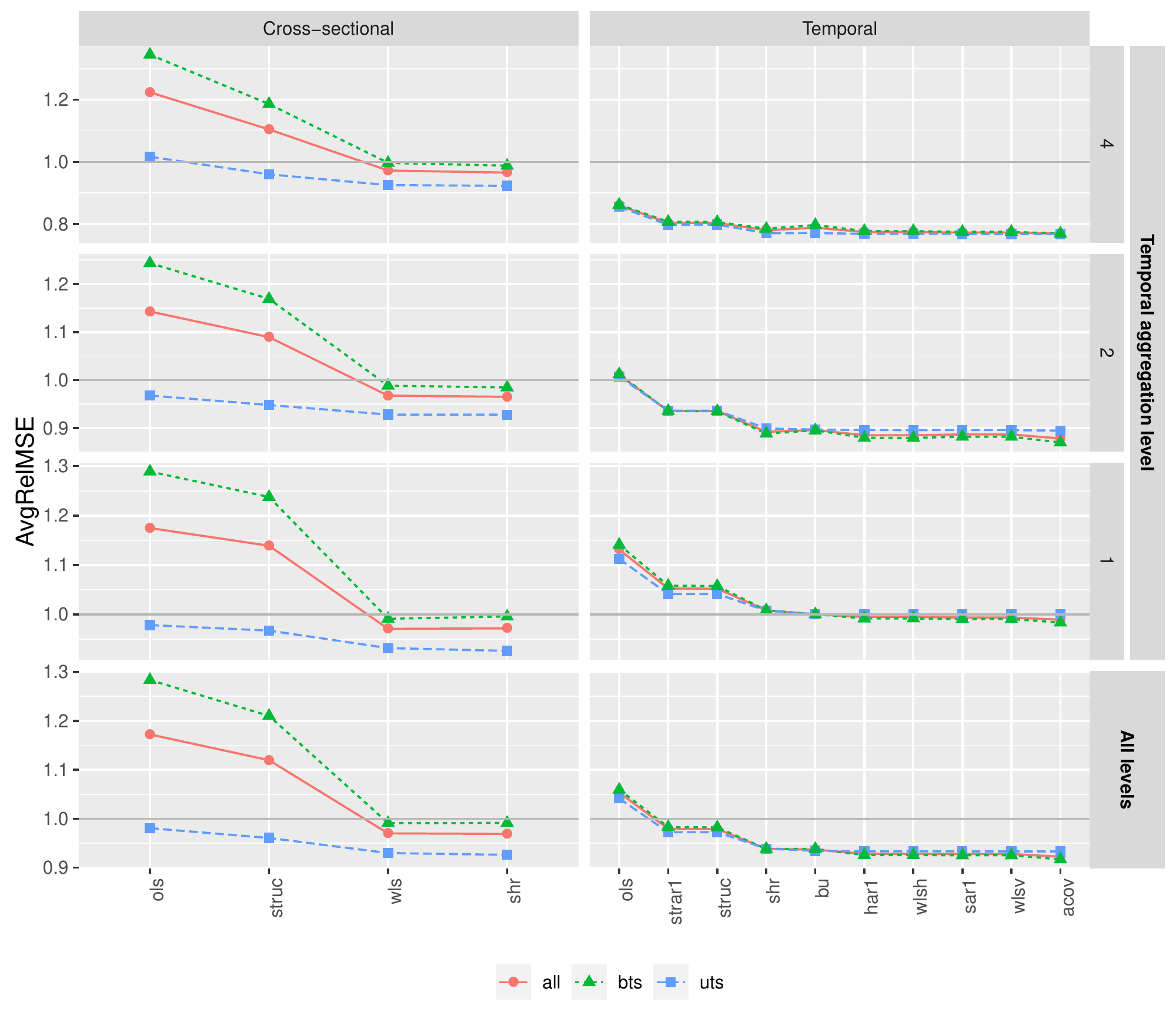}
	\includegraphics{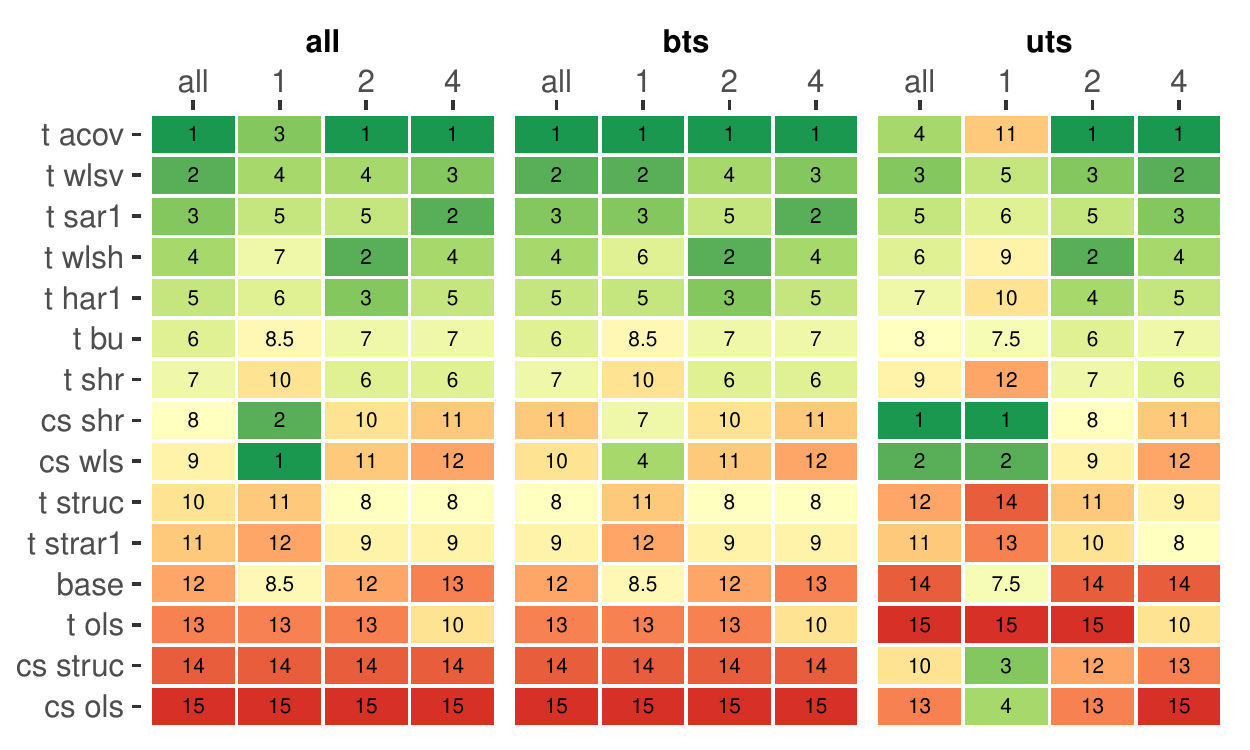}
	\caption{Top panel: Average Relative MSE across all series and forecast horizons, by frequency of observation. Bottom panel: Rankings by frequency of observation and forecast horizon.}
	\label{fig:cs_t_mse}
\end{figure}

\begin{figure}[ht]
	\centering
	\includegraphics[width=\linewidth]{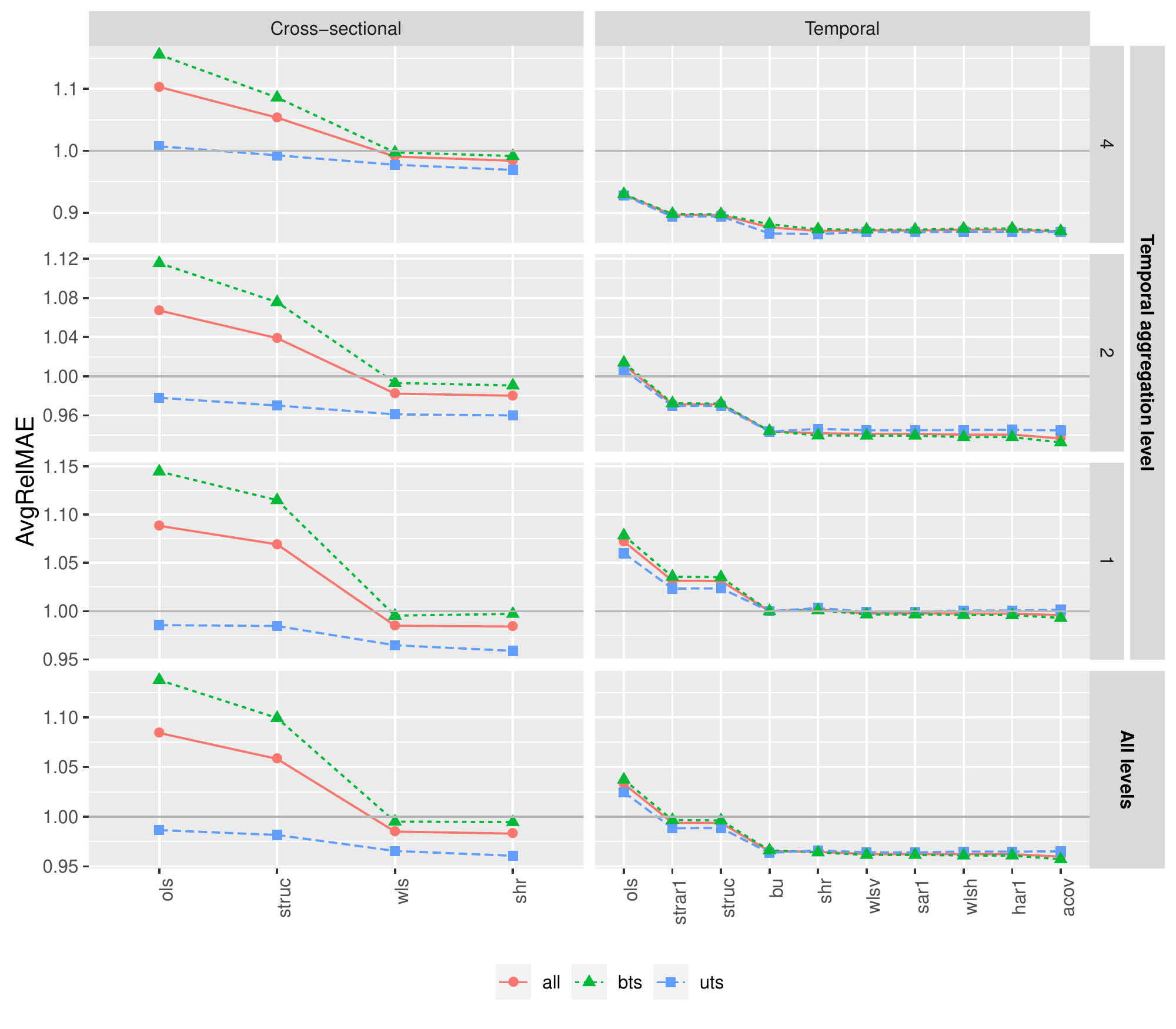}
	\includegraphics{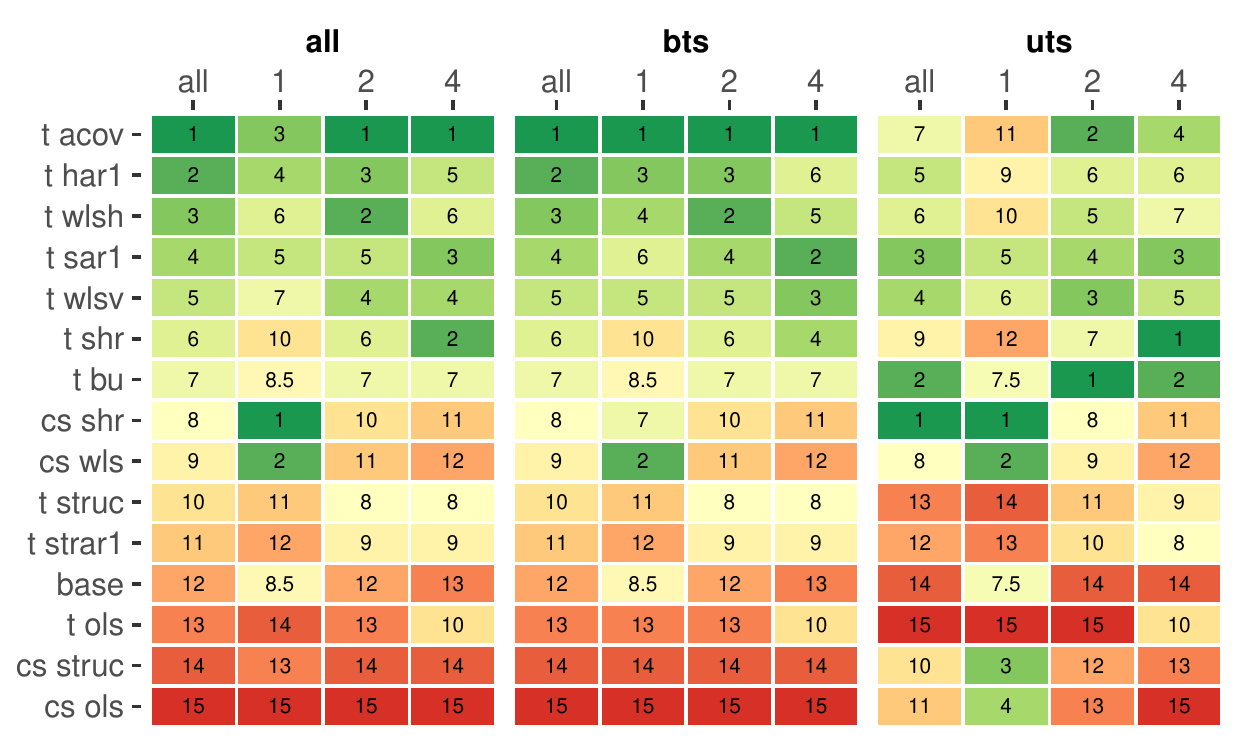}
	\caption{Top panel: Average Relative MAE across all series and forecast horizons, by frequency of observation. Bottom panel: Rankings by frequency of observation and forecast horizon.}
	\label{fig:cs_t_mae}
\end{figure}

\begin{figure}[ht]
	\centering
	\includegraphics[width=\linewidth]{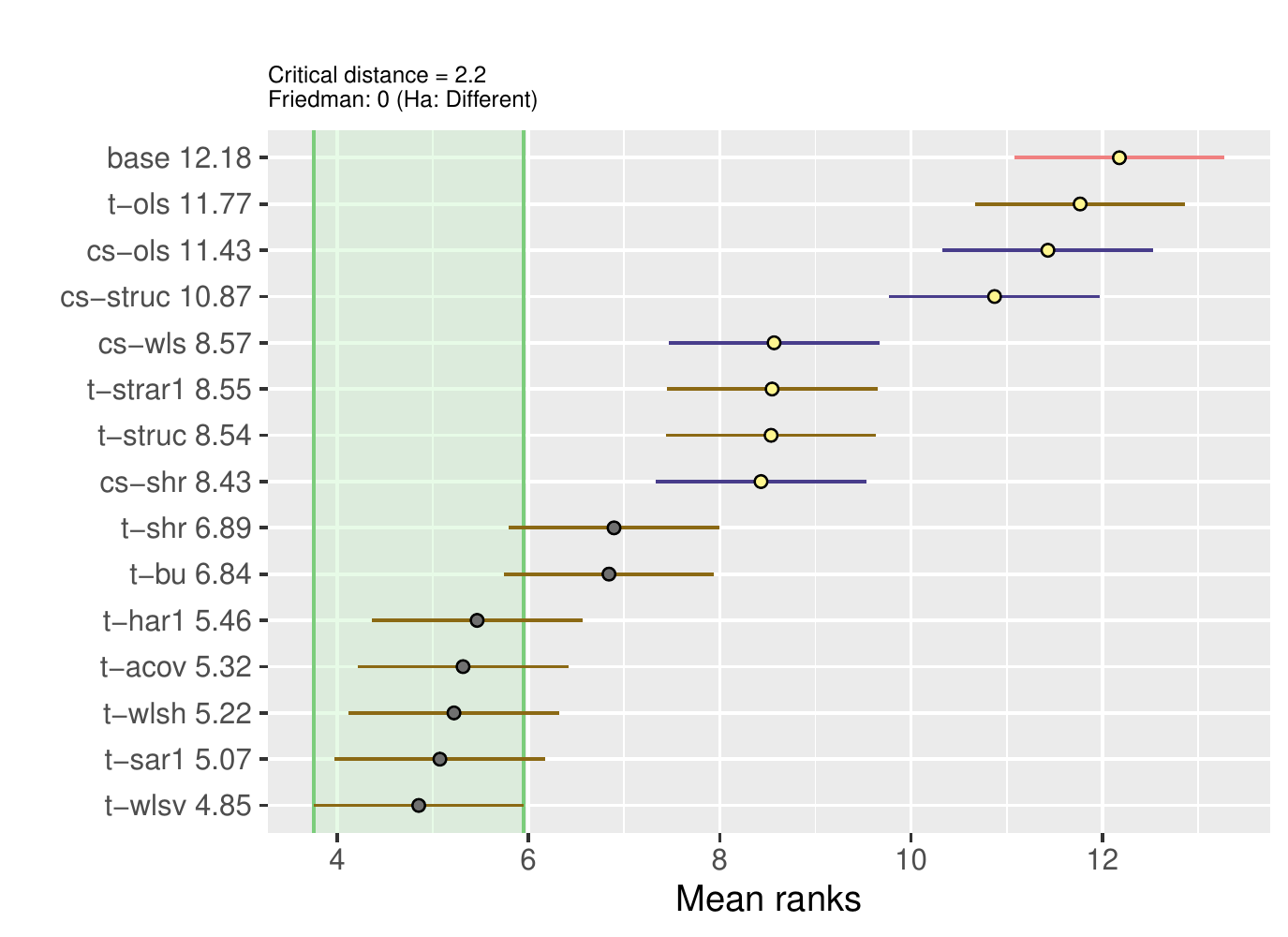}
	\includegraphics[width=\linewidth]{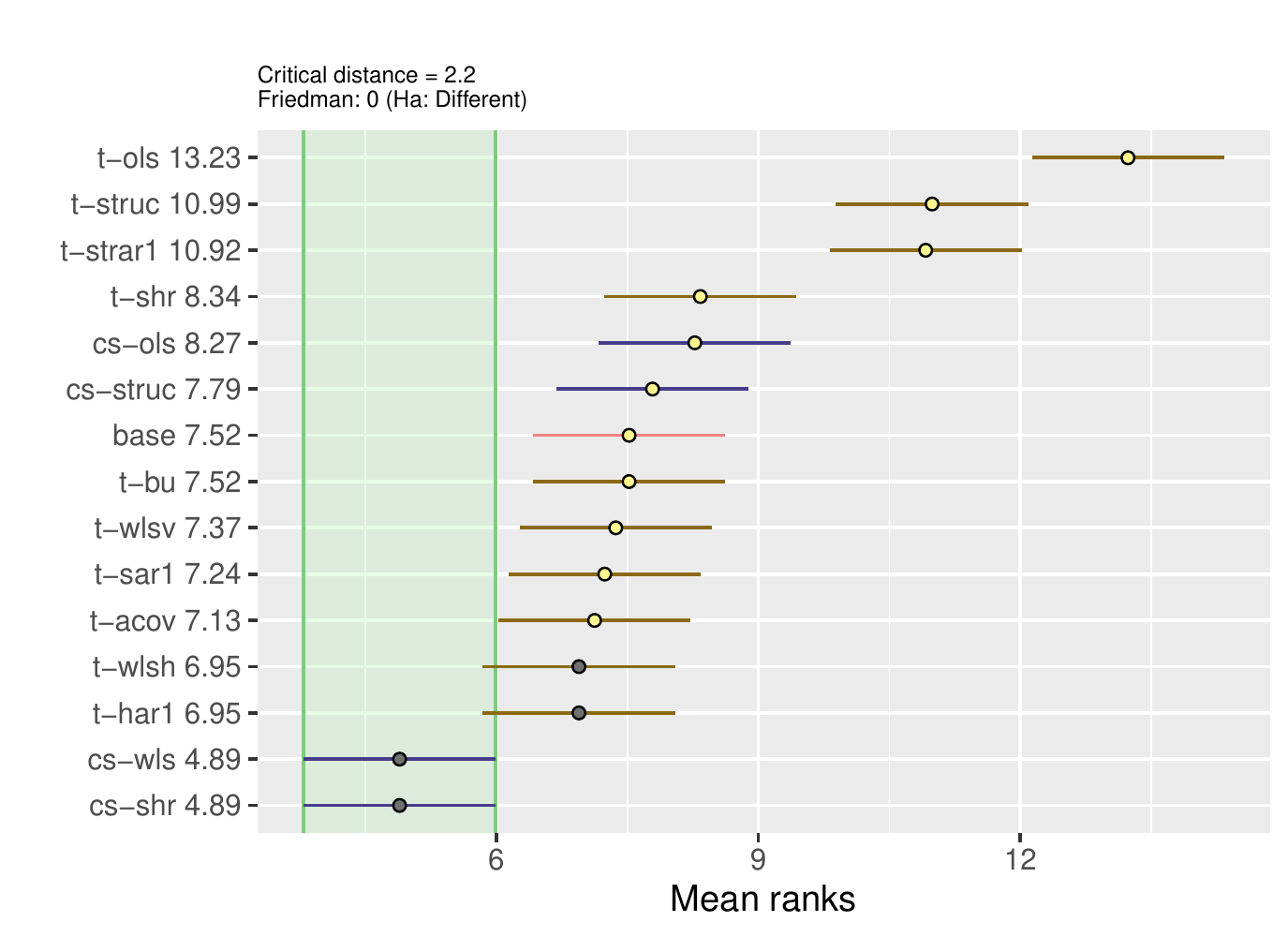}
	\caption{Nemenyi test results at 5\% significance level for all 95 series. The cross-sectional and temporal reconciliation procedures are sorted vertically according to the MSE mean rank (i) across all time frequencies and forecast horizons (top), and (ii) for one-step-ahead quarterly forecasts (bottom).}
	\label{fig:nem_all_mse_univ}
\end{figure}

\begin{figure}[ht]
	\centering
	\includegraphics[width=\linewidth]{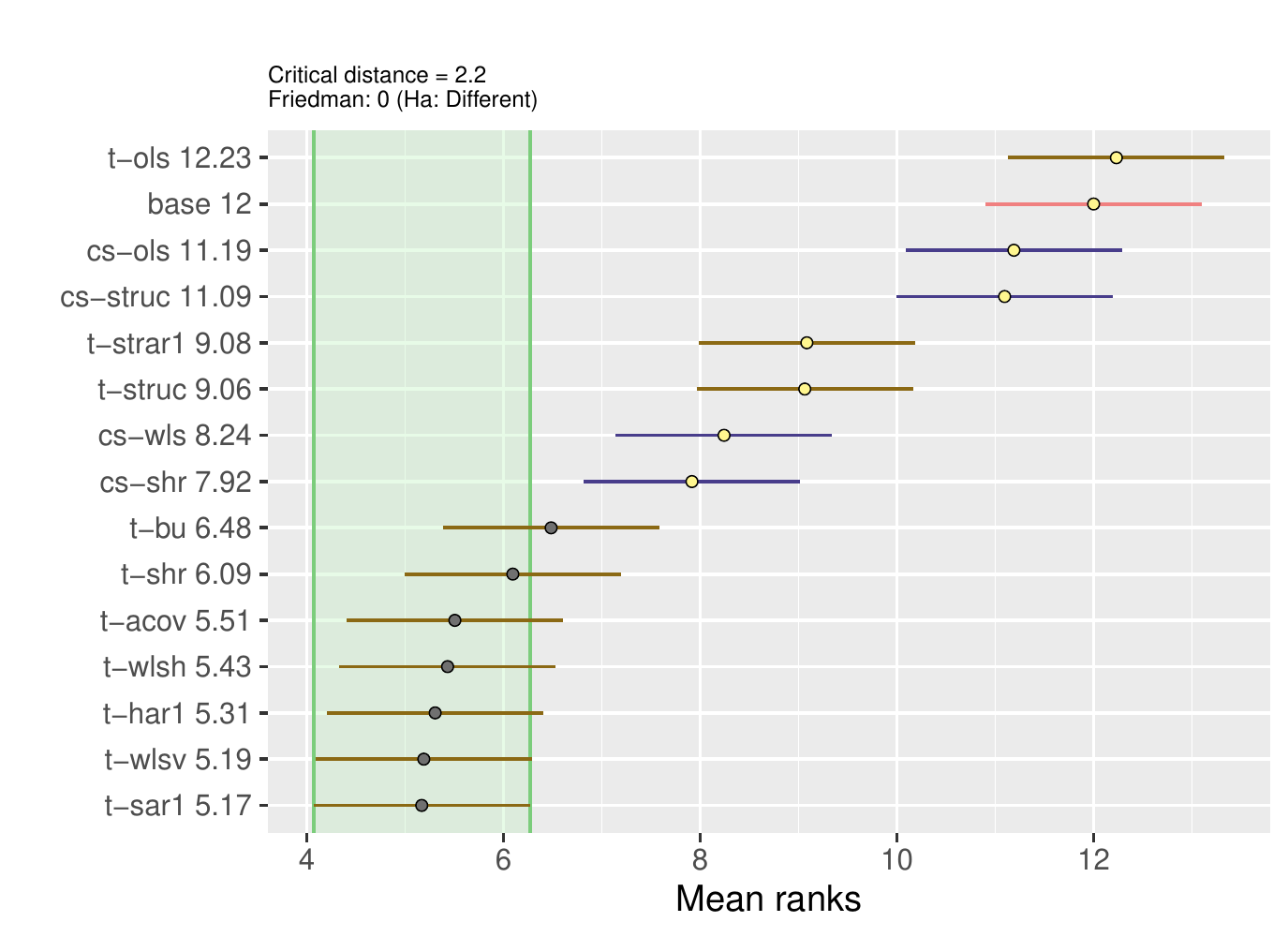}
	\includegraphics[width=\linewidth]{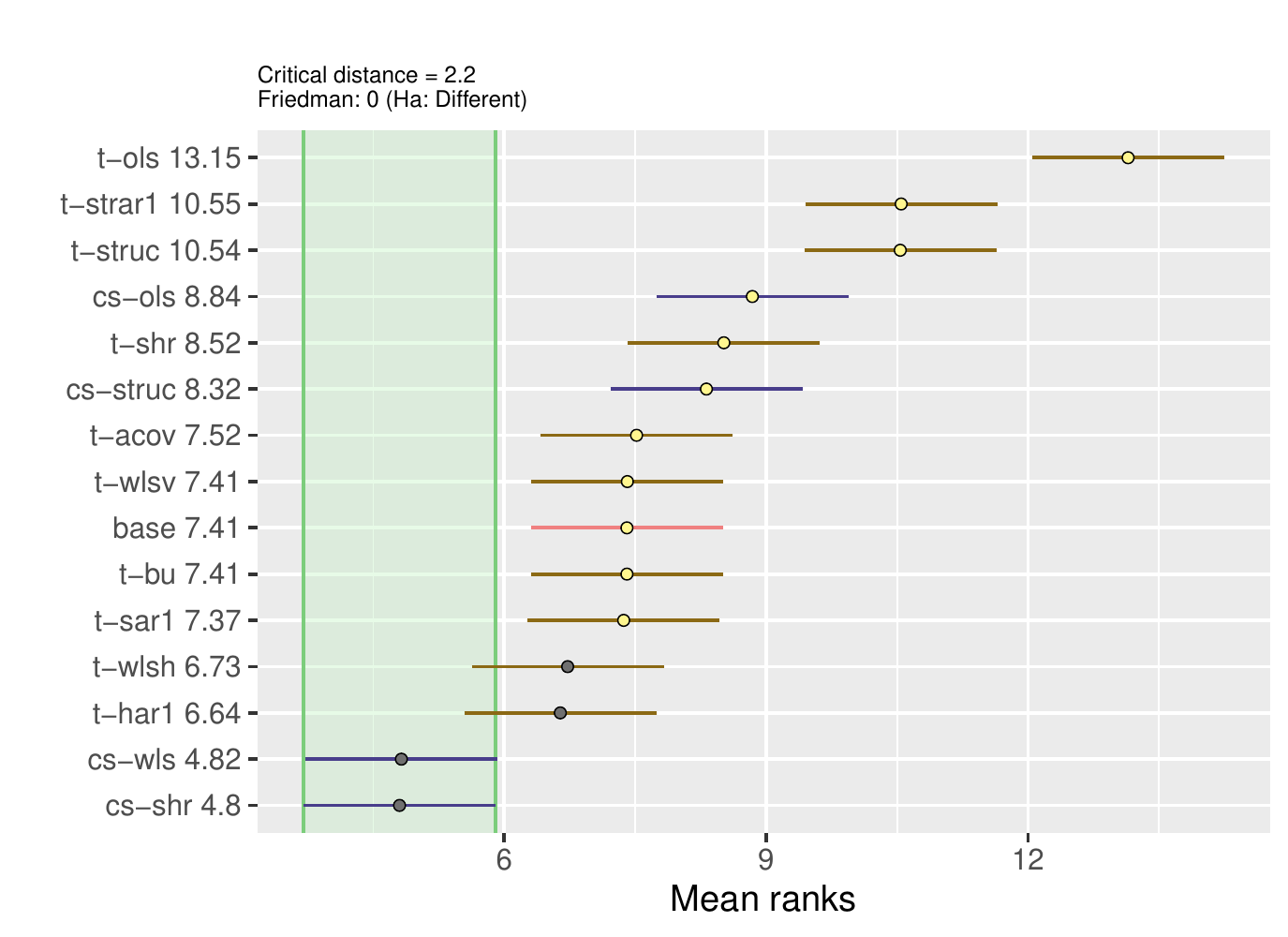}
	\caption{Nemenyi test results at 5\% significance level for all 95 series. The cross-sectional and temporal reconciliation procedures are sorted vertically according to the MAE mean rank (i) across all time frequencies and forecast horizons (top), and (ii) for one-step-ahead quarterly forecasts (bottom).}
	\label{fig:nem_all_mae_univ}
\end{figure}

\clearpage
\subsection*{A.8.3 Heuristic KA, alternatives and iterative cross-temporal reconciliation procedures}

\begin{figure}[ht]
	\centering
	\includegraphics[width=\linewidth]{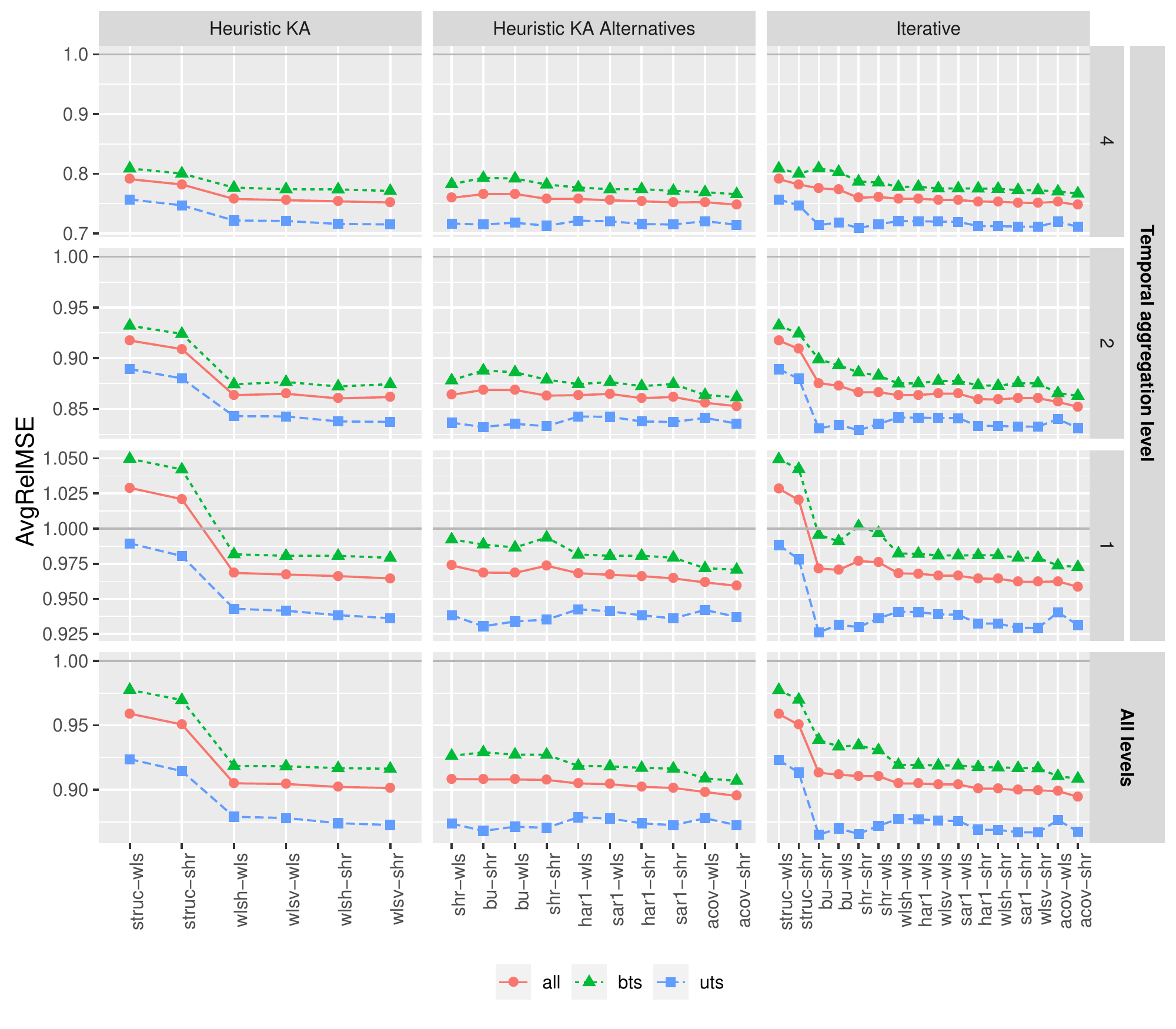}
	\caption{Average Relative MSE across all series and forecast horizons, by frequency of observation (selected procedures).}
	\label{fig:kah_mse}
\end{figure}

\begin{figure}[ht]
	\centering
	\includegraphics[width=\linewidth]{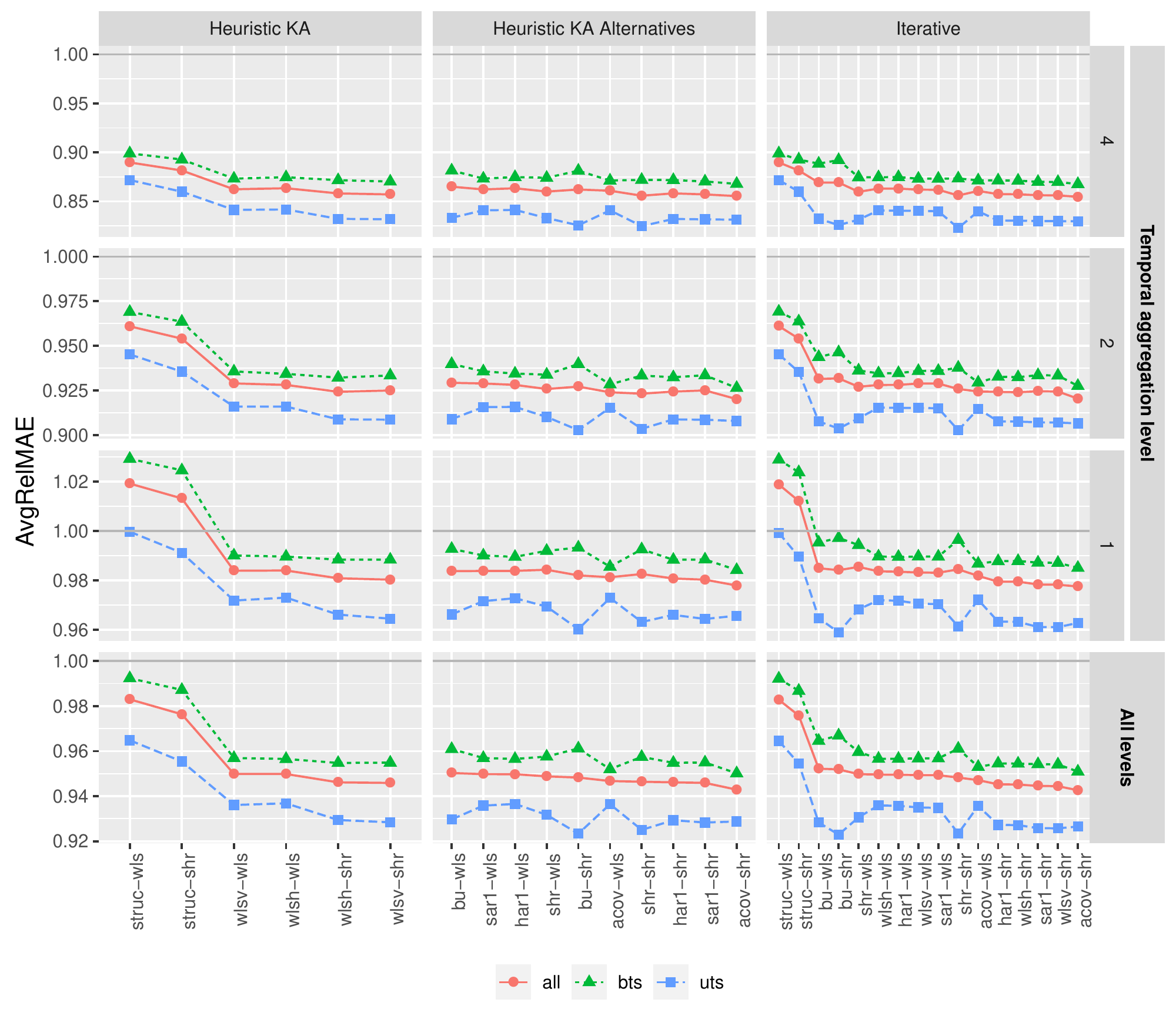}
	\caption{Average Relative MAE across all series and forecast horizons, by frequency of observation (selected procedures).}
	\label{fig:kah_mae}
\end{figure}

\begin{figure}[ht]
	\centering
	\includegraphics[width=\linewidth]{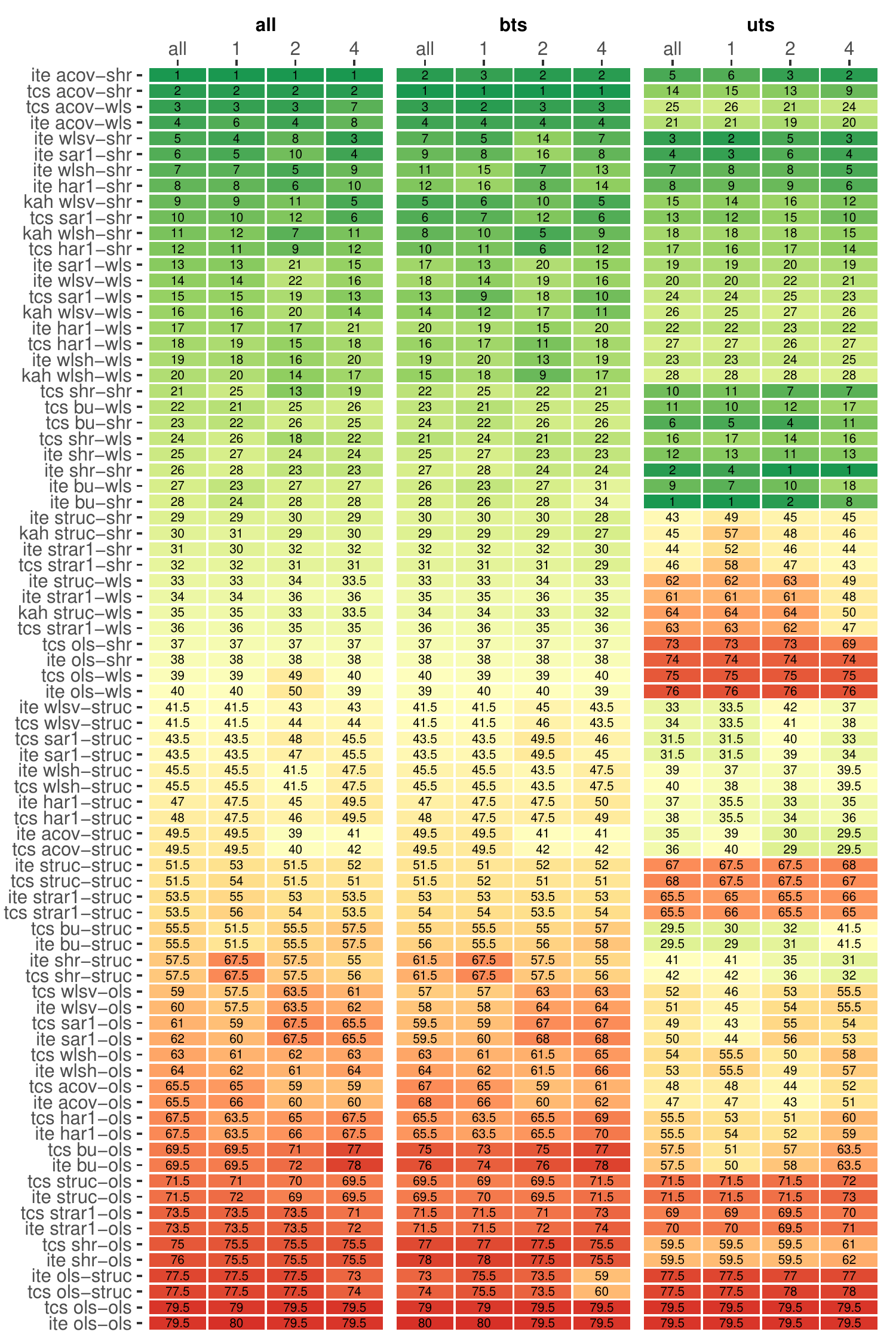}
	\caption{Rankings (Average Relative MSE) by frequency of observation and forecast horizon.}
	\label{fig:kah_RankmSe}
\end{figure}

\begin{figure}[ht]
	\centering
	\includegraphics[width=\linewidth]{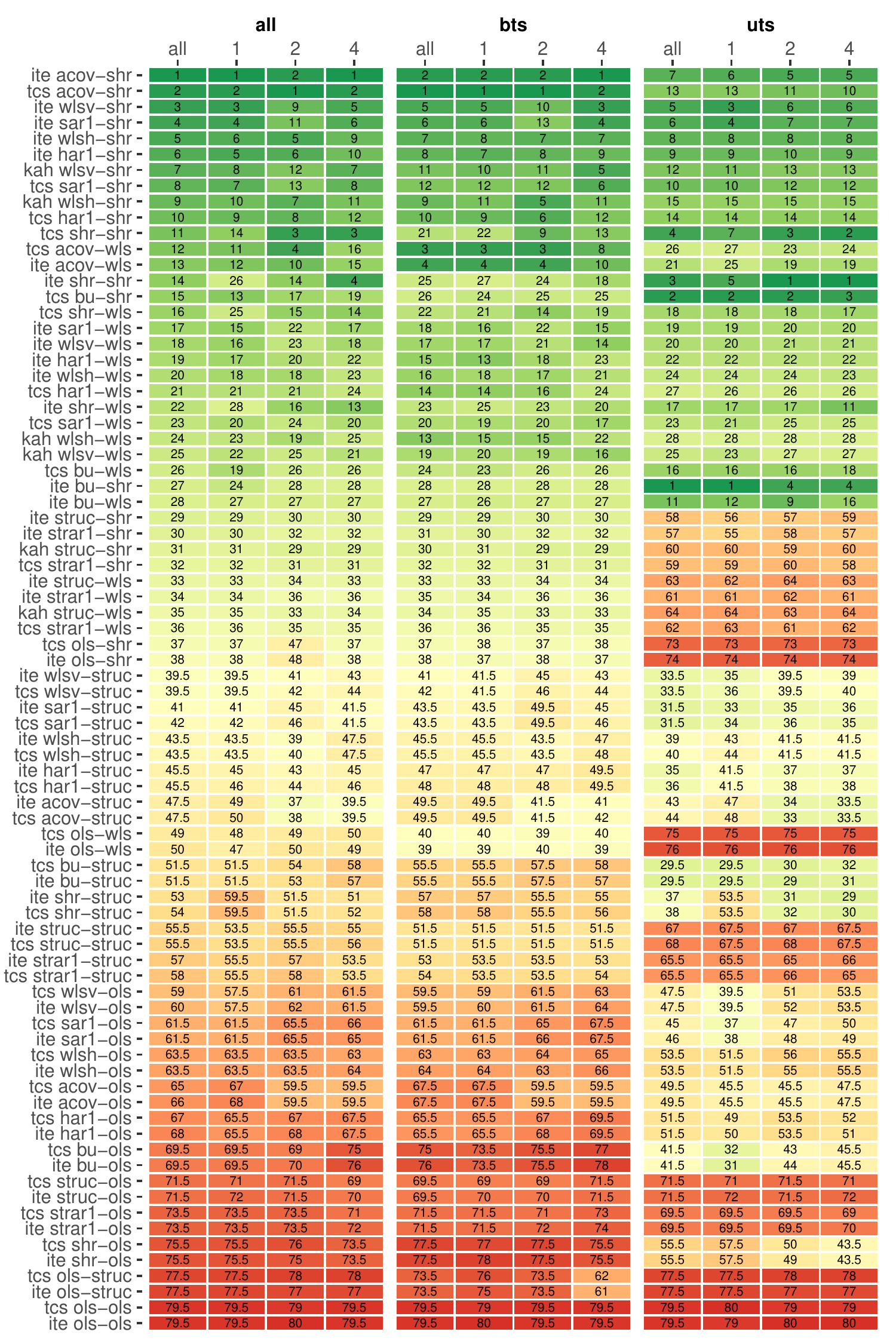}
	\caption{Rankings (Average Relative MAE) by frequency of observation and forecast horizon.}
	\label{fig:kah_Rankmae}
\end{figure}

\begin{table}[ht]
\centering
\caption{AvgRelMSE for the 63 bottom series at any temporal aggregation level and any forecast horizon.}
\resizebox{0.75\linewidth}{!}{
\begin{tabular}{c|c|c|c|c|c|c|c|c|c|c}
  \hline
 & \multicolumn{5}{c|}{\textbf{Quarterly}} & \multicolumn{3}{c|}{\textbf{Semi-annual}} & \multicolumn{1}{c|}{\textbf{Annual}} & \textbf{All}\\
 & \multicolumn{5}{c|}{bts} &
   \multicolumn{3}{c|}{bts} & bts & bts\\
	 \hline
\textbf{Procedure} & 1 & 2 & 3 & 4 & 1-4 & 
            1 & 2 & 1-2 & 1 & \\
     \hline
base & 1 & 1 & 1 & 1 & 1 & 1 & 1 & 1 & 1 & 1 \\ 
  kah-struc-shr & \textbf{1.1655} & \textbf{1.037} & 0.9986 & 0.9774 & \textbf{1.0422} & 0.9146 & 0.9335 & 0.9240 & 0.8004 & 0.9697 \\ 
  kah-struc-wls & \textbf{1.1776} & \textbf{1.044} & \textbf{1.0045} & 0.9832 & \textbf{1.0497} & 0.9241 & 0.9399 & 0.9320 & 0.8088 & 0.9775 \\ 
  kah-wlsh-shr & 0.9665 & 0.9806 & 0.9954 & 0.9802 & 0.9806 & 0.8156 & 0.9327 & 0.8722 & 0.7738 & 0.9168 \\ 
  kah-wlsh-wls & 0.9677 & 0.9812 & 0.9961 & 0.9819 & 0.9817 & 0.8180 & 0.9344 & 0.8743 & 0.7767 & 0.9185 \\
\hline
\end{tabular}}
\end{table}

\begin{table}[ht]
\centering
\caption{AvgRelMSE for the 32 upper series at any temporal aggregation level and any forecast horizon.}
\resizebox{0.75\linewidth}{!}{
\begin{tabular}{c|c|c|c|c|c|c|c|c|c|c}
  \hline
 & \multicolumn{5}{c|}{\textbf{Quarterly}} & \multicolumn{3}{c|}{\textbf{Semi-annual}} & \multicolumn{1}{c|}{\textbf{Annual}} & \textbf{All}\\
 & \multicolumn{5}{c|}{bts} &
   \multicolumn{3}{c|}{bts} & bts & bts\\
	 \hline
\textbf{Procedure} & 1 & 2 & 3 & 4 & 1-4 & 
            1 & 2 & 1-2 & 1 & \\
     \hline
base & 1 & 1 & 1 & 1 & 1 & 1 & 1 & 1 & 1 & 1 \\ 
  kah-struc-shr & \textbf{1.0261} & 0.9976 & 0.9521 & 0.9483 & 0.9805 & 0.863 & 0.8976 & 0.8801 & 0.7471 & 0.9145 \\ 
  kah-struc-wls & \textbf{1.0369} & \textbf{1.0072} & 0.9605 & 0.9556 & 0.9895 & 0.8729 & 0.9060 & 0.8893 & 0.757 & 0.9237 \\ 
  kah-wlsh-shr & 0.9432 & 0.9481 & 0.9295 & 0.9326 & 0.9383 & 0.7995 & 0.8780 & 0.8378 & 0.7159 & 0.8740 \\ 
  kah-wlsh-wls & 0.9469 & 0.9544 & 0.9345 & 0.9359 & 0.9429 & 0.8047 & 0.883 & 0.843 & 0.7217 & 0.8790 \\
\hline
\end{tabular}}
\end{table}

\begin{table}[ht]
\centering
\caption{AvgRelMSE for all 95 series at any temporal aggregation level and any forecast horizon.}
\resizebox{0.75\linewidth}{!}{
\begin{tabular}{c|c|c|c|c|c|c|c|c|c|c}
  \hline
 & \multicolumn{5}{c|}{\textbf{Quarterly}} & \multicolumn{3}{c|}{\textbf{Semi-annual}} & \multicolumn{1}{c|}{\textbf{Annual}} & \textbf{All}\\
 & \multicolumn{5}{c|}{all} &
   \multicolumn{3}{c|}{all} & all & all\\
	 \hline
\textbf{Procedure} & 1 & 2 & 3 & 4 & 1-4 & 
            1 & 2 & 1-2 & 1 & \\
     \hline
base & 1 & 1 & 1 & 1 & 1 & 1 & 1 & 1 & 1 & 1 \\ 
  kah-struc-shr & \textbf{1.1165} & \textbf{1.0235} & 0.9827 & 0.9675 & \textbf{1.021} & 0.8968 & 0.9212 & 0.9090 & 0.7820 & 0.9507 \\ 
  kah-struc-wls & \textbf{1.1282} & \textbf{1.0314} & 0.9894 & 0.9738 & \textbf{1.029} & 0.9065 & 0.9284 & 0.9174 & 0.7909 & 0.9591 \\ 
  kah-wlsh-shr & 0.9586 & 0.9696 & 0.9727 & 0.9639 & 0.9662 & 0.8101 & 0.9139 & 0.8605 & 0.7538 & 0.9021 \\ 
  kah-wlsh-wls & 0.9607 & 0.9721 & 0.9749 & 0.9662 & 0.9684 & 0.8135 & 0.9168 & 0.8636 & 0.7578 & 0.9050 \\
\hline
\end{tabular}}
\end{table}
     
\begin{table}[ht]
\centering
\caption{AvgRelMAE for the 63 bottom series at any temporal aggregation level and any forecast horizon.}
\resizebox{0.75\linewidth}{!}{
\begin{tabular}{c|c|c|c|c|c|c|c|c|c|c}
  \hline
 & \multicolumn{5}{c|}{\textbf{Quarterly}} & \multicolumn{3}{c|}{\textbf{Semi-annual}} & \multicolumn{1}{c|}{\textbf{Annual}} & \textbf{All}\\
 & \multicolumn{5}{c|}{bts} &
   \multicolumn{3}{c|}{bts} & bts & bts\\
	 \hline
\textbf{Procedure} & 1 & 2 & 3 & 4 & 1-4 & 
            1 & 2 & 1-2 & 1 & \\
     \hline
base & 1 & 1 & 1 & 1 & 1 & 1 & 1 & 1 & 1 & 1 \\ 
  kah-struc-shr & \textbf{1.0858} & \textbf{1.0187} & \textbf{1.0046} & 0.9918 & \textbf{1.0246} & 0.9603 & 0.9669 & 0.9636 & 0.8928 & 0.9872 \\ 
  kah-struc-wls & \textbf{1.092} & \textbf{1.0238} & \textbf{1.0082} & 0.9957 & \textbf{1.0293} & 0.9665 & 0.9716 & 0.9690 & 0.8991 & 0.9923 \\ 
  kah-wlsh-shr & 0.9833 & 0.9884 & 0.9960 & 0.9861 & 0.9884 & 0.9054 & 0.9598 & 0.9322 & 0.8717 & 0.9548 \\ 
  kah-wlsh-wls & 0.9840 & 0.9895 & 0.9969 & 0.9882 & 0.9897 & 0.9072 & 0.9622 & 0.9343 & 0.8747 & 0.9565 \\
\hline
\end{tabular}}
\end{table}

\begin{table}[ht]
\centering
\caption{AvgRelMAE for the 32 upper series at any temporal aggregation level and any forecast horizon.}
\resizebox{0.75\linewidth}{!}{
\begin{tabular}{c|c|c|c|c|c|c|c|c|c|c}
  \hline
 & \multicolumn{5}{c|}{\textbf{Quarterly}} & \multicolumn{3}{c|}{\textbf{Semi-annual}} & \multicolumn{1}{c|}{\textbf{Annual}} & \textbf{All}\\
 & \multicolumn{5}{c|}{bts} &
   \multicolumn{3}{c|}{bts} & bts & bts\\
	 \hline
\textbf{Procedure} & 1 & 2 & 3 & 4 & 1-4 & 
            1 & 2 & 1-2 & 1 & \\
     \hline
base & 1 & 1 & 1 & 1 & 1 & 1 & 1 & 1 & 1 & 1 \\ 
  kah-struc-shr & 0.9972 & 0.9988 & 0.9846 & 0.9842 & 0.9912 & 0.9266 & 0.9447 & 0.9356 & 0.8599 & 0.9554 \\ 
  kah-struc-wls & \textbf{1.0058} & \textbf{1.0070} & 0.9925 & 0.9938 & 0.9998 & 0.9351 & 0.9555 & 0.9452 & 0.8718 & 0.9648 \\ 
  kah-wlsh-shr & 0.9584 & 0.9711 & 0.9655 & 0.9697 & 0.9662 & 0.8909 & 0.9271 & 0.9088 & 0.8322 & 0.9294 \\ 
  kah-wlsh-wls & 0.9644 & 0.9777 & 0.9727 & 0.9775 & 0.9731 & 0.8969 & 0.9356 & 0.9160 & 0.8418 & 0.9368 \\
\hline
\end{tabular}}
\end{table}

\begin{table}[ht]
\centering
\caption{AvgRelMAE for the 95 series at any temporal aggregation level and any forecast horizon.}
\resizebox{0.75\linewidth}{!}{
\begin{tabular}{c|c|c|c|c|c|c|c|c|c|c}
  \hline
 & \multicolumn{5}{c|}{\textbf{Quarterly}} & \multicolumn{3}{c|}{\textbf{Semi-annual}} & \multicolumn{1}{c|}{\textbf{Annual}} & \textbf{All}\\
 & \multicolumn{5}{c|}{all} &
   \multicolumn{3}{c|}{all} & all & all\\
	 \hline
\textbf{Procedure} & 1 & 2 & 3 & 4 & 1-4 & 
            1 & 2 & 1-2 & 1 & \\
     \hline
base & 1 & 1 & 1 & 1 & 1 & 1 & 1 & 1 & 1 & 1 \\ 
  kah-struc-shr & \textbf{1.0551} & \textbf{1.0120} & 0.9978 & 0.9893 & \textbf{1.0132} & 0.9488 & 0.9593 & 0.954 & 0.8816 & 0.9763 \\ 
  kah-struc-wls & \textbf{1.0622} & \textbf{1.0181} & \textbf{1.0029} & 0.9951 & \textbf{1.0192} & 0.9558 & 0.9661 & 0.9610 & 0.8898 & 0.9830 \\ 
  kah-wlsh-shr & 0.9749 & 0.9826 & 0.9856 & 0.9805 & 0.9809 & 0.9005 & 0.9487 & 0.9243 & 0.8582 & 0.9461 \\ 
  kah-wlsh-wls & 0.9774 & 0.9855 & 0.9887 & 0.9846 & 0.9840 & 0.9038 & 0.9532 & 0.9281 & 0.8635 & 0.9498 \\
\hline
\end{tabular}}
\end{table}

\begin{figure}[ht]
	\centering
	\includegraphics[width=\linewidth]{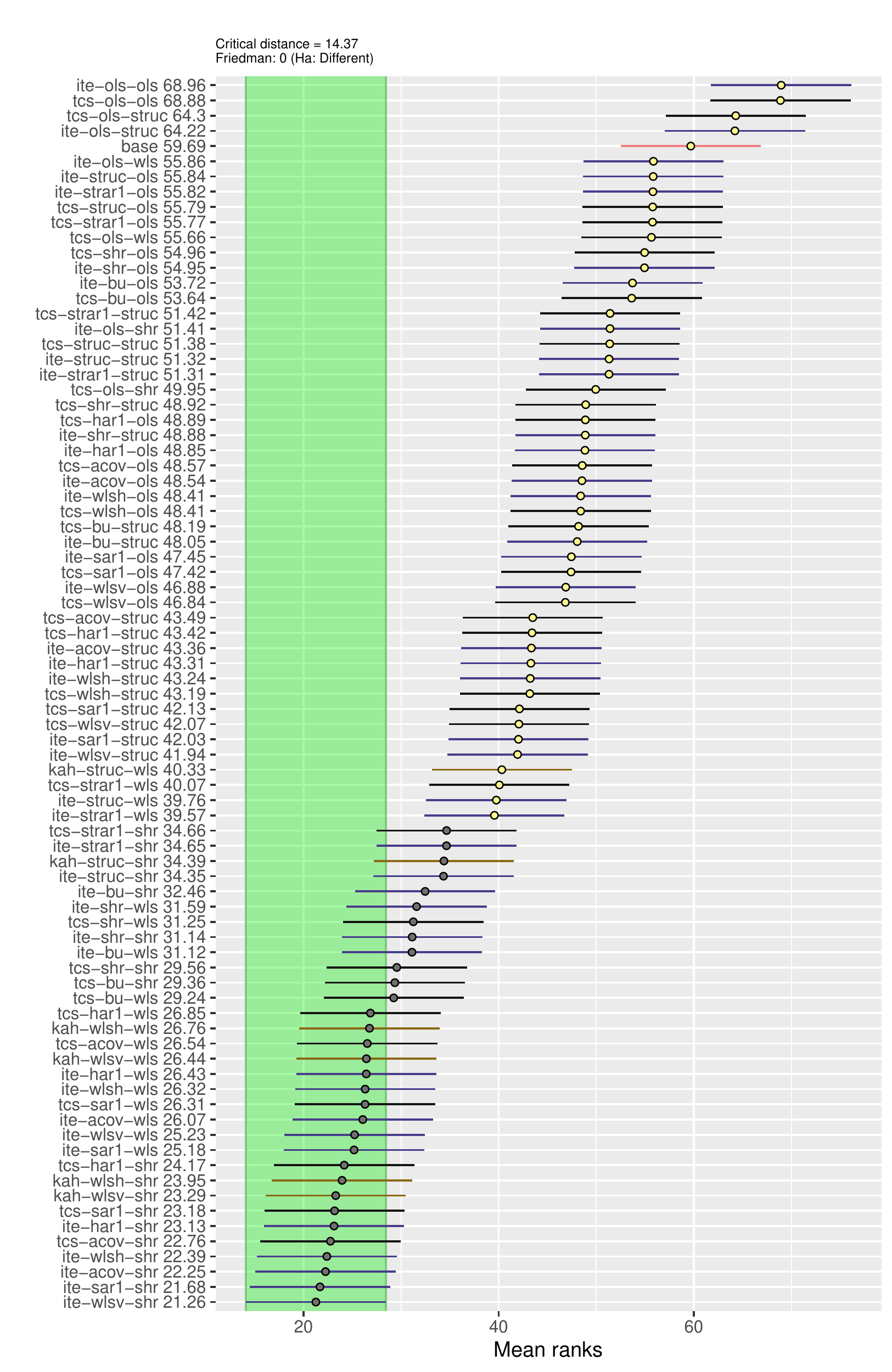}
	\caption{Nemenyi test results at 5\% significance level for all 95 series. The reconciliation procedures are sorted vertically according to the MSE mean rank across all time frequencies and forecast horizons.}
		\label{fig:nem_all_mse_tcs}
\end{figure}

\begin{figure}[ht]
	\centering
	\includegraphics[width=\linewidth]{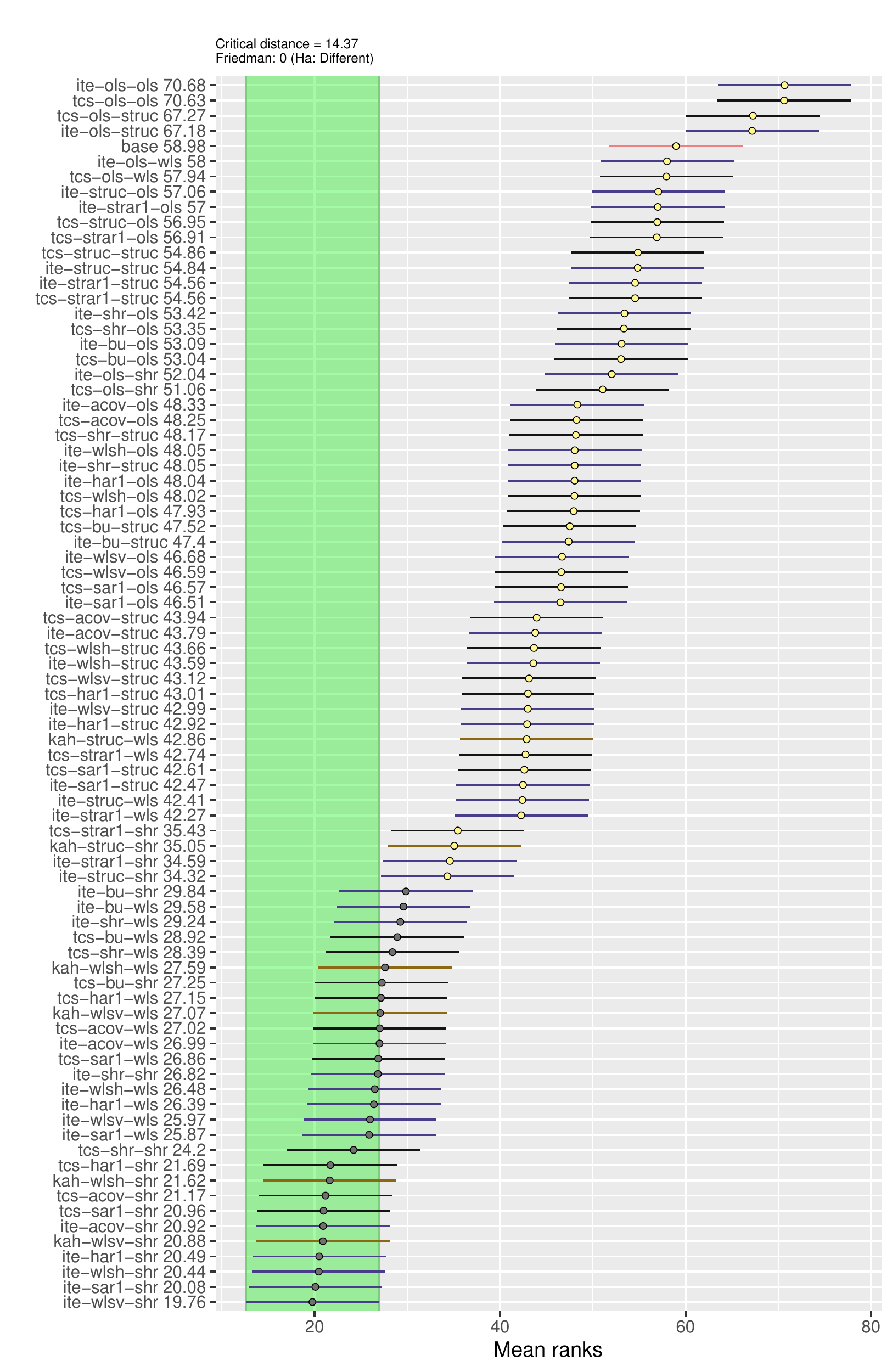}
	\caption{Nemenyi test results at 5\% significance level for all 95 series. The reconciliation procedures are sorted vertically according to the MAE mean rank across all time frequencies and forecast horizons}
		\label{fig:nem_all_mae_tcs}
\end{figure}

\begin{figure}[ht]
	\centering
	\includegraphics[width=\linewidth]{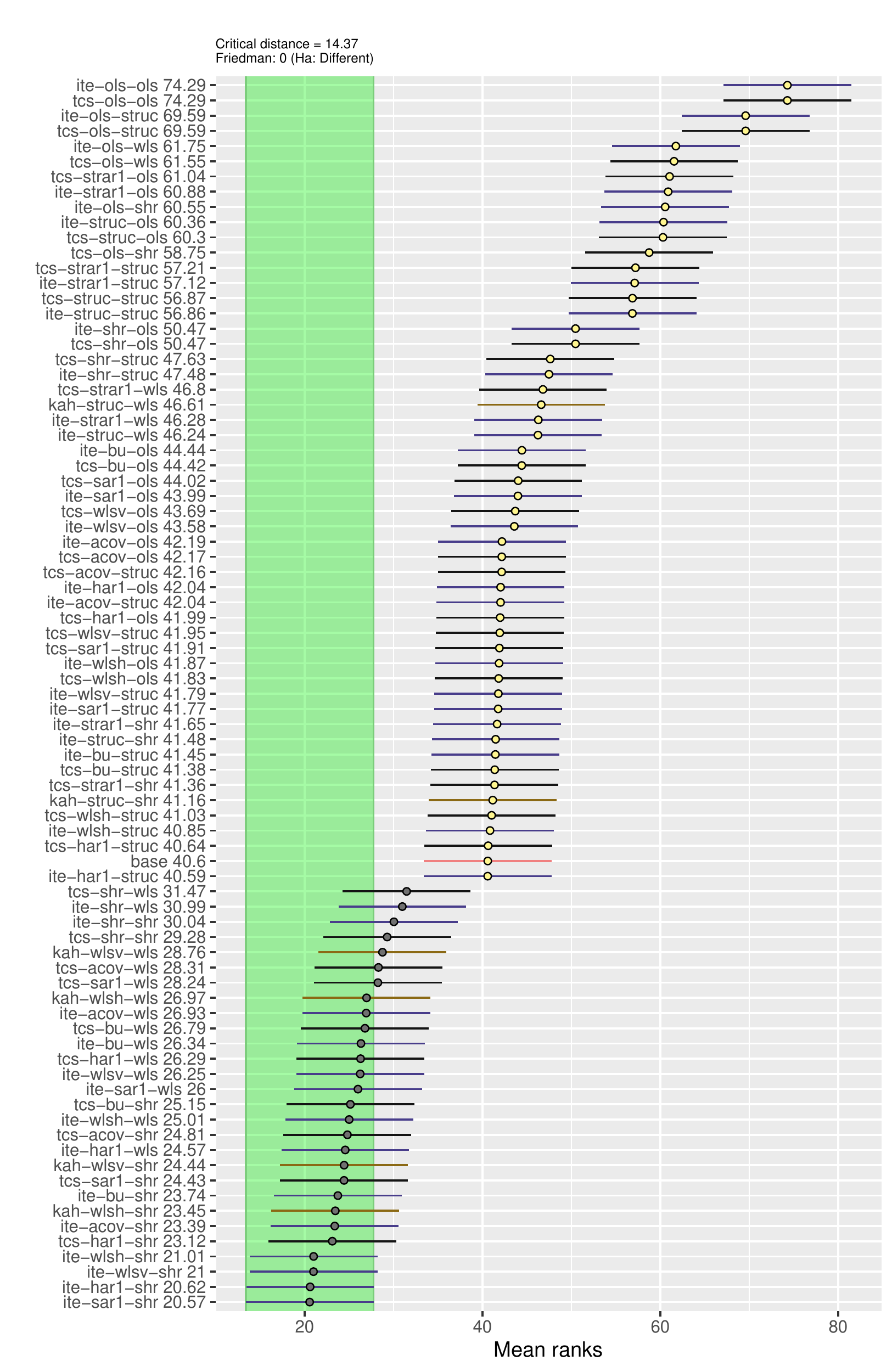}
	\caption{Nemenyi test results at 5\% significance level for all 95 series. The reconciliation procedures are sorted vertically according to the MSE mean rank for the one-step ahead quarterly forecasts.}	\label{fig:nem_k1h1_mse_tcs}
\end{figure}

\begin{figure}[ht]
	\centering
	\includegraphics[width=\linewidth]{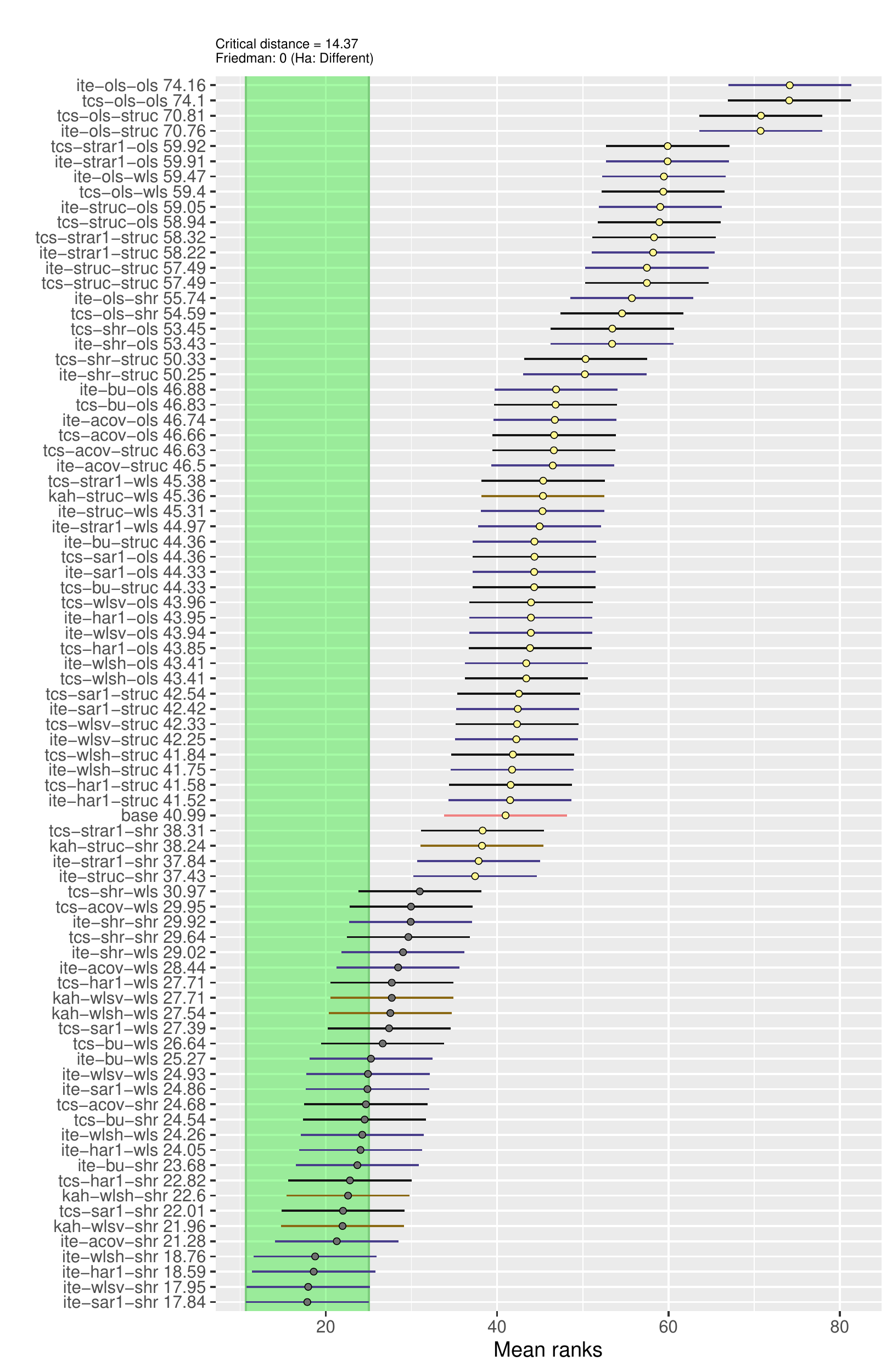}
	\caption{Nemenyi test results at 5\% significance level for all 95 series. The reconciliation procedures are sorted vertically according to the MAE mean rank for the one-step ahead quarterly forecasts.}	\label{fig:nem_k1h1_mae_tcs}
\end{figure}

\clearpage
\subsection*{A.8.4 Optimal combination forecast cross-temporal reconciliation procedures}

\begin{figure}[ht]
	\centering
	\includegraphics[width=\linewidth]{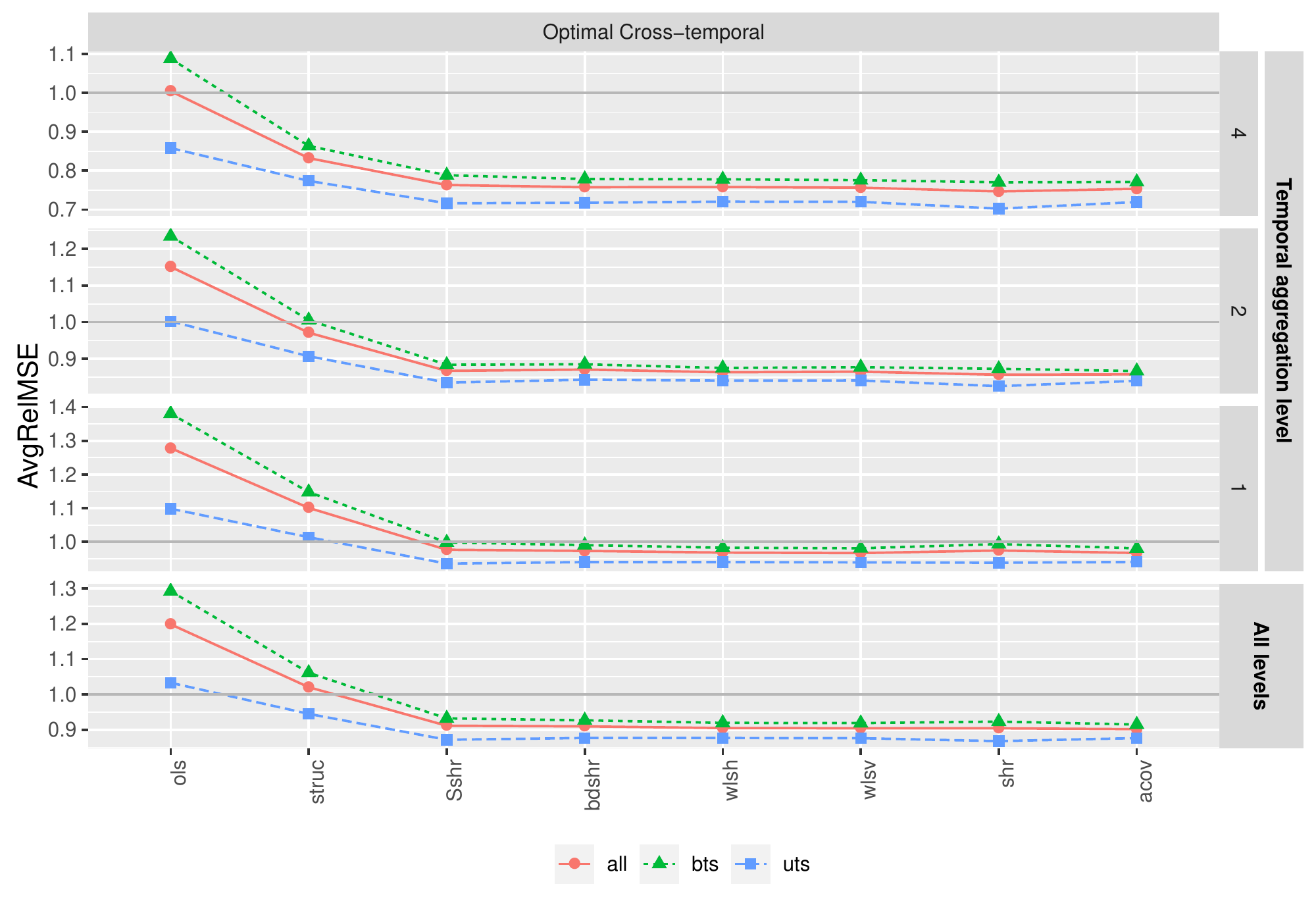}
	\vskip0.5cm
	\includegraphics{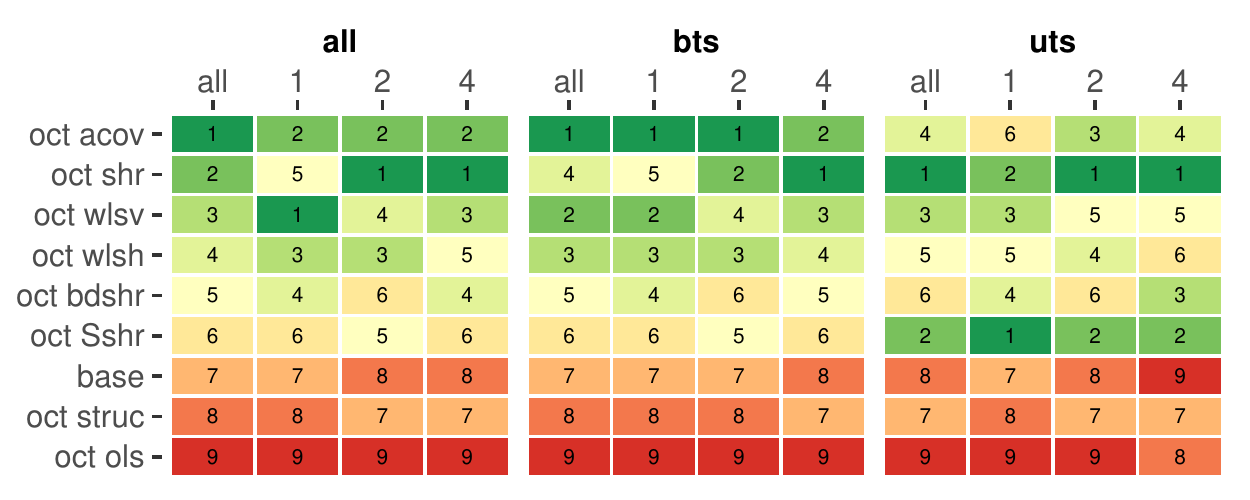}
	\caption{Top panel: Average Relative MSE across all series and forecast horizons, by frequency of observation. Bottom panel: Rankings by frequency of observation and forecast horizon.}
	\label{fig:oct_mse}
\end{figure}

\begin{figure}[ht]
	\centering
	\includegraphics[width=\linewidth]{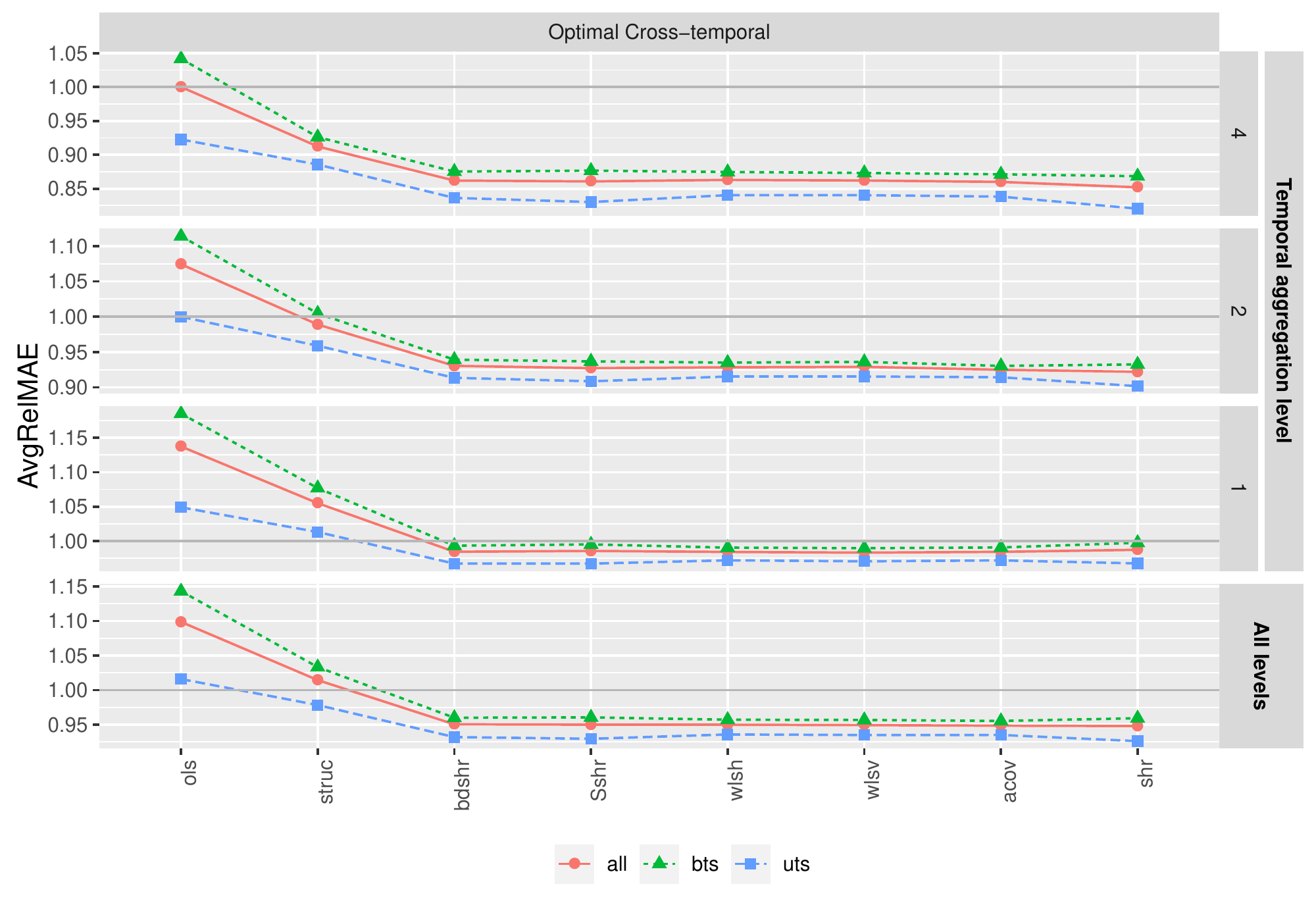}
	\vskip0.5cm
	\includegraphics{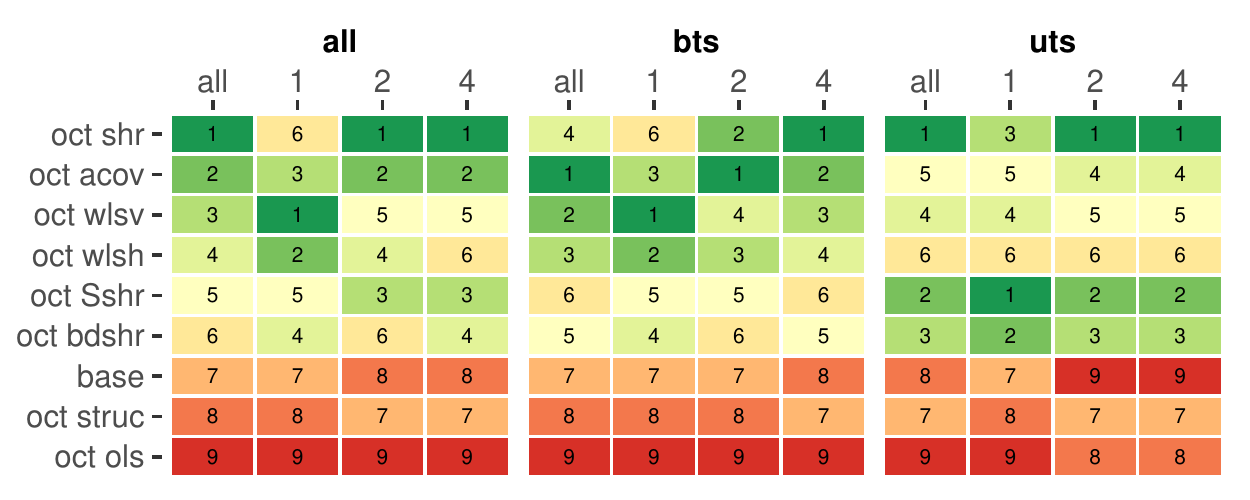}
	\caption{Top panel: Average Relative MAE across all series and forecast horizons, by frequency of observation. Bottom panel: Rankings by frequency of observation and forecast horizon.}
	\label{fig:oct_mae}
\end{figure}

\begin{table}[ht]
\centering
\caption{AvgRelMSE at any temporal aggregation level and any forecast horizon.}
\resizebox{\linewidth}{!}{
\begin{tabular}{c|c|c|c|c|c|c|c|c|c|c}
  \hline
 & \multicolumn{5}{c|}{\textbf{Quarterly}} & \multicolumn{3}{c|}{\textbf{Semi-annual}} & \multicolumn{1}{c|}{\textbf{Annual}} & \textbf{All}\\
	 \hline
\textbf{Procedure} & 1 & 2 & 3 & 4 & 1-4 & 
            1 & 2 & 1-2 & 1 & \\
   \hline
 \multicolumn{11}{c}{\rule{0pt}{4ex}\emph{all 95 series}} \\ [2ex]
base & 1 & 1 & 1 & 1 & 1 & 1 & 1 & 1 & 1 & 1 \\ 
  oct-ols & \textbf{1.5246} & \textbf{1.2871} & \textbf{1.1918} & \textbf{1.1415} & \textbf{1.2782} & \textbf{1.1855} & \textbf{1.1172} & \textbf{1.1508} & \textbf{1.0041} & \textbf{1.1984} \\ 
  oct-struc & \textbf{1.2511} & \textbf{1.1084} & \textbf{1.0430} & \textbf{1.0154} & \textbf{1.1008} & 0.9746 & 0.9682 & 0.9714 & 0.8322 & \textbf{1.0206} \\ 
  oct-wlsh & 0.9548 & 0.9718 & 0.9760 & 0.9696 & 0.9680 & 0.8112 & 0.9194 & 0.8636 & 0.7579 & 0.9048 \\ 
oct-wlsv & 0.9692 & 0.9719 & 0.9622 & \textbf{\textcolor{red}{0.9631}} & \textbf{\textcolor{red}{0.9666}} & 0.8203 & \textbf{\textcolor{red}{0.9125}} & 0.8652 & 0.7562 & 0.9042 \\ 
oct-bdshr & 0.9838 & 0.9798 & \textbf{\textcolor{red}{0.9618}} & 0.9665 & 0.9730 & 0.8297 & 0.9144 & 0.8710 & 0.7573 & 0.9095 \\ 
  oct-acov & 0.9553 & 0.9648 & 0.9767 & 0.9707 & 0.9668 & 0.8013 & 0.9185 & 0.8579 & 0.7531 & \textbf{\textcolor{red}{0.9016}} \\ 
  oct-shr & 0.9652 & \textbf{\textcolor{red}{0.9610}} & 0.9875 & 0.9835 & 0.9742 & \textbf{\textcolor{red}{0.7971}} & 0.9211 & \textbf{\textcolor{red}{0.8569}} & \textbf{\textcolor{red}{0.7465}} & 0.9041 \\ 
  oct-Sshr & \textbf{\textcolor{red}{0.9547}} & 0.9720 & 0.9913 & 0.9884 & 0.9765 & 0.8054 & 0.9343 & 0.8674 & 0.7631 & 0.9113 \\ 
 \multicolumn{11}{c}{\rule{0pt}{4ex}\emph{32 upper series}} \\ [2ex]
base & 1 & 1 & 1 & 1 & 1 & 1 & 1 & 1 & 1 & 1 \\ 
  oct-ols & \textbf{1.2156} & \textbf{1.1122} & \textbf{1.0452} & \textbf{1.0300} & \textbf{1.0984} & \textbf{1.0087} & 0.9962 & \textbf{1.0024} & 0.858 & \textbf{1.0330} \\ 
  oct-struc & \textbf{1.0667} & \textbf{1.0341} & 0.9814 & 0.9756 & \textbf{1.0138} & 0.8898 & 0.9252 & 0.9073 & 0.7737 & 0.9449 \\ 
  oct-wlsh & 0.9387 & 0.9510 & 0.9339 & 0.9360 & 0.9399 & 0.7999 & 0.8842 & 0.8410 & 0.7204 & 0.8766 \\ 
  oct-wlsv & 0.9411 & 0.9506 & 0.9316 & \textbf{\textcolor{red}{0.9326}} & 0.9390 & 0.8032 & 0.8811 & 0.8412 & 0.7198 & 0.8760 \\ 
  oct-bdshr & 0.9453 & 0.9559 & \textbf{\textcolor{red}{0.9246}} & 0.9340 & 0.9399 & 0.8091 & \textbf{\textcolor{red}{0.8791}} & 0.8433 & 0.7174 & 0.8767 \\ 
  oct-acov & 0.9388 & 0.9498 & 0.9353 & 0.9371 & 0.9402 & 0.7984 & 0.8844 & 0.8403 & 0.7193 & 0.8763 \\ 
  oct-shr & 0.9309 & \textbf{\textcolor{red}{0.9237}} & 0.9438 & 0.9532 & 0.9379 & \textbf{\textcolor{red}{0.7691}} & 0.8867 & \textbf{\textcolor{red}{0.8258}} & \textbf{\textcolor{red}{0.7023}} & \textbf{\textcolor{red}{0.8678}} \\ 
  oct-Sshr & \textbf{\textcolor{red}{0.9291}} & 0.9381 & 0.9339 & 0.9398 & \textbf{\textcolor{red}{0.9352}} & 0.7892 & 0.8846 & 0.8355 & 0.7160 & 0.8717 \\  
 \multicolumn{11}{c}{\rule{0pt}{4ex}\emph{63 bottom series}} \\ [2ex]
base & 1 & 1 & 1 & 1 & 1 & 1 & 1 & 1 & 1 & 1 \\ 
  oct-ols & \textbf{1.7104} & \textbf{1.3863} & \textbf{1.2740} & \textbf{1.2026} & \textbf{1.3806} & \textbf{1.2869} & \textbf{1.1842} & \textbf{1.2345} & \textbf{1.0876} & \textbf{1.2924} \\ 
  oct-struc & \textbf{1.3567} & \textbf{1.1481} & \textbf{1.0757} & \textbf{1.0362} & \textbf{1.1479} & \textbf{1.0208} & 0.9908 & \textbf{1.0057} & 0.8636 & \textbf{1.0613} \\ 
oct-wlsh & \textbf{\textcolor{red}{0.9631}} & 0.9825 & 0.9981 & 0.9871 & 0.9826 & 0.8170 & 0.9378 & 0.8753 & 0.7776 & 0.9194 \\ 
oct-wlsv & 0.9837 & 0.9828 & \textbf{\textcolor{red}{0.9782}} & \textbf{\textcolor{red}{0.9789}} & 0.9809 & 0.8292 & \textbf{\textcolor{red}{0.9288}} & 0.8776 & 0.7754 & 0.9188 \\ 
  oct-bdshr & \textbf{1.004} & 0.9922 & 0.9813 & 0.9835 & 0.9902 & 0.8404 & 0.9329 & 0.8854 & 0.7784 & 0.9267 \\ 
  oct-acov & 0.9639 & \textbf{\textcolor{red}{0.9725}} & 0.9984 & 0.9881 & \textbf{\textcolor{red}{0.9806}} & \textbf{\textcolor{red}{0.8028}} & 0.9363 & \textbf{\textcolor{red}{0.8670}} & 0.7709 & \textbf{\textcolor{red}{0.9147}} \\ 
  oct-shr & 0.9830 & 0.9805 & \textbf{1.0105} & 0.9992 & 0.9932 & 0.8117 & 0.9391 & 0.8731 & \textbf{\textcolor{red}{0.7700}} & 0.9231 \\ 
  oct-Sshr & 0.9680 & 0.9897 & \textbf{1.0218} & \textbf{1.0140} & 0.9982 & 0.8138 & 0.9606 & 0.8841 & 0.7881 & 0.9322 \\ 
   \hline
\end{tabular}}
\end{table}

\begin{table}[ht]
\centering
\caption{AvgRelMAE at any temporal aggregation level and any forecast horizon.}
\resizebox{\linewidth}{!}{
\begin{tabular}{c|c|c|c|c|c|c|c|c|c|c}
  \hline
 & \multicolumn{5}{c|}{\textbf{Quarterly}} & \multicolumn{3}{c|}{\textbf{Semi-annual}} & \multicolumn{1}{c|}{\textbf{Annual}} & \textbf{All}\\
	 \hline
\textbf{Procedure} & 1 & 2 & 3 & 4 & 1-4 & 
            1 & 2 & 1-2 & 1 & \\
   \hline
 \multicolumn{11}{c}{\rule{0pt}{4ex}\emph{all 95 series}} \\ [2ex]
base & 1 & 1 & 1 & 1 & 1 & 1 & 1 & 1 & 1 & 1 \\ 
  oct-ols & \textbf{1.2422} & \textbf{1.1377} & \textbf{1.1027} & \textbf{1.0736} & \textbf{1.1373} & \textbf{1.0921} & \textbf{1.0565} & \textbf{1.0741} & \textbf{1.0001} & \textbf{1.0985} \\ 
  oct-struc & \textbf{1.1219} & \textbf{1.0563} & \textbf{1.0298} & \textbf{1.0153} & \textbf{1.0551} & 0.9908 & 0.9870 & 0.9889 & 0.9123 & \textbf{1.0144} \\ 
oct-wlsh & 0.9745 & 0.9863 & 0.9895 & 0.9867 & 0.9842 & 0.9028 & 0.9545 & 0.9283 & 0.8630 & 0.9499 \\ 
oct-wlsv & 0.9813 & 0.9858 & 0.9829 & \textbf{\textcolor{red}{0.9830}} & \textbf{\textcolor{red}{0.9832}} & 0.9078 & 0.9506 & 0.9289 & 0.8620 & 0.9494 \\ 
  oct-bdshr & 0.9858 & 0.9880 & \textbf{\textcolor{red}{0.9809}} & 0.9833 & 0.9845 & 0.9112 & \textbf{\textcolor{red}{0.9499}} & 0.9304 & 0.8620 & 0.9505 \\ 
  oct-acov & 0.9762 & 0.9831 & 0.9904 & 0.9879 & 0.9844 & 0.8965 & 0.9541 & 0.9248 & 0.8600 & 0.9485 \\ 
  oct-shr & 0.9825 & \textbf{\textcolor{red}{0.9804}} & 0.9944 & 0.9920 & 0.9873 & \textbf{\textcolor{red}{0.8939}} & 0.9509 & \textbf{\textcolor{red}{0.9219}} & \textbf{\textcolor{red}{0.8520}} & \textbf{\textcolor{red}{0.9480}} \\ 
  oct-Sshr & \textbf{\textcolor{red}{0.9739}} & 0.9832 & 0.9922 & 0.9936 & 0.9857 & 0.8962 & 0.9591 & 0.9271 & 0.8607 & 0.9500 \\ 
 \multicolumn{11}{c}{\rule{0pt}{4ex}\emph{32 upper series}} \\ [2ex]
 base & 1 & 1 & 1 & 1 & 1 & 1 & 1 & 1 & 1 & 1 \\ 
  oct-ols & \textbf{1.0857} & \textbf{1.0534} & \textbf{1.0331} & \textbf{1.0249} & \textbf{1.0491} & \textbf{1.0027} & 0.9971 & 0.9999 & 0.9226 & \textbf{1.0159} \\ 
  oct-struc & \textbf{1.0173} & \textbf{1.0227} & \textbf{1.0058} & \textbf{1.0069} & \textbf{1.0131} & 0.9467 & 0.9712 & 0.9589 & 0.8857 & 0.9784 \\ 
  oct-wlsh & 0.9604 & 0.9780 & 0.9727 & 0.9773 & 0.9721 & 0.8948 & 0.9362 & 0.9152 & 0.8404 & 0.9358 \\ 
  oct-wlsv & 0.9589 & 0.9773 & 0.9712 & 0.9752 & 0.9706 & 0.8969 & 0.9339 & 0.9152 & 0.8404 & 0.9350 \\ 
  oct-bdshr & \textbf{\textcolor{red}{0.9552}} & 0.9790 & \textbf{\textcolor{red}{0.9632}} & \textbf{\textcolor{red}{0.9719}} & 0.9673 & 0.8983 & \textbf{\textcolor{red}{0.9288}} & 0.9134 & 0.8364 & 0.9320 \\ 
  oct-acov & 0.9631 & 0.9756 & 0.9729 & 0.9764 & 0.9720 & 0.8933 & 0.9356 & 0.9142 & 0.8383 & 0.9351 \\ 
  oct-shr & 0.9591 & \textbf{\textcolor{red}{0.9594}} & 0.9700 & 0.9820 & 0.9676 & \textbf{\textcolor{red}{0.8746}} & 0.9295 & \textbf{\textcolor{red}{0.9016}} & \textbf{\textcolor{red}{0.8205}} & \textbf{\textcolor{red}{0.9262}} \\ 
  oct-Sshr & 0.9587 & 0.9675 & 0.9660 & 0.9769 & \textbf{\textcolor{red}{0.9673}} & 0.885 & 0.9327 & 0.9085 & 0.8303 & 0.9296 \\
 \multicolumn{11}{c}{\rule{0pt}{4ex}\emph{63 bottom series}} \\ [2ex]
base & 1 & 1 & 1 & 1 & 1 & 1 & 1 & 1 & 1 & 1 \\ 
  oct-ols & \textbf{1.3301} & \textbf{1.1831} & \textbf{1.1399} & \textbf{1.0992} & \textbf{1.1850} & \textbf{1.1405} & \textbf{1.088} & \textbf{1.1139} & \textbf{1.0419} & \textbf{1.1430} \\ 
  oct-struc & \textbf{1.1791} & \textbf{1.0737} & \textbf{1.0422} & \textbf{1.0196} & \textbf{1.0770} & \textbf{1.0140} & 0.9951 & \textbf{1.0045} & 0.9262 & \textbf{1.0333} \\ 
oct-wlsh & 0.9818 & 0.9906 & 0.9981 & 0.9916 & 0.9905 & 0.9069 & 0.9639 & 0.9350 & 0.8746 & 0.9571 \\ 
oct-wlsv & 0.9929 & 0.9901 & \textbf{\textcolor{red}{0.9888}} & \textbf{\textcolor{red}{0.9870}} & \textbf{\textcolor{red}{0.9897}} & 0.9133 & \textbf{\textcolor{red}{0.9591}} & 0.9360 & 0.8732 & 0.9568 \\ 
  oct-bdshr & \textbf{1.0018} & 0.9925 & 0.9900 & 0.9892 & 0.9933 & 0.9178 & 0.9609 & 0.9391 & 0.8753 & 0.9600 \\ 
  oct-acov & 0.9830 & \textbf{\textcolor{red}{0.9869}} & 0.9994 & 0.9937 & 0.9907 & \textbf{\textcolor{red}{0.8981}} & 0.9636 & \textbf{\textcolor{red}{0.9303}} & 0.8712 & \textbf{\textcolor{red}{0.9554}} \\ 
  oct-shr & 0.9946 & 0.9913 & \textbf{1.0070} & 0.9972 & 0.9975 & 0.9038 & 0.9619 & 0.9324 & \textbf{\textcolor{red}{0.8684}} & 0.9593 \\ 
  oct-Sshr & \textbf{\textcolor{red}{0.9818}} & 0.9912 & \textbf{1.0059} & \textbf{1.0021} & 0.9952 & 0.9019 & 0.9728 & 0.9367 & 0.8766 & 0.9606 \\ 
   \hline
\end{tabular}}
\end{table}

\begin{figure}[ht]
	\centering
	\includegraphics[width=\linewidth]{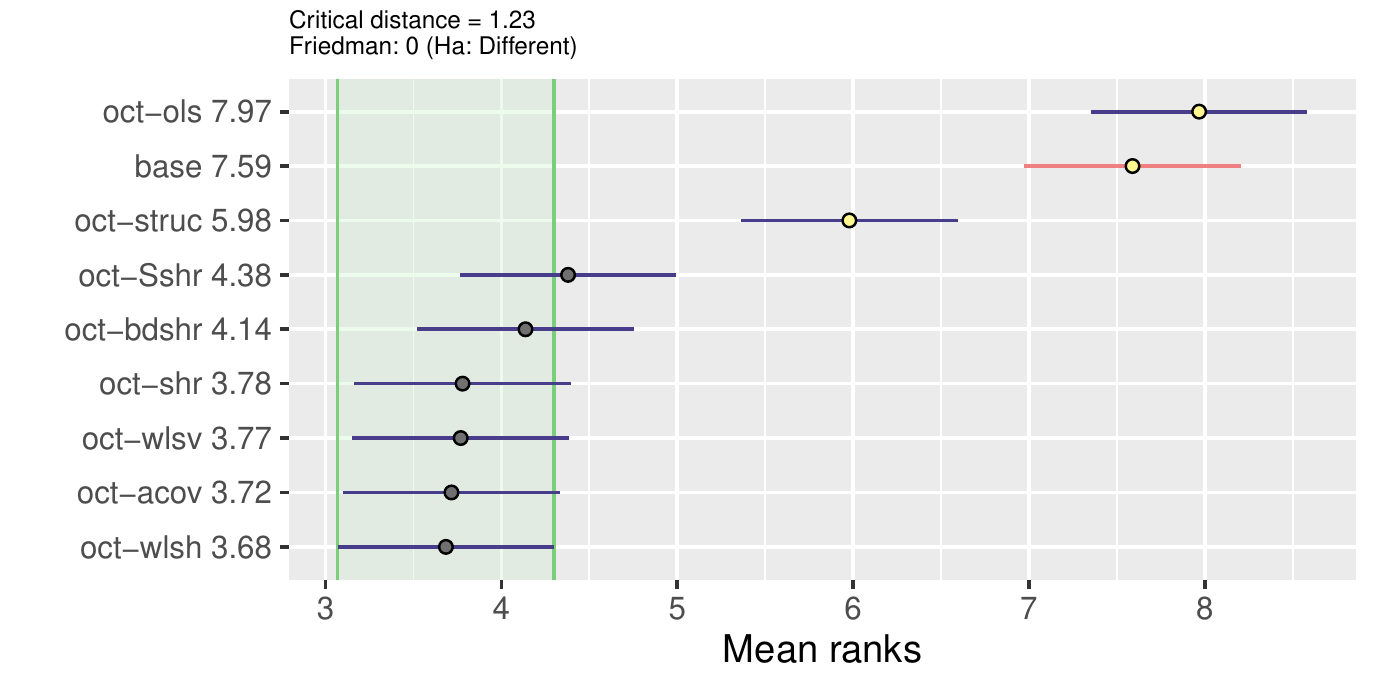}
	\includegraphics[width=\linewidth]{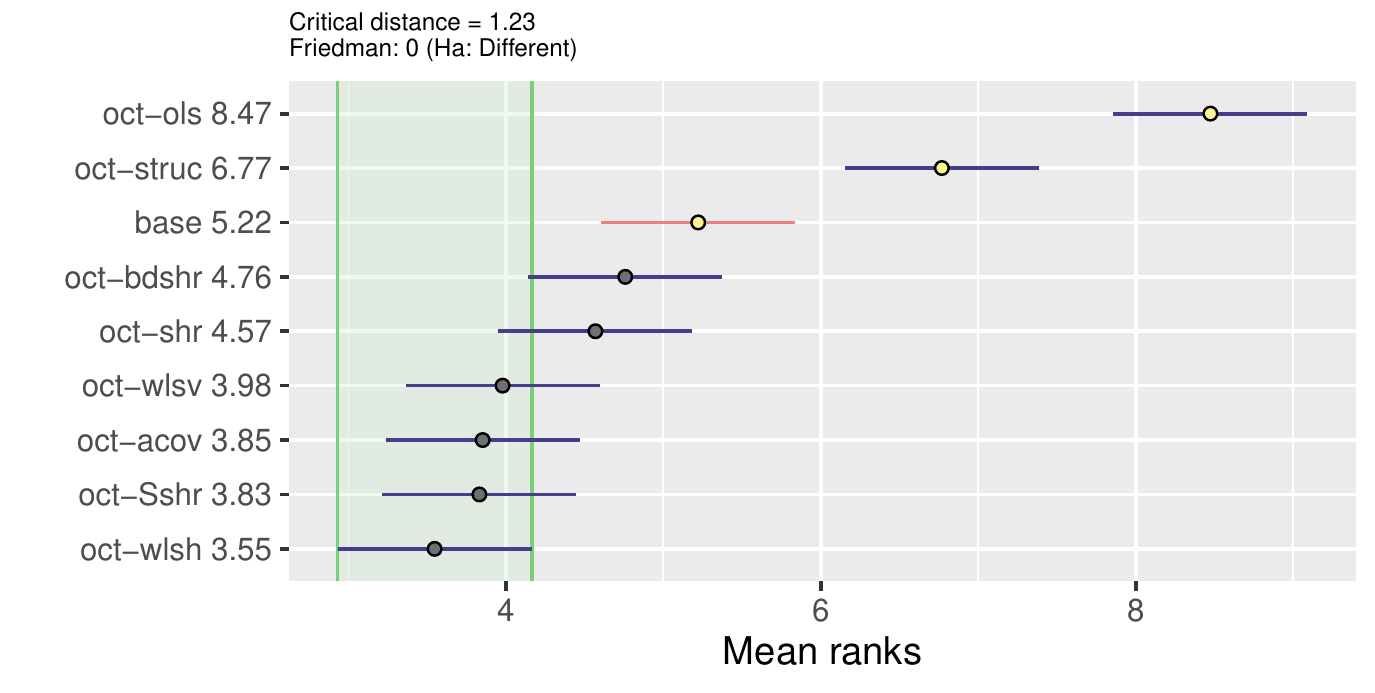}
	\caption{Nemenyi test results at 5\% significance level for all 95 series. The optimal combination reconciliation procedures are sorted vertically according to the MSE mean rank (i) across all time frequencies and forecast horizons (top), and (ii) for 1-step-ahead quarterly forecasts (bottom).}
	\label{fig:nem_all_mse}
\end{figure}

\begin{figure}[ht]
	\centering
	\includegraphics[width=\linewidth]{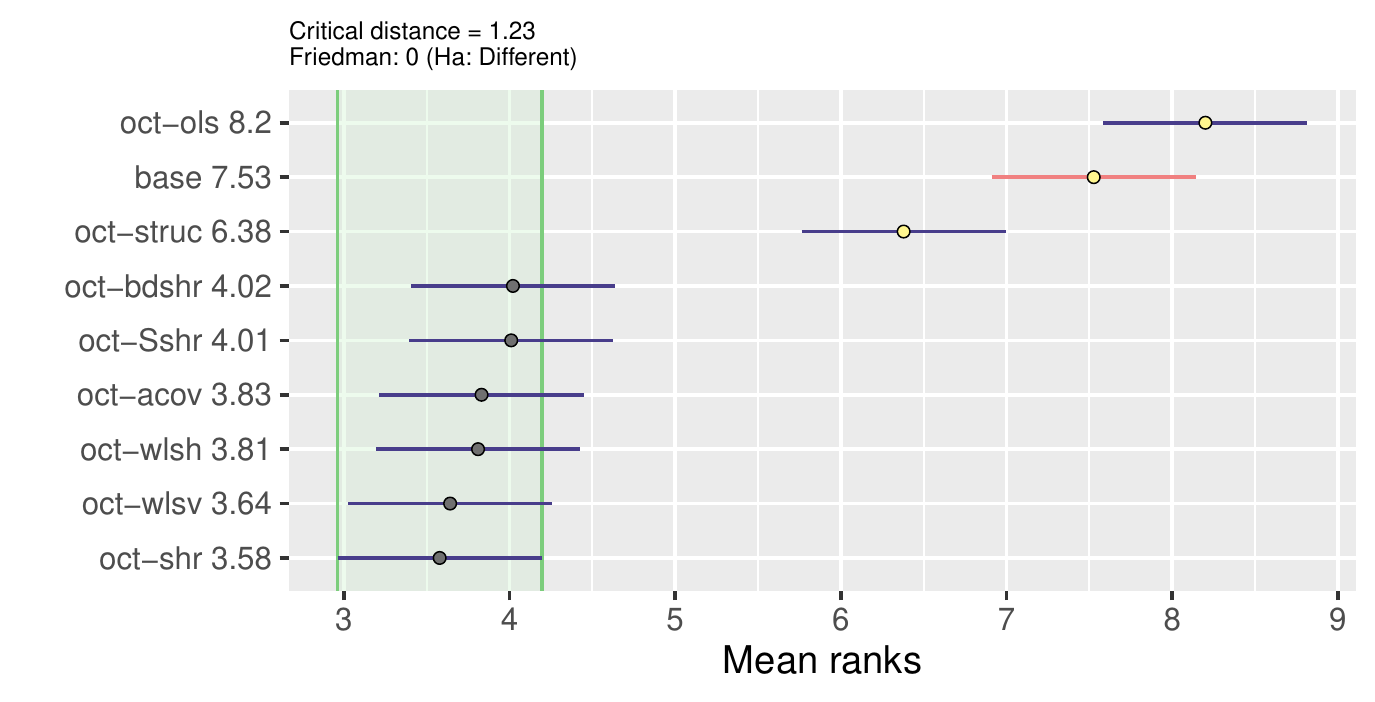}
	\includegraphics[width=\linewidth]{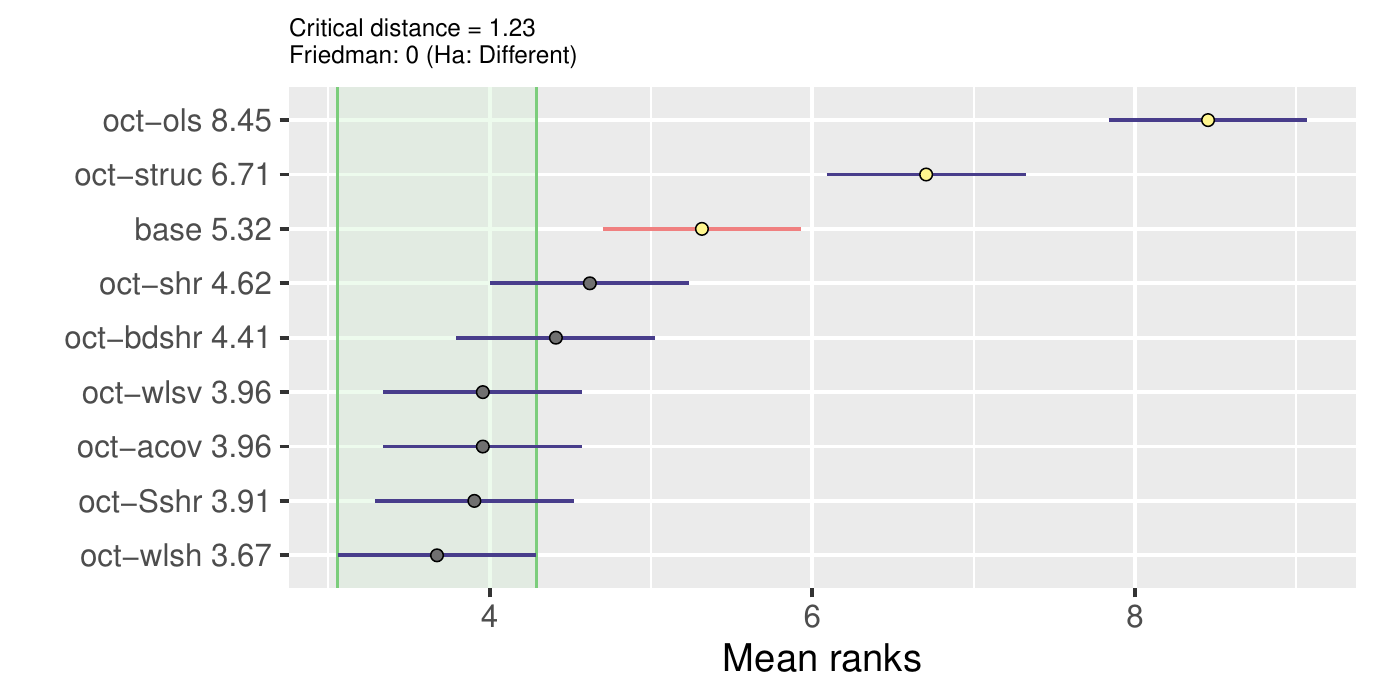}
	\caption{Nemenyi test results at 5\% significance level for all 95 series. The optimal combination reconciliation procedures are sorted vertically according to the MAE mean rank across (i) all time frequencies and forecast horizons (top), and (ii) for 1-step-ahead quarterly forecasts (bottom).}
	\label{fig:nem_all_mae}
\end{figure}


\clearpage
\markboth{Tommaso Di Fonzo, Daniele Girolimetto}{References}
\begin{thebibliography}{99}
\addcontentsline{toc}{section}{References}
\markboth{Tommaso Di Fonzo, Daniele Girolimetto}{References}
\itemsep=1.5pt

%

%
%
%
%
%
\bibitem{hts133}
   Ashouri, M., Hyndman, R.J., Shmueli, G. (2019),
   {\em Fast forecast reconciliation using linear models},
   Department of Econometrics and Business Statistics, Monash University, Working Paper 29/19.
%
\bibitem{hts001}
   Athanasopoulos, G., Ahmed, R.A., Hyndman, R.J. (2009),
   Hierarchical forecasts for Australian domestic tourism,
   {\em International Journal of Forecasting},
   25, 1, 146--166.
%
\bibitem{hts084}
   Athanasopoulos, G., Gamakumara, P., Panagiotelis, A., Hyndman, R.J., Affan, M. (2019),
   Hierarchical Forecasting, in Fuleky, P. (ed.),
   {\em Macroeconomic Forecasting in the Era of Big Data}, pp. 703-733, Cham, Springer.
%
\bibitem{hts005}
   Athanasopoulos, G., Hyndman, R.J., Kourentzes, N., Petropoulos, F. (2017),
   Forecasting with Temporal Hierarchies,
   {\em European Journal of Operational Research},
   262, 1, 60--74.
%
%
%
%
%
%
%
%
%
\bibitem{hts031}
   Ben Taieb, S., Taylor, J.W., Hyndman, R.J. (2020),
   Hierarchical Probabilistic Forecasting of Electricity Demand with Smart Meter Data,
   {\em Journal of the American Statistical Association} (in press).
%
%
%
\bibitem{hts154}
   Bertani, N., Satop\"{a}\"{a}, V., Jensen, S. (2020),
   {\em Joint Bottom-Up Method for Hierarchical Time-Series: Application to Australian Tourism}.
   Available at SSRN: https://ssrn.com/abstract=3542278. Accessed on July 25, 2020.

\bibitem{hts159}
   Bisaglia, L., Di Fonzo, T., Girolimetto, D. (2020),
   {\em Fully reconciled GDP forecasts from Income and Expenditure sides},
   arXiv:2004.03864v2. Accessed on July 25, 2020.
%
%
\bibitem{trimes197}
   Byron, R.P. (1978),
   The estimation of large Social Account Matrices,
   {\em Journal of the Royal Statistical Society A}, 141, 3, 359--367.
%
\bibitem{trimes1206}
   Dagum, E.B., Cholette, P.A. (2006),
   {\em Benchmarking, Temporal Distribution and Reconciliation Methods for Time Series},
   New York, Springer.
%
%
%
%
%
%
%
\bibitem{trimes982}
Danilov, D. and Magnus, J.R. (2008),
On the estimation of a large sparse Bayesian system: The Snaer program,
{\em Computational Statistics \& Data Analysis}, 52, 9, 4203--4224.

%
\bibitem{hts122}
   Davydenko, A., Fildes, R. (2013),
   Measuring forecasting accuracy: The case of judgmental adjustments to SKU-level demand forecasts,
   {\em International Journal of Forecasting}, 29, 3, 510--522.

\bibitem{ras008}
   Deming, E., Stephan, F.F. (1940),
   On a least squares adjustment of a sampled frequency table when the expected marginal totals are known,
   {\em The Annals of Mathematical Statistics}, 11, 4, 427--444.

%
\bibitem{trimes297}
   Di Fonzo, T. (1990),
   The estimation of $M$ disaggregate time series when contemporaneous and temporal aggregates are known,
   {\em The Review of Economics and Statistics}, 72, 1, 188-192.

\bibitem{hts219}
Di Fonzo, T., Girolimetto, D. (2020),
   {\em Package `FoReco'}, R package version 0.1.1 (October 18, 2020).
   
\bibitem{trimes1111}
   Di Fonzo, T., Marini, M. (2011),
   Simultaneous and two-step reconciliation of systems of time series:
   methodological and practical issues,
   {\em Journal of the Royal Statistical Society, Series C}, 60, 2, 143-164.

\bibitem{trimes390}
   Dunn, D.M., Williams, W.H., DeChaine, T.L. (1976),
   Aggregate versus subaggregate models in local area forecasting,
   {\em Journal of the American Statistical Association}, 71, 353, 68--71.

\bibitem{hts097}
   Eckert, F. Hyndman, R.J., Panagiotelis, A. (2019),
   Forecasting Swiss Exports using Bayesian Forecast Reconciliation,
   {\em European Journal of Operational Research}, in press.
%
%
\bibitem{hts012}
    van Erven, T., Cugliari, J. (2015),
    Game-theoretically Optimal Reconciliation of Contemporaneous Hierarchical Time Series Forecasts,
    in Antoniadis, A., Poggi, J.M., Brossat, X.,
    {\em Modeling and Stochastic Learning for Forecasting in High Dimensions},
    Berlin, Springer, 297--317.
%
%
%
\bibitem{hts035}
   Fliedner, G. (2001),
   Hierarchical forecasting: issues and guidelines,
   {\em Industrial Management and Data Systems}, 101, 1, 5--12.

%
%
\bibitem{trimes284}
   Gross, C.W., Sohl, J.E. (1990),
   Disaggregation methods to expedite product line forecasting,
   {\em Journal of Forecasting}, 9, 3, 233--254.

\bibitem{harville}
   Harville, D.A. (2008),
   {\em Matrix algebra from a statistician's perspective},
   New York, Springer-Verlag.

\bibitem{ts051}
Hibon, M., Crone, S.F., Kourentzes, N. (2012),
{\em Statistical significance of forecasting methods}, 
presentation at the 32nd Annual International Symposium on Forecasting, Boston,
available at: https://kourentzes.com/forecasting/wp-content/uploads/2014/04/ISF2012\_Tests\_Kourentzes.pdf.
Accessed on July 25, 2020.

%
\bibitem{hts106}
   Hong, T. and Xie, J. and Black, J. (2019),
   Global energy forecasting competition 2017: Hierarchical probabilistic load forecasting,
   {\em International Journal of Forecasting}, 35, 4, 1389--1399.
%
%
\bibitem{hts002}
   Hyndman, R.J., Ahmed, R.A., Athanasopoulos, G., Shang, H.L. (2011),
   Optimal combination forecasts for hierarchical time series,
   {\em Computational Statistics \& Data Analysis}, 55, 9, 2579--2589.
%
%
\bibitem{hts127}
   Hyndman, R.J., Athanasopoulos, G. (2018),
   {\em Forecasting: principles and practice}, 2nd edition, OTexts: Melbourne, Australia.
   OTexts.com/fpp2. Accessed on November 14, 2019.

\bibitem{hts124}
   Hyndman, R.J., Athanasopoulos, G., Bergmeir, C., Caceres, G., Chhay, L., O'Hara-Wild, M.,
   Petropoulos, F., Razbash, S., Wang, E., Yasmeen, F. (2020),
   {\em Forecast: Forecasting functions for time series and linear models},
   R package version 8.12 (March 31, 2020).

\bibitem{hts126}
   Hyndman, R.J., Kourentzes, N. (2018),
   {\em Package `thief'}, R package version 0.3 (January 24, 2018).

\bibitem{hts010}
   Hyndman, R.J., Lee, A., Wang, E. (2016),
   Fast computation of reconciled forecasts for hierarchical and grouped time series,
   {\em Computational Statistics \& Data Analysis}, 97, 16--32.

\bibitem{hts125}
   Hyndman, R.J., Lee, A., Wang, E., Wickramasuriya, S.L. (2020),
   {\em Package `hts'}, R package version 6.0.1 (August 6, 2020).

\bibitem{hts082}
   Jeon, J., Panagiotelis, A., Petropoulos, F. (2019),
   Probabilistic forecast reconciliation with applications to wind power and electric load,
   {\em European Journal of Operational Research}, 279, 2, 364--379.

\bibitem{ras282}
   Johnston, R.J., Pattie, C.J. (1993),
   Entropy-Maximizing and the Iterative Proportional Fitting Procedure,
   {\em The Professional Geographer}, 45, 3, 317--322.

%
%
\bibitem{ts005}
    Koning, A.J., Franses, P.H., Hibon, M., Stekler, H.O. (2005),
    The M3 competition: Statistical tests of the results,
    {\em International Journal of Forecasting}, 21, 3, 397--409.

%
%
\bibitem{hts175}
   Kourentzes, N. (2017),
   {\em Uncertainty in predictive modelling},
https://kourentzes.com/forecasting/wp-content/uploads/2017/09/OR59\_Kourentzes\_Uncertainty.pdf.
Accessed on July 25, 2020.

%
\bibitem{hts176}
   Kourentzes, N. (2018),
   {\em Model uncertainty in hierarchical forecasting},
https://www.lancaster.ac.uk/media/lancaster-university/content-assets/documents/lums/research/OR60\_Kourentzes.pdf.
Accessed on July 25, 2020.

\bibitem{hts080}
   Kourentzes, N., Athanasopoulos, G. (2019),
   Cross-temporal coherent forecasts for Australian tourism,
   {\em Annals of Tourism Research}, 75, 393--409.

\bibitem{hts168}
Kourentzes, N., Athanasopoulos, G.  (2020),
{\em On the evaluation of hierarchical forecasts},
Department of Econometrics and Business Statistics, Monash University, Working Paper 02/20.

\bibitem{hts167}
   Kourentzes, N., Athanasopoulos, G.  (2021),
   Elucidate structure in intermittent demand series,
   {\em European Journal of Operational Research}, 288, 1, pp. 141--152.

%
%
%
%
\bibitem{ts040}
    Kourentzes, N., Svetunkov, O., Schaer, O. (2020),
    {\em Package `tsutils'. Time Series Exploration, Modelling and Forecasting},
    R package version 0.9.2 (February 6, 2020).

%
%
%

\bibitem{hts193}
   Ledoit, O., Wolf, M. (2004),
   A well-conditioned estimator for large-dimensional covariance matrices,
   {\em Journal of Multivariate Analysis}, 88, 2, 365--411.

%
\bibitem{hts158}
   Li, H., Hyndman, R.J. (2019),
   {\em Assessing longevity inequality in the U.S.: What can be said about the future?}
   Available at 
   http://dx.doi.org/10.2139/ssrn.3550683. Accessed on July 25, 2020.

%
%
%
%
%
%
\bibitem{Lou2002}
   Lou, T.T., Shiou, S.-H. (2002),
   Inverses of 2$\times$2 block matrices,
   {\em Computers and Mathematics with Applications}, 43(1-2), 119--129.

%
\bibitem{Magnus}
   Magnus, J.R., Neudecker, H. (2019),
   {\em Matrix Differential Calculus with Applications in Statistics and Econometrics}, third edition, New York, Wiley.

%
%
\bibitem{hts180}
Mancuso, P., Piccialli, V., Sudoso, A.M. (2020),
{\em A machine learning approach for forecasting hierarchical time series},
Available at https://arxiv.org/abs/2006.00630.
Accessed on June 16, 2020.

\bibitem{ras074}
   Miller, R.E., Blair, P.D. (2009),
   {\em Input-output analysis: foundations and extensions}, 2nd edition,
   New York, Cambridge University Press.

%
%
%
%
%
%
%
%
%
\bibitem{hts091}
   Nystrup, P., Lindstroem, E., Pinson, P., Madsen, H. (2020),
   Temporal hierarchies with autocorrelation for load forecasting,
   {\em European Journal of Operational Research}, 280, 1, 876-888.
%
%
%
\bibitem{hts112}
   Panagiotelis, A., Gamakumara, P., Athanasopoulos, G., Hyndman, R.J. (2020a),
   Forecast reconciliation: A geometric view with new insights on bias correction,
   {\em International Journal of Forecasting}, in press.

\bibitem{hts198}
   Panagiotelis, A., Gamakumara, P., Athanasopoulos, G., Hyndman, R.J. (2020b),
   {\em Probabilistic Forecast Reconciliation: Properties, Evaluation and Score Optimisation},
   Department of Econometrics and Business Statistics, Monash University, Working Paper 26/20.
%
%
%
%

\bibitem{trimes1382}
   Petropoulos, F., Kourentzes, N. (2015),
   Forecast combination for intermittent demand,
   {\em Journal of the Operational Research Society}, 66, 6, 914--924.

%
%
%
%
\bibitem{hts075}
   Roach, C. (2019),
   Reconciled boosted models for GEFCom2017 hierarchical probabilistic load forecasting,
   {\em International Journal of Forecasting}, 35, 4, 1439--1450.
%
%
\bibitem{hts179}
   Sagaert, Y.R., Kourentzes, N., De Vuyst, S., Aghezzaf, E.-H. (2019),
   Incorporating macroeconomic leading indicators in tactical capacity planning,
   {\em International Journal of Production Economics}, 209, 12--19.

%
%
\bibitem{hts132}
   Sch\"{a}fer, J.L., Strimmer, K. (2005),
   A Shrinkage Approach to Large-Scale Covariance Matrix Estimation and Implications for Functional Genomics,
   {\em Statistical Applications in Genetics and Molecular Biology}, 4, 1.

%
\bibitem{hts009}
   Shang, H.L. (2017),
   Reconciling Forecasts of Infant Mortality Rates at National and Sub-National Levels:
   Grouped Time-Series Methods,
   {\em Population Resarch Policy Review}, 36, 55--84.

\bibitem{hts163}
   Shang, H.L. (2019),
   {\em Dynamic principal component regression for forecasting functional time series
   in a group structure}, arXiv:1909.00456.
Accessed on July 25, 2020.

\bibitem{hts013}
   Shang, H.L., Hyndman, R.J. (2017),
   Grouped functional time series forecasting: an application to age-specific mortality rates,
   {\em Journal of Computational and Graphical Statistics}, 26, 2, 330--343.
%
%
%

\bibitem{trimes406}
Solomou, S, Weale, M. (1993),
Balanced estimates of national accounts when measurement errors are autocorrelated,
   {\em Journal of the Royal Statistical Society, A}, 156, 1, 89-105.

%
\bibitem{hts181}
Spiliotis, E., Abolghasemi, M., Hyndman, R.J., Petropoulos, F., Assimakopoulos, V. (2020),
{\em Hierarchical forecast reconciliation with machine learning},
Available at https://arxiv.org/abs/2006.02043.
Accessed on June 16, 2020.

\bibitem{hts089}
   Spiliotis, E., Petropoulos, F., Kourentzes, N., Assimakopoulos, V. (2020),
   Cross-temporal aggregation: Improving the forecast accuracy of hierarchical electricity
   consumption,
   {\em Applied Energy}, 261, 1.
%
%
%
\bibitem{ras015}
   Stone, R., Champernowne, D.G., Meade, J.E. (1942),
   The precision of national income estimates,
   {\em The Review of Economic Studies}, 9, 2, 111--125.
%
%
%
%
%
%
%
%
%
%
%
\bibitem{trimes415}
Weale, M. (1988),
The reconciliation of values, volumes and prices in the National Accounts,
   {\em Journal of the Royal Statistical Society, A}, 151, 1, 211-221.
%
%
%
%
%
\bibitem{hts006}
   Wickramasuriya, S.L., Athanasopoulos, G., Hyndman, R.J. (2019),
   Optimal forecast reconciliation for hierarchical and grouped time series through trace minimization,
   {\em Journal of the American Statistical Association}, 114, 526, 804--819.

\bibitem{hts098}
   Wickramasuriya, S.L., Turlach, B.A., Hyndman, R.J. (2020),
   Optimal non-negative forecast reconciliation,
   {\em Statistics and Computing}, 30, 5, 1167--1182.
%
%
%
%
%
%
\bibitem{hts087}
   Yagli, G.M., Yang, D., Srinivasan, D. (2019),
   Reconciling solar forecasts: Sequential reconciliation,
   {\em Solar Energy}, 179, 391--397.

\bibitem{hts188}
   Yang, D. (2020),
   Reconciling solar forecasts: Probabilistic forecast reconciliation
   in a nonparametric framework,
   {\em Solar Energy}, in press.
%
%
%
%

\bibitem{hts024}
   Yang, D., Quan, H., Disfani, V.R., Rodriguez-Gallego, C.D. (2017),
   Reconciling solar forecasts: Temporal hierarchy,
   {\em Solar Energy}, 158, 332--346.

%
%
%
%
%
%
%
%

\end{thebibliography}
\end{document}